\documentclass[g5paper,10pt,final,openright,coverpage,swedish]{thesis}

\usepackage{epsfig,amsmath,a4wide,amssymb}
\usepackage{epsfig,amsmath,amssymb}
\usepackage{epsfig,wrapfig}
\usepackage{fancyheadings}
\usepackage{varioref}
\usepackage{graphicx}
\usepackage{verbatim,amsfonts}
\usepackage{array}
\usepackage{pstricks}
\usepackage{multirow}
\usepackage{multicol}
\usepackage{hhline}
\usepackage{epsfig}
\usepackage{rotate}
\usepackage{longtable}
\usepackage{supertabular}
\usepackage[square]{natbib}

\newgray{dg}{0.3}
\newgray{mg}{0.8}
\newgray{lg}{0.8}

\title{First Determination of the \\Electric Charge of the \\Top Quark}
\author{Per Hansson}
\date{Oct 2006}
\shortdate{2006}
\type{Licentiate Thesis}
\division{Particle and Astroparticle Physics}
\department{Department of Physics}
\address{SE-106 91 Stockholm, Sweden}
\city{Stockholm}
\country{Sweden}
\publisher{Printed by Universitetsservice US AB 2006}
\copyrightline{\copyright\ Per Hansson, Oct 2006}
\trita{FYS~2006:69}
\issn{0280-316X}
\isrn{KTH/FYS/\mbox{-}\mbox{-}06:69\mbox{-}\mbox{-}SE}
\isbn{91-7178-493-4}

\cplogo{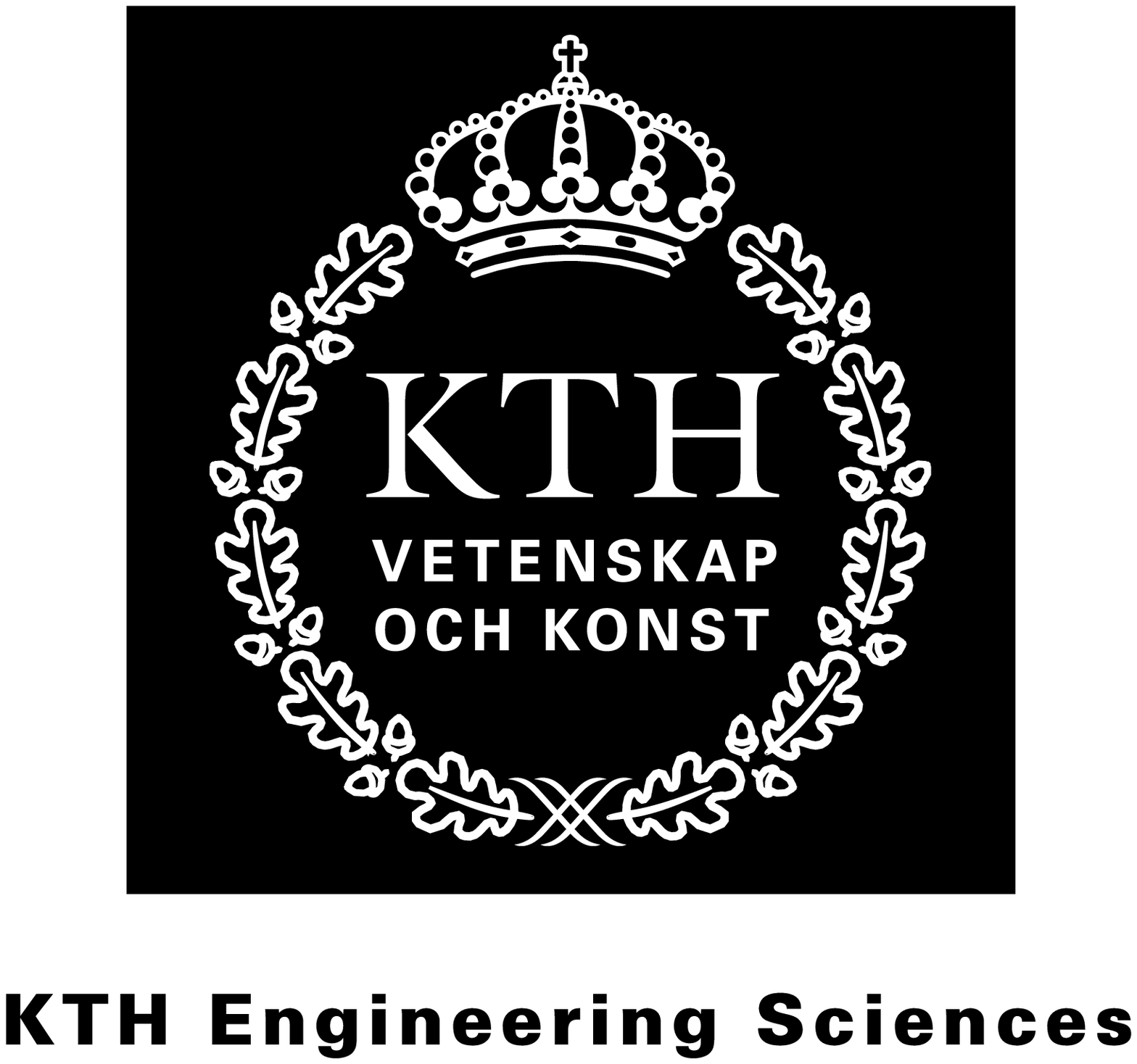}
\cplogonblines{1} 
\innerlogo{\includegraphics[width=32mm]{kth_logo_pics/kth_svv.eps}}
\centercomment{Cover illustration: View of a top quark pair event with an electron and four jets in the final state.\\ Image by D\O\ Collaboration.}
\foregincomment{Akademisk avhandling som med tillst{\aa}nd av Kungliga Tekniska H{\"o}gskolan i Stockholm framl{\"a}gges till offentlig granskning f{\"o}r avl{\"a}ggande av filosofie licentiatexamen fredagen den 24 november 2006 14.00 i sal FB54, AlbaNova Universitets Center, KTH Partikel- och Astropartikelfysik, Roslagstullsbacken 21, Stockholm.\\
\vspace{7mm}\noindent
Avhandlingen f{\"o}rsvaras p{\aa} engelska.}

\newcommand{\fb} {f_{b}}
\newcommand{\fbb}{f_{\bar{b}}}
\newcommand{\fc} {f_{c}}
\newcommand{\fcb}{f_{\bar{c}}}

\newcommand{\fmup} {f_{\mu^+}}
\newcommand{\fmum} {f_{\mu^-}}
\newcommand{\fpmup} {f'_{\mu^+}}
\newcommand{\fpmum} {f'_{\mu^-}}

\newcommand{\xf} {x_{\rm flip}}

\newcommand{\xc} {x_{c}}
\newcommand{\xpc} {x'_{c}}

\newcommand{\pt}        {\mbox{$p_T$}}
\newcommand{\et}        {\mbox{$E_T$}}
\newcommand{\met}       {\mbox{$\not{\!\!\!E_T}$}}
\newcommand{\qqbar}     {\mbox{$q\bar{q}$}}
\newcommand{\ppbar}     {\mbox{$p\bar{p}$}}
\newcommand{\ttbar}     {\mbox{$t\bar{t}$}}
\newcommand{\bbbar}     {\mbox{$b\bar{b}$}}
\newcommand{\Zbbbar}     {\mbox{$Z \rightarrow b\bar{b}$}}
\newcommand{\Zccbar}     {\mbox{$Z \rightarrow c\bar{c}$}}
\newcommand{\ccbar}     {\mbox{$c\bar{c}$}}
\newcommand{\ptrel}     {\mbox{$p_{T,{\rm rel}}$}}

\newcommand{\ljets}     {\mbox{$\ell \text{+jets}$}}

\begin{document}             

\maketitle

\cleardoublepage

\chapter*{Abstract}
\addcontentsline{toc}{chapter}{Abstract}

\smallskip
In this thesis, the first determination of the electric charge of the top quark 
is presented using $370$~pb$^{-1}$ of data recorded by the D\O\ detector at the Fermilab Tevatron 
accelerator. $t\bar{t}$ events are selected with one isolated electron or muon 
and at least four jets out of which two are $b$-tagged by reconstruction of a 
secondary decay vertex (SVT). The method is based on the discrimination
between $b$- and $\bar{b}$-quark jets using a jet charge algorithm applied
to SVT-tagged jets. A method to calibrate the jet charge algorithm with data 
is developed. A constrained kinematic fit is performed to associate the 
$W$ bosons to the correct $b$-quark jets in the event and extract the top quark 
electric charge. The data is in good agreement with the Standard Model top quark 
electric charge of $2e/3$. The scenario where the selected sample is solely composed 
of an exotic quark $Q$ with charge $4e/3$ is excluded at $92$\% confidence level. 
Using a Bayesian approach, an upper limit on the fraction of exotic quarks 
$\rho < 0.80$ at $90$\% confidence level is obtained.





\cleardoublepage

\chapter*{}
\vspace*{2cm}
\centerline{\emph{I de blindas rike, \"ar den en\"ogde kung.}}
\vspace*{1.0cm}
\rightline{Niccol\`o Machiavelli}

\newpage

\tableofcontents

\mainmatter

\chapter*{Introduction}
\addcontentsline{toc}{chapter}{Introduction}


It is widely believed that the new particle discovered at Fermilab in 
1995~\cite{discovery} is the long-sought top quark. Its currently measured 
properties are consistent with the Standard Model (SM) expectations
for the top quark, but many of its properties are still poorly known. 
In particular, the electric charge, which is a fundamental quantity
characterizing a particle, has not yet been measured for this quark.
It still remains not only to confirm that the discovered quark has charge $+2e/3$ and
hence the expected SM quantum numbers, but also to measure
the strength of its electromagnetic (EM) coupling to rule
out anomalous contributions to its EM interactions.
Indeed, one alternative interpretation has not yet been
ruled out: that the new particle is a charge $-4e/3$ quark. In the
published top quark analyses of the CDF and DO collaborations~\cite{top_review}, 
the pairing of the $b$ quarks and the W bosons in 
$p \bar{p} \to t \bar{t}\to W^+W^- b \bar{b}$ processes 
are not determined. As a result, there is a twofold ambiguity in the 
electric charge assignment of the ``top quark''. In addition to the SM 
assignment $t \to W^+ b$, $t \to W^-b$ is also conceivable, in which case
the ``top quark'' would actually be an exotic quark with charge $q = -4e/3$. 
The analysis presented in this thesis is not carried out within the framework 
of any extension to the SM. Nevertheless interpreting the particle found 
at Fermilab as a charge $-4e/3$ quark is consistent with current precision electroweak
data. Current $Z\to \ell^+ \ell^-$ and $Z\to b \bar{b}$ data can be fitted with a top
quark of mass $m_{t} =270$~GeV, provided that the right-handed
$b$-quark mixes with the isospin +1/2 component of an exotic
doublet of charge $-1e/3$ and $-4e/3$ quarks, $(Q1~,Q4)_{R}$~\cite{exotic_top_paper}. 
If the top quark had a mass of $m_{t}=270$~GeV, it would so far 
have escaped detection at the Fermilab Tevatron. The CDF collaboration has 
carried out a search for a heavy $t'$-quark using $760$~pb$^{-1}$ of 
data and excludes masses up to $258$~GeV~\cite{CDF_4th_quark_search}. 
With data sets beyond $1$fb$^{-1}$ and combining D\O\ and CDF, the Tevatron will be 
capable of detecting $t'$-quarks with masses of $270$~GeV and more. It should also be noted 
that a mass of $270$~GeV merely corresponds 
to the best fit to SM precision electroweak data in these models and the 
mass of such a heavy fermion could still be above $300$~GeV.


In this thesis, the first determination of the electric charge of the top
quark using $\sim365$~pb$^{-1}$ of $p\bar{p}$ data collected with the D\O\ 
experiment is presented. The result of the measurement is described 
in the paper
\hspace{2cm}

\begin{quote}
\item[] D\O\ Collaboration, V. M. Abazov {\em et. al}, {\it ''Experimental discrimination between charge $2e/3$ top quark and charge $4e/3$ exotic quark production scenarios''}, hep-ex/0608044, submitted to Phys. Rev. Lett. 

\end{quote}

The thesis is outlined as follows: chapter~\ref{ch:topphysics} gives an 
overview of the SM and the 
top quark. The D\O\ detector is described in chapter~\ref{ch:detector} and 
the object reconstruction is presented in chapter~\ref{ch:reco}. The analysis 
to determine the top quark charge is described in chapter~\ref{ch:topcharge} followed 
by a conclusion and outlook in chapter~\ref{ch:conclusions}.

\section*{Authors Contribution}
In this thesis the result of my work at the D\O\ experiment at Fermilab between 
February 2004 and summer 2006  is presented. Arriving at Fermilab I quickly 
started working in the top quark group with a feasibility study to determine 
the possibility and the amount of data needed for a determination of the 
electric charge of the top quark. The top quark charge had not been measured before and 
was considered very difficult due to the low statistic sample of top quarks. 

I have been responsible for the entire analysis from the first day. This analysis 
was developed through intense collaboration with Dr. Christophe Cl\'ement and 
Dr. David Milstead. At the start, most work went into 
studying various jet charge algorithms and their optimization as described in 
Sec.~\ref{sec:jcttbar}. In autumn 2004, I showed that a measurement should be 
possible with the data that was collected during this period and the work was 
accelerated towards forming a full analysis. During winter 2004 and spring 2005 
most of my work went into defining and validating the jet charge calibration 
discussed in Sec.~\ref{sec:data_calibration} and finding and studying various sources of 
systematic uncertainties. Based on the result of the top 
quark pair cross section, the top quark charge measurement was first presented 
as a preliminary result at the PANIC05 conference in October 2005. During 
winter 2005 and spring of 2006 I worked mostly on refining the data calibration 
method but also to develop the method of a simultaneous measurement of the fraction of exotic 
quarks in the sample. The result was finally submitted to Physical Review Letters 
for publication in the summer of 2006.

During 2005 I was involved in studies of the jet reconstruction efficiency and 
energy calibration, especially studying the out-of-cone radiation correction 
described in chapter~\ref{ch:reco}. During spring 2006 the D\O\ detector was 
upgraded, extending the silicon vertex detector with an additional layer allowing 
for an improved tracking of charge particles. I was responsible for upgrading and 
developing the online software displaying the silicon tracking detector status.

\section*{Notation}
\label{sec:notation}
As mentioned earlier, the particle discovered at Fermilab 
is widely believed to be the SM top quark. To 
this date, many of its parameters are poorly known. 
Until all its properties are determined 
with high precision, exotic scenarios (not included in the 
SM) are not excluded and only measurements such 
as the one presented in this thesis can finally decide if the 
particle is the SM top quark or an exotic quark. 
This thesis has no preconceived opinion on the true nature of 
the particle discovered. From now on, the name ``top'' in this thesis 
is simply a notation chosen for consistency with other papers 
referenced. The ``top'' quark refers to the SM top quark only when 
specifically indicated or when a 
comparison of the exotic quark scenario with the SM 
scenario is carried out.

\cleardoublepage



\chapter{The Standard Model Top Quark}
\label{ch:topphysics}

Elementary particle physics research is the quest for 
understanding the smallest constituents of matter 
and their interactions. The SM is the 
theoretical framework used to describe the known 
elementary particles and their interactions. The 
current view is that all matter is made up of three 
kinds of particles: leptons, quarks and mediators. 
In the SM the particle matter consists of spin-1/2 
quarks and leptons, which, down to a scale of around 
$10^{-18}$~m appear elementary
\footnote{Elementary means here that they don't 
have any internal structure.}. There are six 
``flavors'' of quarks and leptons arranged in 
three generations. There are four fundamental 
forces through which these elementary particles 
interact; gravity, electromagnetic, weak and the 
strong force. The electromagnetic- and weak 
force are manifestations of one single force, 
called the electroweak force, in the 
Glashow-Weinberg-Salam (GWS) model and the number 
of forces are then reduced to three. The SM is a 
quantum field theory (QFT) based on the symmetry 
group $SU(3) \times SU(2)_L \times U(1)_Y$; all particles are 
described as fields and forces between them are 
interpreted as being due to the exchange of mediator particles. 
These particles are known as gauge bosons, which 
are spin-1 particles\footnote{ The graviton is postulated to mediate the gravitational 
force and have spin-2 but is yet to be observed.}. 
The building blocks of the SM are summarized in 
Tab.~\ref{tab:SMparticles}. For a pedagogical introduction 
to elementary particle physics and the Standard 
Model, see e.g. Ref.~\cite{griffiths}.
\begin{table}
\centering
\begin{tabular}{|r|c|c|c|c|}
\hline
& Symbol & Name & Mass & Charge \\
& & &  (MeV) & ($e$) \\
\hline
Quarks & u & up & 1.5 to 4 & +2/3 \\
(spin=1/2)& $d$ & down & 4 to 8 & -1/3 \\
& $s$ & strange & 80 to 130 & -1/3 \\
& $c$ & charm & 1150 to 1350 & +2/3 \\
& $b$ & bottom & 4100 to 4400 & -1/3 \\
& $t$ & top &  172.5~GeV & +2/3 \\
\hline
Leptons & $\nu_e$ & electron neutrino & $<3$~eV & 0 \\
(spin=1/2) & $e$ & electron & 0.511 & -1 \\
& $\nu_\mu$ & muon neutrino & $<$0.19 & 0 \\
& $\mu$ & muon & 105.7 & -1 \\
& $\nu_\tau$ & tau neutrino & $<$18.2 & 0 \\
& $\tau$ & tau & 1777.0 & -1 \\
\hline
Gauge bosons & $\gamma$ & photon & 0 & 0 \\
(spin=1)& $g$ & gluon & 0 & 0 \\
& $W$ & $W$ & (80.425$\pm$0.038)$\times10^3$ & 1 \\
& $Z$ & $Z$ & (91.1876$\pm$0.0021)$\times10^3$ & 0 \\
\hline
Higgs boson & $H$ & Higgs & $>$114~GeV &  \\
(spin=0)    &     &       &                  &  \\ 
\hline
\end{tabular}
\caption{The SM particles~\cite{PDG}. The 
Higgs boson is yet to be observed,  direct searches for the Higgs 
boson puts a lower limit of 114~GeV on its mass~\cite{lep_higgs}.}
\label{tab:SMparticles}
\end{table}
The SM has been extremely successful and agrees 
with nearly all experimental data so far~\cite{PDG}. 
However, the SM is a not a complete theory of particle 
physics. For example, it does not incorporate gravity, nor 
can it account for so-called dark matter and energy. Many 
theoretical extensions of the SM have been postulated 
which predict the existence of hitherto-undiscovered 
fundamental particles including exotic quarks and 
leptons~\cite{frank_mod_article}.


The top quark is the partner to the bottom quark in the 
weak isospin doublet in the SM. 
Unless otherwise specified, in this chapter the term 
``top quark'' refers to the SM top quark, the discovery 
of which was announced by the D\O\ and CDF experiments 
around a decade ago~\cite{discovery}. This is in contrast 
to the notation discussed in the introductory chapter. The 
existence of the top quark was expected since the discovery 
of the bottom ($b$) quark in 1977 implied the existence of 
a further quark to complete the quark sector with a three 
generation structure. This chapter describes the current experimental status of 
the discovered quark and the notation is adapted to simplify 
the discussion. 
The first direct studies of the top quark were performed 
during Run I of the Tevatron at center-of-mass energy of 
$\sqrt{s}=1.8$~TeV and continued during Run II with 
higher-statistic samples. The Tevatron remains to date 
the only top factory. The top quark mass was predicted from precision 
electroweak measurements from LEP, SLD, NuTeV 
and \ppbar~colliders\cite{EWprecLetterB} before its 
discovery. Due to the limited number of top quarks observed 
so far its properties are less well experimentally determined 
than those of other known quarks. However, most existing 
results are consistent with the particle possessing the 
quantum numbers of the SM-top quark~\cite{PDG}.

There are several reasons why the top quark is 
interesting in the framework of the SM 
and possible physics beyond it:
\begin{itemize}
  \item{The top quark production and decay properties 
  are poorly known and provide important tests of the 
  SM at the Tevatron.}
  \item{The short expected lifetime of the top quark implies 
  that it is the only quark that will decay before it 
  hadronizes.}
  \item{The top quark mass is an important parameter in precision 
    electroweak fits and can thus constrain theoretical models of 
    physics beyond the SM~\cite{EWphysicsReport2004}.}
  \item{The top quark may have special dynamics related 
    to new particles beyond the SM due to its 
    large coupling to the Higgs boson.}
  \item{The large mass of the top quark and the increasing 
  production cross section at higher energies implies that 
  top quark production will be one of the principal sources 
  of background when searching for evidence of New Physics 
  processes at the Large Hadron Collider, which will collide 
  protons at $\sqrt{s}=14$~TeV from 2008 onwards.}
\end{itemize}

\section{Production of the Top Quark}
\label{subsec:topquarkprod}
\begin{figure*}[]
\begin{center}
  \includegraphics[width=0.32\textwidth]{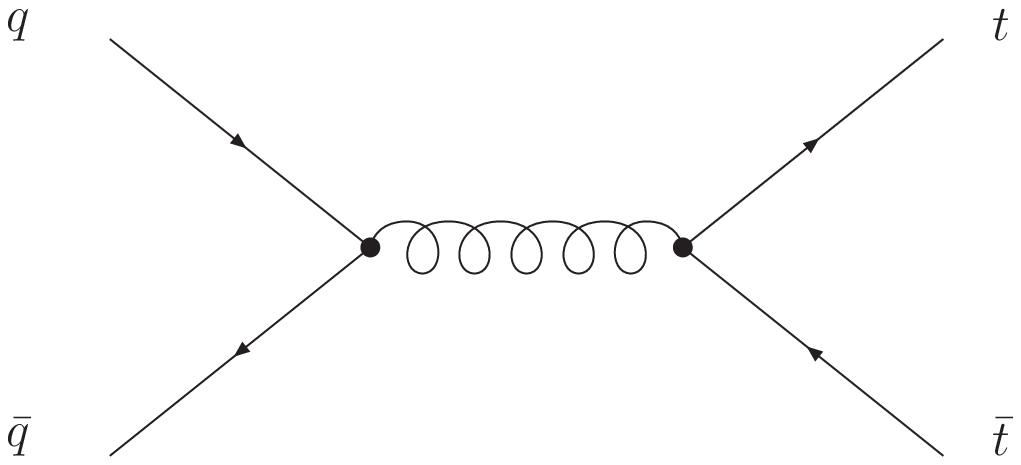}
  \hspace{10.0cm}
  \includegraphics[width=0.9\textwidth]{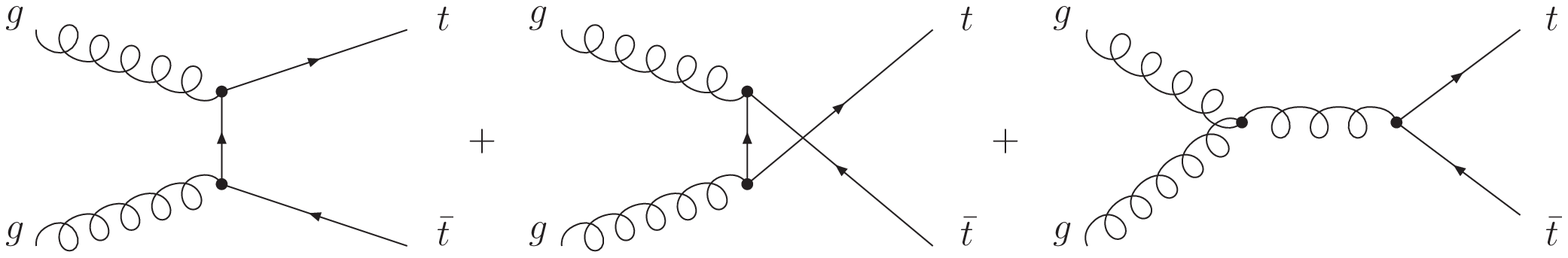}
\caption{Lowest order Feynman diagrams for the top quark pair 
production at the Tevatron.}
\label{fig:top_pair_production}
\end{center} 
\end{figure*}

Evidence for the direct production of the top quark has been 
obtained by the D\O\ and CDF collaboration solely via the 
measurement of \ttbar~pair-production processes. The two main 
processes are $q\bar{q} \to \ttbar$ and $gg \to \ttbar$, as shown 
in Fig.~\ref{fig:top_pair_production}.  
The quark and gluon 
content of the proton is described by so-called parton distribution 
functions. These describe the probability to find a gluon or a quark 
of a certain flavor carrying a fraction $x$ of the proton's 
(or anti-proton's) momentum. 
The value of $x$ required for production of top quarks 
decreases with increasing collision energy. At the Tevatron energy, 
the top quark pairs are produced approximately in 
85\% of the events by quark anti-quark fusion $q\bar{q} \rightarrow \ttbar$ 
and in 15\% from gluon fusion $gg \rightarrow \ttbar$~\cite{top_prod_theory}.
The top quark is also produced singly via the weak interaction via the 
so called $s$- and $t$-channel. Discovering single top production 
is more experimentally challenging due to a less distinctive event 
signature and larger backgrounds. No experimental evidence for 
production of single top quarks has been found so 
far~\cite{top_single_production}. 

The total top quark pair and single production cross section in the SM 
at a center-of-mass energy of $\sqrt{s}=1.96$~TeV is calculated to be 
$\approx 7$~pb and $\approx 3$~pb respectively~\cite{top_single_prod_theory}.

\section{Decay of the Top Quark}
\label{subsec:topquarkdecay}
In the SM the top quark is 
predicted\footnote{Assuming only three families and unitarity of 
the flavor mixing matrix (called Cabibbo-Kobayashi-Maskawa matrix or 
CKM in short) $|V_{tb}|\simeq 1$.} 
to decay to a $W^+$ boson and a $b$-quark with a branching ratio 
of $\sim\> 0.999$~\cite{PDG}. The large decay width 
($\approx1.5$~GeV) corresponds to a lifetime of 
around $5\times10^{-25}$~s. This lifetime is shorter than 
the corresponding time for hadronization and thus no 
bound states with $t$ or $\bar{t}$ exists~\cite{tt_quarkonium}. 

\section{Mass of the Top Quark}
\label{subsec:topquarkmass}
The top quark is heavier than any other 
elementary particle found so far. The mass of the top quark 
have been measured to the best relative precision of 
all the quarks. Combining the results from both experiments 
at the Tevatron the world-average top quark mass is 
$172.5 \pm 2.3$~GeV. More 
information on the techniques and results from the top quark mass 
analyzes can be found in~\cite{topmass_average}. The precision electroweak 
measurements from e.g. LEP can be used to make an indirect 
prediction of the top quark mass. The result, 
$179.4_{-9.2}^{+12.1}$~GeV, is consistent with the direct 
measurements.

\section{Experimental Tests of the Standard Model Top Quark Sector}
\label{subsec:topquarkprop}

To put the work in this thesis into perspective a summary 
of the world measurements in the top quark sector is given. 

\subsection{Top Quark Pair Production Cross Section}
\label{subsec:topprodcrosssection}
Both D\O\ and CDF have measured the \ttbar~production 
cross section. It is extracted by counting the number of observed 
events, estimating the number of background events and measuring 
the integrated luminosity (taking into account the acceptance). 
Any abnormal top quark decay such as $t \rightarrow H^+b$ can result in 
a lower cross section than predicted by the SM. 
A higher than expected cross section would hint at new 
unknown production mechanisms. One example can be found 
in Ref.~\cite{technicolor}. 

\begin{figure}[]
  \centering
  \includegraphics[width=0.8\textwidth]{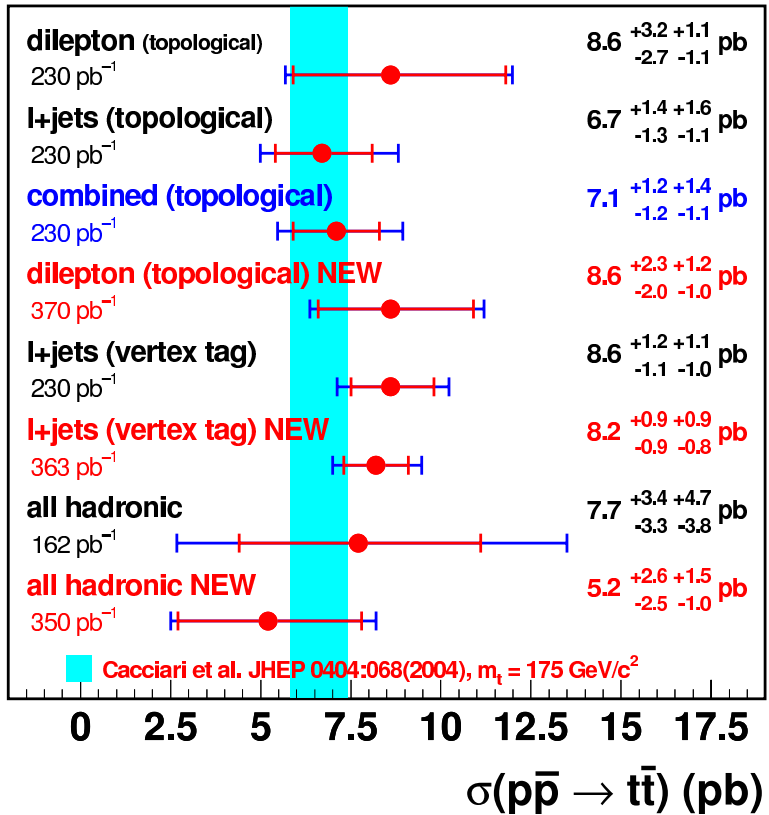}
  \caption{The \ttbar~production cross section measured by 
  the D\O\ collaboration as of fall 2005. The figure contains 
  both published and preliminary 
  results~\cite{d0_collaboration_top_results_online_archive}. 
  The notation of the different measurements is explained in 
  Sec.~\ref{subsec:eventsignature}.}
  \label{fig:measured_topxsec}
\end{figure}

So far all direct measurements of the \ttbar~production 
cross sections are in agreement with the SM 
prediction. Figure~\ref{fig:measured_topxsec} shows the 
measured cross sections from the D\O\ collaboration. 
The full list of cross section measurements at the Tevatron 
can be found in~\cite{PDG}.

\subsection{Top Quark Decay Branching Ratio}
\label{subsubsec:Vtb}
As discussed above, within the SM the dominant decay 
mode for the top quark is $t \to W^+b$. The CKM 
matrix~\cite{CKM_org} element $V_{tx}$ (with $x=b,s,d$) determines 
the coupling between the top quark and other flavors. The $W^+d$ and $W^+s$ decay modes are 
suppressed by the square of the mixing matrix elements. The predicted 
values of the mixing matrix can be tested by determining the ratio $R$ of 
branching ratios $\mathcal{B}$ for the processes,
\begin{equation}
R = \frac{\mathcal{B}(t \rightarrow Wb)}{\mathcal{B}(t \rightarrow Wq)}.
\end{equation}
The SM prediction\footnote{This ratio can be expressed 
in the elements of the mixing matrix elements as 
$R= \left (\left|V_{tb} \right|^2 \right ) / \left ( \left|V_{tb} \right|^2 + \left|V_{ts} \right|^2 + \left|V_{td} \right|^2 \right ) $.} is $0.9980<R<0.9984$ at 
$90$\% confidence level and the current best measurement~\cite{Vtb} is 
$R = 1.03^{+0.19}_{-0.17}$, in good agreement with the SM.

\subsection{$W$ Boson Helicity}
New physics has been searched for in the dominant top quark 
decay vertex $t \to W^+b$ where the helicity of 
the $W$ boson is sensitive to anomalous contributions from 
new physics beyond the SM. In the SM 
the right-handed fraction of $W$ bosons is suppressed compared to 
the longitudinal fraction ($\sim70$\%). By studying the angular distribution 
of the $W$ boson decay products
with respect to the top quark direction, D\O\ puts an upper limit of 
$0.23$ on the fraction of right-handed $W$ bosons at 95\% confidence 
level~\cite{W_hel_d0}. 
Direct measurements of the longitudinal fraction give a value 
of $0.74^{+0.22}_{-0.34}$~\cite{W_hel_cdf_f0} and 
$0.56 \pm 0.31$~\cite{W_hel_d0_f0}.

\subsection{Resonances and Rare Decays}

Due to its large mass there are various physics 
models~\cite{topcondensate,topasstechnicolor} beyond the 
SM in which the top quark plays a central role. 
In these models, a heavy particle decaying to \ttbar~can 
be produced with cross sections large enough to be visible 
at the Tevatron. The D\O\ and CDF collaborations have 
searched for \ttbar~production via an intermediate particle state 
by looking for narrow-width peaks in the spectrum of the invariant mass 
of \ttbar~events. D\O\ and CDF report no evidence for such an intermediate 
state and exclude masses of such a state up to 
$680$~GeV~\cite{ttbarresonance}.

\subsection{Top Quark Spin Correlations}
The top quarks in \ttbar~pairs produced from unpolarized incoming 
particles in \qqbar~annihilation are expected to be unpolarized. 
However, their spin is expected to be highly correlated with a 
higher fraction of events in which the spins are aligned rather 
than anti-aligned. D\O\ measured the spin correlation in a low statistics 
sample in Run I and found no deviation from the SM 
prediction\cite{spincorr}.

\subsection{The Standard Model Higgs Boson}

A key concept of the SM is the so-called gauge invariance, 
which can be interpreted as a transformation of fields in 
such a way that they do not change (they are gauge invariant). 
It can be be shown that to keep a theory like QED gauge invariant 
an additional interaction must be introduced, i.e. the photon. 
From gauge invariance it can be shown that the SM 
predicts massless mediator bosons (and also massless fermions), while 
the $W^{\pm}$ and $Z$ bosons are known to have a large mass. By 
postulating another field, the Higgs field, that all particles interact 
with, this problem is taken into account and all masses are consistent 
with the theory. This additional field implies the existence of the 
Higgs boson as the mediator of the field and is today the only 
undiscovered particle in the SM. For a good introduction 
to gauge theories and spontaneous symmetry breaking, see 
e.g.~\cite{griffiths}. 

\begin{figure}[]
  \centering
  \includegraphics[width=6cm]{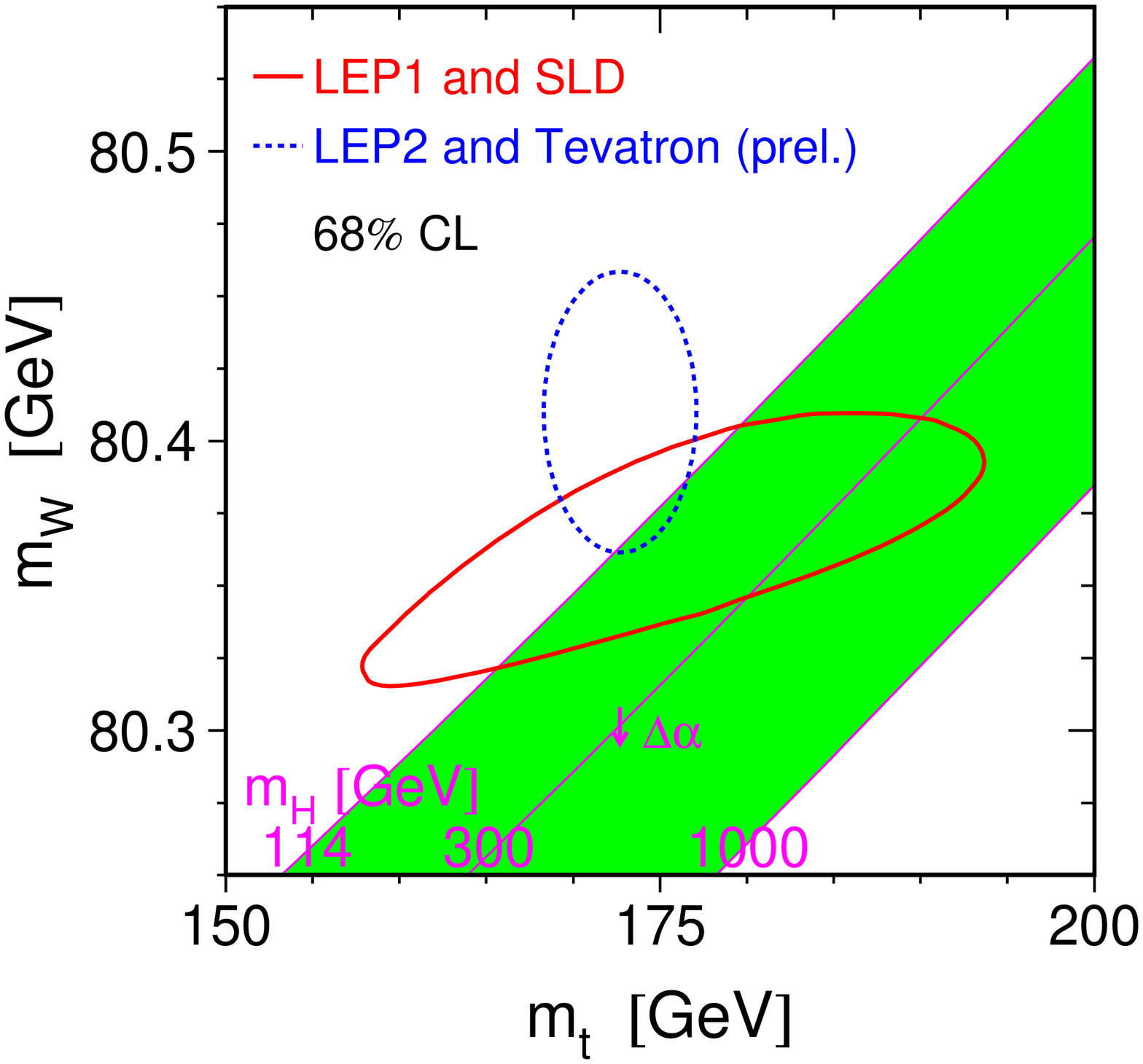}
  \includegraphics[width=6cm]{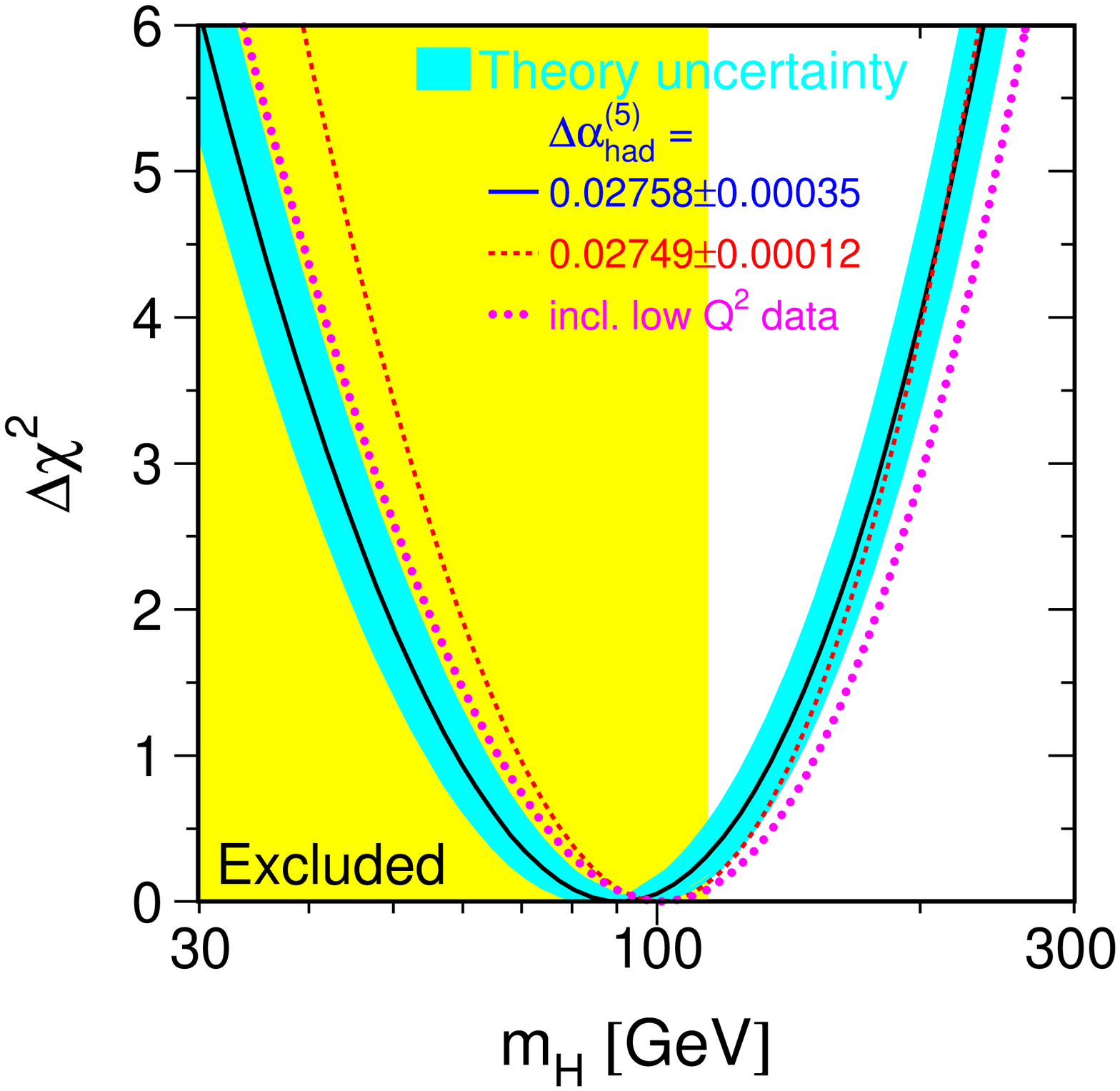}
  \caption{Constraints on the Higgs boson mass as a function of the 
  $W$ boson and the top quark mass (left) and the fit to the electroweak 
  parameters as a function of the mass of the Higgs boson 
  (right)~\cite{EWWG_summer05_combination}.}
  \label{fig:higgsconstraints}
\end{figure}

Although the couplings of the Higgs boson to other particles are predicted the mass 
of the Higgs boson is not. It can however be inferred from high precision 
measurements of the $W$ boson mass where virtual loop corrections involving both 
the Higgs boson and the top quark contribute. The same principle 
used when predicting the top quark mass before its discovery. 
Figure~\ref{fig:higgsconstraints} shows the dependence of the Higgs 
boson mass on the top quark and $W$ boson masses.
Although the uncertainty on the prediction of the Higgs boson mass is large, 
it is evident that the experiments imply a low-mass Higgs 
boson mass. The 95\% lower confidence limit on the Higgs mass from direct searches is 
$114$~GeV and the upper 95\% confidence limit is $166$~GeV 
(including the direct search lower limit increases the upper limit 
to $199$~GeV)~\cite{EWWG_summer05_combination}.

The search for evidence of the existence of the Higgs boson is currently one 
of the largest efforts in the particle physics community and will be 
addressed at the LHC. 

\cleardoublepage

\chapter{The D\O\ Detector}
\label{ch:detector}

The D\O\ detector was proposed in 1983 to study proton 
anti-proton collisions at the Fermilab Tevatron accelerator. The 
purpose was to study a wide range of phenomena focusing on high-mass 
states and high-\pt~processes. The D\O\ detector performed 
well during Run I of the Tevatron, which lasted from 1992 to 1996. 
Among many impressive results was the discovery of the long 
sought top quark and measurements of its mass. During 
Run I, the Tevatron operated with six bunches of protons and 
anti-protons with $3500$~ns between each bunch-crossing. The 
center-of-mass energy was $1.8$~TeV and the peak instantaneous 
luminosity was typically around $1-2 \times 10^{31}$cm$^{-2}$s$^{-1}$. 
The data recorded by the D\O\ experiment in Run I amounted to 
approximately $120$~pb$^{-1}$. Following the completion of the 
Fermilab Main Injector and other substantial Tevatron 
upgrades, the D\O\ experiment was running again in 
2001. In this phase, called Run II, the Tevatron is operated 
with $36$ bunches of protons and anti-protons with $396$~ns between 
each bunch-crossing and a center-of-mass energy of $1.96$~TeV. 
The instantaneous luminosity increased by a factor of ten. 

\begin{figure}[]
  \centering
  \includegraphics[width=1.0\textwidth]{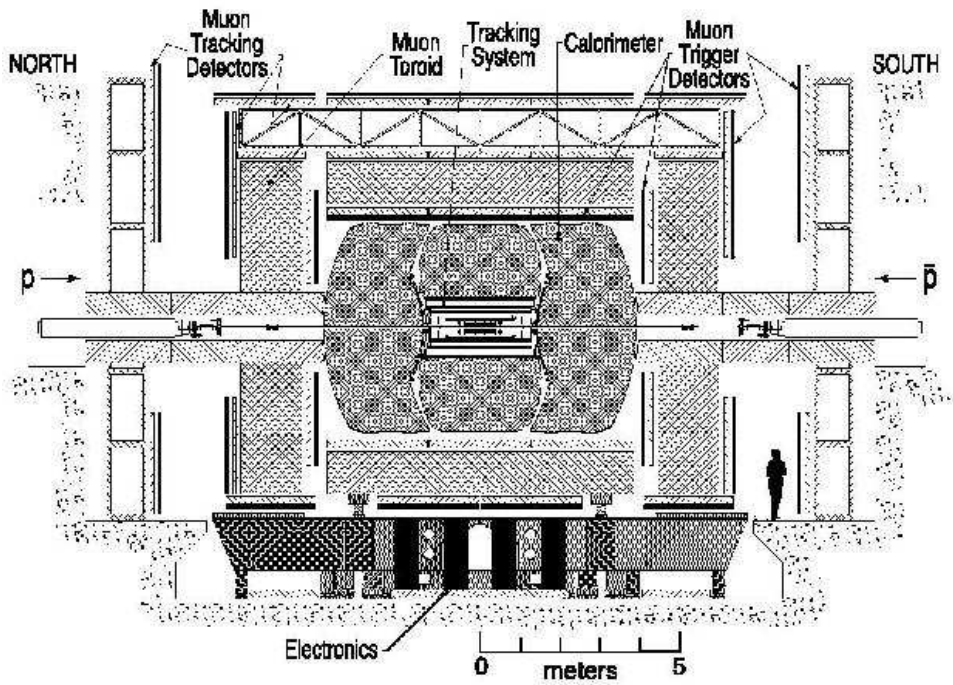} 
  \caption{Diagram showing the upgraded D\O\ detector as seen from the 
  outside of the Tevatron ring. The $+z$ axis is to the right, $+y$ 
  is up and $+x$ is out of the page~\cite{d0_detector}.}
  \label{fig:gen_detector_fig}
\end{figure}

To take advantage of the increased luminosity and center-of-mass 
energy delivered by the Tevatron the D\O\ experiment was greatly 
upgraded during 1996-2001. Among the major upgrades it is important 
to note that the tracking 
system from Run I which lacked a magnetic field and suffered from 
radiation damage was replaced with a silicon microstrip tracker and 
a fiber tracking detector in a $2$~T magnetic field. The detector 
consists of three major subsystems: the central tracking detectors, 
a uranium/liquid-argon calorimeter and a muon spectrometer. A side-view of 
the upgraded D\O\ detector is shown in Fig.~\ref{fig:gen_detector_fig}. 
This chapter gives a brief description of the upgraded D\O\ detector 
and those components most pertinent to the analysis presented in 
this thesis. A more detailed description can be found 
in Ref.~\cite{d0_detector}.

\section{ The D\O\ Coordinate System}
In the detector description and the data analysis, the standard 
D\O\ collaboration coordinate system is used where the positive 
$z$-axis points in the direction of the proton beam, 
the positive $x$-axis points radially outward from the Tevatron 
center and the positive y-axis is pointing upwards. To specify a 
direction in the detector, the polar and azimuthal angles 
$\theta$ and $\phi$ can be used. Since the angle $\theta$ is 
not invariant under Lorentz transformations along the $z$-axis 
it is common to use the pseudorapidity $\eta = -ln(tan \frac{\theta}{2})$ 
instead. $\eta$ approximates the true rapidity 
$ \mathbf{y} = \frac{1}{2} ln(\frac{E + p_z}{E - p_z})$ in the kinematic 
region where the mass is negligible, i.e. when $E \approx p$. The 
separation between two objects labeled 1 and 2 can be expressed 
as the distance $\Delta R$ between them in the ($\eta,\phi$) 
plane, defined as $\Delta R = \sqrt{\Delta \eta^2 + \Delta \phi^2}$. 
The term ``forward'' is commonly used to describe regions of the 
detector at large $|\eta|$. Since the initial momentum along the 
beam axis is unknown and some particles escape detection close to the beam 
axis the measured variables are in general quantities transverse 
to the beampipe direction, such as transverse momentum ($p_T$) or 
energy (\et), and missing transverse energy, \met, from neutrinos 
escaping the detector.

\section{The Central Tracking System}
\label{sec:trackingsystem}

\begin{figure}[]
  \centering
  \includegraphics[width=1.0\textwidth]{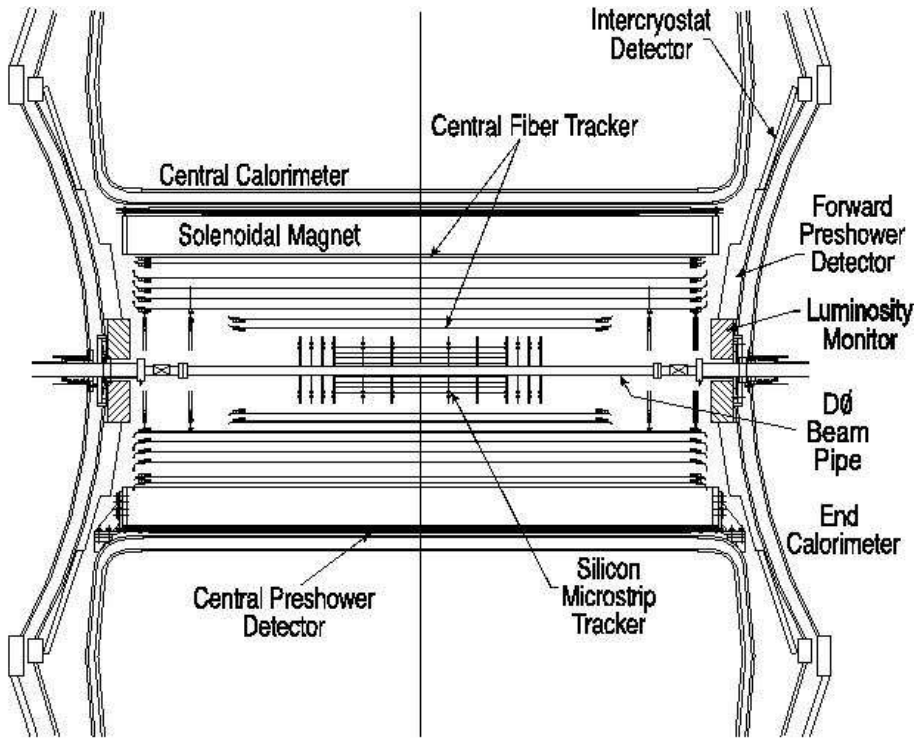}
  \caption{Schematic view of the central tracking system. The 
  preshower detectors, calorimeter and luminosity monitors are 
  also shown~\cite{d0_detector}.}
\label{fig:general_tracking_system}
\end{figure}

The measurement of tracks of charged particles and the reconstruction 
of a production or decay vertex is an important part of experimental 
studies at collider experiments. A precisely determined primary 
interaction vertex allows accurate measurements of lepton $p_T$, 
jet $E_T$ and \met. Using the tracking information 
it is possible to identify jets containing decay products of a 
$b$-quark by finding tracks emanating from a secondary vertex 
which is displaced with respect to the primary interaction vertex. 
This is especially important for top quark 
physics were the dominant top quark decay is to a $b$-quark 
and a $W$ boson. The central tracking system in D\O\ was 
completely replaced after Run I. The new system 
consists of two parts: The Silicon Microvertex Tracker (SMT) and 
the Central Fiber Tracker (CFT) enclosed in a magnetic field 
oriented along the beam axis. The $2$~T magnetic field is 
provided by a $2.8$~m long superconducting solenoid magnet with a 
radius of approximately $60$~cm. Charged particles produced 
in the collision are bent around the field lines in a magnetic 
field of strength $B$. The radius $r$ of the particle trajectory 
can be used to calculate the \pt~through~\cite{kleinknecht}:
\begin{equation} 
\pt[{\rm GeV}] = 0.3 \times r[{\rm m}] \times B[{\rm T}].
\end{equation} 
Combined, the SMT and CFT locates the primary 
interaction vertex with a resolution of $\approx35~\mu$m along 
the beam direction. They provide an impact 
parameter\footnote{The impact parameter is defined as the distance of 
closest approach ($d_{ca}$) of the track to the primary vertex in 
the plane transverse to the beamline. The impact parameter significance 
is defined as $d_{ca}/\sigma_{d_{ca}}$, where $\sigma_{d_{ca}}$ is 
the uncertainty on $d_{ca}$.} resolution of about $15~\mu$m in 
the $r-\phi$ plane for particles with $p_T>10$~GeV in the central 
region~\cite{d0_detector}. In addition, they also provide information on track $p_T$ 
to the trigger system for fast event decisions. A schematic view of the 
central tracking system is shown in 
Fig.~\ref{fig:general_tracking_system}.

\subsection{The Silicon Microvertex Tracker}
\label{subsec:smt}
\begin{figure}[]
  \centering
  \includegraphics[width=10cm]{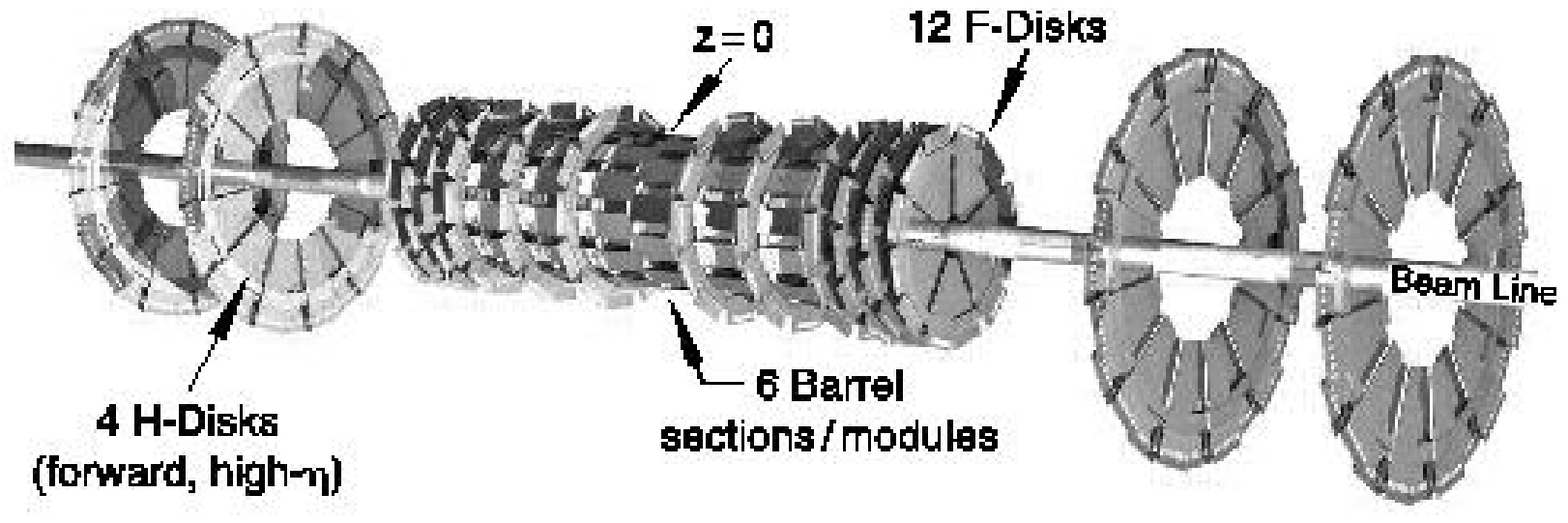}
  \caption{Design view of the Silicon Microvertex Tracker~\cite{d0_detector}.}
  \label{fig:smt_tracking_detector}
\end{figure}

The Silicon Microvertex Tracker (SMT) is the innermost part 
of the D\O\ detector. Its purpose is to provide both high-quality 
vertex finding and high resolution tracking. 
Its design is primarily dictated by the accelerator environment. 
For example the length of the device is determined based on the length of 
the interaction region, $\sim25$~cm. 
Since the SMT has to cover a significant solid angle it is 
difficult to ensure that the detector planes are always 
perpendicular to the outgoing particle trajectories. Therefore, the 
SMT has a barrel design interspersed with discs in the central 
region, while the forward region consists primarily of disks. There 
are six barrels, each with four silicon readout 
layers. Each barrel is attached (at the high $|z|$ side) to a 
disk with wedge detectors. At the outside of the barrel-disk 
assembly three disks are mounted. In the forward region four larger 
disks provide tracking capabilities up to $|\eta|=3$, see 
Fig.~\ref{fig:smt_tracking_detector}.
Particles with low pseudorapidities are mainly measured 
by the barrels while particles with larger pseudorapidities are 
also measured by the disks. 

There are several different types of silicon sensors. Both the
disks and barrels uses a combination of single-sided and 
double-sided sensors depending on the location in the SMT (varying 
with both layer and $|z|$ for the barrel). The SMT has in total 
912 readout modules, with 792,576 channels. Most of the sensors 
have a pitch of $50~\mu$m and the hit resolution is approximately 
$10~\mu$m (improving from the $1/\sqrt{12}$ dependence due to the pulse height 
information). The resolution in $z$-direction varies depending on the 
detector type in the various part of the SMT ranging from around 
$35~\mu$m to $450~\mu$m for $90^\circ$ and $2^\circ$ stereo angle 
detectors respectively. The $p_T$ resolution for central tracks 
with $|\eta|<2$ varies with momentum from $2-5$\% at track momentum 
of around $1$~GeV to $5-10$\% for tracks with approximately 
$10$~GeV momentum. The resolution degrades fast in the forward 
region up to $30$\% for tracks around $10$~GeV at $|\eta|\approx3$.

%
%


\subsection{The Central Fiber Tracker}
\label{subsec:cft}
\begin{figure}[]
  \centering
  \includegraphics[width=10cm]{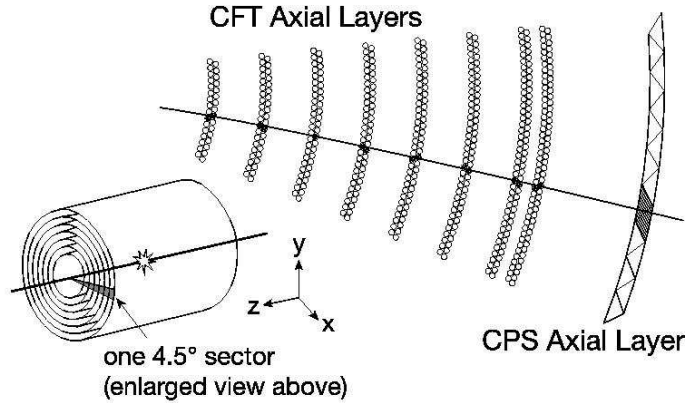}
  \caption{View of the axial layers of the CFT and the Central Preshower 
  (CPS) detectors 
  with a hypothetic track overlaid. Each of the axial doublet layer 
  has an associated additional doublet layer with a stereo angle 
  of $\pm 3^\circ$ not shown in this picture.}
  \label{fig:cft_layer}
\end{figure}

The Central Fiber Tracker (CFT) surrounds the SMT and covers 
the radial space from $20$ to $52$~cm from the center of the 
beampipe as shown in Fig.~\ref{fig:general_tracking_system}. 
The essential part of the CFT is the scintillating fiber 
system. Each fiber is $835~\mu$m in diameter 
(including cladding which is approximately $50~\mu$m thick) 
and oriented along the beam pipe in 
doublet layers on eight concentric cylinders. The innermost 
two cylinders are $1.66$~m long and the outer six are $2.52$~m 
long. The fibers in 
each doublet layer are separated by half the fiber diameter 
to achieve total coverage. Each cylinder supports one axial 
(oriented along the beam axis) doublet layer, see 
Fig.~\ref{fig:cft_layer}, and a second 
doublet layer oriented with $\pm 3^\circ$ stereo angle. 
The scintillating fibers are arranged in a multiclad structure using polystyrene 
as core material and paraterphenyl 
as the light-emitting material. To get the light out a second 
wavelength-shifter material is added and the light is transported via 
a clear fiber to the Visible Light Photon Counters (VLPC's) 
connected to one end of the fibers 
where the light is converted to an electric pulse and read out. 
The CFT has in total 76,800 channels of VLPC read out and the 
hit resolution is around $100~\mu$m. 
Approximately $200$~km of scintillating and $800$~km of 
clear fiber is used in the CFT in total.
 
The CFT's axial layers are part of the fast Level 1 trigger 
which aid in finding the interesting collisions discussed in 
more detail in Sec~\ref{sec:trigger}.

The $p_T$ resolution achieved combining SMT and CFT is studied 
using $Z\to\mu^+\mu^-$ events and resolutions of 
$\sigma/p_T^2 \approx 0.002$ have been obtained~\cite{pt_res}.

\section{The Preshower Detectors}
\label{sec:preshower}
The preshower detectors provide an early energy sampling and good 
position measurement. The detectors are designed to help in electron 
identification and to correct for the energy lost in the 
upstream material (mainly the solenoid). The fast response 
also allows the preshower detectors to be part of the event 
trigger. 

The design consists of two similar detectors, the Central Preshower 
Detector (CPS) and the Forward Preshower Detector (FPS). The CPS 
(FPS) consists of three (two) layers of triangular strips of 
scintillating material interleaved to remove any gaps. 

The central preshower detector (CPS) is placed in the $5$~cm 
gap between the solenoid magnet and the central calorimeter 
 covering $|\eta|<1.3$ as shown in 
Fig.~\ref{fig:general_tracking_system}. Inside the 
CPS, a lead radiator about one radiation length, $X_0$, 
thick (corresponding to $\approx0.55$~cm) and $244$~cm long is inserted. The 
solenoid and the lead radiator together comprise about two radiation 
lengths (the solenoid is $0.9X_0$ thick) for normal incident 
particles increasing to about four radiation lengths at maximum CPS 
coverage. Electrons and photons are converted into showers in the upstream 
material and this provides a discrimination between electrons 
or photons and pions, where the latter mostly passes through without 
showering.

The two (north and south) forward preshower detectors are 
mounted on the inner part of the end cap calorimeter 
(see Fig.~\ref{fig:general_tracking_system}) covering 
$1.5<|\eta|<2.5$. Each detector consists of 
a two radiation lengths thick stainless steel radiator 
sandwiched between two layers of scintillating strips. This 
design allows for position measurements as well as 
possible discrimination between electrons or photons 
and pions. 


\section{The Calorimeter}
\label{sec:calorimeter}
\begin{figure}[]
  \centering
  \includegraphics[width=1.0\textwidth]{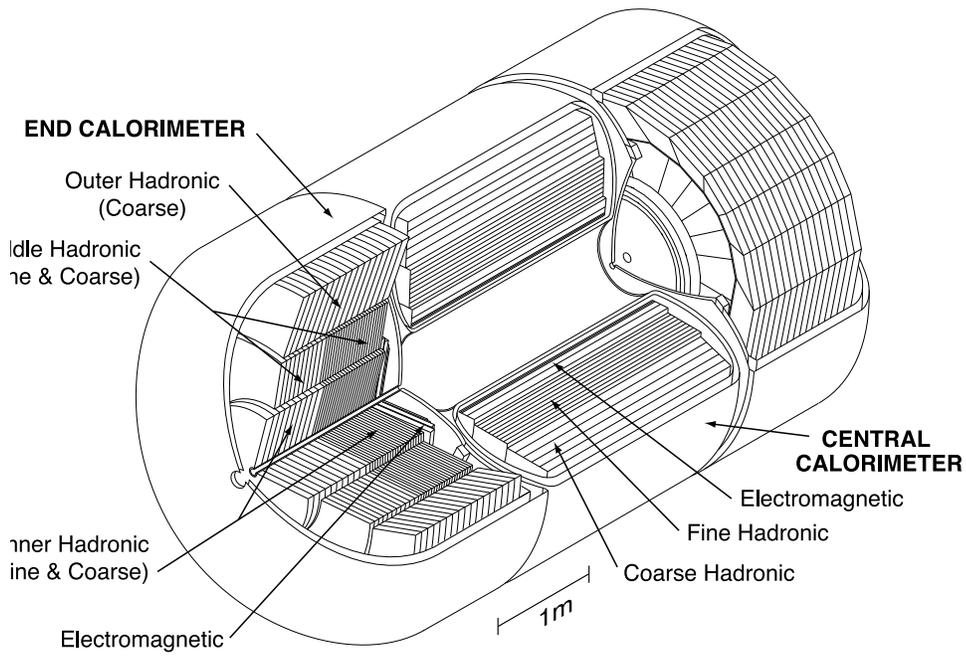}
  \caption{View of the central and two end cap 
    calorimeters~\cite{d0_detector}.}
  \label{fig:general_calorimeter_detector}
\end{figure}

The calorimeter absorbs and measures particle energy and 
the position of the deposited energy. It consists of a 
central calorimeter (CC) and 
two (north and south) end cap calorimeters (EC), see 
Fig.~\ref{fig:general_calorimeter_detector}. The Run~II 
calorimeter is essentially the same calorimeter as in Run~I 
but with upgraded electronics adapted to the new accelerator 
environment, i.e. the higher bunch crossing frequency.
\begin{figure}[]
  \centering
  \includegraphics[width=1.0\textwidth]{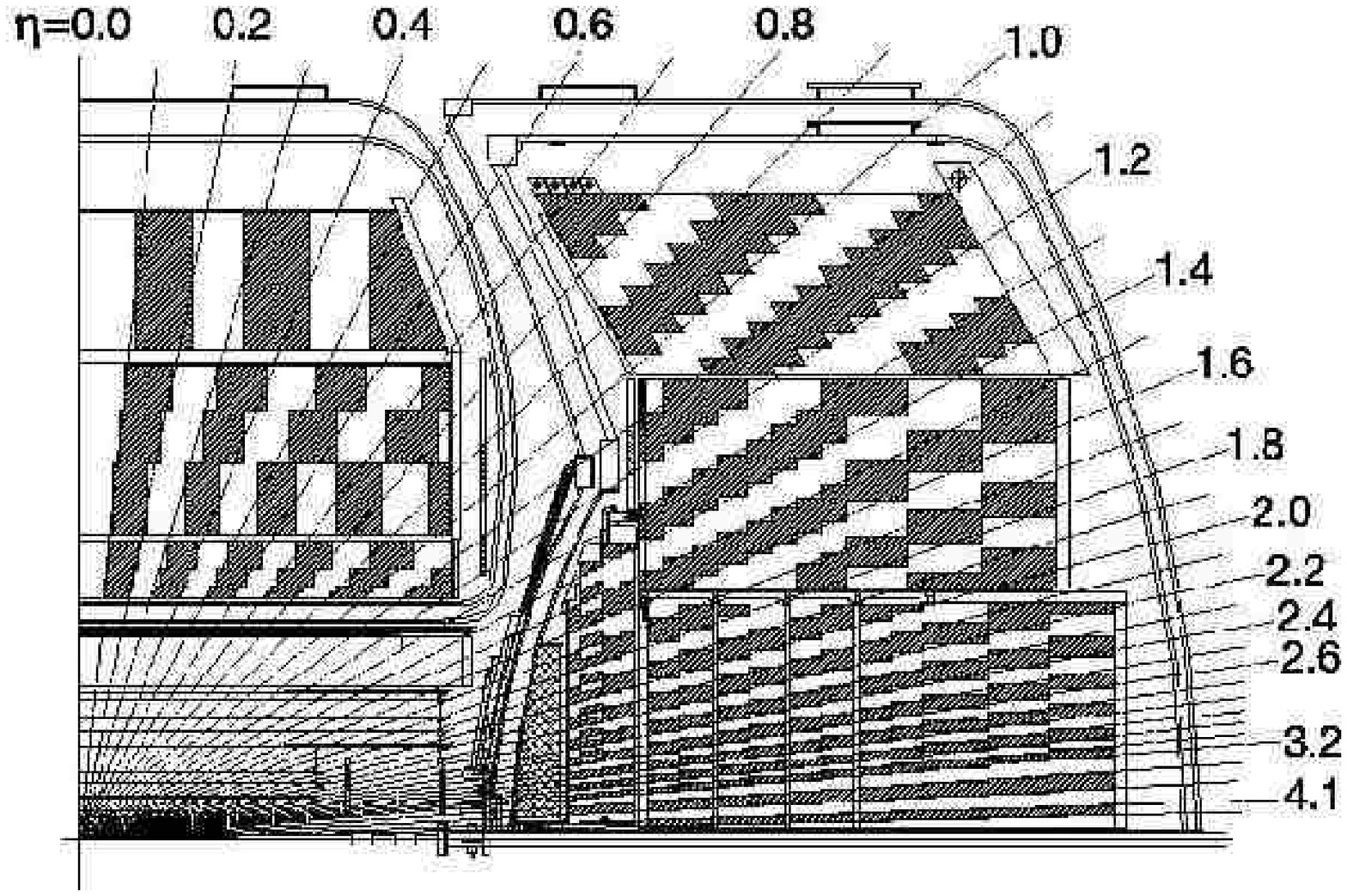}
  \caption{Schematic view of a portion of the D\O\ calorimeter 
  showing the transverse and longitudinal segmentation pattern. 
  The shaded areas corresponds to cells grouped together for 
  readout. The lines indicate values of psuedorapidity, as measured with
  respect to the centre of the detector~\cite{d0_detector}.}
  \label{fig:cal_zoom}
\end{figure} 
The CC covers a region up to $|\eta|<1.0$ and the two end cap 
calorimeters extend the coverage to $|\eta|=4$, as shown in 
Fig.~\ref{fig:cal_zoom}. The CC 
and EC are constructed in three parts; the electromagnetic 
section (EM) closest to the beam pipe followed by the 
fine hadronic section (FH) and the coarse hadronic (CH) section. 
The active medium for all the calorimeters is liquid 
argon and the three calorimeters are enclosed in a 
cryostat at a temperature of approximately $80$~K. Different 
locations have different absorber plates. The main absorber used 
in the EM calorimeter is nearly pure depleted uranium 
assembled into thin plates ($\approx3$~mm) in both CC 
and EC. The fine hadronic section uses $6$~mm thick 
uranium alloy plates (both in the CC and EC) and the 
coarse hadronic section uses $46.5$~mm thick copper 
(stainless steel) plates in the CC (EC). 

The readout cells of the calorimeter are arranged in 
sizes such that each cell covers 
$\Delta\phi \times \Delta\eta=0.1\times0.1$, which is 
comparable to the transverse sizes of showers: $1-2$~cm 
for EM showers and about $10$~cm for hadronic showers. 
Longitudinal depth segmentation is important when 
distinguishing between electrons or photons and hadrons. In the 
EM calorimeter there are four depth layers (in both 
EC and CC). The third layer is placed at the expected shower 
maximum and is twice as finely segmented in the lateral 
direction for increased spatial resolution.
The amount of material (tracking, cryostats, solenoid, etc.) 
between the interaction region and the first active 
gap in the EM calorimter at amounts to approximately 
$4X_0$ in the CC and $4.4X_0$ in the EC. The EM calorimeter contains uranium 
comparable to approximately 20 radiation lengths 
($X_0^{\rm U}=3.2$~mm) to capture the overwhelming part 
of the electromagnetic shower. As the nuclear interaction 
length is much larger than the radiation length 
($\lambda_I^{\rm U} \approx 10.5~{\rm cm} \approx 30X_0^{\rm U}$) 
the hadronic particles typically deposits most of its 
energy in the hadronic part of the calorimeter.

The calorimeter provides trigger information to all three 
trigger levels. The Level 1 and Level 2 triggers are based on 
analog sums of energy in special trigger towers. 

The energy resolution of a sampling calorimeter can 
be parametrized by 
\begin{equation}
  \frac{\sigma(E)}{E} = \sqrt{ C^2 + 
                              \left ( \frac{S}{\sqrt{E}} \right )^2 + 
							  \left ( \frac{N}{E} \right )^2  }.
  \label{eq:res}
\end{equation}
The parameter $C$ is called the ``constant term'' and comes 
from calibration errors or other systematic effects, $N$ is an 
energy independent ``noise term'' including contributions from uranium 
decays and electronic noise. The largest contribution comes from the 
``sampling term'', $S$, which is the statistical error 
in the sampling procedure. 
For the Run II detector, preliminary studies shows a 
degradation of the calorimeter resolution from several 
sources e.g. worse noise characteristics of 
the detector electronics, shorter pulse shaping due to the increased 
bunch crossing frequency, large cell-to-cell miscalibrations 
and more upstream material from the new tracking system, 
degrading the sampling term. The jet energy resolution 
is described in Sec.~\ref{subsec:jer}.

\subsection{The Inter-Cryostat Detector}
\label{subsec:icd}
Due to the fact that the calorimeter is contained in three 
separate cryostats it has incomplete coverage in the region 
$0.8<|\eta|<1.4$. Therefore, scintillation counters 
with a cell size matching the calorimeter as well 
as single cell structured scintillation counters are 
inserted in this region. These detectors allow for 
a sampling of the inter-cryostat region improving the 
energy resolution.

\section{The Muon Spectrometer}
\label{sec:muonsystem}

\begin{figure}[]
  \centering
  \includegraphics[width=\textwidth]{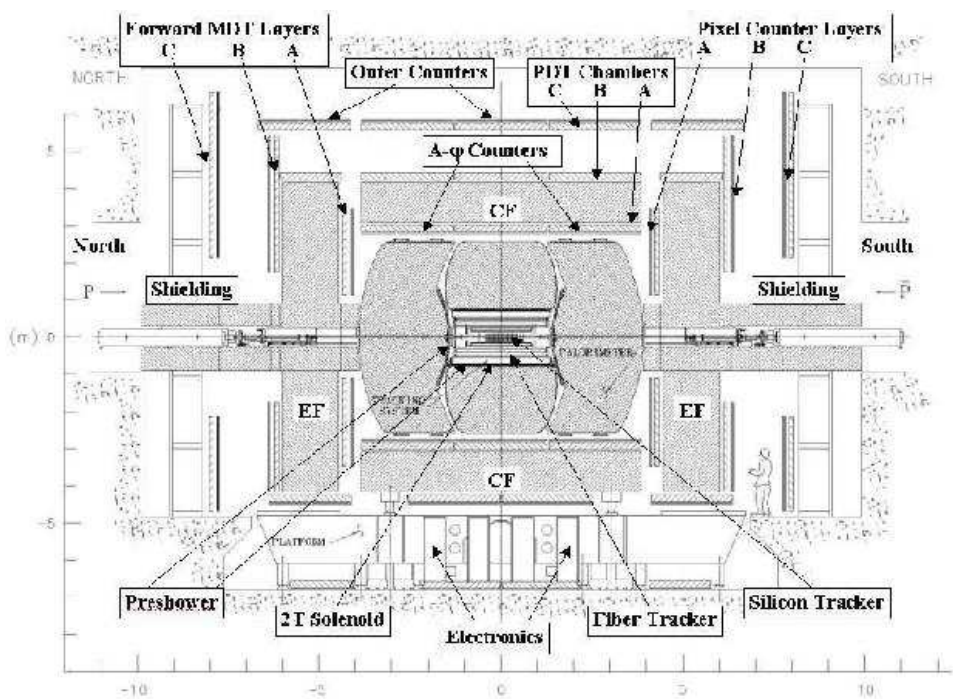}
  \caption{Schematic view of the muon system. CF and EF denote 
    the toroid magnets~\cite{muon_detector}.}
  \label{fig:muon_system}
\end{figure}

Muons lose only a small fraction of their energy in 
the central tracking system and calorimeter. The D\O\ muon 
system~\cite{muon_detector} is located around the calorimeter 
and is used to trigger and to measure muon $p_T$ and charge 
independently of the tracking system. An overview of the 
muon system is shown in Fig.~\ref{fig:muon_system}.
The system is divided into a central and forward detector. 
A $1.8$~T magnetic field is supplied by a $109$~cm thick 
iron toroid magnet, built in three sections to allow for 
easier access to the inner part of the detector. The 
central magnet is located at a radial distance of $318$~cm from 
the beam line covering the region $|\eta|<1.0$. The forward 
toroids are located at $454<|z|<610$~cm. The muon detectors consist 
of proportional drift tubes (PDT's), mini drift tubes (MDT's) and 
scintillation counters. The PDT's are rectangular volumes filled 
with gas and cover $|\eta|<1.0$. 
A charged particle ionizes the gas and the electrons are amplified 
at the $50~\mu$m thick anode wire. The Vernier cathode pads above 
and below the wire are segmented to provide information on the 
ionization position along the wire. The maximum electron drift 
velocity is $450$~ns and gives a single wire resolution of around 
$1$~mm in the radial direction of the wire for $10$~cm wide 
drift cells. The MDTs extend the coverage up to $|\eta|<2.0$ and 
consist of drift tubes with shorter electron drift times ($40-60$~ns) 
than the PDTs (the MDT cell width is $9.4$~mm and the length 
ranges from $1$ to $6$~m). The radial resolution for single wires 
is $\approx0.7$~mm. 

\begin{figure}[]
  \centering
  \includegraphics[width=10cm]{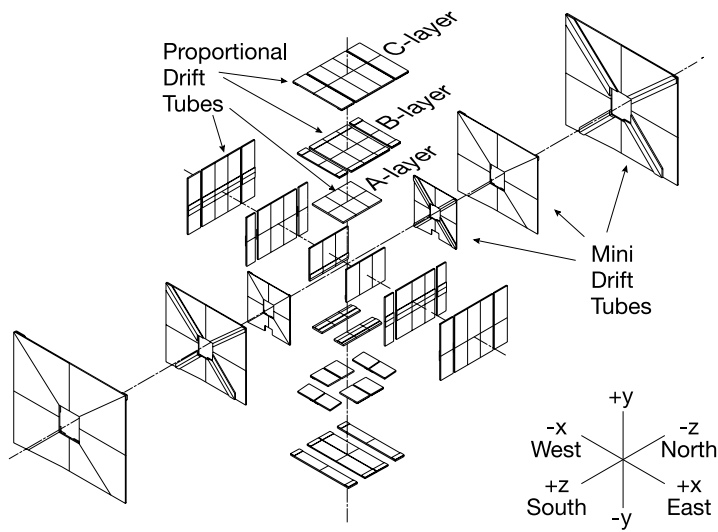}
  \caption{Exploded view of the wire chambers in the muon system~\cite{muon_detector}.}
  \label{fig:muon_system_wirechambers_exploded}
\end{figure}

Both the central and forward drift chambers consist of three layers, 
A, B and C. The A layer is located inside the toroid magnet while the B 
and C layers are outside. Each layer also has a sheet of scintillating 
pixels (except layer B in the central region) used for triggering, 
cosmic muon (and other background) rejection and track reconstruction. 
The scintillator geometry is matched to the central fiber tracker 
trigger read out to provide matching of tracks from the central 
tracking system to the muon system at the first trigger level. The 
scintillation counters allow for triggering on muons with $p_T$ down 
to $3$~GeV (the A layer drift tubes and scintillation counters also 
allow for triggering on muons that do not penetrate the toroid magnet). 
The muon system drift tubes are shown in Fig.~\ref{fig:muon_system_wirechambers_exploded}. 

\begin{figure}[]
  \centering
  \includegraphics[width=8cm]{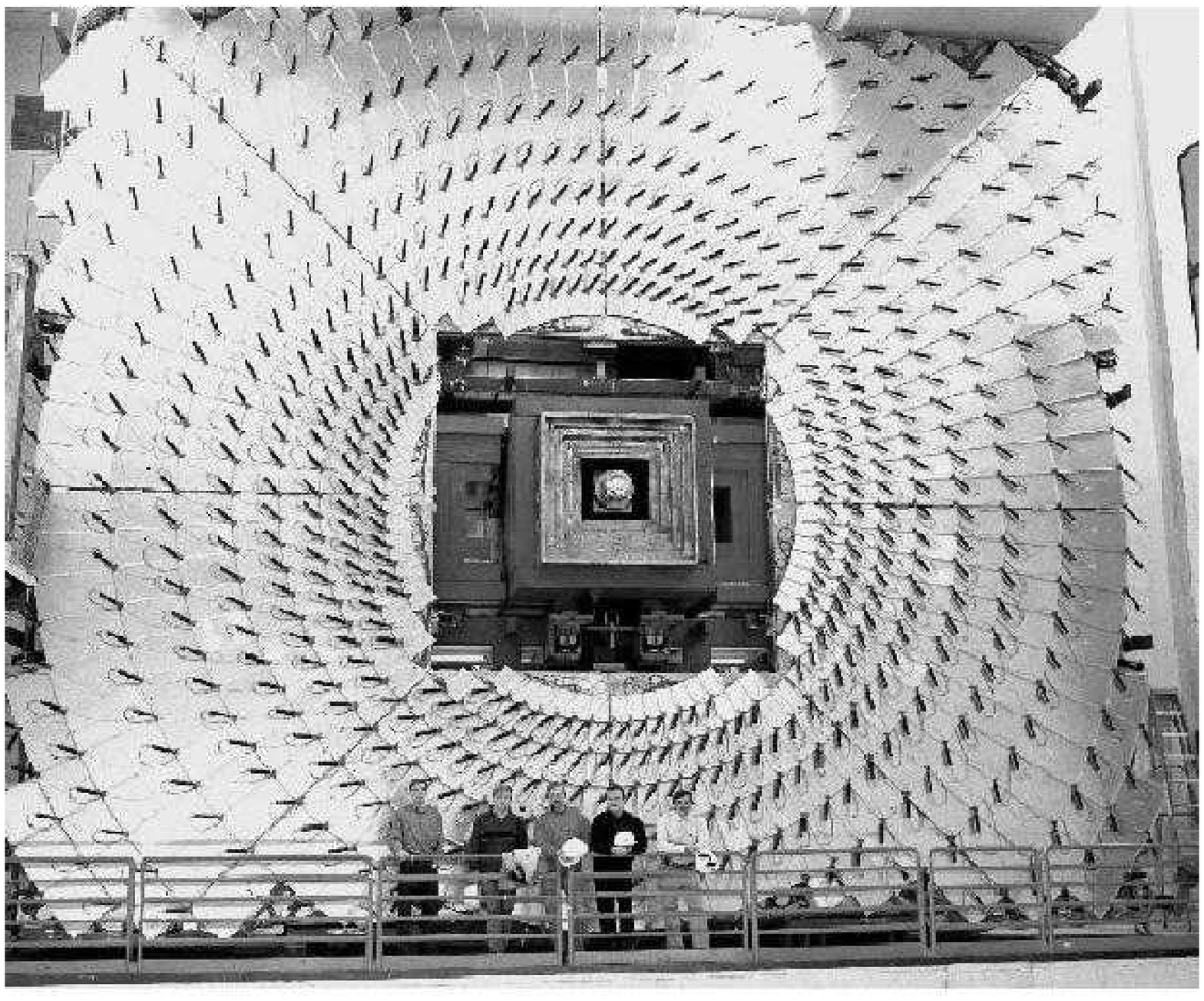}
  \caption{Photograph of the forward C layer scintillator 
    counter~\cite{muon_detector}.}
  \label{fig:c_layer}
\end{figure}

Directly below the D\O\ detector, the support 
structure and the readout electronics causes the muon system to 
have only partial coverage in this region. The forward C layer of 
scintillation detectors are shown in Fig.~\ref{fig:c_layer}.
The overall momentum resolution, including information from the 
silicon microvertex tracker and central fiber tracker, is 
defined by the central tracking system for muons with momentum 
up to approximately $100$~GeV. The muon spectrometer 
improves the resolution only for very high energy muons.



\section{Luminosity Monitoring}
\label{sec:lm}

The number of observed events $N_{\rm class}$ for a certain class of process 
in a collider is given by, 
\begin{equation}
  N_{\rm class} = \epsilon A \sigma_{\rm class} \int\mathcal{L}dt,
  \label{eq:eventrate}
\end{equation}
where $A$ is the acceptance, $\sigma_{\rm class}$ is the 
cross section for the process and $\epsilon$ the probability to 
record the event if it is within the acceptance region (i.e. the efficiency). 
For a cross section measurement e.g. \ttbar~production, 
the efficiency and acceptance for \ttbar~events (and background) 
is calculated and the only unknown in Eq.~\ref{eq:eventrate} is 
$\sigma_{{\rm class}=\ttbar}$ and the proportionality factor called 
{\sl instantaneous luminosity} $\mathcal{L}$. The luminosity 
is defined by the beam parameters of the Tevatron accelerator e.g. 
the number of protons and anti-protons in each bunch, the bunch 
crossing frequency, the lateral bunch size, the bunch overlap in the 
collision region etc. 

\begin{figure}[]
  \centering
  \includegraphics[width=10cm]{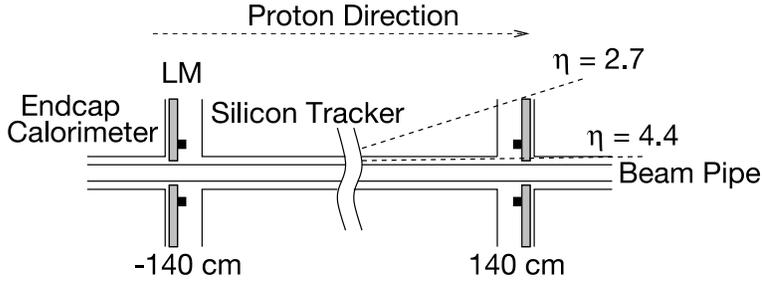}
  \caption{Schematic view of the location of the luminosity 
  monitoring detectors.}
  \label{fig:lm_detector}
\end{figure}

From Eq.~\ref{eq:eventrate} the instantaneous luminosity can be calculated by 
counting the number of observed events for a process with 
known cross section. At D\O\ \cite{d0_lumi}, the process used is the inelastic 
\ppbar~cross section $\sigma_{\ppbar}$, i.e.
\begin{equation}
  \mathcal{L} = \frac{1}{\epsilon A \sigma_{\ppbar}} \frac{dN}{dt}.
  \label{eq:lumi}
\end{equation}
The inelastic \ppbar~cross section has been measured by several 
experiments~\cite{inel_ppbar_xsec} to be 
$\sigma_{\ppbar} = 60.7 \pm 2.4$~mb. The detectors used for 
counting
the interaction rate are the 
Luminosity Monitors (LM) consisting of two arrays of 24 plastic 
scintillators located at $z = \pm 140$~cm from the center of 
the detector covering the region $2.7 < |\eta| < 4.4$ as 
shown in Fig.~\ref{fig:lm_detector}. They are attached to the 
inner part of the cryostat housing the end cap calorimeters, 
see Fig.~\ref{fig:general_tracking_system}.
In addition to measuring the luminosity, the LM is used to 
provide a fast measurement of the position of the primary interaction 
vertex, used by the fast Level 1 trigger. The time-of-flight 
difference between particles in the \ppbar~collision is calculated 
with a resolution of $\approx0.3$~ns and the resolution of the 
vertex position in the $z$-direction is $<10$~cm~\cite{d0_detector}.

\section{The Trigger System}
\label{sec:trigger}

Bunch crossings at the Tevatron occur at 2.5MHz rate. This 
immense rate is needed as the overwhelming majority of 
\ppbar~encounters result in collisions of little interest. The 
production of heavy objects like the top quark, $W$ or $Z$ 
bosons or collisions that could indicate the existence of New Physics 
processes occur at very low rate. The trigger system allows for a fast 
event-by-event decision on whether or not the collision was 
interesting and reduces the output rate to $50$~Hz, more 
suitable for writing to disk. 

The trigger is a 3-tiered system where each tier (Level 1, 
Level 2 and Level 3) 
investigates the event in larger detail than the preceding one 
and restricts the amount of data sent to the next 
level. An event can fail the trigger because: It did not 
fulfill the trigger requirements and was declared uninteresting, 
it was mistaken for an uninteresting event (trigger 
inefficiency) or the trigger system was busy processing 
previous events (dead time).

The Level 1 (L1) Trigger is built from specialized hardware 
investigating every event for interesting features. 
Pipelines mounted on the front-end electronic boards allows for an 
event decision in $4.2~\mu$s. The L1 trigger receives input from several 
subdetectors: The calorimeter L1 trigger looks for patterns 
of large transverse energy deposits in special trigger 
towers\footnote{Due to noise considerations 
not all trigger towers are used in the L1 calorimeter trigger.}, 
the Central Fiber Tracker L1 trigger looks for tracks 
exceeding predetermined thresholds in transverse momentum, 
the L1 muon trigger searches for muon candidates with a 
matched track from the CFT, an indication of 
the collision point is given by the luminosity monitors. 
The L1 reduces the rate from $2.5$~MHz to about $2$~kHz.

The level 2 (L2) trigger stage has an event decision 
time of approximately $100~\mu$s and further analyzes 
the event by two stages upon a L1 trigger accept: A 
preprocessing stage that analyzes 
the data into simple physics objects e.g. track clusters, and a global 
stage that combines trigger information 
from different subdetectors e.g. matching tracks from 
the inner detector to electromagnetic clusters in the 
calorimeter. The output rate of the L2 trigger is about 
$1$~kHz. 


The last trigger level, Level 3 (L3), is a software trigger 
which reduces the event rate from to $50$~Hz to allow 
for writing the interesting events to disk for later offline 
processing. The L3 trigger is fully programmable using 
algorithms based on complete physics objects which are as 
sophisticated as those available during the offline reconstruction 
phase. A L3 decision is based on full event information 
with complex variables such as spatial separation between jets 
and electrons, invariant masses of objects, displaced tracks 
from the primary vertex, etc. 

\cleardoublepage


\chapter{Event Reconstruction}
\label{ch:reco}

The data in a single event collected from the D\O\ detector
is the immediate detector response from nearly a million
detector readout channels. To find
evidence for the products of the collision and measure
their properties these signals needs to be processed
carefully.
To reduce the huge amount of data from the experiment
the information is handled by a chain of sophisticated
software algorithms which create and define physics
objects that represent the particles originating from 
the \ppbar~interaction.
Each algorithm is designed to identify a particular object
often based on the required efficiency and purity. The
analysis presented here is based on the 
\ttbar$\rightarrow l \nu jjb \bar{b}$ final state which
requires identification of the primary vertex, tracks,
leptons (electrons and muons), jets and their flavors and
missing transverse energy \met.

This chapter describes the event signature of top quark 
pair production and the most important background processes. 
A short description of the identification and reconstruction 
of the different physics objects is also given.

\section{Event Signatures}
\label{sec:topquarksigbkg}

\subsection{Experimental Signature of \ttbar~Production}
\label{subsec:eventsignature}
\begin{figure}[]
  \centering
  \includegraphics[width=10cm]{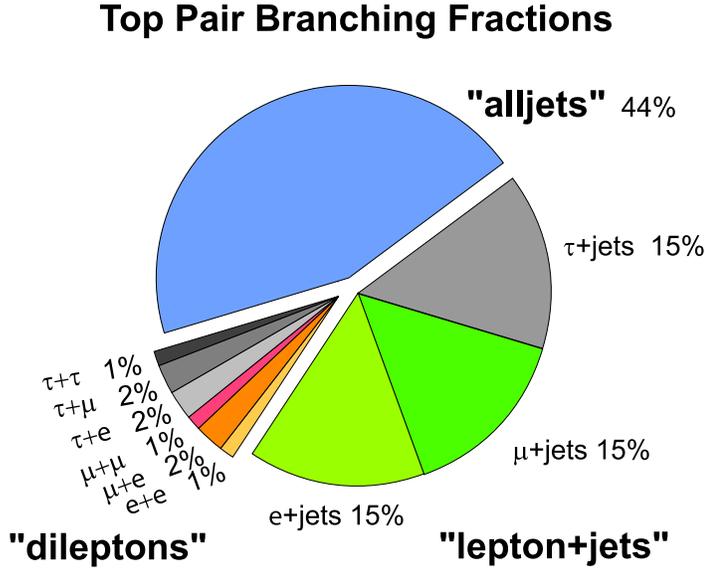}
  \caption{Schematic view of the characterisation of the top quark pair decay channels and 
    their branching fractions~\cite{top_group}.}
  \label{fig:br_ratios}
\end{figure}

Since the top quark decays almost exclusively through
$t \rightarrow Wb$, the final state of the top quark pair production
can be characterized by the decay of the two $W$ bosons. The 
$W$ boson decays leptonically via $W \to e \nu$, $W \to \mu \nu$ 
or $W \to \tau \nu$ with a branching fraction of $\approx 11$\% 
each or to hadrons with $\approx 67$\%. The decay modes of the $W$ bosons 
are reflected in the experimental search channels:
\begin{itemize}

  \item{{\bf All jets channel}\\
  Both $W$ bosons decay hadronically into $q\bar{q}$ pairs and the
  final state is characterized by two $b$-quark jets and at least four jets
  from the hadronization of the $q\bar{q}$ pairs. No significant 
  \met~is expected. This channel has the largest 
  branching fraction but suffers from large multijet backgrounds.}

  \item{{\bf lepton-plus-jets channels}\\
  One $W$ boson decays hadronically and the other leptonically. The
  final state is characterized by two $b$-quark jets, at least two
  jets from the $q\bar{q}$ pair, one charged lepton and significant \met~due 
  to the neutrino from the leptonically decaying $W$ boson. This decay 
  chain provides a clean signature of a single isolated lepton with high transverse
  momentum and large \met. Together with the large branching fraction this channel
  is most promising for measurements of top quark properties and is also the one
  used to determine the electric charge of the top quark in the the 
  present analysis. This channel is also referred to 
  as $e$+jets and $\mu$+jets separately depending on the flavor of the charged high 
  transverse momentum lepton or $\ttbar \to \ljets$~collectively.}

  \item{{\bf dilepton channels} \\
  Both $W$ bosons decay leptonically. The final state is
  characterized by two $b$-quark jets, two charged leptons and
  large \met. These channels has an excellent signal-to-background ratios but
  suffer from small branching fractions.}

\end{itemize}
The top quark pair decay channels and their branching ratios are summarized 
in Fig.~\ref{fig:br_ratios}.
Note that the top quark pair analysis in D\O\ includes the
leptonically decaying $\tau$ in the \ljets~and dilepton
channels since this gives a similar experimental
signature. 

\begin{figure}[]
  \centering
  \includegraphics[width=10cm]{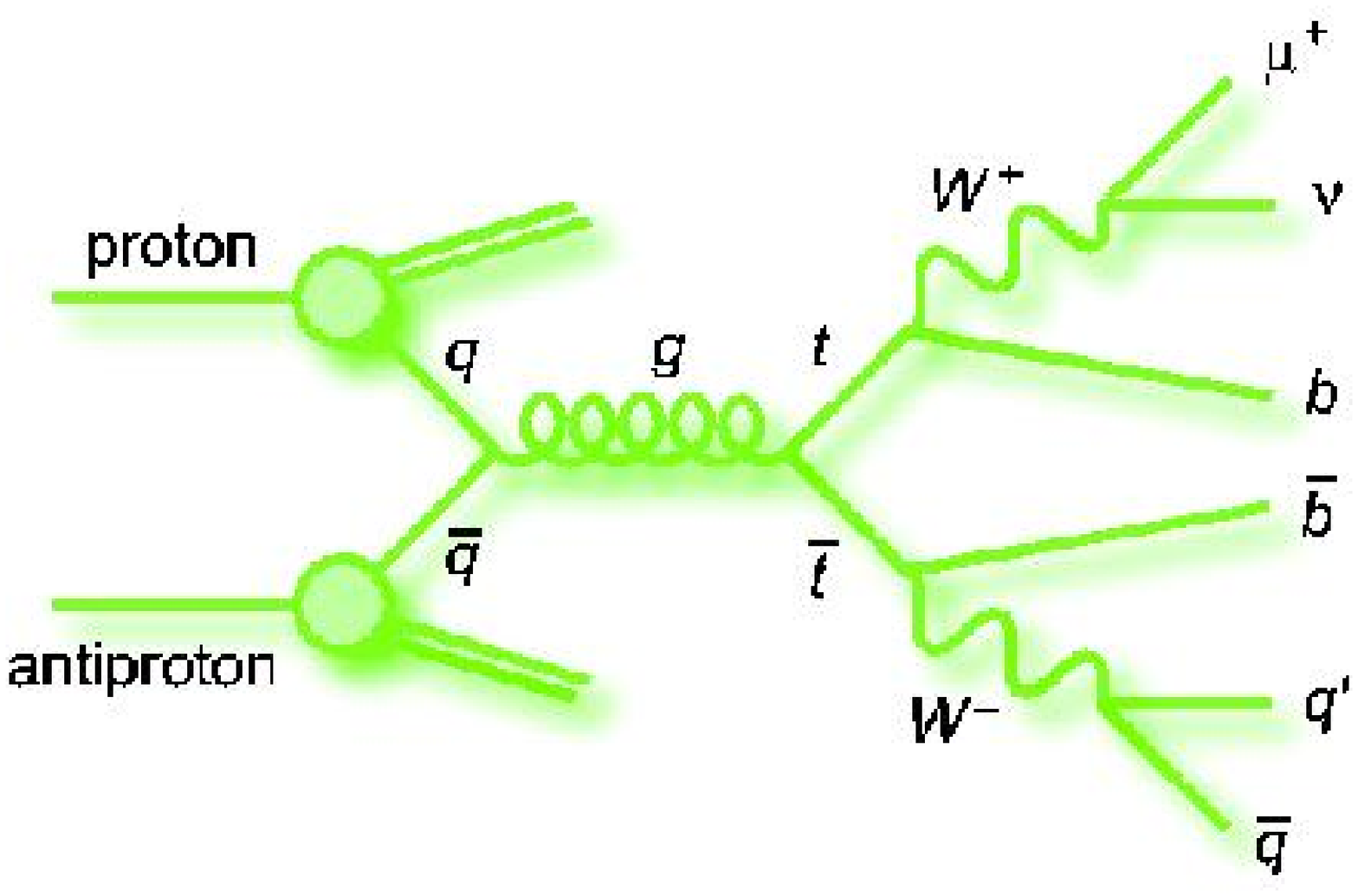}
  \caption{A sketch of a $\ttbar \to \mu \text{+jets}$ event~\cite{top_group}.}
  \label{fig:sketch_full_ttbar}
\end{figure}

Figure~\ref{fig:sketch_full_ttbar} shows a schematical view of a $\mu \text{+jets}$
event. Additional jets can be produced in all channels due to initial
(ISR) and final state radiation (FSR) discussed below.

In summary, the main feature of a $\ttbar \to \ljets$ event is the presence
of one charged isolated lepton produced centrally in the detector with high
transverse momentum, a neutrino with comparable momentum giving
rise to a significant \met~and several jets.

\subsection{Background Signature}
\label{subsec:bkgsignature}

At hadron colliders, QCD multijet production has a large cross section 
and is initially the largest background. This strong production of jets 
contain no genuinely isolated leptons nor missing
transverse energy. However, these can be faked by instrumental effects. A 
jet fluctuating to a high electromagnetic content can fake an electron. 
In addition, semi-leptonic decay modes of $b$- and $c$-quarks 
can give rise to fake isolated leptons if the associated jet is not
reconstructed, either due to a low energy deposition in the calorimeter
or inefficiency of the jet reconstruction. In combination, fake missing
transverse energy can arise due to unreconstructed jets, the neutrino
from the heavy flavor decay or a mismeasurement of the lepton momentum. 
However, even an unreconstructed jet deposits a small amount of energy 
in the calorimeter and gives signal in the tracking detectors. Thus, a 
good handle to suppress this background is to
require that the lepton is isolated from other activity in the detector.

\begin{figure}[]
  \centering
  \includegraphics[width=10cm]{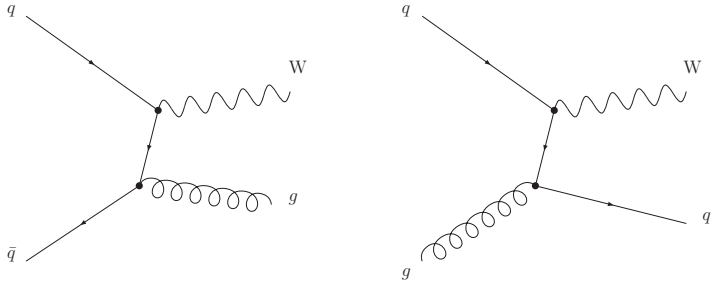}
  \caption{Examples of Feynman diagrams for the production of $W$ bosons with
  one additional parton in the final state, called $W\text{+1jet}$.}
  \label{fig:W_plus_1}
\end{figure}

The main source of $W$ boson production at the Tevatron is due to quark 
anti-quark fusion. 
In this process gluon radiation from the incoming quark lines (ISR) can
give the $W$ boson transverse momentum and add more partons to the final
state. Partons are a common used notation, remaining from the 70's, for 
the constituents of a hadron (quarks and gluons). 
Figure~\ref{fig:W_plus_1} shows examples of $W$ production with
an additional parton in the final state, called $W\text{+1jet}$\footnote{Here
the additional partons in the final state are assumed to hadronize into
jets.}.
Processes leading to even higher number of partons in the final state can
be produced by gluon radiation of the quark and gluon lines in the
initial or final state. Computation of $W$ boson production in 
association with up to four partons in the final state has been performed to leading
order~\cite{alpgen_manual} with different techniques to handle the immense
amount of Feynman diagrams contributing to the process. 

Due to the similar experimental signature as $\ttbar \to \ljets$ events, the 
production of $W$ (with subsequent leptonic decay) in association with four partons 
in the final state is the dominant
background in this analysis after standard preselection of $W$ boson candidates.
Even though the objects in the final state are the same as in a $\ttbar \to \ljets$ 
event, there are significant differences in several aspects of the event that can be
exploited to separate \ttbar~events from this background. The analyzes measuring
the \ttbar~production cross section utilize the fact that; ({\it i}) $W$ bosons 
produced from the decay of top quarks have on average larger transverse momentum 
and are produced at lower $|\eta|$ 
({\it ii}) the additional jets are 
mainly produced by gluon radiation, resulting in jet with lower transverse momenta 
and at higher $|\eta|$~\cite{xsec_240_ljets_topo_publ}. 
Another way to discriminate between a \ttbar~and a $W \text{+jets}$ event is to exploit 
the fact that a \ttbar~event has a higher fraction of $b$-quarks in the final state. 
The $b$-quarks in a \ttbar~event hadronize into $B$ hadrons and the event is 
expected to have at least two heavy 
flavor jets originating from $b$-quarks (see Sec.~\ref{sec:bjets}). 
As discussed in more detail in Sec.~\ref{sec:preselection}, the dominant 
background in the $\ttbar \to \ljets$ channel after the requirement of at least two jets 
identified as $b$-quark jets is the production of a $W$ boson in association 
with four jets out of which two are $b$-quark jets (denoted as $Wb\bar{b}jj$). 



\section{Tracks}

Charged particles traversing the inner detector deposit energy 
in the silicon layers of the SMT and produce scintillation light 
in the CFT\footnote{In reality, tracks do not always have hits in all 
layers of the SMT and CFT.}. 
The hits in the different inner detector layers together 
with the bending in the magnetic field allows for the reconstruction 
of the particle's trajectory. The track reconstruction 
algorithm groups together hits in different detectors into clusters 
which are then fitted to find a possible physical path of the 
particle~\cite{kalman_fitter}.

\section{Primary Vertex}
\label{sec:pv}
In proton anti-proton collisions 
several interactions between the constituent partons are possible. The primary 
interaction vertex (PV) are the point were the hard scattering (high transverse 
momentum) interaction took place. The spread 
of the interaction point in the ($x-y$) plane, transverse to the 
beam line, is small due to the 
transverse size of the Tevatron beam which is of the order of $30~\mu$m. 
In $z$-direction, the spread of the PV position 
extends up to $60$~cm following a Gaussian
distribution with a width of approximately $25$~cm. Finding the
primary vertex (PV) is crucial for all $b$-tagging algorithms 
and in order to determine if a lepton originates from the PV. 

The PV algorithm~\cite{pv_alg_p14} used by D\O\ starts by fitting all 
reconstructed tracks to a common vertex and removes bad track fits 
until a predefined value of the goodness of fit is reached. The 
same procedure is repeated for the tracks that were removed until 
all tracks are assigned to a PV. There are two similar implementations 
of the PV algorithm ,{\sl D\O\ reco} and {\sl D\O\ root}, 
with the difference that {\sl D\O\ root} has an additional step of 
clustering tracks in the $z$-direction and slightly tighter 
track selection criteria (the $d_{ca}$ significance is required to 
be $\leq 3.0$ compared to $\leq 5.0$). In both algorithms, only tracks 
with $p_T > 0.5$~GeV are considered and at least two hits in the 
SMT detector.
If more than one PV is found, the hard scatter vertex is selected 
by observing that hard scatter vertices have on average tracks with 
larger transverse momentum associated to it than minimum bias 
vertices~\cite{pv_prob_p14}.

The performance of the PV selection algorithms are comparable. There are 
on average 20 tracks in a generic QCD multijet event and the average PV 
reconstruction efficiency is 98\%. This efficiency is about 100\% in the
central $|z|$ region of the SMT fiducial region ($|z|<36~cm$ for the barrel) 
and drops quickly outside of this region due to the requirement of at least 
two SMT hits for tracks forming the PV.

\section{Muons}
\label{sec:muons}

Muons are identified in the drift chambers and scintillation counters 
by matching hits in the layers on either side of the toroid magnet. 
The D\O\ muon group has established a set of 
standard muon identification criteria applied to the candidate 
muon~\cite{p14muonid}:
\begin{itemize}
  \item 
        At least two A layer wire hits,
  \item 
        at least one A layer scintillator hit,
  \item 
        at least two BC layer wire hits,
  \item 
        at least one BC scintillator hit(except for central muons with
        less than four BC wire hits),
  \item
        to be inconsistent with a cosmic muon based on timing information 
        from the scintillator hits.
\end{itemize}

A muon identified in the above fashion is the basis for the muon reconstruction. 
The superior track resolution of the central tracker (SMT and
CFT) is used to improve the muon's momentum resolution. Therefore, in 
addition to the above criteria, a track consistent 
with originating from the PV is required to be spatially matched to 
the muon candidate. 

Muon tracks with no hits in the SMT (which have been shown to have a worse 
fit and thus a worse resolution) are re-fitted constraining the muon track 
to the PV in order to improve their momentum resolution. 

The muon momentum scale and resolution was determined by reconstructing the
$Z$ boson invariant mass peak in $Z \to \mu^+ \mu^-$ events. Comparison of the invariant 
mass peak in data and simulation reveals a significantly better resolution 
in the simulation than in data as well as a shifted peak position. This is accounted for by 
smearing the reconstructed muon momenta in simulated events to match the 
resolution in data~\cite{xsec_note240_mujets_topological}. 

\section{Electrons}
\label{sec:electrons}

The ability to identify and reconstruct high $p_T$ electrons is essential
for many analyzes, including top quark measurements, electroweak processes
and searches for New Physics. Being charged particles, electrons deposit
energy in the central tracking detectors before showering predominantly in
the EM section of the calorimeter. 
The main backgrounds to reconstructed true electrons (so-called ``fake'' 
electrons) are:
\begin{itemize}
  \item{$\pi^0$ showers overlapping with a track from a charged particle,}
  \item{photons which convert to $e^+e^-$ pairs,}
  \item{$\pi^{\pm}$ which undergo charge exchange in the detector material,}
  \item{fluctuations of hadronic showers.}
\end{itemize}
At D\O\ electron identification involves three steps. First, electron
candidates are searched for by looking for clusters in the EM calorimeter. 
Secondly, a track in the central tracking system that is spatially matched to the 
EM cluster is searched for and finally 
the electron has to pass a likelihood test based on shower shape variables. 
To handle all the sources of backgrounds while keeping the efficiency to reconstruct
real electrons high, several variables are used:
\begin{itemize}
  \item 
        The fraction of energy deposited in the electromagnetic part of the 
        calorimeter is required to be above 90\% of the total deposited energy 
		in the calorimeter inside the cone of $\Delta R< 0.2$.
		
  \item 
        Electrons tend to be isolated from other activity in the calorimeter. 
        Therefore, 
        at most 15\% of the energy of the cluster is allowed to be deposited 
        in a hollow cone ($0.2 < \Delta R < 0.4$) around the electron's 
        direction.
		
  \item 
        The shower shape of candidate EM clusters is compared to the
        expected shape from electrons~\cite{emlhood_p14_org}. 
		
  \item 
        A track is required to point to the EM cluster.

  \item 
        A seven parameter likelihood is built that rejects background-like EM 
		candidates~\cite{emlhood_p14}.

%
  \item 
        The electron candidate is required to be in the central calorimeter, since 
		the fake electron background is not completely understood in the end 
        cap calorimeters.
		
\end{itemize}

The electron momentum scale and resolution was studied by reconstructing
the $Z$ boson invariant mass peak in $Z \to e^+e^-$ events. The comparison of 
data with simulation further revealed a higher resolution in the simulation and a
corresponding scale factor and smearing is applied to simulated electrons to reproduce the
measured quantities. Detailed information on the selection criteria, 
electron momentum scale and resolution can be found 
in Ref.~\cite{xsec_note240_ejets_topological}.

\section{Jets}
\label{sec:jets}

In the analysis presented in this thesis, jets form 
an essential ingredient in the event selection. Each event is required to 
have at least four jets out of which two are identified as $b$-quark jets. 
This section describes the identification 
and energy calibration of jets and explains the identification of jets 
originating from the hadronization of $b$-quarks. 

\subsection{Jet Identification}
\label{sec:jetid}

Jets are reconstructed based on finding calorimeter towers with an 
energy above a predefined threshold of $E_T>0.5$~GeV which are further 
collected into clusters which form candidate jets~\cite{cone_algorithm}. The jet 
cone size is $\Delta R < 0.5$ and the uncalibrated transverse energy reconstruction 
threshold is $8$~GeV. Before any calorimeter physics object (jets, 
electrons, photon or \met) are reconstructed a special calorimeter algorithm 
is applied to remove measured cell energies likely to arise from noise~\cite{T42_1,T42_2,T42_3}. 

After jets have been reconstructed it is important to reject those 
that are poorly reconstructed or are electrons or photons mis-identified
as jets. Therefore, several additional requirements based on the expected 
properties of jets are applied: The probability for a jet to deposit a large 
fraction of its energy in the coarse hadronic section of the calorimeter is 
low (and the coarse hadronic section is subject to a higher noise level	
due to its larger cell size) and the energy deposition in this section 
is required to be below 40\%.
Electrons and photons typically deposits all of their energy in the EM 
section and therefore the fraction of energy deposited in the EM part for 
jets is required to be between 5\% and 95\%. Jets 
reconstructed from few or single cells 
containing a large fraction of the jets total energy are likely to be 
fake jets due to noise in the cell. Therefore, jets are rejected if one 
cell contains above 90\% 
of the total energy or the ratio of the cell with the highest energy 
to the next-to-highest is above $10$. Since the Level 1 calorimeter 
trigger has an independent readout chain it has been proven to be a 
powerful handle to reject fake jets due to noise in the precision readout 
electronics. Therefore, the sum of energy in the Level 1 calorimeter 
trigger towers is required to be above 24-40\% (12-20\%) in the central 
 and end cap calorimeter  (inter-cryostat) regions depending on the fraction 
of total energy deposited in the coarse hadronic section.
If a jet overlaps with EM candidates (which are also reconstructed as 
jets if their energy is above $8$~GeV) an ambiguity 
appears about which energy scale that should be applied. The solution 
is to reject jets that overlap with electrons or photons within $\Delta R < 0.5$.
The minimum transverse momentum after scale and corrections described below 
is required to be $15$~GeV.

\subsection{Jet Energy Scale}
\label{subsec:jes}
\begin{figure}[]
  \centering
  \includegraphics[width=0.9\textwidth]{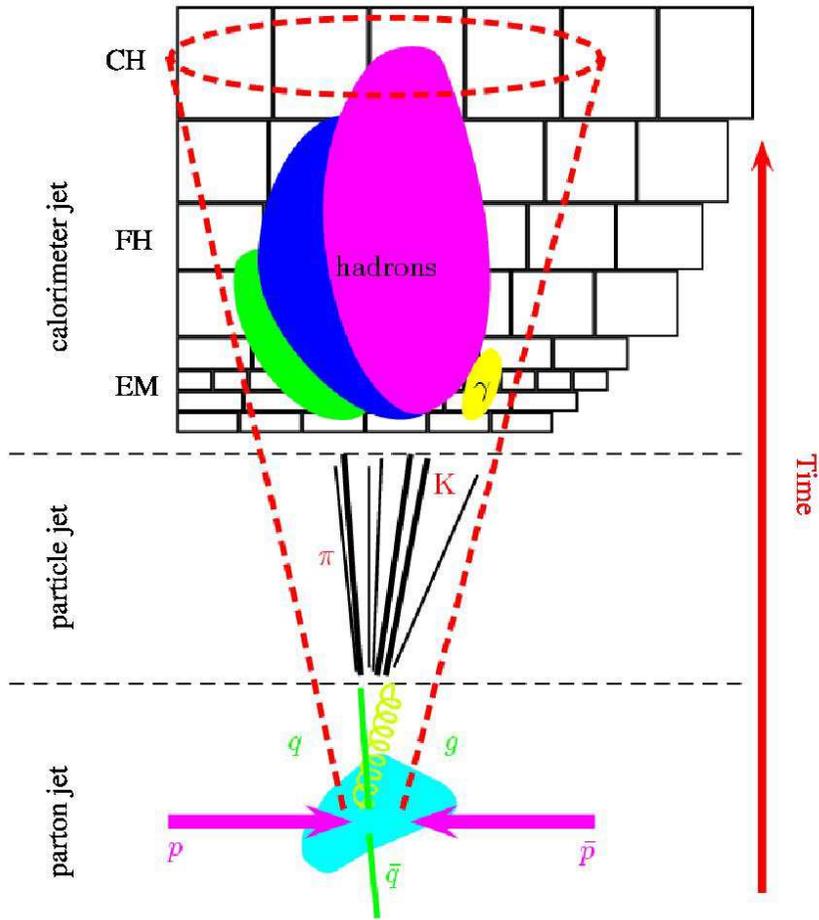}
  \caption{Schematic view of the process taking a parton from the hard 
  scattering to energy deposited in the calorimeter~\cite{jes_group}.}
  \label{fig:jes_view}
\end{figure}

Quarks and gluons from the hard interaction hadronize into jets and deposits 
energy in the calorimeter as shown schematically in Fig.~\ref{fig:jes_view}. 
The goal of the Jet Energy Scale (JES) is to correct the measured jet energy 
in the calorimeter back to the stable-particle 
energy before interacting with the calorimeter. Various effects cause the 
measured energy to be different from the particle jet e.g. the use of a 
sampling calorimeter, noise and dead-material. 

The JES is derived using $\gamma \text{+jet}$ events in a back-to-back 
configuration~\cite{jes_nim}. Using the fact that the electromagnetic 
energy scale is known to high precision\footnote{The EM energy scale can be calibrated by 
calculating the invariant mass of electrons or photons in inclusive samples 
such as $Z \rightarrow e^+e^-$, $J/\Psi \rightarrow e^+e^-$ and 
$\pi^0 \rightarrow \gamma\gamma$.}
the JES can be extracted from the transverse momentum imbalance in such an event. 

The JES is divided into different subcorrections applied in order of 
appearance to jets: 
The {\bf offset correction} corrects for energy not part of the hard 
scatter (detector and electronic noise, pile-up and energy from the underlying 
event, see Sec.~\ref{subsec:topquarkmcsim} for a description of pile-up and 
underlying event). This luminosity dependent offset correction is calculated 
in data using events triggered by the luminosity monitors, signaling a possible 
inelastic \ppbar~collision (minimum-bias events). 
The {\bf response correction} is a correction for the non-uniform response of 
parts of the detector (dominated by the ICD region) and an absolute response 
for a uniform treatment of jet energies derived from well defined 
$\gamma\text{+jet}$ events in a back-to-back configuration. The 
last correction is the {\bf showering correction} which attempts to correct 
for particles inside the jet that deposits their energy outside the jet cone 
(or the reverse process). This correction is extracted from jet profiles in 
$\gamma \text{+jet}$ events. 

\begin{figure}[]
  \centering
  \includegraphics[width=0.65\textwidth,angle=270]{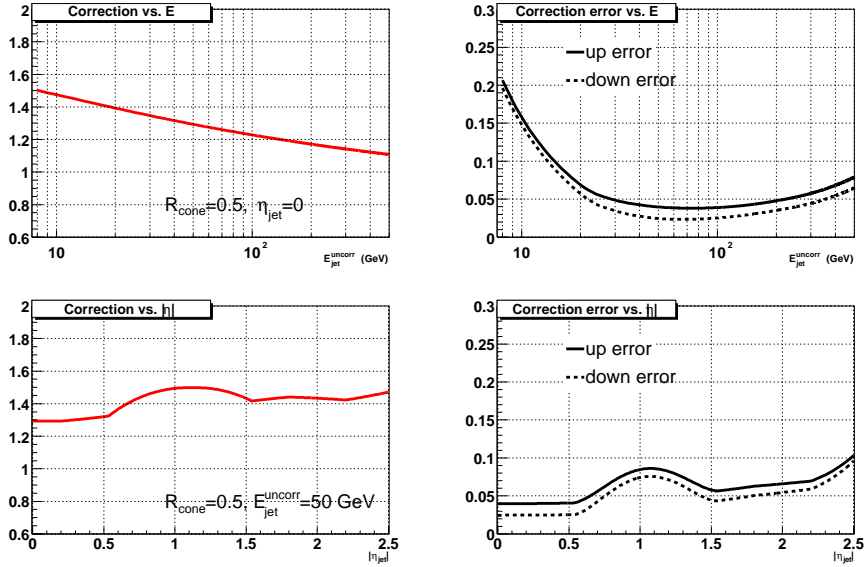}
  \caption{The JES corrections for data (left) and the error (right) 
  as a function of the measured jet energy (top) and $\eta$ (bottom). 
  The ``up'' and ``down'' error are the combined undertainties from all the 
  sub-corrections~\cite{jes_group}.}
  \label{fig:jes_data}
\end{figure}

\begin{figure}[]
  \centering
  \includegraphics[width=0.65\textwidth,angle=270]{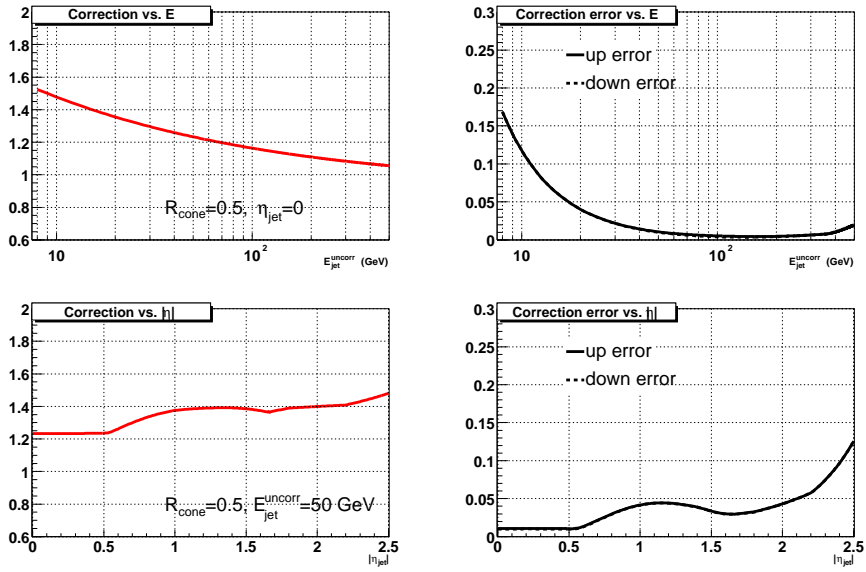}
  \caption{The JES corrections for simulated jets (left) and the error (right) 
  as a function of the measured jet energy (top) and $\eta$ (bottom). 
  The ``up'' and ``down'' error are the combined undertainties from all 
  the sub-corrections~\cite{jes_group}.}
  \label{fig:jes_mc}
\end{figure}

The final result is a JES correction factor shown in 
Fig.~\ref{fig:jes_data} and~\ref{fig:jes_mc} for data and simulation 
respectively.

\subsection{Jet Energy Resolution}
\label{subsec:jer}
The Jet Energy Resolution (JER) can be determined by studying the $p_T$
imbalance in dijet events~\cite{jer_jetcorr5.3}. The asymmetry variable
$A=\left ( p_T^{{\rm jet}_1} - p_T^{{\rm jet}_2} \right ) / \left( p_T^{{\rm jet}_1} + p_T^{{\rm jet}_2} \right)$ 
is calculated and the jet resolution can be inferred from the width of 
the asymmetry $\sigma_A$ as,
\begin{equation}
\frac{\sigma_{p_T}^{\rm jet}}{p_T^{\rm jet}} = \sqrt{2}\sigma_A.
\end{equation}
At lower energies ($<50$~GeV) $\gamma + jet$ events are used.
\footnote{This because the triggers used to select dijet 
events are inefficient for jet energies below $50$~GeV.}. The asymmetry 
variable $A_{pj}$ is calculated as 
$A_{pj} = \left ( p_T^{\gamma} - p_T^{\rm jet} \right ) / p_T^{\gamma}$ and the
JER can be expressed as,
\begin{equation}
\frac{\sigma_{p_{T}}^{\rm jet}}{p_{T}^{\rm jet}} = \sigma_{A_{pj}} \times \frac{p_T^{\gamma}}{p_T^{\rm jet}}.
\end{equation}
The JER measured using the procedure above is then parametrized 
as (see Sec.~\ref{sec:calorimeter}),
\begin{equation}
\frac{\sigma_{p_T}^{\rm jet}}{p_T^{\rm jet}} =
\sqrt{ \frac{N^2}{p_T^2} + \frac{S^2}{p_T} + C^2}, 
\end{equation}
and the results for data and simulation are shown in Tab.~\ref{tab:jer_data} 
and Tab.~\ref{tab:jer_mc} respectively. 
\begin{table}
\centering
\begin{tabular}{c|ccc}
\hline
\hline
$|\eta_{\rm det}|$ range & $N$ & $S$ & $C$ \\
\hline
$0.0 <|\eta_{\rm det}| < 0.5$ & $5.05$ & $0.753$  & $0.0893$ \\
$0.5 <|\eta_{\rm det}| < 1.0$ & $0.0$ & $1.20$ & $0.0870$ \\
$1.0 <|\eta_{\rm det}| < 1.5$ & $2.24$ & $0.924$ & $0.135$ \\
$1.5 <|\eta_{\rm det}| < 2.0$ & $6.42$ & $0.0$ & $0.0974$ \\
\hline
\hline
\end{tabular}
\caption{The JER parameters for data.}
\label{tab:jer_data}
\end{table}
\begin{table}
\centering
\begin{tabular}{c|ccc}
\hline
\hline
$|\eta_{\rm det}|$ range & $N$ & $S$ & $C$ \\
\hline
$0.0 <|\eta_{\rm det}| < 0.5$ & $4.26$ & $0.658$  & $0.0436$ \\
$0.5 <|\eta_{\rm det}| < 1.0$ & $4.61$ & $0.621$ & $0.0578$ \\
$1.0 <|\eta_{\rm det}| < 1.5$ & $3.08$ & $0.816$ & $0.0729$ \\
$1.5 <|\eta_{\rm det}| < 2.0$ & $4.83$ & $0.0$ & $0.0735$ \\
\hline
\hline
\end{tabular}
\caption{The JER parameters for simulated jets.}
\label{tab:jer_mc}
\end{table}
More details can be found in~\cite{jer_jetcorr5.3}. 
The resolution is clearly better in the simulation and simulated 
jets are therefore smeared to match the resolution in data.

\subsection{$b$-Quark Jets}
\label{sec:bjets}
\begin{figure}[]
  \centering
  \includegraphics[width=0.7\textwidth]{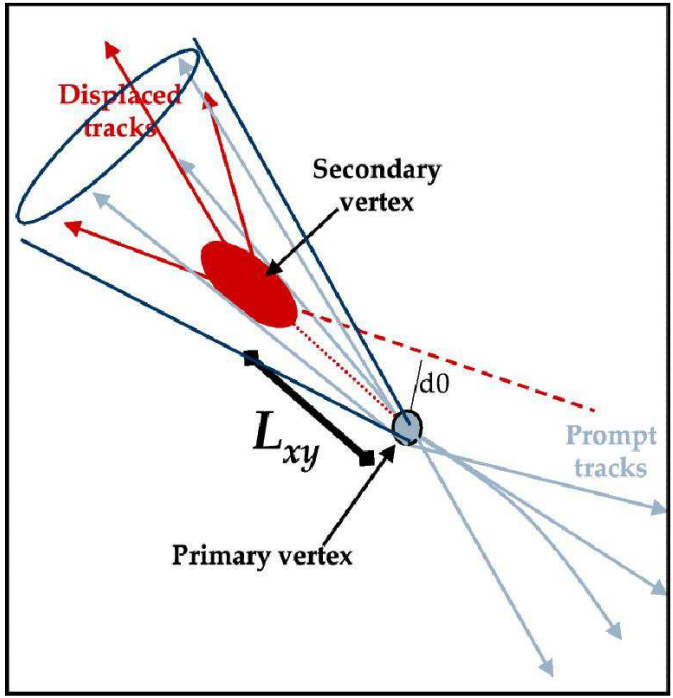}
  \caption{A quantitative description of a jet with a secondary vertex arising 
  from the long lifetime of a $B$ hadron. Several displaced tracks are fitted 
  to the secondary vertex which can be used to identify $b$-quark jets. $d_0$ 
  is the distance of closest approach of a track with respect to the primary 
  vertex.}
  \label{fig:secvtx}
\end{figure}

Jets can further be classified by their flavor. A jet originating from 
the hadronization of a gluon ($g$), $u$-, $d$- or $s$-quark or from 
the a $c$- or $b$-quark is referred to as a light jet 
or a heavy flavor jet respectively. There are two techniques 
to distinguishing between light and heavy flavor jets: {\bf Soft Lepton Tagging}
uses the presence of a lepton within a jet as a signature for a semi-leptonic 
decay of a heavy flavor hadron. The branching fraction for a semi-leptonic 
($\mu^{\pm}$ or $e^{\pm}$) $B$ hadron decay is $\approx 10$\% for each lepton 
mode~\cite{PDG}. {\bf Lifetime Tagging} uses the the fact that hadrons 
containing a $b$-quark have a lifetime of approximately $1.6$~ps. A 
$B$ hadron originating from the primary vertex will therefore travel a 
significant distance from the primary vertex before decaying (the average flight length is 
around $3$~mm for $40$~GeV hadron).
Approximately 70\% of the $B$ mesons with a decay length~\cite{dl} greater than 
$1$~mm have more than $2$ displaced tracks from its decay products with an impact 
parameter significance above three~\cite{bid_group}. The lifetime 
tagging algorithm therefore searches for tracks significantly displaced from 
the primary vertex as shown schematically in Fig.~\ref{fig:secvtx}.
In this analysis both ways to identify heavy flavor jets are used. 
A jet tagged by the lifetime tagging algorithm is called a ``$b$-tagged jet'' 
or simply a ``tagged jet''. A jet tagged by the soft lepton tagging algorithm 
is called a ``$\mu$-tagged jet''. Due the difficulty of identifying an 
electron within a jet, the soft lepton tagging algorithm uses only the muonic 
semi-leptonic decay mode. Note that a $b$-tagged jet is 
not necessarily a jet originating from the hadronization of a $b$-quark 
or even a $c$-quark. The tagging technique may wrongly tag light jets as discussed 
below in more detail. In this thesis a jet from a $x$-quark ($x=b,c,s,u,d$) 
is called $x$-quark jet\footnote{Here the notation for a jet originating from e.g. a $b$ or 
$\bar{b}$-quark is called $b$-quark jet collectively. In later sections 
the difference between a $b$ or $\bar{b}$-quark jet is exploited and the 
notation will be obvious.} or gluon jets for gluon initiated jets.

To separate the performance of the different algorithms from detector 
effects, such as calorimeter noise and tracking inefficiencies, the 
probability for a jet to be tagged using lifetime tagging is broken down 
into two components: {\it (i)} The probability for a jet to be taggable (also called 
{\sl taggability}) and {\it (ii)} the probability for a taggable jet to be tagged 
(also called {\sl tagging efficiency}). The D\O\ $b$-identification 
group~\cite{bid_group} defines a taggable calorimeter jet as a jet 
reconstructed in the calorimeter 
matched to a track-based jet (track-jet) within $\Delta R<0.5$. A track-jet is 
defined by the following track requirements: $p_T > 0.5$~GeV (where 
at least one track has $p_T> 1$~GeV), at least one hit in the central part 
of the SMT detector and matched to the primary vertex 
within $0.5$ in $r-\phi$ space and within $2$~cm in the $z$-direction. 

\begin{figure}[]
  \centering
  \includegraphics[width=1.0\textwidth]{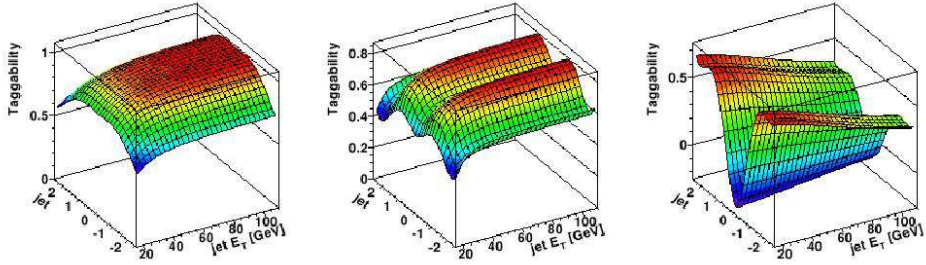}
  \caption{Taggability parameterizations as a function of jet $E_T$ and
  $\eta$ for the combined preselected data sample in three different
  regions of $z_{PV}$: $|z_{PV}|<30$(left), $30<|z_{PV}|<45$(middle) and 
  $45<|z_{PV}|<60$(right).}
  \label{fig:tagga}
\end{figure}

The taggability is determined directly from data and parameterized as a
function of the jet $p_T$ and $\eta$. Due to the requirement of hits in the
central region of the SMT the taggability is expected to have a large
dependence on the $z$-position of the primary vertex. The taggability is shown in
Fig.~\ref{fig:tagga} for the preselected data sample (see 
Sec.~\ref{sec:preselection} for the sample definition).

There are several ways of tagging a jet using lifetime tagging~\cite{bid_group}. 
In this analysis the Secondary Vertex Tagger (SVT) algorithm is used. 
The SVT algorithm 
finds tracks with large impact parameters within the matched track-jets and 
reconstructs secondary vertices that are within $\Delta R < 0.5$ of the 
calorimeter jet direction. 
A calorimeter jet is tagged by the SVT algorithm (SVT-tagged) if it has 
at least one secondary vertex with decay length significance~\cite{dl} 
greater than $7.0$. More information on the
exact definition of the SVT tagging algorithm can be found 
in Ref.~\cite{SVT} and on the selections used in this analysis 
in Ref.~\cite{xsec_240_ljets_btag_publ}. The simulation is unable to describe 
the details of the tracking and is thus overestimating the tagging efficiency 
for simulated jets. Therefore the tagging efficiency is measured on data 
with as little input as possible from simulation. 

\begin{figure}[]
  \centering
  \includegraphics[width=1.0\textwidth]{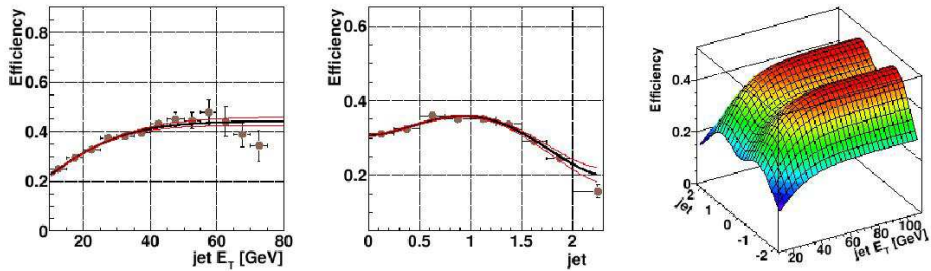}
  \caption{The semi-leptonic tagging efficiency in data as a function
  of jet $p_T$ (left), $\eta$ (middle) and the combined two-dimensional
  parametrization (right)~\cite{xsec_note360_ljets_btag}.}
  \label{fig:btag_eff}
\end{figure}

To determine the SVT tagging efficiency for $b$-quark jets in data, a sample 
enriched in $b$-quark jets is selected by requiring at least one $\mu$-tagged 
jet in the event. The measured SVT tagging efficiency of jets in this sample 
is called ``semi-leptonic''  and is parametrized as 
a function of jet $p_T$ and $\eta$ as shown in Fig.~\ref{fig:btag_eff}. 
In order to find the tagging efficiency for inclusive $b$-quark jets in data 
a scale factor is derived. This scale factor is the ratio of the measured 
semi-leptonic tagging efficiency in data and the semi-leptonic tagging 
efficiency in simulated events. The tagging efficiency for inclusive 
jets in the simulation is then multiplied with the scale factor to find 
the tagging efficiency for inclusive $b$-quark jets in data. Assuming that 
the scale factor for $b$- and
$c$-quark jets are the same, the $c$-quark jet tagging efficiency in data 
can be calculated in a similar way and varies between 7-12\%. Due to the 
limited track resolution and/or mis-reconstructed tracks, light jets can 
be wrongly tagged. 
The probability for a light jet to be tagged (``mistag rate'') is derived 
on a QCD data sample (which dominantly consists of light jets) and 
parametrized again as function of jet $p_T$ and $\eta$. The mistag rate is 
of the order of 1\%.

\section{Missing Transverse Energy}
In the final state of a $\ttbar \to \ljets$ event the neutrino can only be detected by 
the measurement of the imbalance of momentum in the transverse plane. 
This imbalance is measured by the vector sum of all energy depositions 
in calorimeter cells. 
Cells in the coarse hadronic section of the calorimeter are excluded if they are not part of 
a reconstructed jet. This has been shown to improve the missing transverse 
energy resolution due to the higher noise level in those cells. 
The vector opposite to this total visible momentum is the raw missing transverse 
energy ($\met_{raw}$).

The calorimeter response to electromagnetic particles such as electrons and photons
is different than for hadronic particles, and in particular jets. This
difference propagates directly to the $\met_{\rm raw}$ if the energy depositions 
are not calibrated correctly. All jets are corrected 
for the jet energy scale and the correction is also propagated to the
$\met_{\rm raw}$. The same procedure is used for the EM calibration. 
After all corrections, the resulting missing transverse energy is called 
the ``calorimeter missing transverse energy''.

Muons are minimal ionizing particles and traverse the whole detector 
if the muon momentum exceeds a few GeV and their measured transverse momentum 
must be added to the missing transverse energy. The muon deposits 
only a small fraction of its total energy in the calorimeter which 
is estimated and subtracted from the missing transverse energy vector. 
This fully corrected missing transverse energy is denoted simply as \met.

\section{Monte Carlo Simulation}
\label{subsec:topquarkmcsim}

The detailed study of proton anti-proton collisions requires a detailed 
understanding of all aspects of the event. Monte Carlo simulation of such 
collisions are necessary to provide this knowledge.These simulations includes 
the hard scattering interaction, hadronization, 
detector response and digitization, allowing for a detailed comparison between 
simulated events and data.

{\bf The hard scatter interaction} is modeled using 
{\sc alpgen 1.3.3}~\cite{alpgen_manual} that calculates the leading order 
matrix element. The set of parton distribution functions used is 
{\sc CTEQ5L}~\cite{pdf_fit_cteq}.

The complexity of \ppbar~collisions is due to the fact that the colliding (anti-)protons 
are composite states of many partons. As a consequence, an interesting hard 
(high transverse momentum) interaction is 
accompanied by what is called the {\bf underlying event}. 
It consists of the {\it beam-beam remnants}, 
which are what is left over after a parton has been knocked out of the initial incoming 
(anti-)proton, and {\it multiple parton interactions} where in addition to the hard 
scattering interaction there are one or more semi-hard interaction in the same event.
A study of the transverse momentum distribution of charged particles has led to the so-called 
``Tune A''~\cite{tuneA} tuning of the Pythia underlying event model parameters 
to better describe data.
\begin{figure}[]
\begin{center}
\epsfig{file=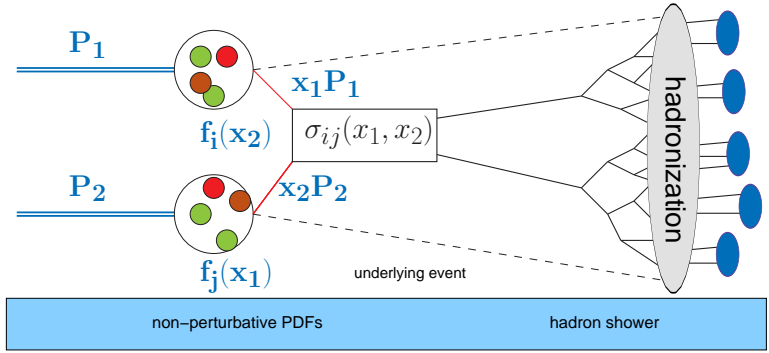,width=0.8\textwidth}
\caption{A schematic view of a \ppbar~ collision.}
\label{fig:ppbar_full}
\end{center}
\end{figure}

In each bunch crossing of protons and anti-protons more than one hard scatter 
\ppbar~collision may occur. These {\bf multiple interactions} are modeled by superimposing 
simulated events with minimum bias data. This data is collected by selecting 
real collider events with minimum activity in the detector i.e. not triggered by a high 
transverse momentum lepton, jet or missing transverse momentum.
In addition \ppbar~collisions from consecutive bunch-crossings can sometimes 
be reconstructed in the same event. This process, called {\bf Pile-up}, is 
taken into account in the simulation. 
At long distances, QCD becomes non-perturbative and in this domain the colored 
partons are transformed 
into colorless hadrons, a process called {\bf fragmentation}\footnote{The fragmentation 
and decay of particles are often collectively called hadronization.}. There exist several 
phenomenological models describing this process, {\sc PYTHIA} uses the Lund string 
model~\cite{lund_model} for fragmentation. Heavy quark fragmentation is 
an important aspect in this analysis and two models ({\sc Bowler}~\cite{bowler} and 
{\sc Peterson}~\cite{peterson} fragmentation) with alternative heavy quark fragmentation 
schemes are used as a cross-check in the analysis presented in this thesis. The decay 
of $B$ mesons are handled 
by {\sc EVTGEN}~\cite{ref_evtgen} and other decays are handled by {\sc PYTHIA}. The 
important simulation of the {\bf detector response} to charged and neutral particles 
is simulated using {\em D\O\ gstar}~\cite{d0gstar}, a {\sc GEANT3}~\cite{ref_geant} 
model describing the material and geometry of the D\O\ detector. The simulated signal 
produced by the detector is digitized using the software 
package {\em D\O\ sim}. From this point onwards events from the 
simulattion and data can be treated in a uniform way. The D\O\ event reconstruction 
software packages ({\em D\O\ reco}) transform the detector 
signals into reconstructed physics objects such as electromagnetic clusters, muon 
candidates, tracks, etc. Finally, the {\em TopAnalyze}~\cite{topanalyze} program 
processes the reconstructed events further with algorithms and object 
identification selections specific for the D\O\ top quark working group and produces 
{\sc ROOT}~\cite{root} files. Throughout the analysis presented in this thesis, 
{\sc ROOT} is used as analysis tool.

\subsection{Simulated Samples}
\label{subsubsec:tt_sig_MC}
\begin{table}
  \begin{tabular}{l|cccc}
    \hline
    Process                    & Generator     &  PDF   & Comment \\
    \hline
    $\ttbar \to \ell$+ jets     & Alpgen+Pythia & CTEQ5L &     \\
    $\ttbar j \to \ell$+ jets    & Alpgen+Pythia & CTEQ5L &     \\
    $\ttbar \to \ell \ell$     & Alpgen+Pythia & CTEQ5L &     \\
    $\ttbar \to \ell$+ jets     & Alpgen+Pythia & CTEQ5L & Bowler fragmentation~\cite{bowler} \\
    $\ttbar \to \ell$+ jets     & Alpgen+Pythia & CTEQ5L & Peterson fragmentation~\cite{peterson} \\
    \hline
    $Wb\bar{b}jj$              & Alpgen+Pythia & CTEQ5L & \\
    \hline
    $Z \to b\bar{b}$           & Pythia        & CTEQ5L & \\
    $Z \to b\bar{b} \to \mu X$ & Pythia        & CTEQ5L & \\
    $Z \to c\bar{c}$           & Pythia        & CTEQ5L & \\
    \hline
    $c\bar{c}$                 & Pythia        & CTEQ5L & \\
    \hline
  \end{tabular}
  \caption{The Monte Carlo samples used in the analysis.}
  \label{tab:mcsamples}
\end{table}
The Monte Carlo samples used in the analysis are listed in 
Tab.~\ref{tab:mcsamples}. The samples are generated with parameters given in 
Ref.~\cite{topphysics_winter2004_note} using {\sc ALPGEN} as primary generator. 
Due to the uncertainty in the modeling of initial state radiation in \ttbar~events, 
a $\ttbar j \to \ljets$ sample with a jet in the initial state is produced 
to estimate this effect.


\chapter{Determination of the Electric Charge of the Top Quark}
\label{ch:topcharge}


\section{Overview of the Method}
\label{sec:overview}

\begin{figure}[]
  \centering
  \includegraphics[width=0.8\textwidth]{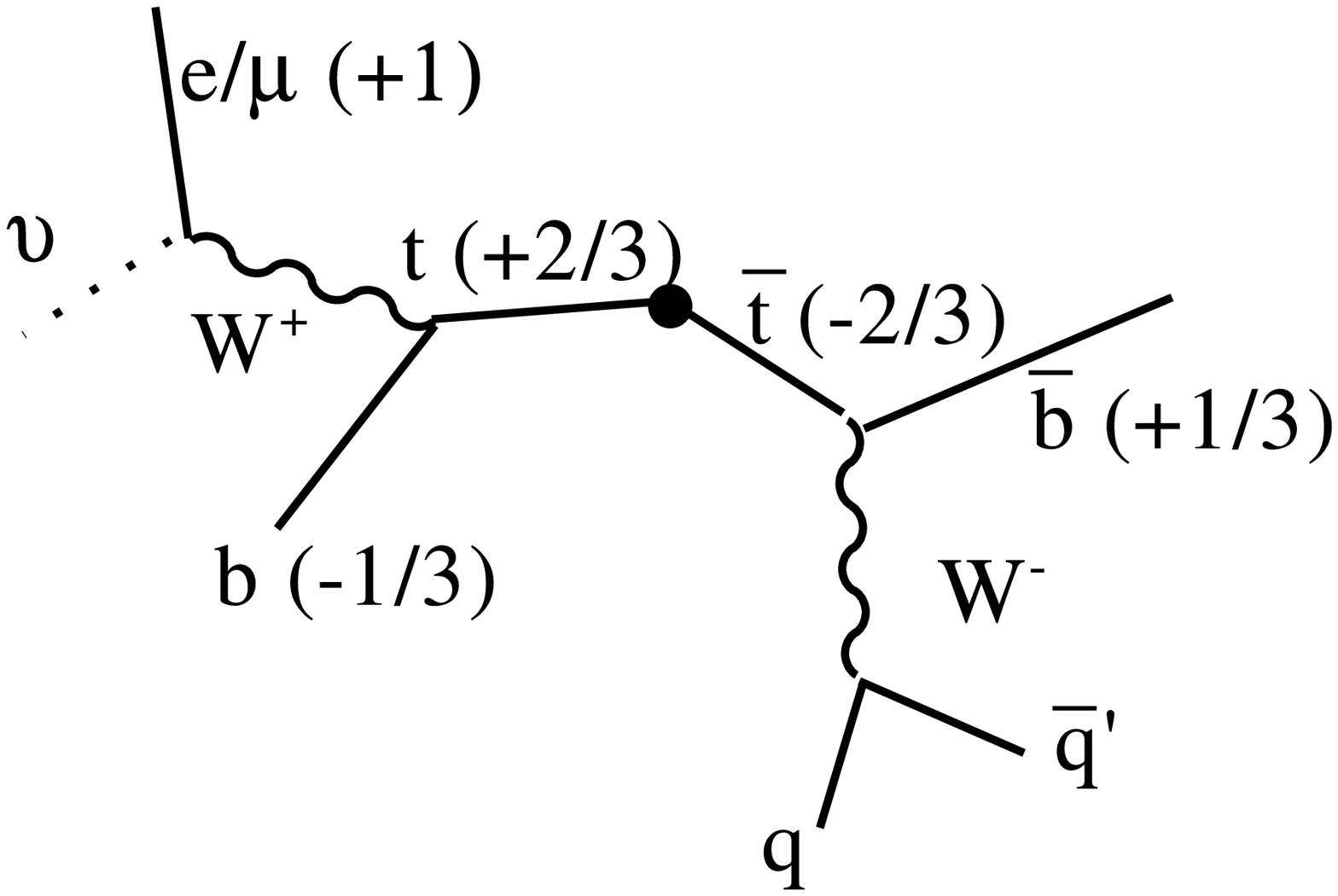}
  \includegraphics[width=0.8\textwidth]{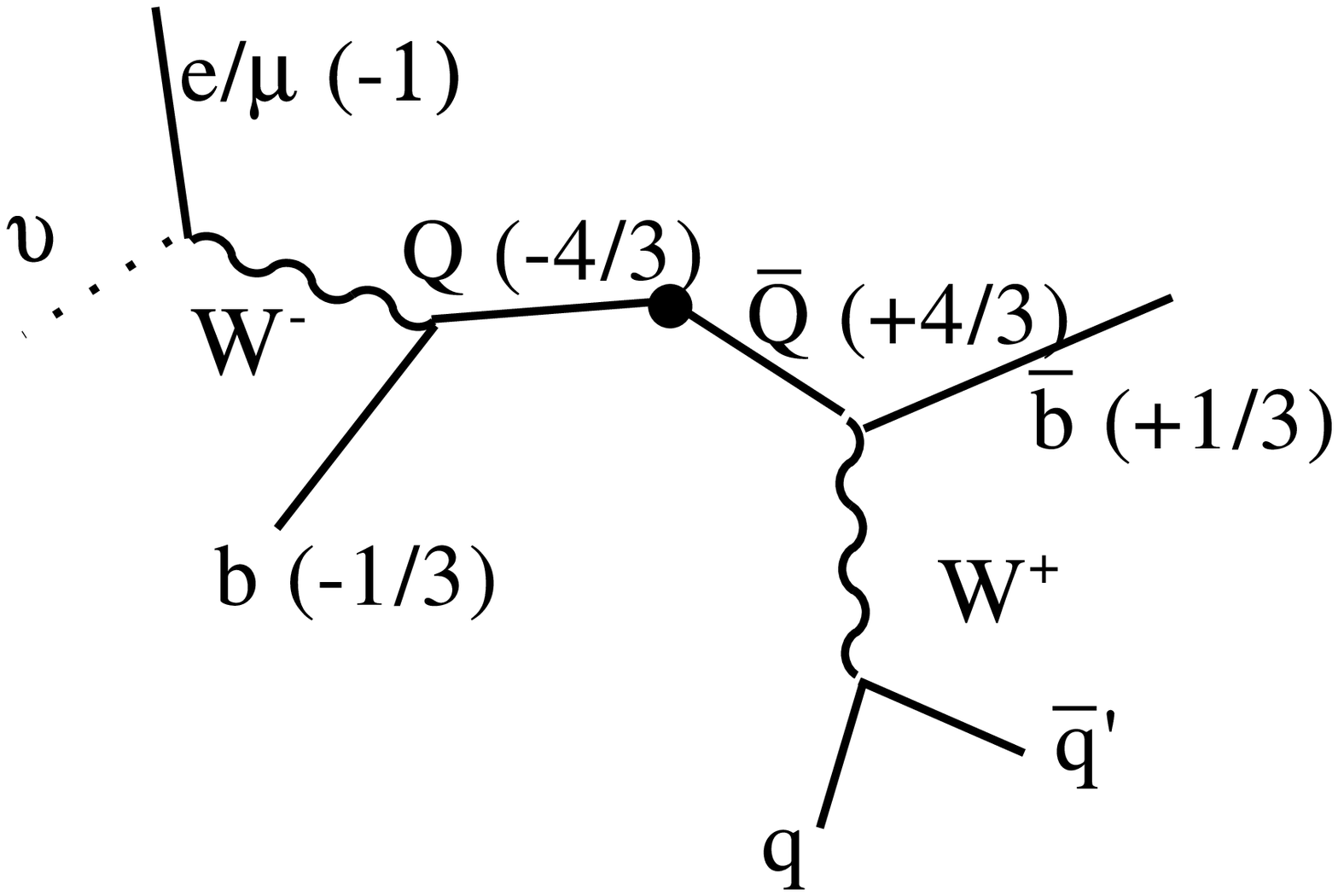}
  \caption{Illustration of the decay of the two different hypotheses, the SM 
  top quark pair (top) and an exotic quark pair decay (below). Process where the lepton 
  appears on the $\bar{t}$ or $\bar{Q}$ are also possible.}
  \label{fig:diagram1}
\end{figure}

The determination of the electric charge of the top quark proceeds in three steps. First 
a pure sample of $\ttbar$ events in the \ljets~channel is selected in data. This 
final state provides a sample with a signal-to-background ratio of $\sim 1$. The 
jet charge algorithm is sensitive to poorly simulated tracking, and is 
therefore only applied to jets tagged by the SVT algorithm. Only events 
with four or more reconstructed jets are considered, two of which are required to be 
SVT-tagged further increasing the purity of the sample (signal-to-background ratio 
$\sim 11$). 
Each of the selected $\ttbar \to \ljets $ events have two ``legs'', one with a 
leptonically decaying $W$ boson ($t \to W b \to \ell \nu b$) and one with a hadronically 
decaying $W$ boson ($t \to W b \to q q^{'} b$), see Fig.~\ref{fig:diagram1}. 

The second step of the analysis consists of assigning the correct jets and leptons 
to the two ``legs'' of the event. To make this assignment the same constrained kinematic 
fit package as for measurements of the top quark mass is used~\cite{hitfit}. The goal 
of the analysis presented in this thesis is to discriminate between two hypotheses: 
The SM top quark charge of $+2e/3$ or an exotic quark with charge $-4e/3$. 
In fact, the analysis is only sensitive to the modulus of the quark's charge ($|2e/3|$ 
or $|4e/3|$). This limitation does not lead to any loss of information since it is 
always assumed that charge is conserved and every event contains one quark and one 
anti-quark. It also allows 
for an equivalent treatment of the quark and anti-quark.
In each \ttbar~event, the absolute value of the charge of the quark and anti-quark 
are computed, which are assumed to be the same. Thus, each \ttbar~candidate event 
has two observables $Q$ which are built in the 
following way: The first combines the charged of the lepton from the $W$ decay and 
the charge of the $b$-quark jet associated to the leptonic side of the event by the 
kinematic fit. The charge of the $b$-quark jet is computed using a jet charge algorithm 
and simply added to the lepton charge and the absolute value of this number is 
taken. The second observable is the charge of the second ``leg'' of the event. 
It is obtained by taking the charge of the second $b$-quark jet and subtracting the 
charge of the lepton. This procedure is almost equivalent to the doubling of the dataset 
size in terms of statistical sensitivity. Any mistake in the assignment of the lepton and the 
$b$-quark jet due to the kinematic fit will automatically propagate to both legs 
of the event, therefore the two charge measurements are not uncorrelated. Nevertheless, 
the measurement of the $b$-quark jet charge is applied twice since it improves 
the sensitivity of the measurement. 

The jet charge distributions for $b$- and $\bar{b}$-quark jets are extracted from data and used 
to derive the expected distribution for the SM and the exotic 
scenarios. The distribution of $Q$ observed in the selected \ttbar~data sample 
is then compared to the SM and exotic expectations for $Q$. 

In the next section the \ttbar~event selection and the sample composition are presented. 
Section~\ref{sec:jcttbar} is devoted to the description of the jet charge algorithm. 
A method is developed to derive the performance of the jet charge algorithm for 
SVT-tagged jets using data. This is presented in Sec.~\ref{sec:data_calibration}. 
Section~\ref{sec:topcharge} gives a detailed description of the method to discriminate between 
the $2e/3$ and $4e/3$ scenarios by combining the kinematic fit with the jet charge algorithm. 
The systematic uncertainties are reviewed in Sec.~\ref{sec:systematics}. 
The final result is presented in Sec.~\ref{sec:results} together with an upper 
limit on the fraction of exotic quark pairs in the data set.

\clearpage

\section{Signal Sample}
\label{sec:preselection}

As discussed earlier only a fraction of the collisions delivered by the Fermilab 
Tevatron is recorded. Furthermore, well known variables in the data are continuously 
monitored by detector experts and only the data marked by these experts as good 
is used. This analysis uses data collected by the D\O\ experiment in a period from June 2002 
to August 2004. The total integrated luminosity amounts to approximately 
$363$~pb$^{-1}$ and 
$366$~pb$^{-1}$ 
in the $\mu \text{+jets}$ and $e \text{+jets}$ channels respectively. 
The difference between the two channels is due to the different signal triggers used.
In this section the selection of a data sample enriched in $\ttbar \to \ljets$ events 
is described. The signal event signature consists of:
\begin{itemize}

  \item{One charged high transverse momentum lepton (electron or muon, either prompt or 
  from a leptonically decaying $\tau$),}

  \item{large missing transverse energy,}
  
  \item{two jets from the hadronization of the $b$-quarks from the top and anti-top 
  quark decay,}

  \item{two jets from the hadronically decaying $W$ boson.}

\end{itemize}



The requirement of one high transverse momentum electron or muon and 
large missing transverse energy rejects most multijet backgrounds. Other physics 
processes, like $W\text{+jets}$, have the same signature.

There are three stages of the event selection: The first 
stage is to make sure that the interesting events fulfilling the $\ttbar \to \ljets$ 
event signature are written to disk. Therefore a set of trigger requirements is 
defined. Secondly, 
a set of selection criteria is defined to select a sample enriched in events with isolated 
high $p_T$ leptons (composed primarily of $W$+multijet and \ttbar~events). This second 
stage is referred to as the preselection and is mostly concerned with selections 
based on lepton requirements. In the third and last stage the \ttbar~events are 
separated from most of the backgrounds by requiring the presence of two $b$-tagged 
jets in the event.


\subsection{Trigger Selection}
\label{subsec:trigger_presel}

The triggers used to record the signal sample require a lepton and at least one jet 
and are different for the $e\text{+jets}$ and $\mu \text{+jets}$ 
channels\footnote{The specific trigger 
requirements are divided into well defined trigger lists. These trigger lists 
change with time to accommodate detector and luminosity changes.}. 

The signal trigger for the $e \text{ +jets channel}$ requires at Level 1, 
at least one electromagnetic 
calorimeter (EM) tower with transverse energy \et~above $10$~GeV and one 
additional calorimeter tower (EM+H) with \et~above $5$~GeV. At Level 2 
an EM candidate with electromagnetic fraction above 85\% and \et~above $10$~GeV 
is required. Level 3 requires an EM candidate with \et~above $15$~GeV, passing 
a transverse shower shape criteria and one Level 3 jet with \et~above $15$~GeV.

For the $\mu \text{+jets channel}$ the signal trigger requires 
a calorimeter trigger tower above $5$~GeV at Level 1. At Level 2, the calorimeter 
trigger varies depending on the period when the data was collected and changes from 
no requirement to at least one Level 2 jet with \et~above $10$~GeV. The Level 3 calorimeter 
trigger requirement is one jet candidate with \et~above $20$ or $25$~GeV depending 
on the period of data taking. The muon trigger uses information from both 
Level 1 and Level 2. At Level 1, a candidate is required two have a coincidence 
between at least two layers of scintillators and similar requirements for 
Level 2 with the additional requirement of hits in the drift tubes. 

The probability for a \ttbar~event to pass all the trigger requirements is 
expressed in the \ttbar~trigger efficiency. The per muon, electron and jet probabilities 
to fire the trigger are derived on data. The \ttbar~trigger efficiency is then 
obtained by folding per lepton and per jet efficiencies with the $\eta$ and 
$E_T/p_T$ of the lepton and jets in simulated events. 
The \ttbar~trigger efficiency in the $e \text{+jets}$ ($\mu \text{+jets}$) channel 
is $92.82 \pm 0.08$\% ($91.65 \pm 0.91$\%) in events with four or more 
reconstructed jets.

More detailed information on the specific trigger requirements used and the 
measurement of the trigger efficiency can be found 
in~\cite{xsec_note240_mujets_topological,xsec_note240_ejets_topological,top_trigger_package}.
%
%

\subsection{Preselection}
\label{subsec:ljets_presel}

Apart from requirements on the charged lepton the preselection in both channels are 
identical. The preselection in this analysis are the same as 
in Ref.~\cite{xsec_note360_ljets_btag} where a more detailed discussion can be found. 

The common event preselection criteria for both $\mu \text{+jets}$ and $e \text{+jets}$ 
channel are:

\begin{itemize}

	\item 
	At least four jets with $p_T>15$~GeV and $|\eta|<2.5$. All 
	additional jets in the event are subject to the same $p_T$ and $\eta$ 
	requirements,

	\item 
	missing transverse energy $\met > 20$~GeV, 
	
	\item 
	a primary vertex with at least three tracks fitted to it and a $z$ coordinate 
	$z_{\rm PV}$ within the fiducial volume of the SMT detector 
	($|z_{\rm PV}|\le60$~cm) ,
	\item 
	the distance in the $z$-direction between the primary vertex and the 
	lepton track has to be less than $1$~cm.  

\end{itemize}

\subsubsection{Preselections specific to the $\mu \text{+jets}$ channel}
\label{subsec:mujets_presel}

The event preselection criteria specific for the $\mu \text{+jets}$ channel are:

\begin{itemize}

	\item One muon with $p_T > 20$~GeV and $|\eta| <2.0$. 
	

  \item
        The muon is required to be separated ($\Delta R(\mu,{\rm jet}) >0.5$) from 
		reconstructed jets and also isolated from activity in the calorimeter 
		by requiring that the scalar sum of $E_T$ of calorimeter clusters in a hollow cone
		between $\Delta R=0.1$ and $\Delta R=0.4$ away from the muon is less than 
		$8$\% of the muon $p_T$.
		
  \item
        The matched muon track is required to be isolated from other activity in 
		the central tracker by requiring that the sum of $p_T$ of all tracks inside a cone 
		of $\Delta R=0.5$ around the muon is less than $6$\% of the muon $p_T$.

	\item The missing transverse energy direction must be separated from the direction 
	of the muon in $\phi$. The missing transverse energy in QCD events passing the muon 
	isolation requirements is found mostly in or opposite the muon direction. This 
	can be explained by $b\bar{b}$ production where one or both $B$ hadrons decay 
	semi-leptonically and the jet is not reconstructed.

	\item Reject events with an electron with $p_T>15$~GeV in the central or the 
	end-cap calorimeter ensuring orthogonality between the 
	$\ttbar \rightarrow \mu$+jets, $\ttbar \rightarrow e$+jets and
	$\ttbar \rightarrow e\mu$ channel.

	\item Reject events with a second muon with $p_T>15$~GeV. This ensures 
	orthogonality between the $\ttbar \rightarrow \mu$+jets channel and 
	the $\ttbar \rightarrow \mu^+\mu^-$ channel and also rejects 
	$Z \rightarrow \mu^+\mu^-$ events.

\end{itemize}

\subsubsection{Preselections specific to the $e \text{+jets}$ channel}
\label{subsec:ejets_presel}

The event preselections specific to the $e \text{+jets}$ channel are:

\begin{itemize}

	\item The presence of an electron with $p_T > 20$~GeV and $|\eta|<1.1$.

	\item The missing transverse energy direction must be separated from the direction 
	of the electron in $\phi$ in order to eliminate events in which a jet 
	was misidentified as an electron.

	\item Reject events with a second electron with $p_T>15$~GeV ensuring 
	orthogonality between the $\ttbar \rightarrow e$+jets and 
	$\ttbar \rightarrow e^+e^-$ channels. This also rejects 
	$Z \rightarrow e^+e^-$ events.

	\item Reject events with a muon with $p_T>15$~GeV 
	thus ensuring orthogonality between 
	$\ttbar \rightarrow e \text{+jets}$, $\ttbar \rightarrow e\mu$ and 
	$\ttbar \rightarrow \mu \text{+jets}$ channels.

\end{itemize}

\subsubsection{Preselected Sample}
\label{subsec:preselected_sample}

The number of preselected events in the $\mu \text{+jets}$ ($e \text{+jets}$) channel 
are shown in Tab.~\ref{mujets_presel_excl} (\ref{ejets_presel_excl}) together with the 
expected contribution of QCD and $W \text{+jets}$ events. 
\begin{table}[htp]
\centering
\begin{tabular}{lcccc}
\hline\hline
{} & {\rm 1jet} & {\rm 2jets} & {\rm 3jets} & {\rm $\geq$4jets} \\
\hline\hline
$N^{presel}_{\mu+jets}$ & 10101 & 3863 & 933 & 231 \\
\hline 
$N^{presel}_{(W \mu)+jets}$ & 9726$\pm$103 & 3669$\pm$63 & 874$\pm$31 & 215$\pm$16 \\
$N^{presel}_{QCD \mu+jets}$ & 375$\pm$12 & 195$\pm$7 & 60$\pm$3 & 16$\pm$2 \\

\hline\hline
\end{tabular}
\caption{Number of preselected events in the $\mu \text{+jets}$ channel and 
expected contribution from QCD and $W$-like events categorized by the number 
of (exclusive) jets in the events, extracted from Ref.~\cite{xsec_note360_ljets_btag}.}
\label{mujets_presel_excl}
\end{table}
\begin{table}[htp]
\centering
\begin{tabular}{lcccc}
\hline\hline
{} & {\rm 1jet} & {\rm 2jets} & {\rm 3jets} & {\rm $\geq$4jets} \\
\hline\hline
$N^{presel}_{e+jets}$ & 12668 & 4587 & 1078 & 277 \\
\hline 
$N^{presel}_{(W e)+jets}$ & 12130$\pm$173 & 4142$\pm$101 & 909$\pm$41 & 221$\pm$19 \\
$N^{presel}_{QCD e+jets}$ & 538$\pm$127 & 445$\pm$72 & 169$\pm$22 & 56$\pm$7 \\
\hline\hline
\end{tabular}
\caption{Number of preselected events in the $e \text{+jets}$ channel and 
expected contribution from QCD and $W$-like events categorized by the number 
of (exclusive) jets in the events, extracted from Ref.~\cite{xsec_note360_ljets_btag}.}
\label{ejets_presel_excl}
\end{table}
To estimate the fraction of QCD events a sample with a ``looser'' selection criteria 
is defined\footnote{In the $\mu \text{+jets}$ channel, the muon isolation is loosened 
by only requiring the muon to be isolated from jets in a cone of $\Delta R =0.5$. In 
the $e \text{+jets}$ channel, the requirement on the electron likelihood is dropped.}. 
By calculating the probability for events that pass the ``looser'' selection 
to also pass the normal preselection requirement (for both QCD events and events with 
a true isolated lepton e.g. $W \text{+jets}$ and $\ttbar$), the fraction of QCD events 
in the preselected sample can be estimated. More information on how to estimate the QCD 
background can be found in Ref.~\cite{xsec_240_ljets_topo_publ}.

The preselection efficiency for $\ttbar \to \ljets$ events is calculated from simulated 
events and scale factors are applied to take into account significant data-to-simulation 
discrepancies. 
%
%
%
%
The efficiency to select $\ttbar \to \ljets$ events is 
$13.4 \pm 1.8$\% ($13.0 \pm 1.5$\%) in the $\mu \text{+jets}$ 
($e \text{+jets}$) channel requiring at least four or more jets. Other backgrounds than 
$W \text{+jets}$ and QCD have preselection efficiencies less than 1\% except single top 
quark and \ttbar~dilepton backgrounds which have preselection efficiencies around $1.5$\%.

\subsection{Final Event Selection}

The majority of the backgrounds does not contain $b$-quark jets and thus the identification 
of $b$-quark jets can be used to preferentially select \ttbar~events while removing background events. 
This is used to obtain a pure sample of $\ttbar \to \ljets$ events 
in the last stage of the event selection:
\begin{quote}
  \item[] The event must contain at least two jets tagged by the SVT algorithm.
\end{quote}

\subsubsection{Sample Composition}
\label{subsec:sample_compostion}

The composition of the data sample after double-tagging (called signal sample) 
is determined in Ref.~\cite{xsec_note360_ljets_btag} 
starting from the preselected sample and the predicted number of $W$+jets and 
QCD events. This is further multiplied by the probability for an event to 
contain jets that are $b$-tagged. 
To obtain an estimate of the number and type of events after applying $b$-tagging 
on the preselected sample, the probability for an event to have two or more $b$-tagged 
jets must be calculated. This event tagging probability depends strongly on 
the topology and jet flavors in the event. It is determined from 
Monte Carlo simulations, applying to each jet the per jet $b$-tagging efficiency 
derived from data (as described in Sec.~\ref{sec:bjets}). The QCD background 
contribution to the signal sample is estimated to be $<0.01$ events. 
To estimate the $W$+jets contribution to the signal sample, the predicted 
number of $W$+jets events in the preselected sample is split into its expected 
flavor composition as predicted by {\sc ALPGEN}. 
The $W \text{+jets}$ background in the preselected sample is dominated by non-heavy 
flavor jets (81\%) due to the low production cross section of a $W$ boson in association 
with heavy flavor jets. The low probability to tag light jets leads to an effective 
rejection of $W \text{+ light jets}$ in the signal sample ($<0.01$ events). 
The only sizable $W$+jets background remaining is the production of a $W$ boson 
in association with a $b\bar{b}$ pair which has an event tagging probability 
($\approx 10$\%) on the same order as a $\ttbar \to \ljets$ event ($\approx15$\%).

Other physics backgrounds that contributes to the signal sample are 
diboson production
($WW \rightarrow \ell \text{+jets},
~WZ\rightarrow \ell\text{+jets},
~WZ\rightarrow jj\ell\bar{\ell},
~ZZ\rightarrow \ell\bar{\ell}jj$), 
single top production 
and 
$Z\text{+jets}$($Z\rightarrow \tau \tau \rightarrow \ljets$ and 
$Z\rightarrow \ell\text{+jets}$). The $Z$+jets background has a similar event 
signature as \ttbar~but is suppressed due to no physical source 
of \met. For a given process $i$, the number of events $N_i^{\text{presel}}$ in 
the preselected sample is determined by 
$N_i^{\text{presel}} = \sigma_i \epsilon^{\text{presel}}_i \mathcal{BR}_i \int\mathcal{L}{\rm dt}$,  
where $\sigma_i$, $\epsilon^{\text{presel}}_i$, $\mathcal{BR}_i$ and 
$\int\mathcal{L}{\rm dt}$ are, respectively, the cross section, the preselection efficiency, 
the branching fraction for the specific process and the integrated luminosity. The 
expected number 
of events in the signal sample is estimated by multiplying $N_i^{\text{presel}}$ 
with the event tagging probability for that specific process.
The only non-negligible backgrounds are the single top quark production and the 
$\ttbar \rightarrow \text{dileptons}$ processes. 

After all selections, 21 data events remain. 
Table~\ref{svt_summary_table_2_ejets} and~\ref{svt_summary_table_2_mujets} summarizes 
the sample composition according to predicted signal and background contributions 
in the $e \text{+jets}$ and $\mu \text{+jets}$ channel respectively. 
It should be noted that in the analysis presented in this thesis, only events 
with four or more jets are used.
\begin{table}[]
\centering
\begin{tabular}{l|ccc}
\hline\hline
&\multicolumn{3}{c}{$e \text{+jets}$} \\
\hline
 & 2 jets & 3 jets & $\ge$4 jets \\
\hline
$W$+light & 0.015$\pm$0.002 & 0.012$\pm$0.001 & $<0.01$  \\
$W(c\bar{c})$ & 0.021$\pm$0.003 & 0.014$\pm$0.001 & $<0.01$ \\
$W(b\bar{b})$ & 0.29$\pm$0.04 & 0.12$\pm$0.01 & 0.03$\pm$0.01 \\
$Wc$ & 0.037$\pm$0.002 & 0.015$\pm$0.001 & $<0.01$ \\
$Wc\bar{c}$ & 0.36$\pm$0.02 & 0.15$\pm$0.01 & 0.05$\pm$0.01 \\
$Wb\bar{b}$ & 4.77$\pm$0.17 & 1.46$\pm$0.10 & 0.49$\pm$0.06 \\
\hline
$W$+jets & 5.5$\pm$0.2 & 1.77$\pm$0.10 & 0.59$\pm$0.06 \\
\hline
QCD & $<0.01$ & 0.29$\pm$0.34 & $<0.01$ \\
\hline
$tb$ & 0.96$\pm$0.01 & 0.52$\pm$0.01 & 0.16$\pm$0.01 \\
$t\bar{t}\rightarrow ll$ & 1.76$\pm$0.02 & 0.92$\pm$0.02 & 0.20$\pm$0.01 \\
diboson & 0.47$\pm$0.02 & 0.06$\pm$0.01 & $<0.01$ \\
$Z\rightarrow\tau^+\tau^-$ & 0.03$\pm$0.03 & 0.02$\pm$0.02 & $<0.01$ \\
\hline
Background & 8.3$\pm$0.2 & 3.58$\pm$0.37 & 0.76$\pm$0.36 \\
\hline
Syst. uncert. (bkg) & +1.42-1.42 & +0.49-0.51 & +0.17-0.15 \\
\hline
$t\bar{t}\rightarrow l$+jets & 0.94$\pm$0.03 & 5.8$\pm$0.1 & 8.6$\pm$0.1 \\
\hline
Sum pred. & 9.2$\pm$0.2 & 9.3$\pm$0.4 & 9.4$\pm$0.4 \\
\hline
Syst. uncert. & +1.49-1.46 & +0.87-0.87 & +1.22-1.42 \\
\hline\hline
Observed & 11 & 7 & {\bf 13} \\
\hline
\hline
\end{tabular}
\caption{Summary of observed and predicted number of events with 
two SVT-tagged jets in the $e \text{+jets}$ channel. The individual 
contributions from the various backgrounds processes are shown. Unless 
explicitly stated, uncertainties are statistical only.}
\label{svt_summary_table_2_ejets}
\end{table}
\begin{table}[]
\centering
\begin{tabular}{l|ccc}
\hline\hline
&\multicolumn{3}{c}{$\mu \text{+jets}$}\\
\hline
 & 2 jets & 3 jets & $\ge$4 jets\\
\hline
$W$+light & 0.028$\pm$0.003 & 0.016$\pm$0.002 & $<0.01$ \\
$W(c\bar{c})$ & 0.027$\pm$0.004 & 0.015$\pm$0.001 & $<0.01$ \\
$W(b\bar{b})$ & 0.30$\pm$0.04 & 0.09$\pm$0.01 & 0.04$\pm$0.01 \\
$Wc$ & 0.044$\pm$0.003 & 0.017$\pm$0.001 & $<0.01$ \\
$Wc\bar{c}$ & 0.35$\pm$0.01 & 0.13$\pm$0.01 & 0.05$\pm$0.01 \\
$Wb\bar{b}$ & 4.69$\pm$0.15 & 1.56$\pm$0.09 & 0.48$\pm$0.05 \\
\hline
$W$+jets & 5.4$\pm$0.2 & 1.83$\pm$0.09 & 0.59$\pm$0.05 \\
\hline
QCD & 0.02$\pm$0.14 & 0.03$\pm$0.10 & $<0.01$ \\
\hline
$tb$ & 0.85$\pm$0.01 & 0.48$\pm$0.01 & 0.15$\pm$0.01 \\
$t\bar{t}\rightarrow ll$ & 1.51$\pm$0.02 & 0.84$\pm$0.02 & 0.18$\pm$0.01 \\
diboson & 0.46$\pm$0.02 & 0.07$\pm$0.01 & $<0.01$ \\
$Z\rightarrow\tau^+\tau^-$ & $<0.01$ & $<0.01$ & $<0.01$ \\
\hline
Background & 8.3$\pm$0.2 & 3.25$\pm$0.15 & 0.77$\pm$0.12 \\
\hline
Syst. uncert. (bkg) & +1.40-1.39 & +0.49-0.51 & +0.15-0.16 \\ 
\hline
$\ttbar \to \ljets$ & 0.76$\pm$0.02 & 5.1$\pm$0.1 & 8.8$\pm$0.1 \\
\hline
Sum pred. & 9.0$\pm$0.2 & 8.4$\pm$0.2 & 9.6$\pm$0.1 \\
\hline
Syst. uncert. & +1.44-1.44 & +0.92-0.94 & +1.33-1.50 \\ 
\hline\hline
Observed & 11 & 4 & {\bf 8} \\
\hline
\hline
\end{tabular}
\caption{Summary of observed and predicted number of events with 
two SVT-tagged jets in the $\mu \text{+jets}$ channel. The individual 
contributions from the various backgrounds processes are shown. Unless 
explicitly stated, uncertainties are statistical only.}
\label{svt_summary_table_2_mujets} 
\end{table}

\clearpage

\clearpage

\section{Jet Charge Algorithms}
\label{sec:jcttbar}

In this section a method to use tracks inside jets to form a variable 
which is capable to discriminate between jets arising from the hadronization 
of $b$- and $\bar{b}$-quarks is presented.

\subsection{Jet Charge Algorithm Definition}
\label{subsec:definition}
Various algorithms to discriminate between $b$- and $\bar{b}$-quark jets have been 
used in the 
past~\cite{jet_charge_aleph,jet_charge_cdf,jet_charge_delphi,jet_charge_l3,jet_charge_opal}.
The idea is to use the particle tracks reconstructed by the central
tracker, associate them with a reconstructed jet in the calorimeter and 
compute the collected charge of the jet. In the hadronization, the initial quark 
transverse momentum is shared in a multi-stage process between many particles. 
A common feature of the modeling of this process is 
that in most jets arising from the hadronization of a $b$- or $\bar{b}$-quark, the 
tracks with highest $p_T$ are the decay products of the $B$ hadron in the jet. 
A weighted sum of the charges of the tracks are therefore used in most algorithms. 
The weights are usually functions
of the track momentum or its projection along a certain direction. In designing such a
jet charge algorithm one needs to decide:
\begin{itemize}
  \item What tracks are associated with the jet, within a cone in $\Delta R$ or in a whole 
  hemisphere,
  \item the quality criteria of the tracks to be considered,
  \item what weight to give each track.
\end{itemize} 

In this analysis, only tracks that fulfill the following criteria 
are considered:

\begin{enumerate}
  \item Track $p_T > 0.5$~GeV,
  \item distance between the track and jet axis must be less than 0.5 in $\Delta R$,
  \item a distance to the primary vertex in the $z$-direction of less than $1$~cm unless 
  the track is fitted to a secondary vertex.
\end{enumerate}

At least four jets are present in the $\ttbar \to \ljets$ events and it is therefore important to consider 
only tracks that are associated with the corresponding jet. Also, in this analysis the interest 
is to calculate only the jet charge for the jets originating from the $b$- and $\bar{b}$-quarks 
in the $t \to Wb$ decay. The SVT algorithm determines what jets should be considered 
and only tracks within a $\Delta R$ cone (studied below) of the jet axis are considered.

\begin{figure}[] 
\centering 
  \includegraphics[width=0.9\textwidth]{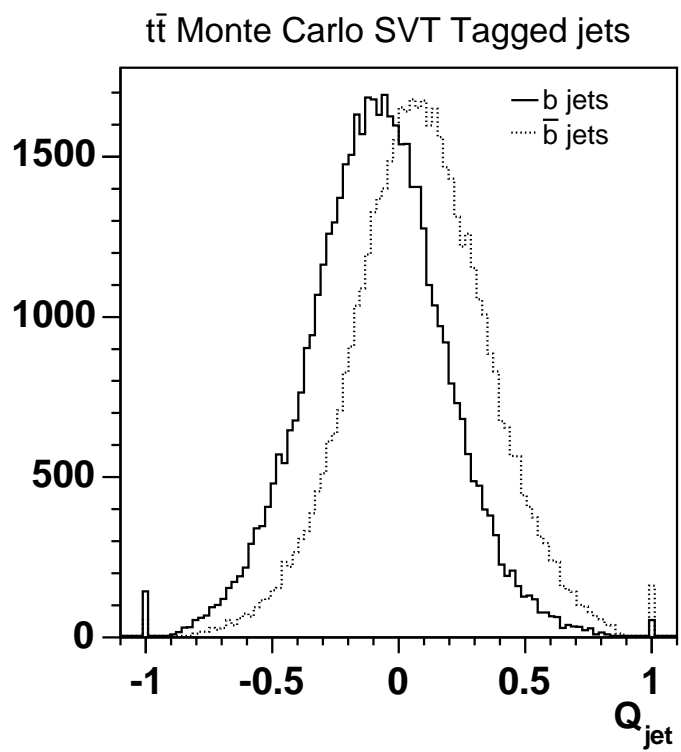} 
  \caption{Jet charge distribution for SVT-tagged $b$ and $\bar{b}$-quark jets in 
  simulated $\ttbar \to \ljets$ events using algorithm I with $a=0.6$ and a jet 
  cone size of $\Delta R = 0.5$.}
  \label{fig:ttbar_jet_algo_performance1}
\end{figure}

The performance of two algorithms is evaluated, algorithm I:
\begin{equation}
  Q_{\text{jet}}=\frac{\sum_{i}q_i \cdot p_{T_i}^a}{\sum_{i}p_{T_i}^a},
\label{eq:jc1}
\end{equation}
and algorithm II:
\begin{equation}
  Q_{\text{jet}}=\frac{\sum_{i}q_i \cdot |p_{||_i|}|^a}{\sum_{i}|p_{||_i}|^a},
\label{eq:jc2}
\end{equation}

where the subscript $i$ runs over all charged tracks passing the track quality cuts 
described above and that are located within a cone of $\Delta R$ from the jet axis.
Each track has a charge $q_i$ and a transverse momentum $p_{T_i}$, while
$p_{||_i}$ represent the projection of the track momentum along the jet axis.
The parameter $a$ is an arbitrary number which is optimized from simulated 
$\ttbar \to \ljets$ events. 
For $a=0$, the weight given to each track is equal to one and hence, $p_T$ independent, 
while $a=\infty$ is equivalent to considering solely the highest $p_T$ track. 
Figure~\ref{fig:ttbar_jet_algo_performance1} shows the jet charge 
distribution for SVT-tagged $b$- and $\bar{b}$-quarks in simulated $\ttbar \to \ljets$ 
events.

\subsection{Optimization}
\label{subsec:optimization}
A simple optimization of the jet charge algorithm parameters using 
$b$- and $\bar{b}$-quark jets in simulated $\ttbar \to \ljets$ events 
is performed. In order to quantify the separation between $b$- and 
$\bar{b}$-quark jets, a variable called discriminating (or discriminant) power $D$ 
was defined as:
\begin{equation}
  D=\frac{|a_b-a_{\bar{b}}|}{\sqrt{V_b+V_{\bar{b}}}},
\label{eq:dpower}
\end{equation}
where $a_{b},V_{b}$ and $a_{\bar{b}},V_{\bar{b}}$ are the mean and variance of 
the jet charge distributions obtained for $b$- and $\bar{b}$-quark jets respectively. 

The purpose of the algorithm is to discriminate between $b$- and $\bar{b}$-quark 
jets from the decay of top quarks in $\ttbar \to \ljets$ events. The optimization 
is carried out using simulated $\ttbar \to \ljets$ 
events. In the optimization, a reconstructed jet is labeled as a true 
$b$-quark ($\bar{b}$-quark) jet if a $b$-quark ($\bar{b}$-quark) at parton level 
in the Monte Carlo history is found within a cone of $\Delta R=0.5$ with respect 
to the jet axis. If more than one heavy flavor ($b$ or $\bar{b}$) parton was 
inside the cone, the closest one is used. Only SVT-tagged jets are considered. To 
avoid biases from the event topology, the events were required to have at least 
four reconstructed jets with $p_T>15$~GeV and $|\eta|<2.5$ (in the simulated 
\Zbbbar~events used for comparison below, exactly two jets were required).

\begin{figure}[]
  \centering
  \includegraphics[width=1.2\textwidth]{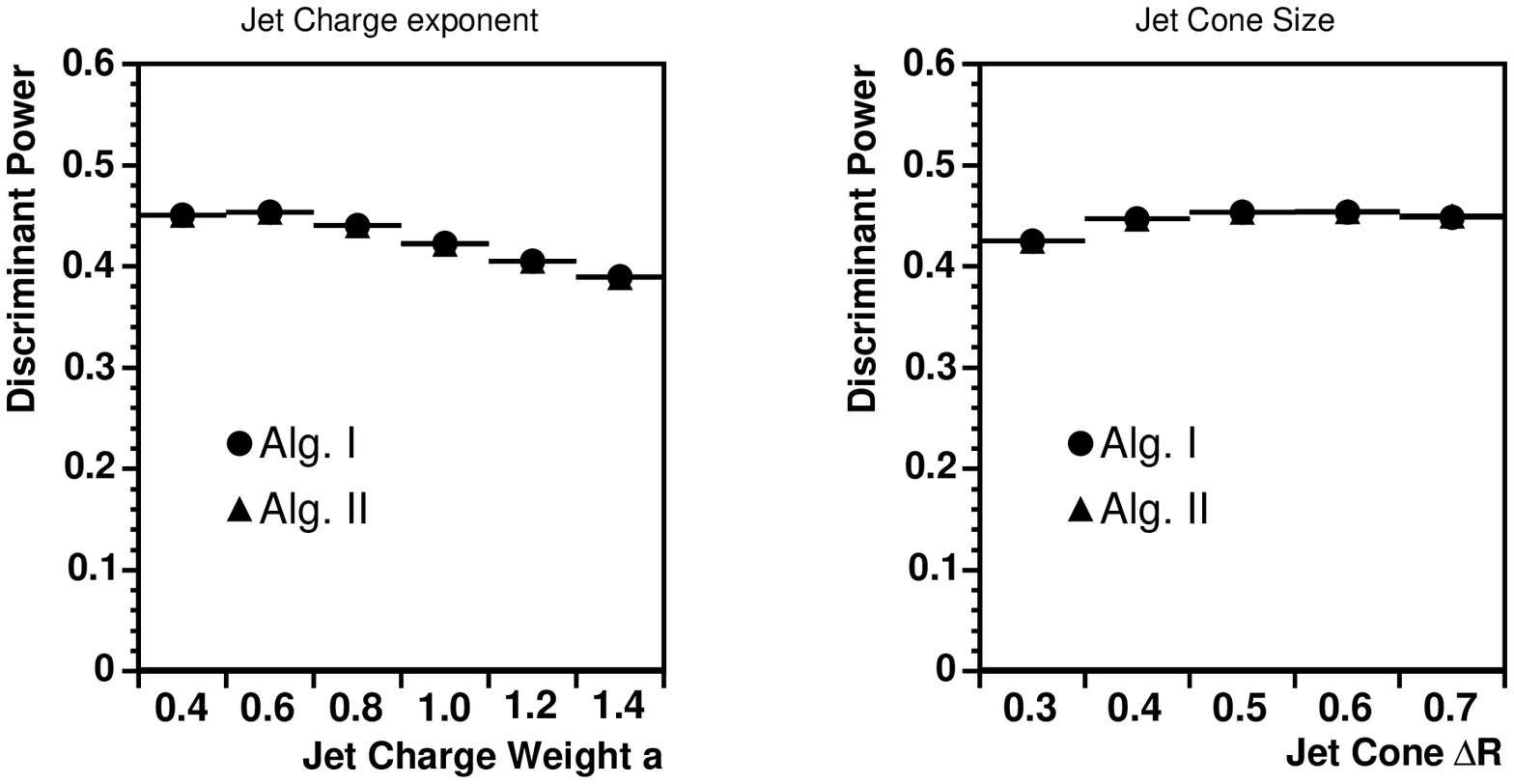}
  \caption{Comparison of the discriminating power for the jet charge algorithms I and 
II as function of $a$ (left) and $\Delta R$ (right).}
  \label{fig:dpoweropt}
\end{figure}

In the optimization, the parameter $a$ was varied between $0.4$ and $1.6$ in steps 
of $0.2$ and the size of the jet cone was varied between $0.3$ and $0.7$ 
in steps of $0.1$. Figure~\ref{fig:dpoweropt} shows the discriminating power for the
algorithms I and II as function of $a$ and $\Delta R$. Little difference is found
between the two algorithms and algorithm I is chosen for the remaining of this 
analysis. It is found that the highest discriminating power is obtained for
$\Delta R =0.5$ and $a=0.6$. 

In general, tracks originating from the hadronization are expected to come from 
the primary vertex unless they come from the decay products of the $B$ hadron and 
consequently displaced. Therefore, the maximum distance of the tracks to the primary 
vertex in the $z$-direction was studied. 
Tracks determined 
to originate from a displaced $B$ hadron vertex are not included. In this 
study using simulated events, the discriminating power increases when loosening this 
requirement. This is expected to be a variable heavily affected by the detailed modeling of the 
tracking which is known to be poor. Therefore, a conservative requirement of $<0.1$~cm is 
used to ensure that tracks only close to the primary vertex are considered. 

\begin{figure}[]
  \centering
  \includegraphics[width=0.7\textwidth]{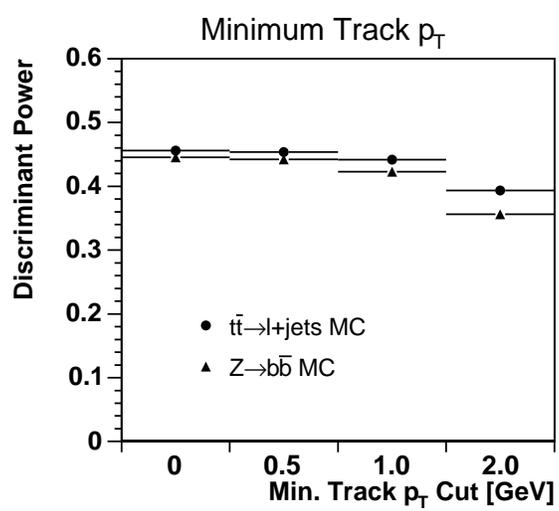}
  \caption{Discriminating power as a function of the minimum track $p_T$ requirement.}
  \label{fig:jcvsmintrkpt}
\end{figure}

In Fig.~\ref{fig:jcvsmintrkpt} the discriminating power is shown as function of the 
minimum requirement of the $p_T$ of the tracks. 
This variable is also expected to exhibit the same critical dependence on the correct 
modeling of the tracking. Therefore, the minimum $p_T$ requirement was chosen to be $0.5$~GeV to ensure 
the quality of the tracks. The requirements on the minimum track $p_T$ and maximum distance 
from the track to the primary vertex are kept in the rest of this analysis.





%

\clearpage


\clearpage

\section{Jet Charge Calibration on Data}
\label{sec:data_calibration}

%
%
The Monte Carlo description of the D\O\ tracking system includes the detailed 
geometry of the detector and a modeling of the electronic noise. Nevertheless, the Monte Carlo 
simulations are 
unable to describe fine details of track quality distributions such as $\chi^2$, 
hit multiplicities and tracking efficiency within jets. 
Physics analyzes in D\O\ that use the tracking detectors for $b$-tagging 
determine the performance of the $b$-tagging algorithms on data, or with as little 
input from simulation as possible. A similar approach is chosen in this analysis.

The goal of the present section is to show that the jet charge distribution templates 
for SVT-tagged $b$-, $\bar{b}$-, $c$- and $\bar{c}$-quark jets can be extracted 
from data. The jet charge distributions are then used 
to derive the expected distributions of the charge observables in the 
SM top and exotic quark scenarios.

The difficulty is to find the true flavor of the particle 
initiating a jet in data. Therefore, a data sample enriched in \bbbar~pairs 
is selected. In such an event with exactly two jets back-to-back 
in azimuth, called dijet event, one of the jets is required to contain a muon from 
the semi-leptonic decay of a $B$ hadron and is referred to as the {\sl tag}-jet, the 
muon associated with the {\sl tag}-jet is referred to as the {\sl tag}-muon or simply 
tagging muon. Studies~\cite{rick_field,cdf_b_production_runI,dzero_thesis} have shown that the 
mechanism for \bbbar~production at the Tevatron depends heavily on the azimuthal 
distance between the jets, with large distances dominated by flavor creation. In such an event, 
the charge of the muon can be used to find the type of quark initiating the 
{\sl tag}-jet and, consequently, also the other jet in the event (referred to 
as the {\sl probe}-jet) as given in Tab.~\ref{tab:data_calib1}.
\begin{table}[h]
\centering
\begin{tabular}{|l|c|c|}
\hline
Muon 	    & Quark initiating the & Quark initiating the   \\
                    & {\sl tag}-jet              & {\sl probe}-jet              \\
\hline
		    & 			   &                        \\
\Large $\mu^-$      & \Large $b$           & \Large $\bar{b}$ 	    \\
\Large $\mu^+$      & \Large $\bar{b}$     & \Large $b$ 	    \\
\hline
\end{tabular}
\caption{\label{tab:data_calib1} Charge of the muon in the {\sl tag}-jet 
and the corresponding type of $b$-quark in the {\sl tag}-jet and the {\sl probe}-jet, 
here assuming no $B$ mixing and only direct semi-leptonic $b$-decays.}
\end{table}
The method above is called the 
{\sl tag}-and-{\sl probe} method.

\begin{figure}[h]
  \centering
  \includegraphics[width=1.0\textwidth]{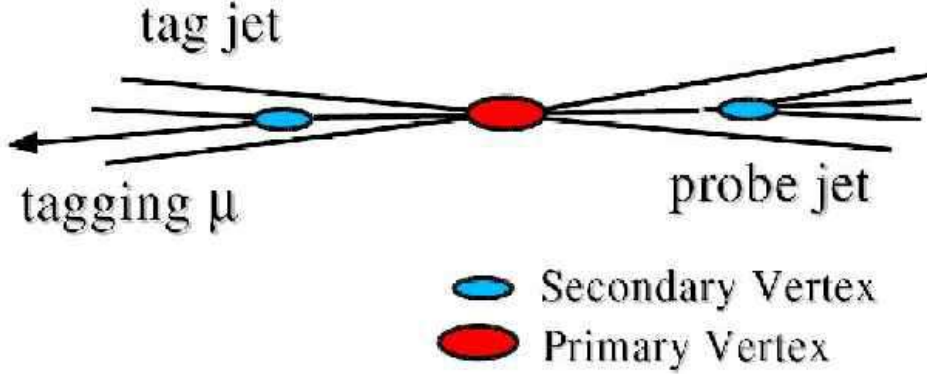}
  \caption{Illustration of the {\sl tag}-and-{\sl probe} method in the tight dijet sample 
  (defined in Sec~\ref{subsubsec:tripletag_sample}).
  The events contain exactly two SVT-tagged jets in a back-to-back configuration in 
  azimuth, one of which is also $\mu$-tagged.}
  \label{fig:triple_tag_illustration}
\end{figure}

The starting point are the dijet samples (defined below in 
Sec.~\ref{subsubsec:tripletag_sample}) dominated by $b$- and $\bar{b}$-quark jets. 
The dijet samples (illustrated in Fig.~\ref{fig:triple_tag_illustration}) contain a jet 
with a $\mu$- and a SVT-tag on one side and a SVT-tagged jet on the other side of the 
event.  
The goal is to find the jet charge templates for $b$- and $c$-quark jets in general, not 
necessarily $\mu$-tagged jets and therefore the jet charge distribution is 
extracted from  the {\sl probe}-jet.
 
In the simplified case, when there is no $B$ mixing~\cite{PDG}, nor contamination of 
the sample by $c$-quark jets or light-jets, and all the tagging muons were coming 
from a direct $B$ hadron decay, the charge of the tagging muon would reliably 
tell if the {\sl tag}-jet was initiated by a $b$- or a $\bar{b}$-quark. According 
to this procedure, the two distributions of jet charge for jets that are 
believed to be $b$-quark jets and those that are believed
to be $\bar{b}$-quark jets can be extracted by plotting separately $Q_{\rm jet}$ 
for the $\mu^+$ and $\mu^-$ samples. This ideal scenario is complicated 
by a number of issues in data. First the
sample is not pure $b\bar{b}$ production, but contains a small fraction of $c\bar{c}$
production. Therefore the observed $b$- and $\bar{b}$-quark jet charge distributions 
from the {\sl probe}-jet are a mixture of the jet charge for $b$-quark jets and for 
$c$-quark jets. The contamination by $c$-quark jets in the dijet samples is determined 
in Sec.~\ref{subsec:triple_tag_composition}.


The tagging muon can also arise from a decay which is not a direct $B$ hadron decay.
This sort of muon is referred to as a cascade muon. It can either arise from
the decay of a $D$ meson or lighter hadrons. In this case, the relation 
between the charge of the tagging muon and the type of $b$-quark in the {\sl probe}-jet is 
not given by Tab.~\ref{tab:data_calib1} anymore. $B$ mixing on the side of 
the {\sl tag}-jet can also destroy the correlation between the muon sign and the type 
of $b$-quark initiating the {\sl probe}-jet. The $B$ meson on the {\sl probe}-jet side 
can also mix, but this does not matter, since in the end the 
data-derived $b$-quark jet charge distributions is applied to the $b$-quark from the 
decay of the top quarks in $t\bar{t}$ events, not the $B$ meson. Another effect that 
can destroy the correlation between the muon sign and the type of $b$-quark initiating the 
{\sl probe}-jet is an incorrectly measured the muon charge. 

\begin{figure}[]
  \centering
  \includegraphics[width=0.95\textwidth]{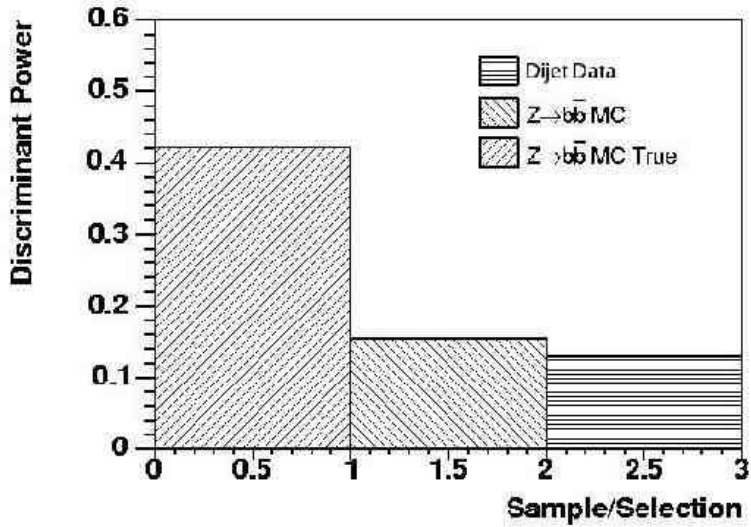}
  \caption{Comparison of the discriminating power using the {\sl tag}-and-{\sl probe} 
  method in the tight dijet sample (rightmost bin) and in simulated \Zbbbar~events 
  (middle bin). In the leftmost bin, the discriminating power for simulated 
  \Zbbbar~events is shown using the true charge of the $b$-quark jet from the Monte Carlo history 
  to sort the $b$- and $\bar{b}$-quark jets.}
  \label{fig:svtmethod_datamc_comp}
\end{figure}

To illustrate the effect of the $B$ mixing, cascade decay and $c$-contamination,
Fig.~\ref{fig:svtmethod_datamc_comp} shows a comparison of the discriminating power 
of the jet charge distributions between:
\begin{itemize}
  \item[{\it i})]  
  Simulated \Zbbbar~events where the true $b$-type ($b$ or $\bar{b}$) is extracted from 
  the Monte Carlo truth to sort between the $b$- and $\bar{b}$-quark jet charge. Note 
  that we look for the $b$- or $\bar{b}$-quark in the Monte Carlo history that is closest 
  in $\Delta R$ to the jet axis.
  \item[{\it ii})] 
  simulated \Zbbbar~events using the {\sl tag}-and-{\sl probe} method, 
  \item[{\it iii})] 
  a dijet sample (defined in Sec.\ref{subsubsec:tripletag_sample}) using the {\sl tag}-and-{\sl probe} method.
\end{itemize}
As expected the discriminating power in case {\it i}) is much 
better than in either case {\it ii}) and {\it iii}) which both suffer from cascade
decays and $B$ mixing. Data also suffers from contamination by \ccbar~events and from 
poorer tracking than the simulated events.

\begin{figure}[]
  \centering
  \includegraphics[width=0.85\textwidth]{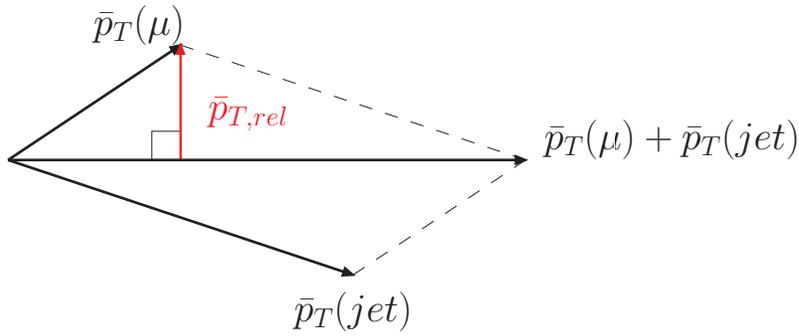}
  \caption{Definition of \ptrel~for a muon within a jet.}
  \label{fig:def_ptrel}
\end{figure}

\begin{figure}[]
  \centering
  \includegraphics[width=1.0\textwidth]{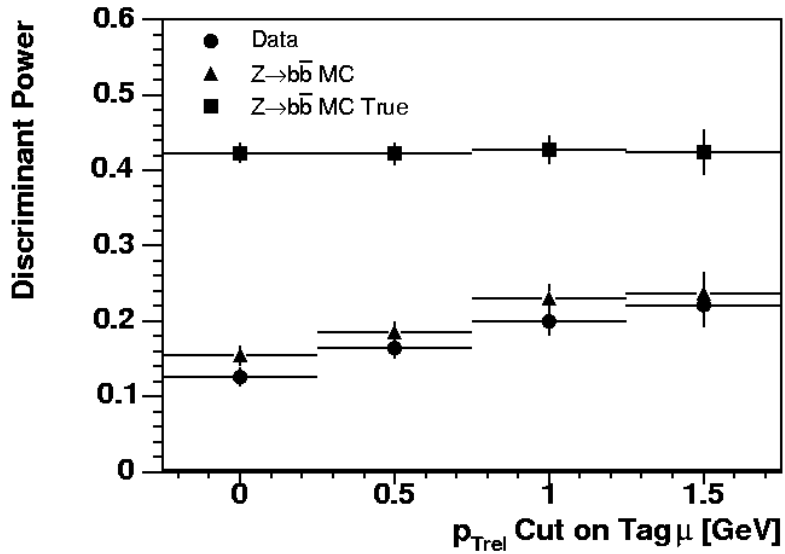}
  \caption{The effect of the requirement of a minimum \ptrel~of the tagging muon in the 
  dijet sample and simulated \Zbbbar~events using the {\sl tag}-and-{\sl probe} 
  method. For comparison the discriminating power in simulated \Zbbbar~events is 
  shown when using the Monte Carlo truth to find the quark initiating the jet.}
  \label{fig:DPvsPtrelDatabbMC}
\end{figure}

The amount of contamination from cascade decay and $c$-quark jets can be changed by 
requiring a high \ptrel, defined as the relative momentum of the tagging muon with respect 
to the jet axis, see Fig.~\ref{fig:def_ptrel}. \ptrel~is expected to be larger for 
$b$-quark jets than from $c$-quark or light jets due to the larger $b$-quark mass.
This effect is illustrated in Fig.~\ref{fig:DPvsPtrelDatabbMC}.
The goal is to extract the jet charge distributions of the {\sl probe}-jet. The 
tagging muon only serves as the source to find the true type of quark initiating the 
jet. Thus, after taking into account all sources that affect the charge of the tagging 
muon (that can destroy the correlation between the charge and the type of quark) the 
extracted jet charge distributions from the {\sl probe}-jet should be independent of 
the specific requirement on the minimum \ptrel~of the tagging muon. As a cross-check, 
the extraction of the jet charge distributions are therefore performed with different 
requirements of the minimum \ptrel~of the tagging muon. The different requirements 
are however later shown to give similar results. The final jet charge 
distributions are derived without a \ptrel~requirement on the tagging muon in order to 
minimize the statistical uncertainty.

\subsection{Dijet Data Samples}
\label{subsubsec:tripletag_sample}
To calibrate the jet charge algorithm using data a sample enriched in 
$b$-quark jets is selected as discussed above. The data sample is based on a 
generally selected sample requiring a muon matched to a jet. The following additional 
requirements are used to further enhance the sample in events with \bbbar~production:
\begin{itemize}
  \item	The event must have exactly two jets $j_1$ and $j_2$ with $p_T>15$~GeV and $|\eta|<2.5$.
  \item The azimuthal distance ($\Delta \phi$) between $j_1$ and $j_2$ is larger than $3.0$. 
  \item $j_1$ and $j_2$ must be $b$-tagged using the SVT algorithm, 
  \item $j_1$ must be associated with a muon within $\Delta R(\mu,{\rm jet})<0.5$ from the jet axis.
\end{itemize}
In summary, this sample has exactly two jets in a back-to-back configuration in $\phi$, 
one jet $j_1$ being both SVT-tagged and $\mu$-tagged and the second jet $j_2$ being 
SVT-tagged as shown schematically in Fig.~\ref{fig:triple_tag_illustration}. 
This sample is referred to as the ``{\bf tight dijet sample}''.



Similarly, the ``{\bf Loose Dijet sample}'' is defined in the same way as the 
``tight dijet sample'' apart from the requirement of the SVT-tag for 
$j_1$ which is removed. Thus, the ``tight dijet sample'' is a subset of the 
``loose dijet sample''.


\subsection{Extraction of Jet Charge Templates from Dijet Data}
\label{subsec:jetcharge_extraction}

In the ideal case where 
the tagging muon comes from a direct $B$ hadron decay without $B$ mixing, the 
{\sl tag}-and-{\sl probe} method would in fact yield the true jet charge distributions. 
In practice when the tag muons
are for example of positive sign the {\sl probe}-jets are in majority from $b$-quark jets
but mixed with a certain fraction of $\bar{b}$-, $c$-, and $\bar{c}$-quark jets.

The jet charge distributions of the {\sl probe}-jet obtained in 
the tight dijet sample are denoted: $f_{\mu+}$ and $f_{\mu-}$ where the subscript 
$\mu\pm$ indicates the sign of the tagging muon.
Similarly, $f_b$ and $f_{\bar{b}}$ denotes the jet charge distributions for
$b$ and $\bar{b}$-quark jets, these are the distributions we want to extract from the
data.

In absence of $B$ mixing, if the fraction of $c$-quark jets were zero and if 
all tagging muons were coming from a direct $B$ decay, then the jet charge 
distributions for $b$- and $\bar{b}$-quark jets would simply be given by: 
$f_b = f_{\mu+}$  and $f_{\bar{b}} = f_{\mu-}$.

In reality, care has to be taken to the processes which change the sign of 
the tagging muon. In the tight dijet sample $x_{\rm flip}$ is defined as the 
fraction of tagging muons which do not have the same sign as the parent $B$ hadron 
due to cascade decay, or because it originates from a kaon or pion decay.

If this is the only process that affects the correlation between the sign 
of the tagging muon and the quark that initiated the {\sl probe}-jet, $f_{\mu+}$ and 
$f_{\mu-}$ can be written as:
\begin{gather} 
  f_{\mu^+} = x_{\rm flip}\times f_{\bar{b}} + x_{\rm noflip}\times f_b \\
  f_{\mu^-} = x_{\rm flip}\times f_b + x_{\rm noflip}\times f_{\bar{b}}
\label{eq:example_jetchargecorrelation}
\end{gather}
where $x_{\rm noflip}$ is simply $1-x_{\rm flip}$.

The tight dijet sample contains also a fraction of $c$- and $\bar{c}$-quark jets.
Therefore the jet charge templates $f_{\mu+}$ and $f_{\mu-}$ contain also
a fraction of the jet charge distributions $f_{c}$ and $f_{\bar{c}}$  for
$c$- and $\bar{c}$-quark jets respectively. The fraction of $c$-quark 
jets in the tight dijet sample is denoted $x_c$. 
Equation~\ref{eq:example_jetchargecorrelation} can therefore be rewritten as
\begin{eqnarray}
  f_{\mu^+} & = & (1-x_c) \left( x_{\rm flip}\times f_{\bar{b}} + x_{\rm noflip}\times f_b \right) + x_c\times f_{\bar{c}} \nonumber \\ 
  f_{\mu^-} & = & (1-x_c) \left( x_{\rm flip}\times f_b + x_{\rm noflip}\times f_{\bar{b}} \right) + x_c\times f_c ,
\label{eq:example_jetchargecorrelation2_tripletag}
\end{eqnarray}
 Note that the fraction of $c$ ($\bar{c}$) jets contributing to $f_{\mu^+}$ 
($f_{\mu^-}$) are neglected. These contributions are at most of the order of 
$x_c \times x_{\rm flip}$ relative to the flipped muon contributions from direct $B$ 
decays. In addition, the muons from $D$ meson
decays have significantly lower momenta and are less likely to pass the muon momentum
requirement in the dijet samples. Therefore the 
component of $c$ ($\bar{c}$) jets contributing to $f_{\mu^+}$ ($f_{\mu^-}$)
is further suppressed.

Equations~\ref{eq:example_jetchargecorrelation2_tripletag} provides two equations with 
four unknowns ($f_b$, $f_{\bar{b}}$, $f_c$, $f_{\bar{c}}$). To be able to extract these
four jet charge distributions another two equations are needed. This is achieved by using
the loose dijet sample, which has a different fraction of $c$-quark jets, that we denote 
$x'_{c}$ because of the relaxation of the SVT-tag on the {\sl tag}-jet.

In the same fashion as for the tight dijet sample, $f'_{\mu^+}$ and $f'_{\mu^-}$  are the 
jet charge distributions observed for the {\sl probe}-jet in the loose dijet data when 
the tag muon is positive or negative respectively. The equivalent of 
equations~\ref{eq:example_jetchargecorrelation2_tripletag} in the loose dijet sample are:
\begin{eqnarray}
    f'_{\mu^+} & = & (1-x'_c) \left( x'_{\rm flip}\times f_{\bar{b}} + x'_{\rm noflip} \times f_b \right)  + x'_c \times f_{\bar{c}} \nonumber \\
    f'_{\mu^-} & = & (1-x'_c) \left( x'_{\rm flip}\times f_b + x'_{\rm noflip} \times f_{\bar{b}}\right)  + x'_c \times f_c,
    \label{eq:example_jetchargecorrelation2_musvt}
\end{eqnarray}

Using Eq.~\ref{eq:example_jetchargecorrelation2_musvt} 
and~\ref{eq:example_jetchargecorrelation2_tripletag} the system of equation 
can be solved for $f_b$, $f_{\bar{b}}$, $f_c$ and $f_{\bar{c}}$ provided that 
the fraction of $c$-quark jets and the fraction of times the measured tagging muon charge 
is changed with respect to the quark initiating the jet. The procedure to determine the fraction of 
$c$-quark jets in both samples is described in Sec.~\ref{subsec:triple_tag_composition} 
and the calculation of fraction of events with changed  muon charge sign $x_{\rm flip}$ and $x_{\rm flip}^{'}$ is described in 
Sec.~\ref{subsec:muflip}. 

The solutions to Eq.~\ref{eq:example_jetchargecorrelation2_tripletag} 
and are~\ref{eq:example_jetchargecorrelation2_musvt}:

\begin{eqnarray}
  \fbb & = & \frac{ \fpmum \xc - \fpmum \xc \xf -\fpmup \xc \xf -\fmum \xpc  + \fmum \xf \xpc +\fmup \xf \xpc}{(-1+2\xf)(\xpc-\xc)}   \nonumber \\
  \fb  & = & \frac{ \fpmup \xc - \fpmum \xc \xf -\fpmup \xc \xf -\fmup \xpc  + \fmum \xf \xpc +\fmup \xf \xpc}{(-1+2\xf)(\xpc-\xc)}   \nonumber \\
  \fcb & = & \frac{-\fmup + \fpmup -\fpmup \xc + \fmup \xpc}{\xpc - \xc}   \nonumber\\
  \fc  & = & \frac{-\fmum + \fpmum -\fpmum \xc + \fmum \xpc}{\xpc - \xc}   \label{eq:sol} 
\end{eqnarray}

\subsection{Fraction of $c$-Quark Jets in the Dijet Samples}
\label{subsec:triple_tag_composition}

The difference in mass between the $b$-quark, $c$-quark and 
light quarks implies that the momentum distribution of muons within jets from 
semi-leptonic decays of $B$-, $D$- and light mesons are different. This property 
is utilized to extract the respective fraction of $b$-quark, $c$-quark and light 
jets in the dijet samples which were used to extract the jet charge 
distributions for $b$- and $c$-quark jets above.

The expected \ptrel~spectra from $b$- and $c$-quark jets found from Monte Carlo 
simulation are used to fit the observed \ptrel~spectra in the dijet samples. The 
light flavor contribution is assumed to be negligible. This is also confirmed 
a posteriori by the fact that the fitted fraction of $c$-quark jets is small and 
that the light jet tagging efficiency is $\approx15$ times lower than than the 
$c$-quark tagging efficiency (see Sec.~\ref{sec:bjets}). 

The expected $b$-quark jet \ptrel~spectrum is determined using simulated 
\Zbbbar~events including cascade decays ($b\rightarrow (c\rightarrow)~\mu$) 
and the expected $c$-quark jet \ptrel~spectrum is determined from simulated 
\Zccbar~events. \ptrel~templates in three bins of muon $p_T$: $4$~GeV$<p_{T\mu}<8$~GeV, 
$8$~GeV$<p_{T\mu}<10$~GeV and $p_{T\mu} >10$~GeV and in three bins of jet 
$p_T$: $15$~GeV$<p_T<35$~GeV, $35$~GeV$<p_T<55$~GeV and $p_T>55$~GeV are constructed. 
The jet flavor in the Monte Carlo simulated events is determined by matching the direction of the 
reconstructed jet to the heaviest hadron flavor within a cone of $\Delta R<0.5$. 
If there is more than one hadron found within the cone, the jet is considered to 
be a $b$-quark jet if the cone contains at least one $B$ hadron. It is called a 
$c$-quark jet if there is at least one $D$ meson in the cone and no $B$ hadron. 
The fitted parameter is the fraction of $c$-quark jets among the {\sl tag}-jets 
of the dijet samples.

\begin{figure}[]
  \centering
  \includegraphics[width=0.6\textwidth]{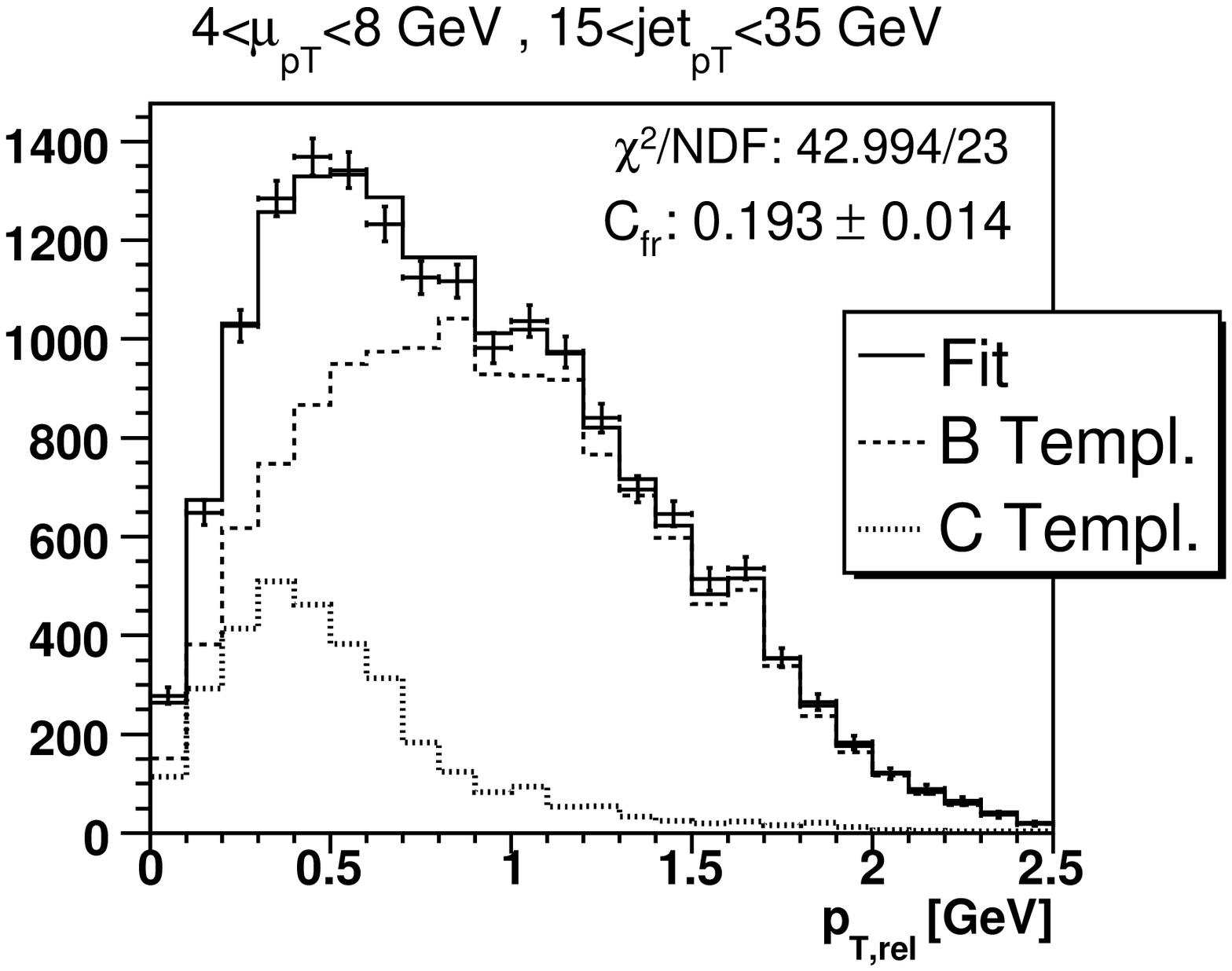}
  \includegraphics[width=0.6\textwidth]{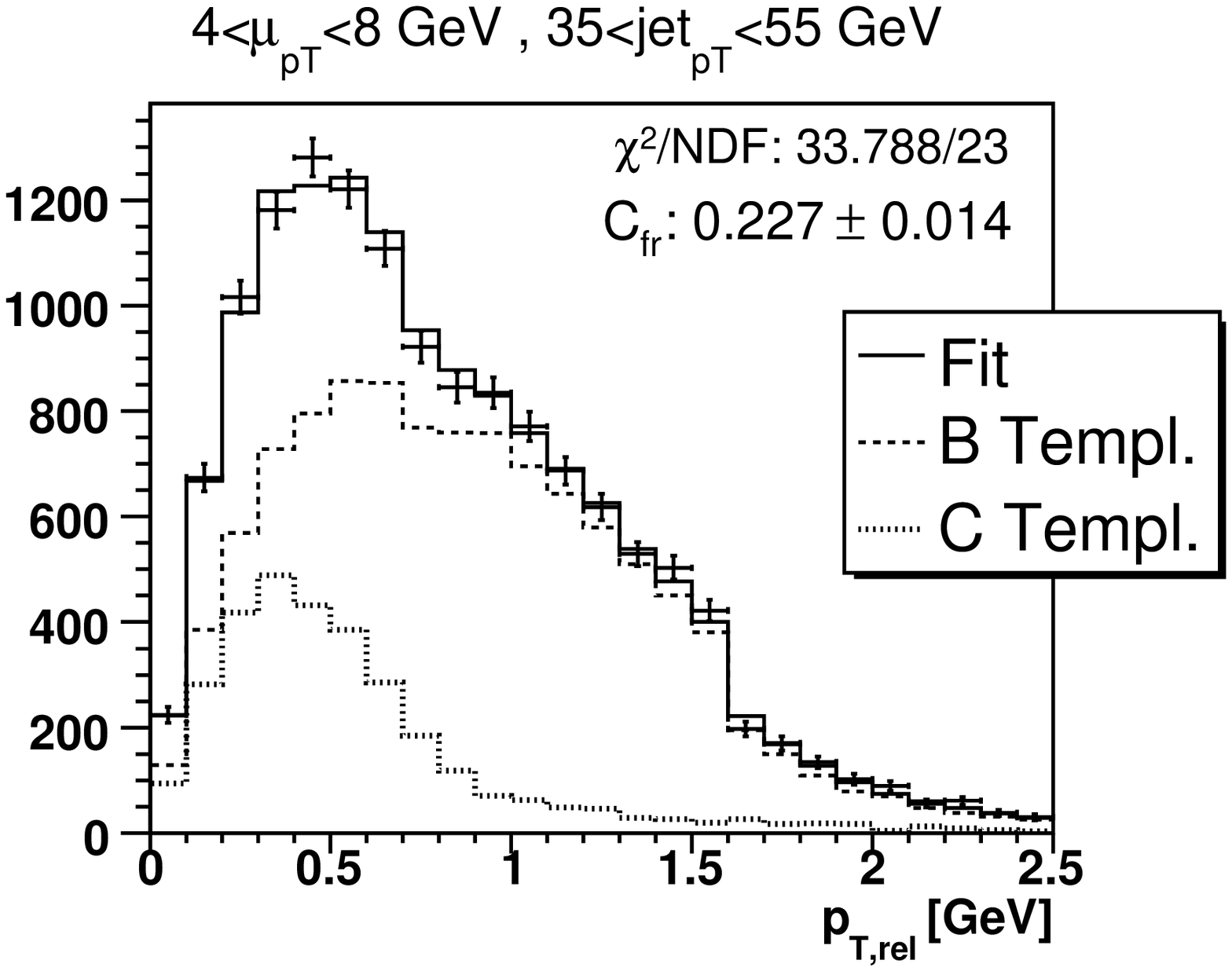}
  \includegraphics[width=0.6\textwidth]{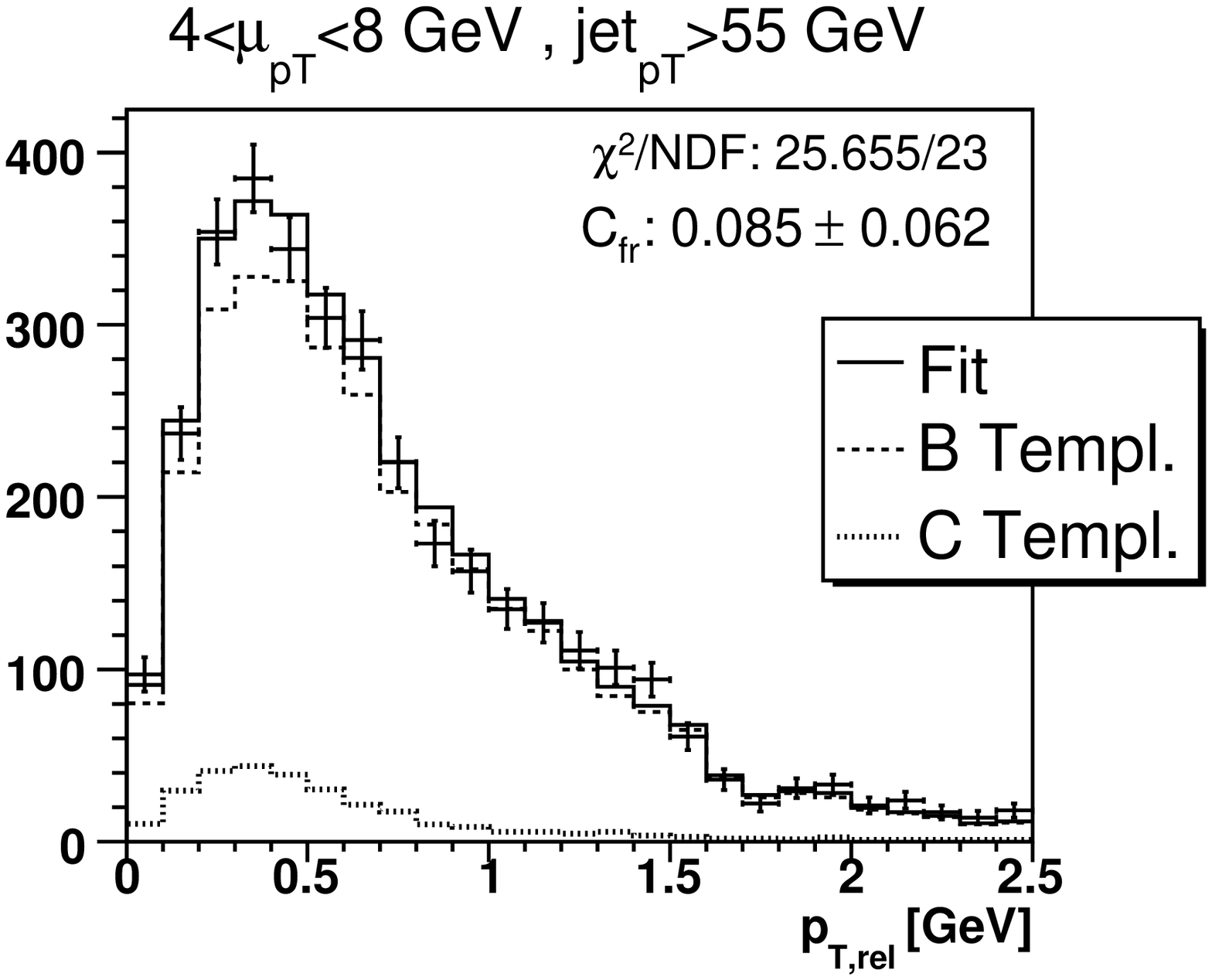}
  \caption{The \ptrel~template fit in the loose dijet sample for muon $p_T$ between $4$~GeV 
  and $8$~GeV and three bins of jet $p_T$. Similarly for the other two muon $p_T$ bins.}
  \label{fig:bfraction_template_doubletag}
\end{figure}

\begin{figure}[]
  \centering
  \includegraphics[width=0.6\textwidth]{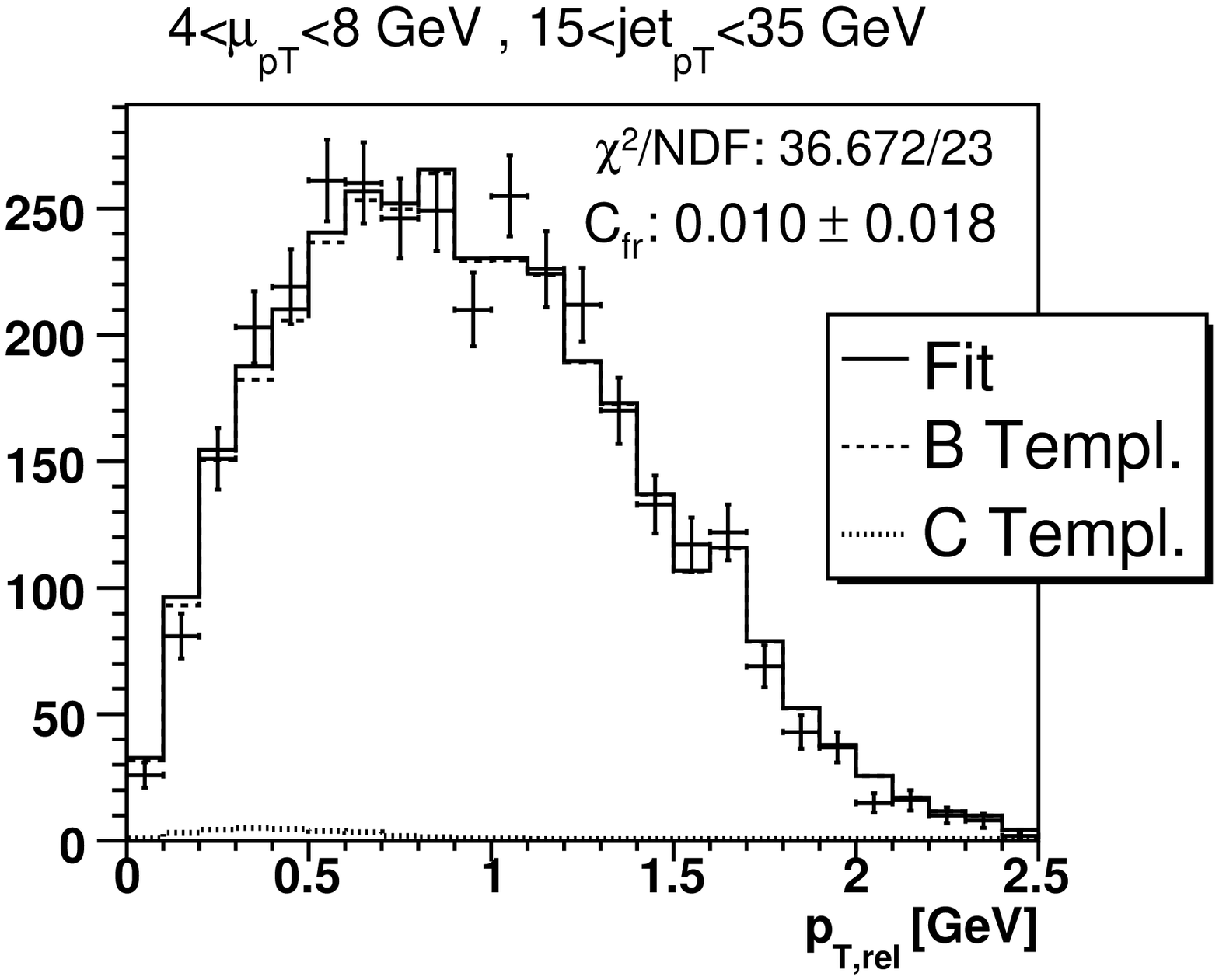}
  \includegraphics[width=0.6\textwidth]{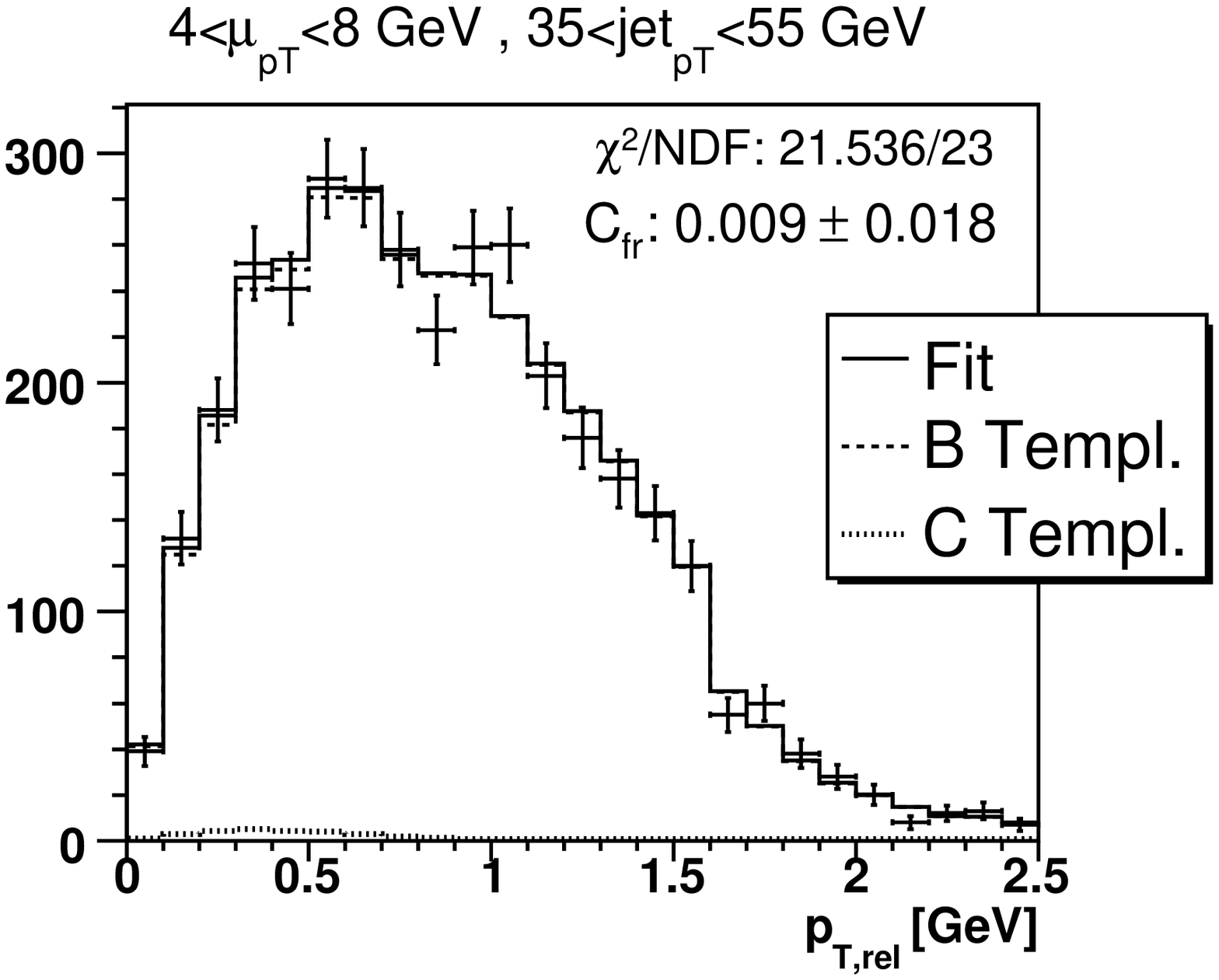}
  \includegraphics[width=0.6\textwidth]{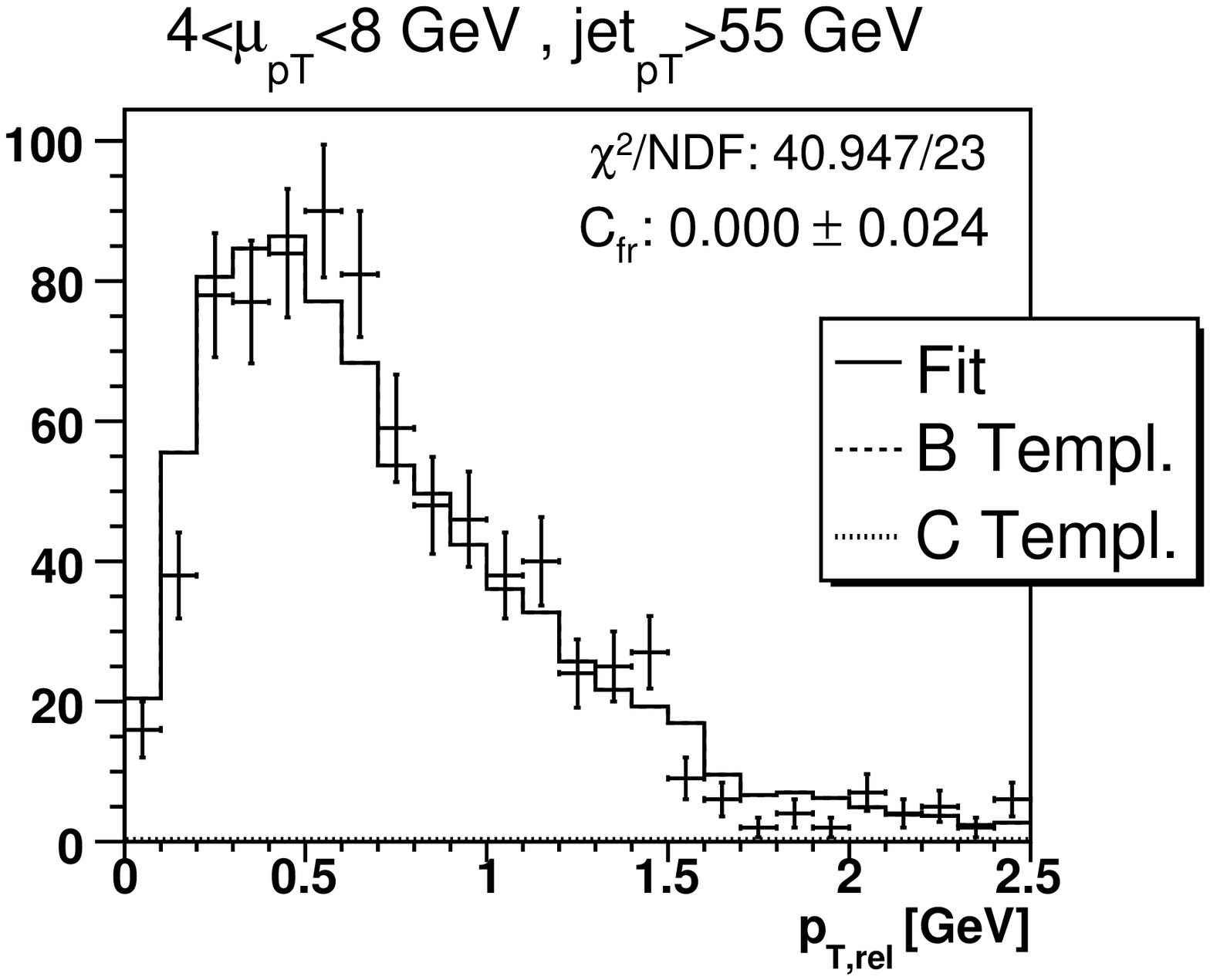}
  \caption{The \ptrel~template fit in the tight dijet sample for muon $p_T$ between $4$~GeV 
  and $8$~GeV and three bins of jet $p_T$. Similarly for the other two muon $p_T$ bins.}
  \label{fig:bfraction_template_tripletag}
\end{figure}

The \ptrel~template fits are shown in Fig.~\ref{fig:bfraction_template_doubletag} 
and Fig.~\ref{fig:bfraction_template_tripletag} for the loose and tight dijet 
samples respectively.
The fitted fractions of $b$- and $c$-quark jets are presented in 
Tab.~\ref{tab:cfraction_doubletag} and~\ref{tab:cfraction_tripletag} respectively. The 
fit also takes into account the statistical uncertainty on the \ptrel~spectra 
obtained from simulated events as well as the uncertainty from the observed spectrum. 
\begin{table}
\caption{Result of the \ptrel~fit for the loose dijet sample for the nine different 
combinations of muon $p_T$ and jet $p_T$.}
\begin{center}
\begin{tabular}{ccccccc} \\ \hline
   Muon $p_T$ & Jet $p_T$ & Fitted $c$-fraction
   \\ \hline
   $4 < p_{T\mu} < 8 $ GeV & $15 < p_T < 35 $ GeV  & 0.19 $\pm$ 0.01
   \\ 
   $4 < p_{T\mu} < 8 $ GeV & $35 < p_T < 55 $ GeV  & 0.23 $\pm$ 0.01
   \\ 
   $4 < p_{T\mu} < 8 $ GeV & $p_T > 55 $ GeV       & 0.09 $\pm$ 0.06
   \\
   $8 < p_{T\mu} < 10 $ GeV & $15 < p_T < 35 $ GeV  &  0.15 $\pm$ 0.03
   \\ 
   $8 < p_{T\mu} < 10 $ GeV & $35 < p_T < 55 $ GeV  & 0.24 $\pm$ 0.02
   \\ 
   $8 < p_{T\mu} < 10 $ GeV & $p_T > 55 $ GeV       & 0.07 $\pm$ 0.06
   \\ 
   $p_{T\mu} > 10 $ GeV & $15 < p_T < 35 $ GeV      & 0.18 $\pm$ 0.04
   \\ 
   $p_{T\mu} > 10 $ GeV & $35 < p_T < 55 $ GeV      & 0.21 $\pm$ 0.02
   \\ 
   $p_{T\mu} > 10 $ GeV & $p_T > 55 $ GeV  & 0.14 $\pm$ 0.03
\\   \hline
   Weighted sample average:    &                   & 0.19 $\pm$ 0.02
\\   \hline
   \end{tabular} 
\label{tab:cfraction_doubletag}
\end{center}
\end{table}
\begin{table}
\caption{Result of the \ptrel~fit for the tight dijet sample for the nine different 
combinations of muon $p_T$ and jet $p_T$.}
\begin{center}
\begin{tabular}{ccccccc} \\ \hline
   Muon $p_T$ & Jet $p_T$ & Fitted $c$-fraction
   \\ \hline
   $4 < p_{T\mu} < 8 $ GeV & $15 < p_T < 35 $ GeV  & 0.01 $\pm$ 0.02
   \\ 
   $4 < p_{T\mu} < 8 $ GeV & $35 < p_T < 55 $ GeV  & 0.01$\pm$ 0.02
   \\ 
   $4 < p_{T\mu} < 8 $ GeV & $p_T > 55 $ GeV       & 0.00 $\pm$ 0.02
   \\
   $8 < p_{T\mu} < 10 $ GeV & $15 < p_T < 35 $ GeV  & 0.01 $\pm$ 0.04
   \\ 
   $8 < p_{T\mu} < 10 $ GeV & $35 < p_T < 55 $ GeV  & 0.02 $\pm$ 0.03
   \\ 
   $8 < p_{T\mu} < 10 $ GeV & $p_T > 55 $ GeV       & 0.00 $\pm$ 0.03
   \\ 
   $p_{T\mu} > 10 $ GeV & $15 < p_T < 35 $ GeV      & 0.02$\pm$ 0.05
   \\ 
   $p_{T\mu} > 10 $ GeV & $35 < p_T < 55 $ GeV      & 0.01 $\pm$ 0.02
   \\ 
   $p_{T\mu} > 10 $ GeV & $p_T > 55 $ GeV           & 0.000 $\pm$ 0.004
\\   \hline

   Weighted sample average:    &                   & 0.01 $\pm$ 0.02
\\   \hline
   \end{tabular} 
\label{tab:cfraction_tripletag}
\end{center}
\end{table}

The total fraction of of $c$- and $b$-quark jets in the dijet samples 
can be extracted by the sum of the measured fraction in each bin 
weighted by the event population in each bin. As expected, removing 
the requirement of a SVT-tag on the {\sl tag}-jet enhances the fraction of 
$c$-quark jets.


\subsection{Determination of the Tagging Muon Charge Flip Fraction}
\label{subsec:muflip}

\begin{figure}[]
  \centering
  \includegraphics[width=0.8\textwidth]{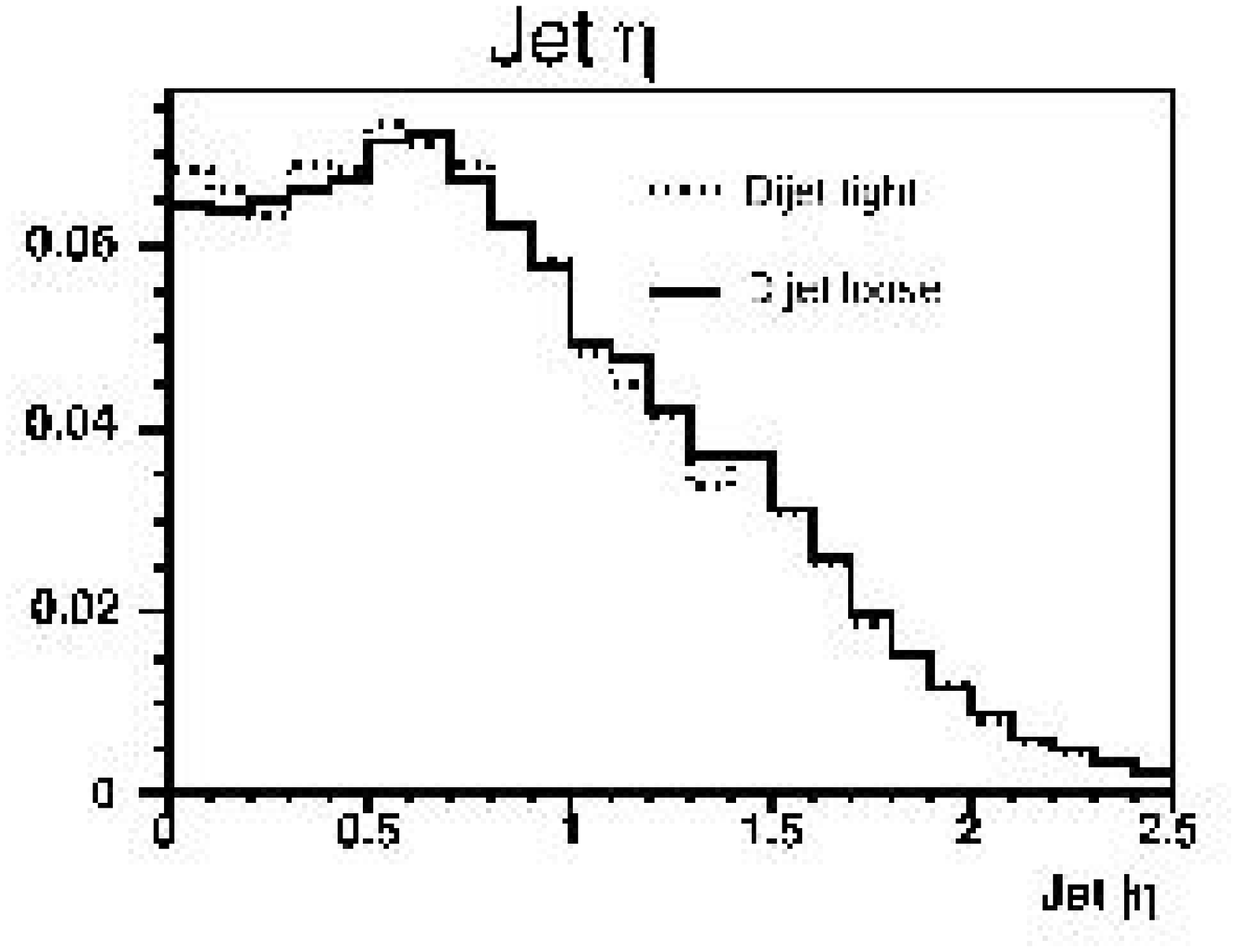}
  \includegraphics[width=0.8\textwidth]{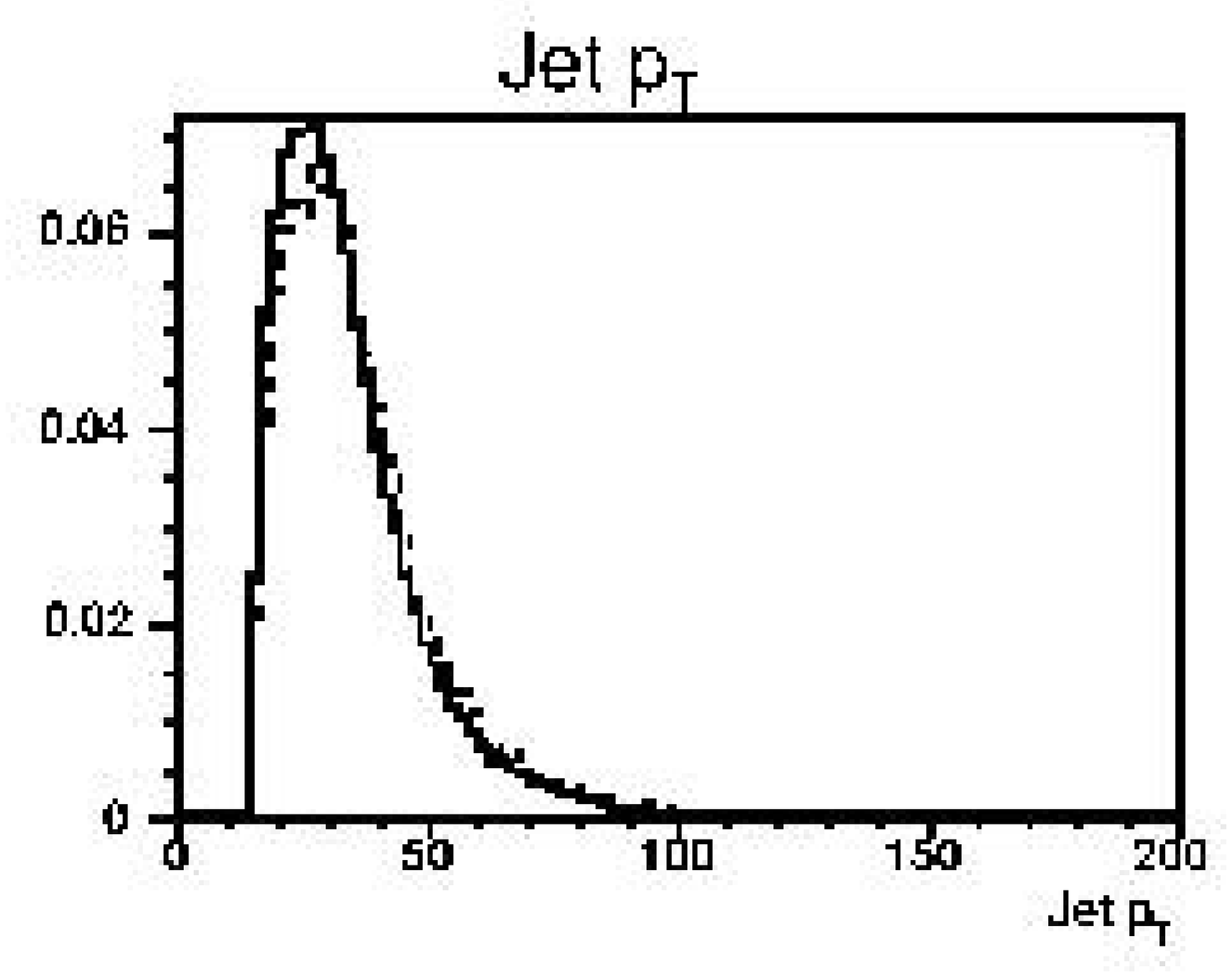}
  \caption{Comparison of jet $\eta$ (top) and $p_T$ (bottom) 
  for {\sl tag}-jets in the loose (solid) and tight (dashed) dijet sample.}
  \label{fig:tripletag_musvt_kincomp}
\end{figure}

The total fraction of times the tagging muon charge sign is ``flipped'' compared to 
the quark that initiated the jet in the tight (loose) dijet 
sample is given by the variable $x_{\rm flip}$ ($x'_{\rm flip}$). The value of 
$x_{\rm flip}$ is measured from simulated \bbbar~events 
where the Monte Carlo truth information allows for a determination of the 
correlation between the tagging muon and the quark sign. The 
fraction of times the tagging muon changes sign depends on 
the $p_T$ spectrum of the jets. At lower jet $p_T$, the probability 
that a cascade muon is found above the reconstruction threshold is lower and the 
fraction of cascade muons thus increases as a function of increasing 
jet $p_T$. The fraction of times the tagging muon 
changes sign determined from simulated events is therefore weighted to 
the $p_T$ spectrum observed in the tight dijet sample. The result 
is shown in Tab~\ref{tab:muflip}. 
\begin{table}
\caption{Fraction of times the tagging muon changes sign 
$x_{\rm flip}$ in four different bins of {\sl tag}-jet $p_T$. The value of 
$x_{\rm flip}$ plugged into Eq.~\ref{eq:sol} is the weighted 
average to take into account the {\sl tag}-jet $p_T$ spectra of the dijet 
samples.}
\begin{center}
\begin{tabular}{ccccc} \\ \hline
   Jet $p_T$ & $x_{\rm flip}$
   \\ \hline
   $15 < p_T < 25$~GeV      & $0.257 \pm 0.064$
   \\ 
   $25 < p_T < 35$~GeV      & $0.290 \pm 0.013$
   \\ 
   $35 < p_T < 45$~GeV      & $0.305 \pm 0.009$ 
   \\ 
   $p_T > 45$~GeV  & $0.317 \pm 0.009$
   \\   \hline
   Weighted sample average    &                    $0.303 \pm 0.006$ &
   \\   \hline
   \end{tabular} 
\label{tab:muflip}
\end{center}
\end{table}
Any topological differences between the loose and tight dijet 
sample such as the jet $p_T$ spectrum could lead to a 
difference between $x_{\rm flip}$ and $x'_{\rm flip}$. Therefore, 
the {\sl tag}-jet $p_T$ and $\eta$ spectra are compared (see 
Fig.~\ref{fig:tripletag_musvt_kincomp}) but no significant 
differences were found and $x_{\rm flip}=x'_{\rm flip}$ is thus used.


\subsubsection{Cross-check of tagging muon charge flip}
\label{subsec:crosscheckmuflip}
The fraction of times the tagging muon in the loose and tight dijet samples 
changes sign calculated above can be cross-checked with data by 
requiring that the {\sl probe}-jet in the tight dijet sample contain a muon 
track (with the same quality as the tagging muon). If $N_{b\bar{b}}$ 
is the total number of events in this sample, $N_{os}$ is the number 
of events with two opposite sign muons and $N_{ss}$ the number 
of events with two same sign muons, the 
following relationships can be used to predict the number of $N_{ss}$ and 
$N_{os}$ in this sample given the flip fraction estimated above, 
\begin{equation}
N_{b\bar{b}} = N_{ss} + N_{os}
\label{eq:quadfr}
\end{equation}
\begin{equation}
N_{ss} = 2N_{b\bar{b}}x_{\rm flip}(1-x_{\rm flip}).
\label{eq:quadfr2}
\end{equation}
The prediction gives $N_{ss}=76\pm6$ and $N_{os}=105\pm15$ while the observation 
is $N_{ss}=79$ and $N_{os}=103$ which is consistent within the statistical 
uncertainty. The caveat of this cross-check is that a rather broad range of 
$x_{\rm flip}$ is allowed before conflicting with the statistical uncertainties.

\subsection{Correction for Kinematical Differences in the Signal and Dijet Samples}
\label{subsec:kincorr}
\begin{figure}[]
  \centering
  \includegraphics[width=0.7\textwidth]{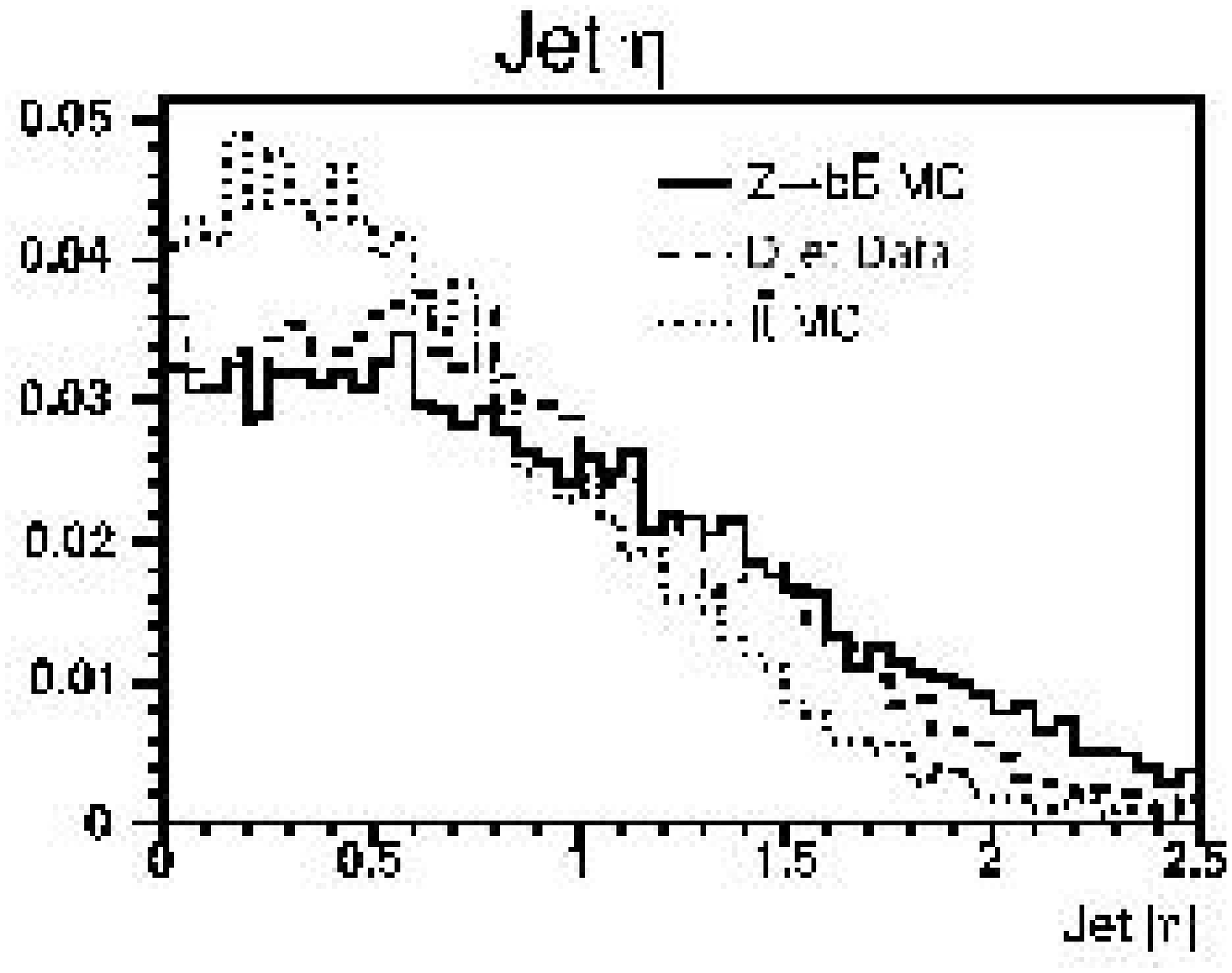}
  \includegraphics[width=0.7\textwidth]{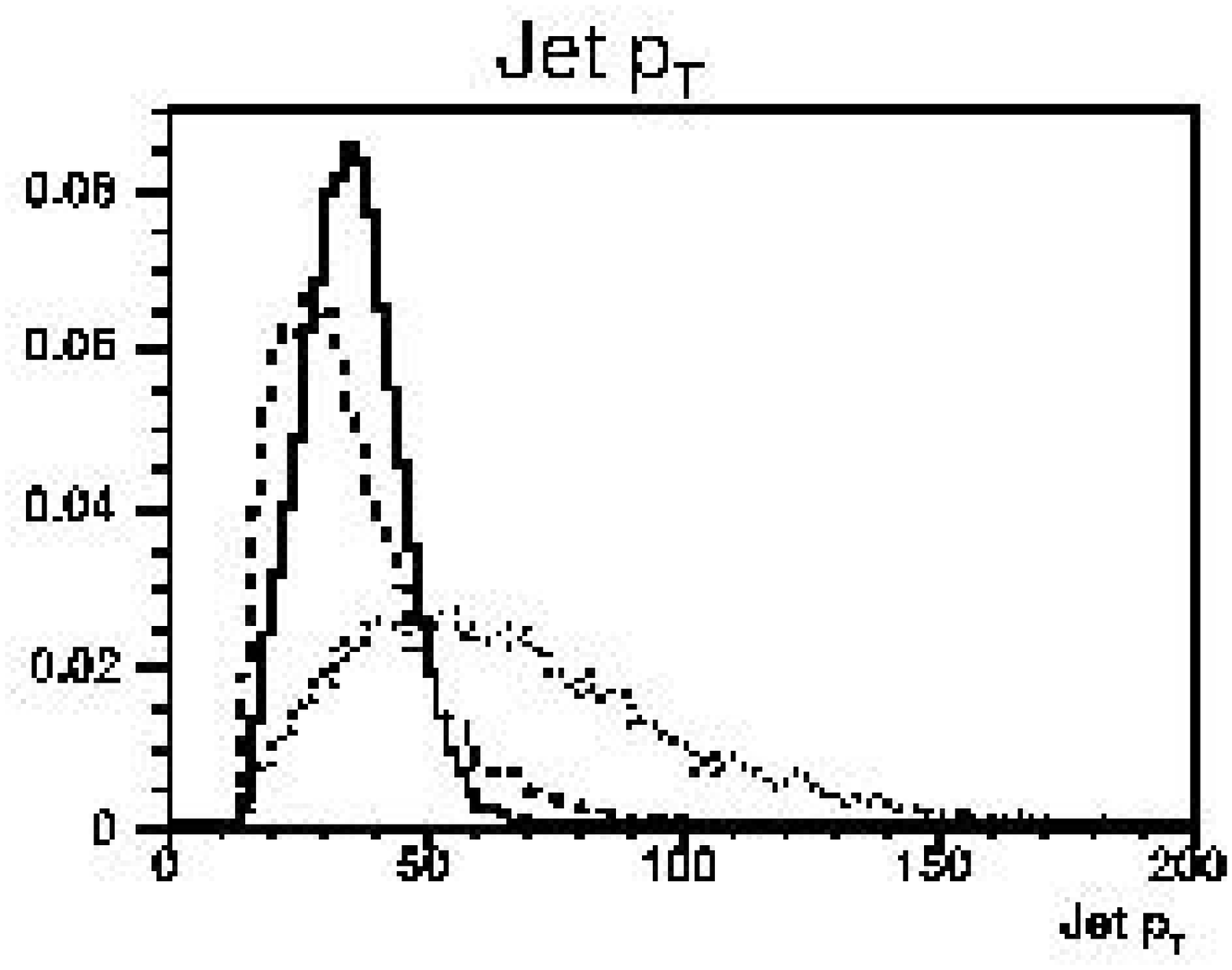}
\caption{Jet $\eta$ (top) and $p_T$ (bottom) for the {\sl probe}-jet in 
the tight dijet sample, in simulated \Zbbbar~events and for SVT-tagged 
jets in simulated $\ttbar \to \ell\text{+jets}$ events.}
\label{fig:svtmethodkinprop_data_MC}
\end{figure}

\begin{figure}[]
  \centering
  \includegraphics[width=1.0\textwidth]{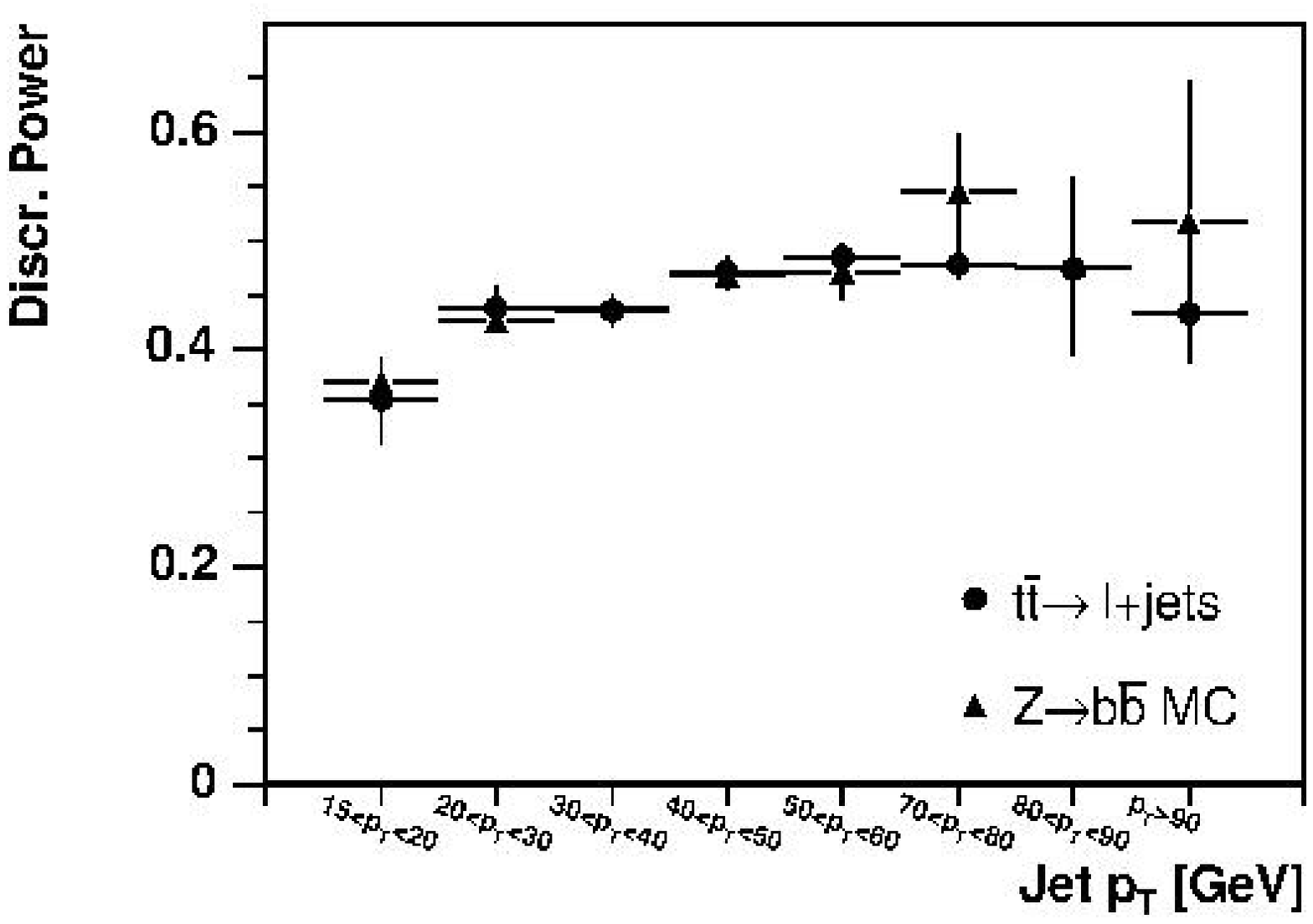}
  \includegraphics[width=1.0\textwidth]{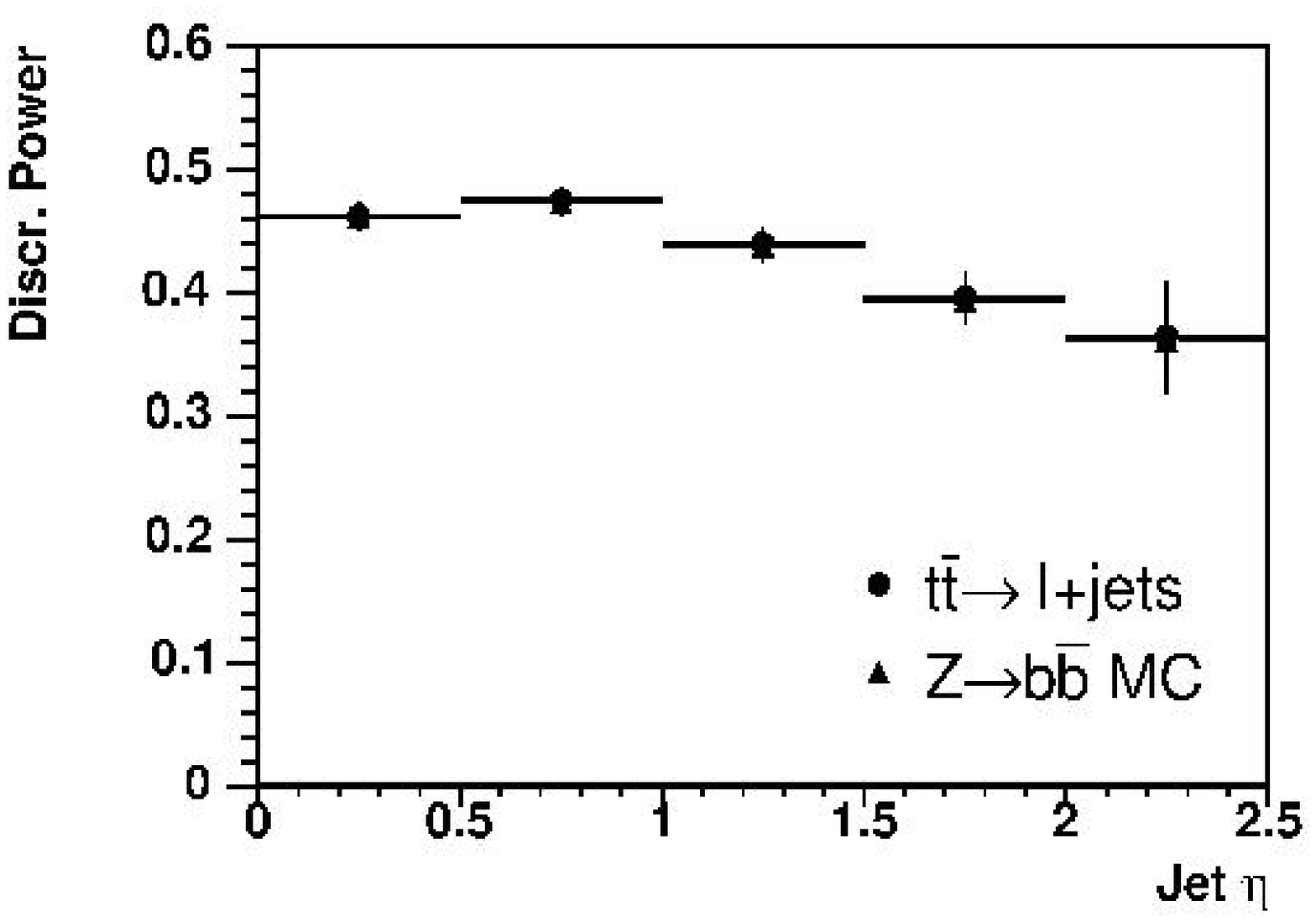}
  \caption{Discriminating power for SVT-tagged $b$-quark jets as a function of 
  jet $p_T$ (top) and $\eta$ (bottom) for simulated $\ttbar \to \ljets$ (circles) and 
  \Zbbbar~events (triangles).}
  \label{fig:dp_ttbar_bbbar}
\end{figure}

The jet charge templates were derived from the dijet data samples 
above. The aim is to extract the jet charge templates for $b$- and $c$-quark jets 
in $\ttbar \to \ell\text{+jets}$~events. Figure~\ref{fig:svtmethodkinprop_data_MC} 
shows a comparison of the jet $p_T$ and $\eta$ for the {\sl probe}-jet 
in the tight dijet sample and the SVT-tagged $b$-quark jets in 
simulated $\ttbar \to \ljets$ events. The $b$-quark jet from the top 
quark decay has as expected harder $p_T$ and more central $\eta$ spectra. 
From the correlation between jet $p_T$ and the number of tracks associated 
with the jet one can expect that the discriminating 
power depends on the jet $p_T$. A dependence on jet $|\eta|$ is also expected 
due to the geometry of the inner tracking detector where particles around 
$|\eta|=1.0$ traverses more layers in the SMT ( a similar behavior exists 
for the tracking efficiency in Fig.~\ref{fig:btag_eff} in Sec.~\ref{sec:bjets}). 
The jet $p_T$ and $|\eta|$ 
dependences of the jet charge algorithm are shown in Fig.~\ref{fig:dp_ttbar_bbbar}. 
As the jet charge 
algorithm performance is improving with increasing jet $p_T$ and mostly decreasing 
with increasing $|\eta|$ the conclusion is that the jet charge templates 
derived on the dijet samples are underestimating the jet charge algorithm 
discriminating power when applied to $b$-quark jets in $\ttbar \to \ljets$ events.

This section describes how to correct for this by weighting simulated \ttbar~events 
to have the same $b$-quark jet $p_T$ and $\eta$ spectrum as the dijet samples. A 
correction function is then extracted based on the deviation 
of the weighted compared to the unweighted jet charge templates. This correction 
is subsequently applied to the jet charged templates derived from the dijet samples 
to extract the expected jet charge templates for $b$- and $c$-quark jets in 
$\ttbar \to \ljets$ events.

\subsubsection{Jet Kinematical Weighting}
\label{subsubsec:kincorr_der}
\begin{figure}[]
  \centering
  \includegraphics[width=1.0\textwidth]{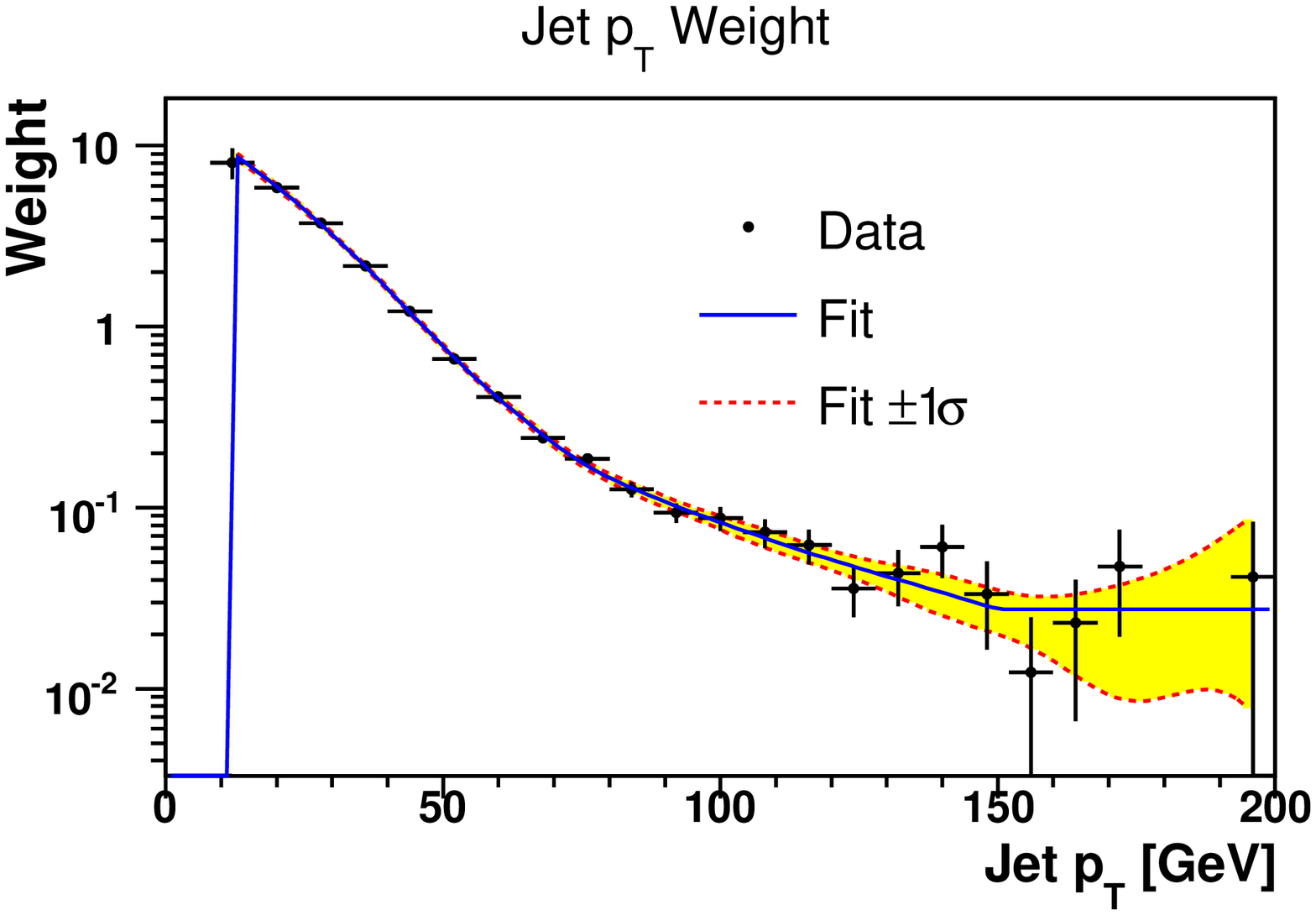}
  \caption{The weight applied to the $b$-quark jet $p_T$ spectrum of 
  simulated $\ttbar \to \ell\text{+jets}$ events to obtain the same $p_T$ spectrum as the 
  {\sl probe}-jet in the tight dijet sample.}
  \label{fig:ptreweightstep}
\end{figure}

\begin{figure}[]
  \centering
  \includegraphics[width=0.8\textwidth]{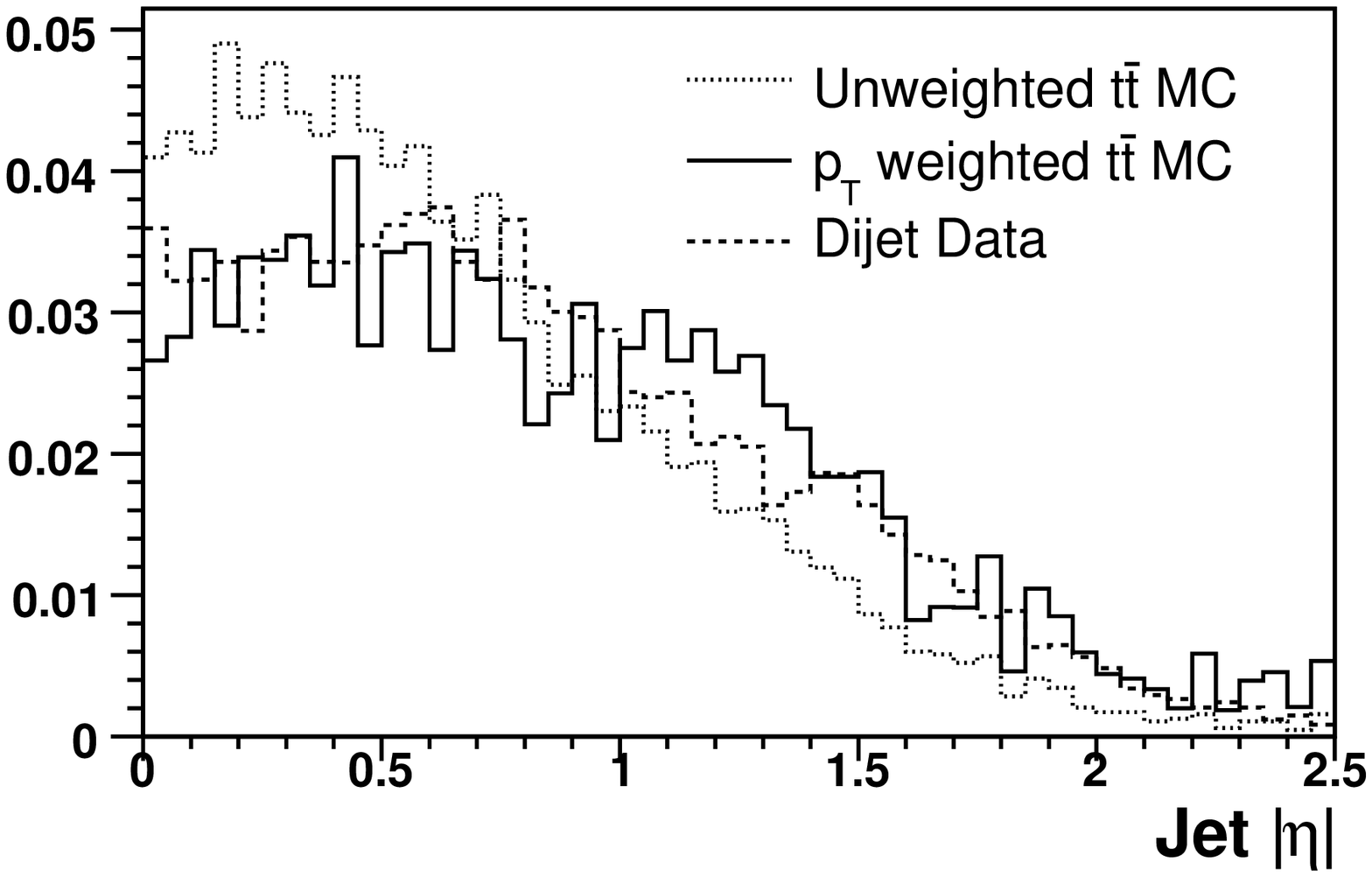}
  \includegraphics[width=0.8\textwidth]{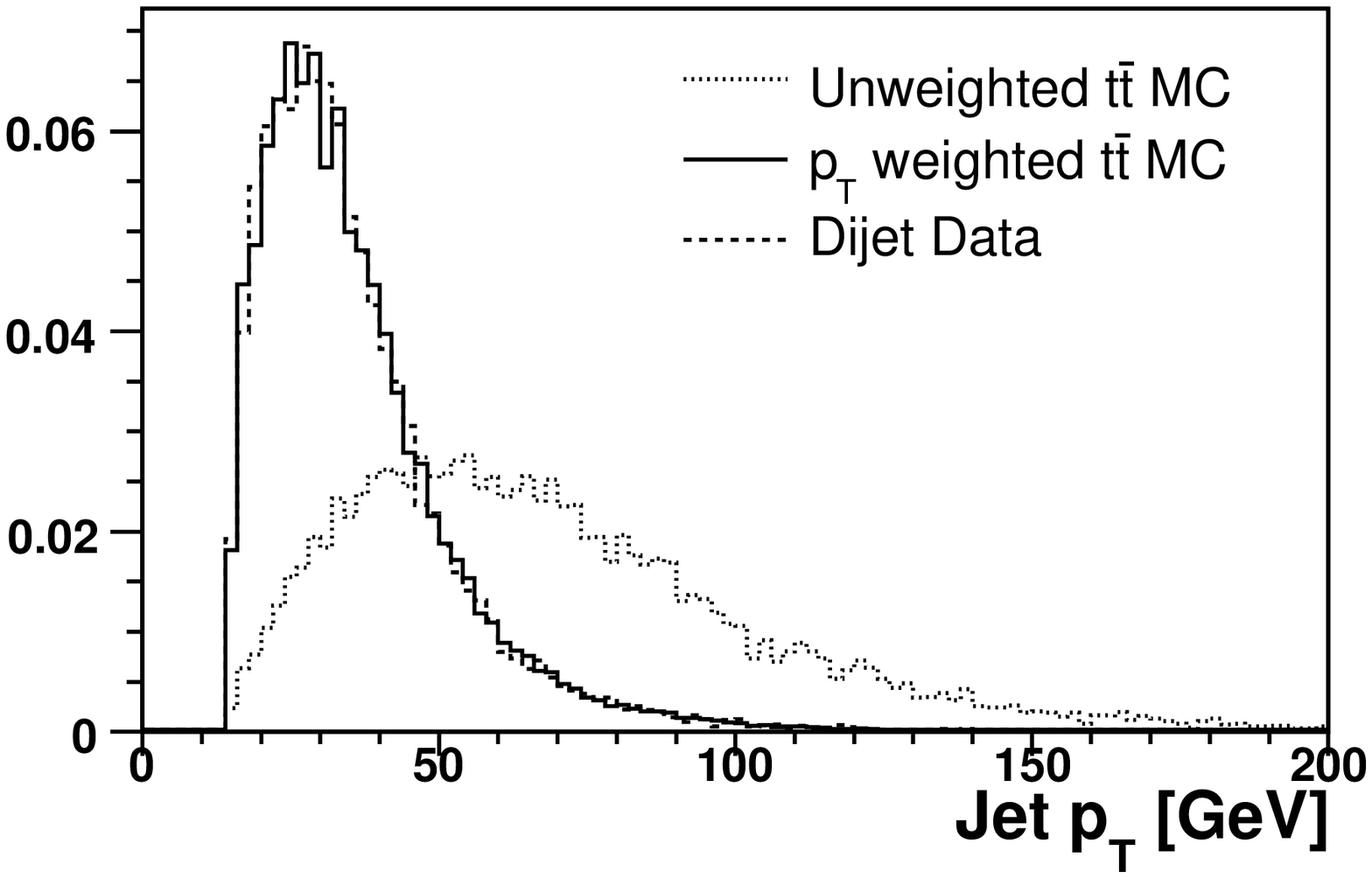}
  \caption{Comparison of jet $\eta$ (top) and $p_T$(bottom) for SVT-tagged 
  $b$-quark jets in simulated $\ttbar \to \ljets$ events passing signal 
  selections before and after weighting of the $p_T$ spectrum. The $p_T$ spectrum 
  of {\sl probe}-jets in the tight dijet sample is shown for comparison.}
  \label{fig:ptreweightstep_comp}
\end{figure}

\begin{figure}[]
  \centering
  \includegraphics[width=0.9\textwidth]{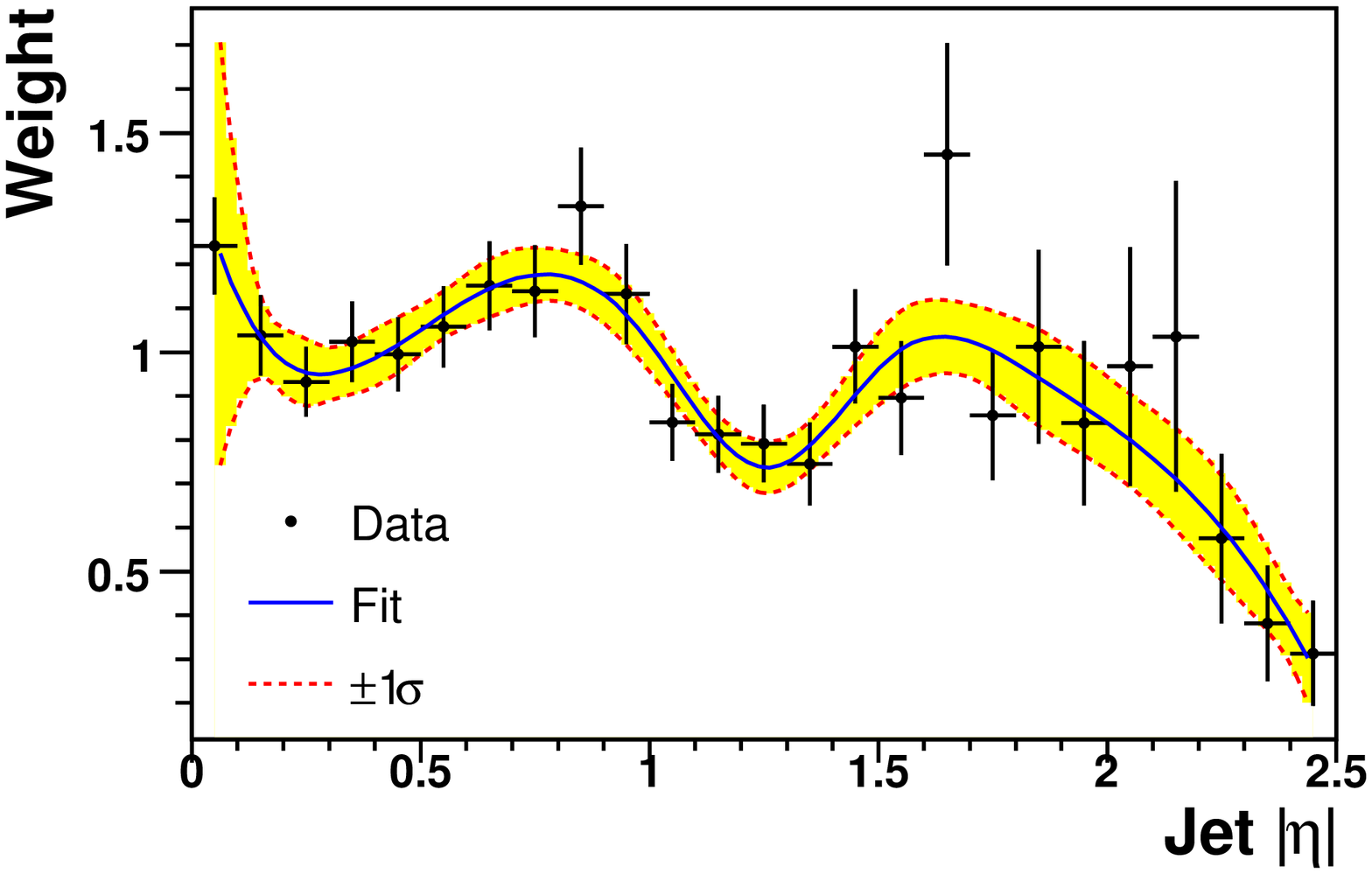}
  \caption{The $\eta$ weight applied to the $p_T$ weighted $b$-quark jets in 
  simulated $\ttbar \to \ljets$ events.}
  \label{fig:eta_ptreweightstep}
\end{figure}

\begin{figure}[]
  \centering
  \includegraphics[width=0.8\textwidth]{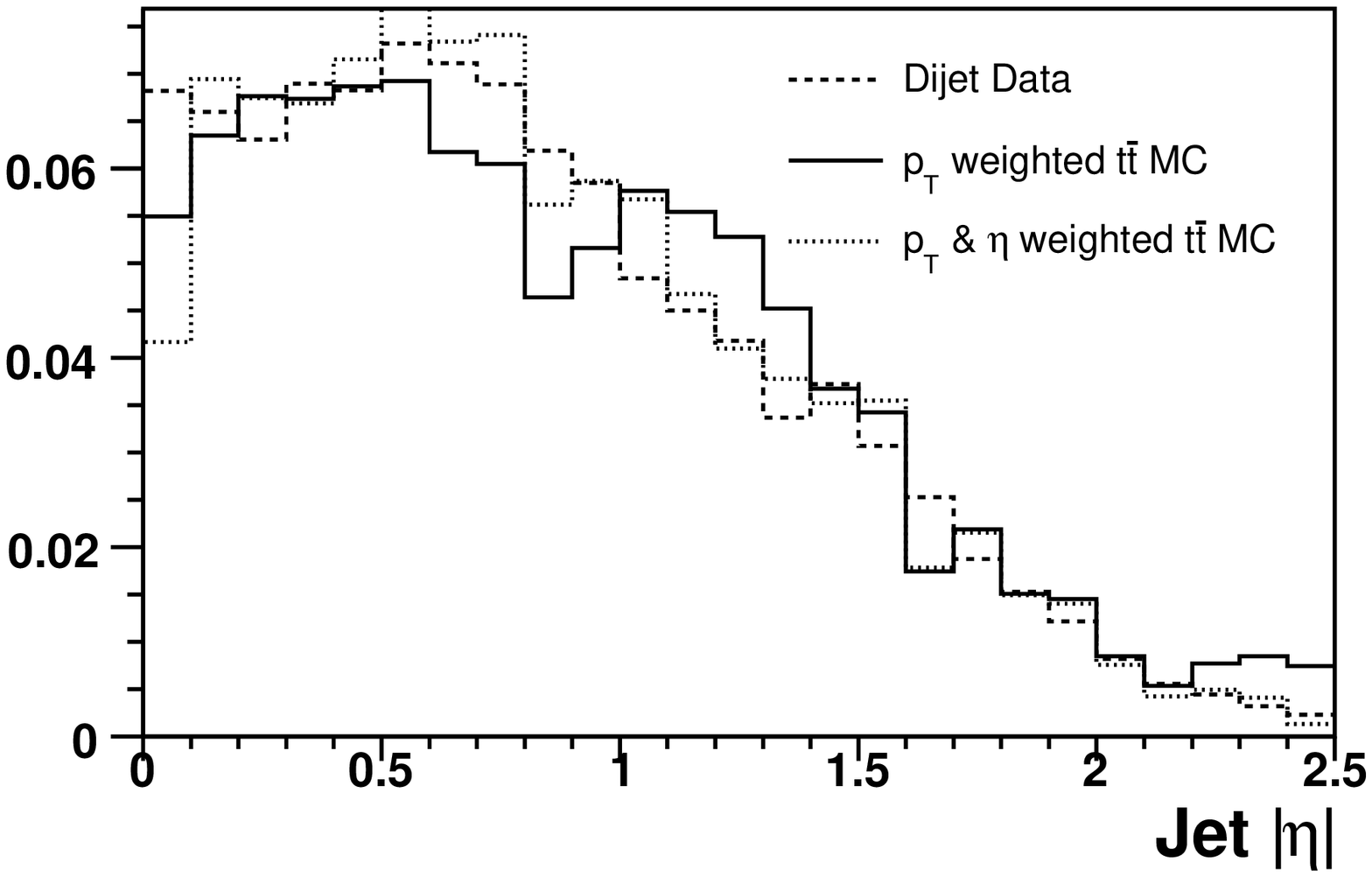}
  \includegraphics[width=0.8\textwidth]{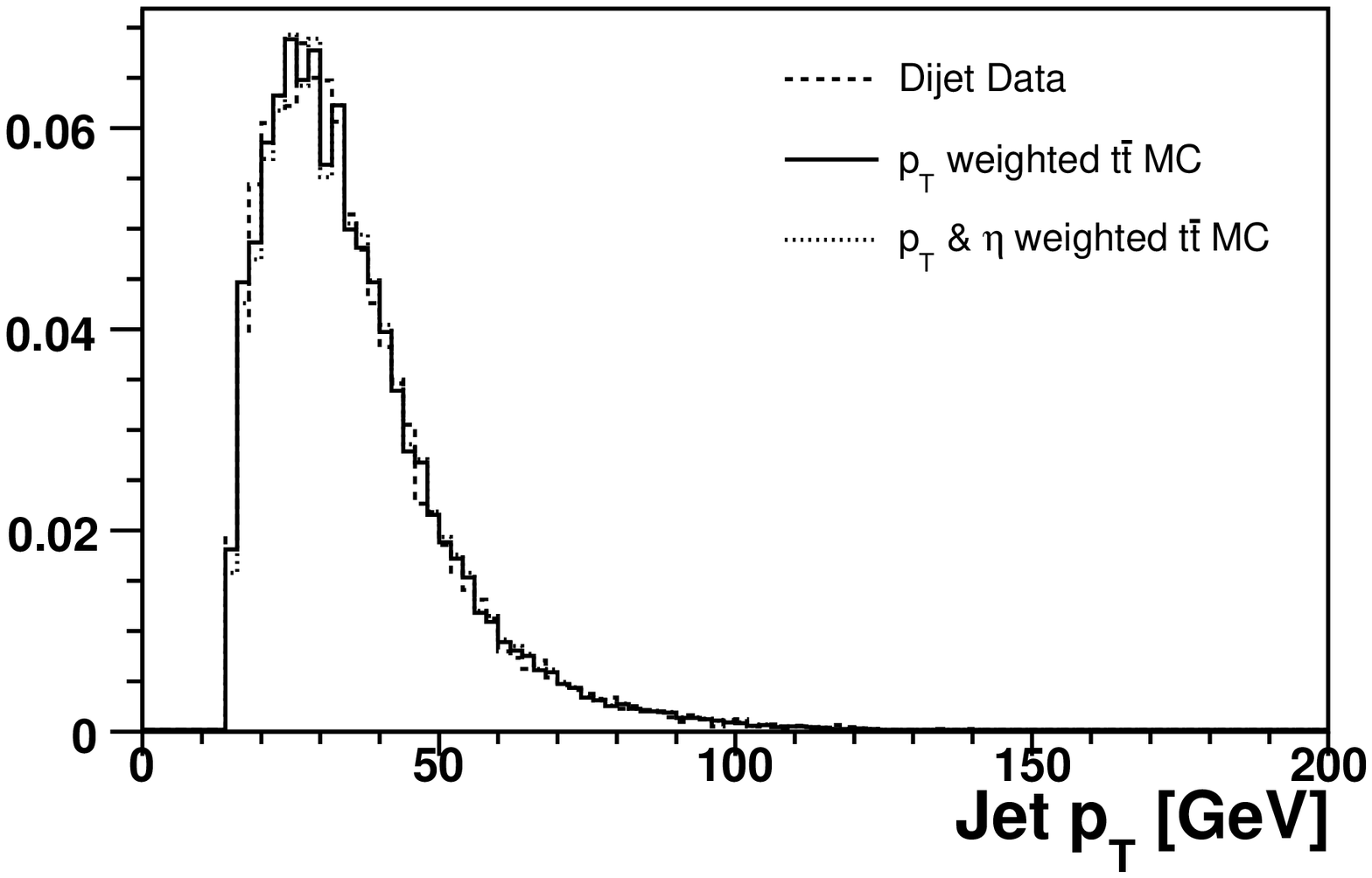}
  \caption{Comparison of jet $\eta$ (top) and $p_T$(bottom) for 
  $b$-quark jets in simulated $\ttbar \to \ljets$ events passing signal 
  selections before and after weighting. The {\sl probe}-jet 
  $p_T$ and $\eta$ spectrum in the tight dijet sample is shown for comparison.}
\label{fig:eta_ptreweightstep_comp}
\end{figure}

To fully take into account the kinematical differences for $b$-quark jets in 
the dijet samples and $\ttbar \to \ell\text{+jets}$ both the difference in 
jet $p_T$ and $\eta$ and their correlations have to considered.
This can be achieved by weighting the $b$-quark jet $p_T$ spectrum of 
$\ttbar \to \ljets$ to the {\sl probe}-jet $p_T$ spectrum in the 
tight dijet sample (this direction of weighting is the only possible due to 
the low statistics at high $p_T$ in the dijet samples). 
Figure~\ref{fig:ptreweightstep} shows the weight and the result is shown 
in Fig.~\ref{fig:ptreweightstep_comp}.
The weighting of the $p_T$ spectrum does not take into account the differences 
in $\eta$ spectrum between the two samples. This difference is treated in a similar 
fashion. The simulated $\ttbar \to \ell\text{+jets}$ events are weighted  
again but this time with respect to the $b$-quark jet $\eta$ spectrum. The weight as a 
function of $\eta$ is shown in Fig.~\ref{fig:eta_ptreweightstep}. 
After applying both weights the jet $p_T$ and $\eta$ spectrum of 
the different samples agree as can be seen in 
Fig.~\ref{fig:eta_ptreweightstep_comp}, confirming the validity of the 
assumption of uncorrelated $p_T$ and $\eta$.

\subsubsection{Derivation of the Kinematical Correction}
\label{subsubsec:kincorr_der_func}

\begin{figure}[]
  \centering
  \includegraphics[width=1.0\textwidth]{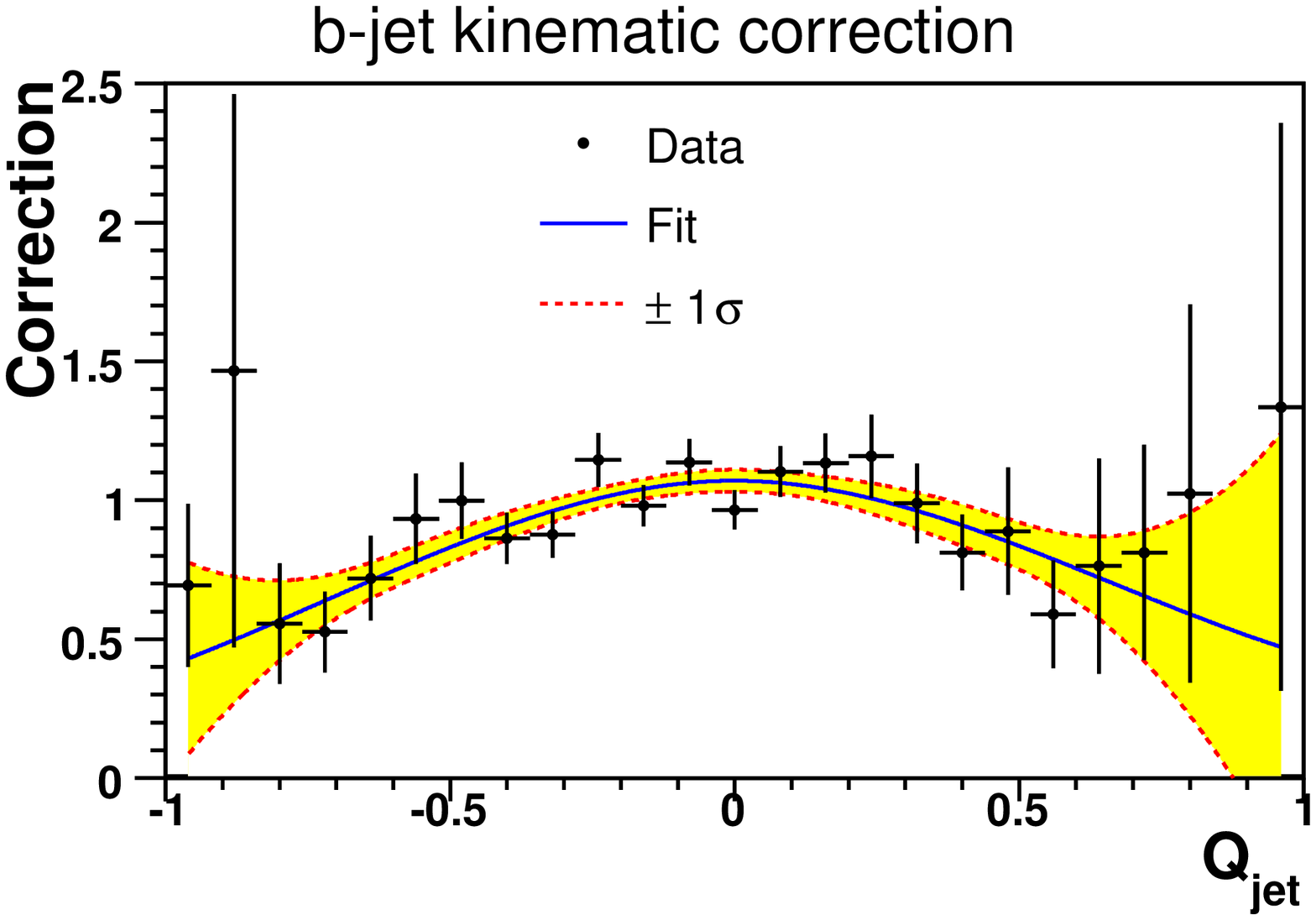}
  \includegraphics[width=1.0\textwidth]{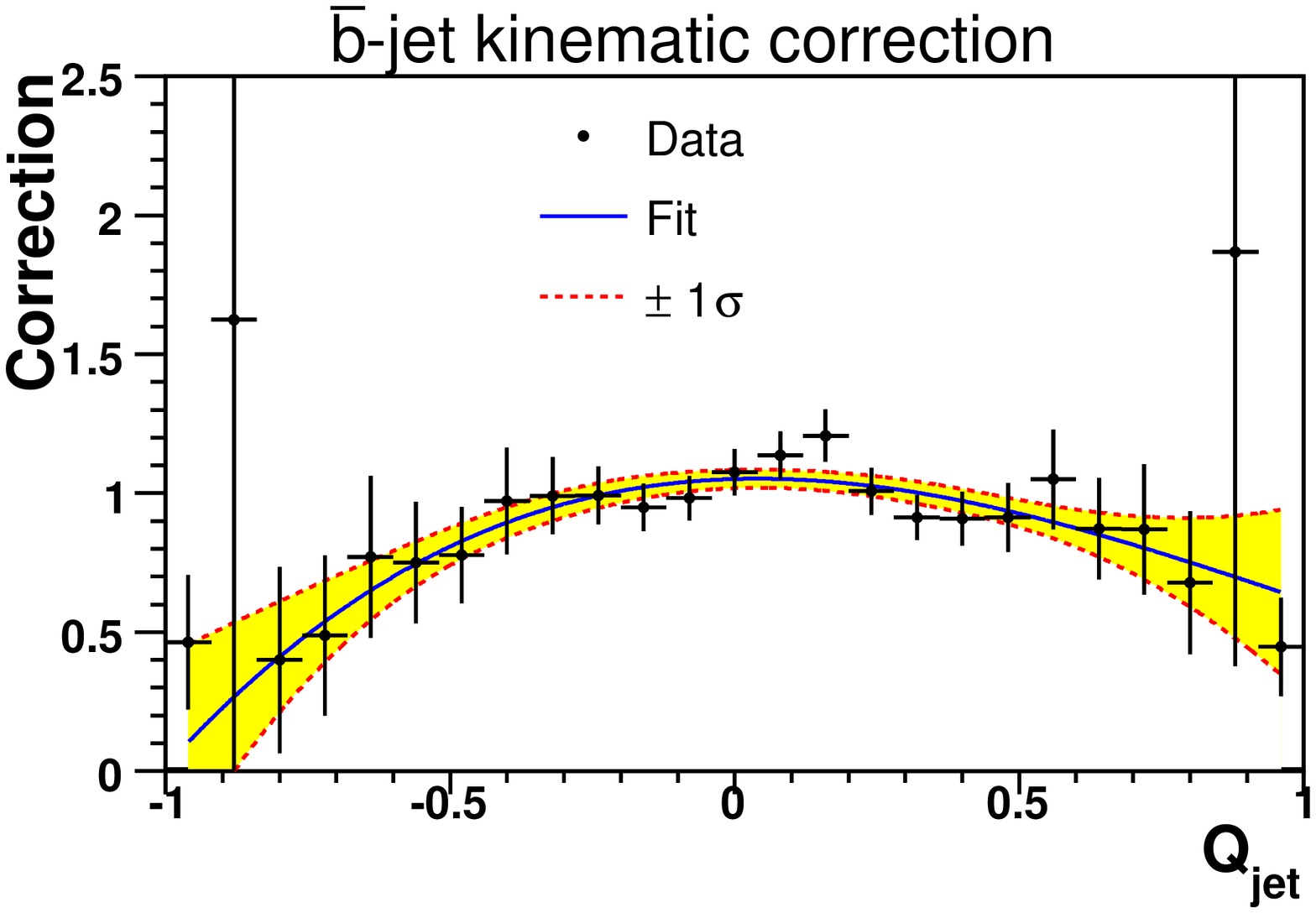}
  \caption{The kinematical correction function for $b$- (top) and $\bar{b}$-quark 
  jets (bottom).} 
  \label{fig:kincorr_b_c_jets}
\end{figure}

\begin{figure}[]
  \centering
  \includegraphics[width=1.0\textwidth]{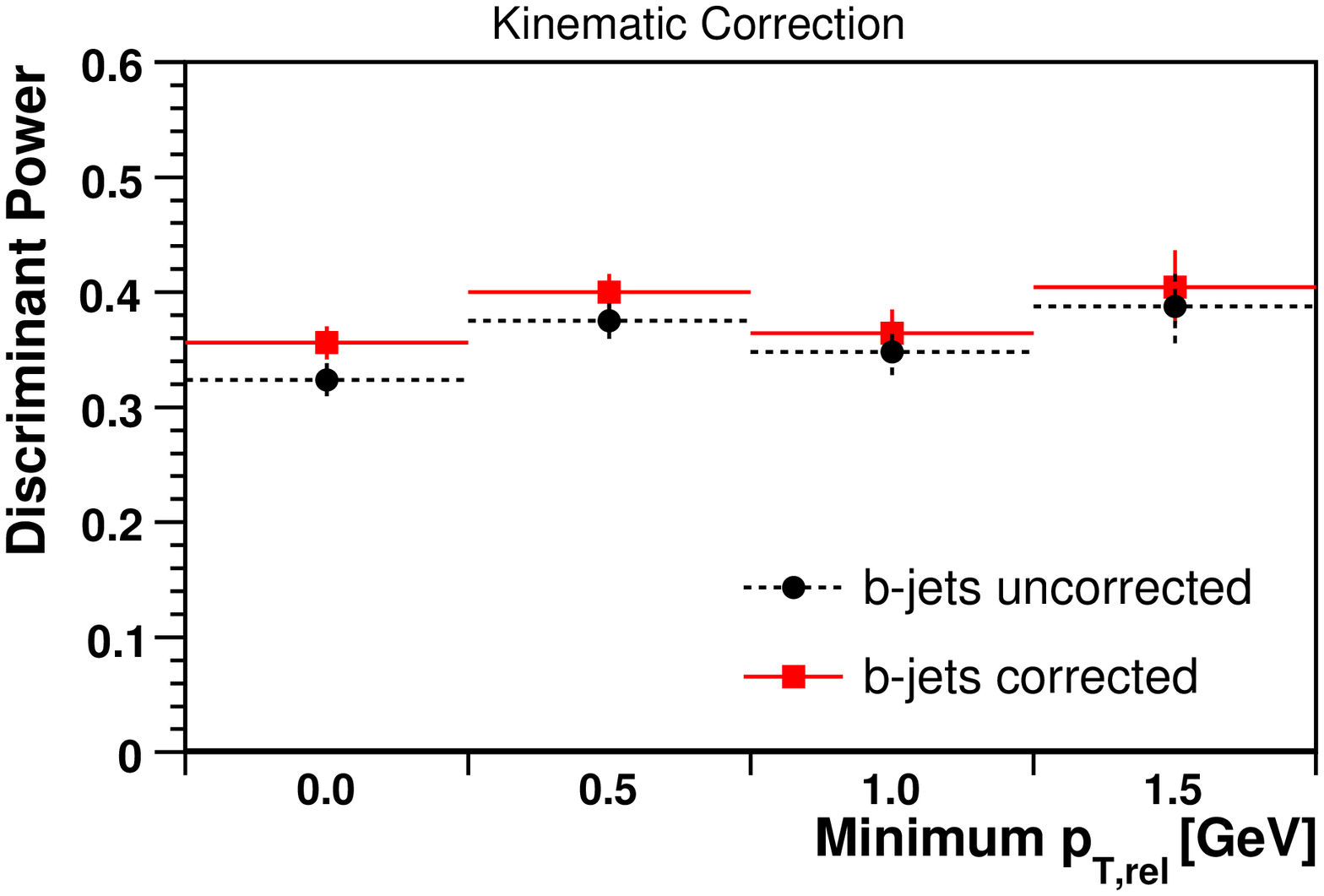}
  \caption{Discriminating power for $b$-quark jet charge templates before and after 
  the kinematical correction is applied as a function of the \ptrel~cut. As expected 
  the kinematical correction increases the discriminating power.}
\label{fig:DP_kincorr_b_c_jets}
\end{figure}

The correction is extracted as the ratio between the $b$-quark jet charge 
templates in normal and re-weighted simulated events. The jet charge templates 
can be seen as probability densities 
to observe a certain jet charge given the true charge and type of quark. 
In the following, the jet charge templates are modeled as functions of 
the measured jet charge $Q_{\rm jet}$ defined in Eq.~\ref{eq:jc1}. Further, we denote:
\begin{itemize}
  \item $f_{b}^{\text{data}}(Q_{\rm jet})$ the $b$-quark jet charge template obtained directly 
  from dijet data (as described in Sec.~\ref{subsec:jetcharge_extraction}) before any correction,
  \item $f_b^{\ttbar}(Q_{\rm jet})$ the $b$-quark jet charge template from 
  simulated $\ttbar \to \ell\text{+jets}$ events passing signal selections,
  \item $f_{b}^{\ttbar,\omega}(Q_{\rm jet})$ the $b$-quark jet charge template 
  from simulated $\ttbar \to \ell\text{+jets}$ events passing signal selections 
  re-weighted as discussed in the previous section,
  \item $f_{b}(Q_{\rm jet})$ the jet charge distribution from the tight dijet data, corrected 
  to reproduce the jet charge distribution for jets with the same $p_T$ and 
  $\eta$ as the simulated \ttbar~events.
\end{itemize}
The corrected distribution is obtained in the following way:
\begin{equation}
  f_{b}(Q_{\rm jet})  =  f_{b}^{\text{data}}(Q_{\rm jet}) \cdot \frac{f_b^{\ttbar}(Q_{\rm jet})}{f_{b}^{\ttbar\omega}(Q_{\rm jet})}
\label{eq:kin_corr}
\end{equation}
The correction for the $\bar{b}$-quark jet charge template is derived in a similar way.  
The correction functions are shown in Fig.~\ref{fig:kincorr_b_c_jets} for both $b$- 
and $\bar{b}$-quark jets. 
As expected, the corrections decrease the width of the 
jet charge templates since a higher jet $p_T$ give a larger probability to observe 
a higher number of tracks associated with the jet. The more tracks associated with the 
jet means less probability to observe $|Q_{\rm jet}|$ close to one. 
Figure~\ref{fig:DP_kincorr_b_c_jets} shows the discriminating power for $b$-quark jets 
before and after the kinematical correction.

\subsection{Final Jet Charge Distributions Extracted from Data}
\label{subsec:jcdata}

\begin{figure}[]
  \centering
  \includegraphics[width=1.0\textwidth]{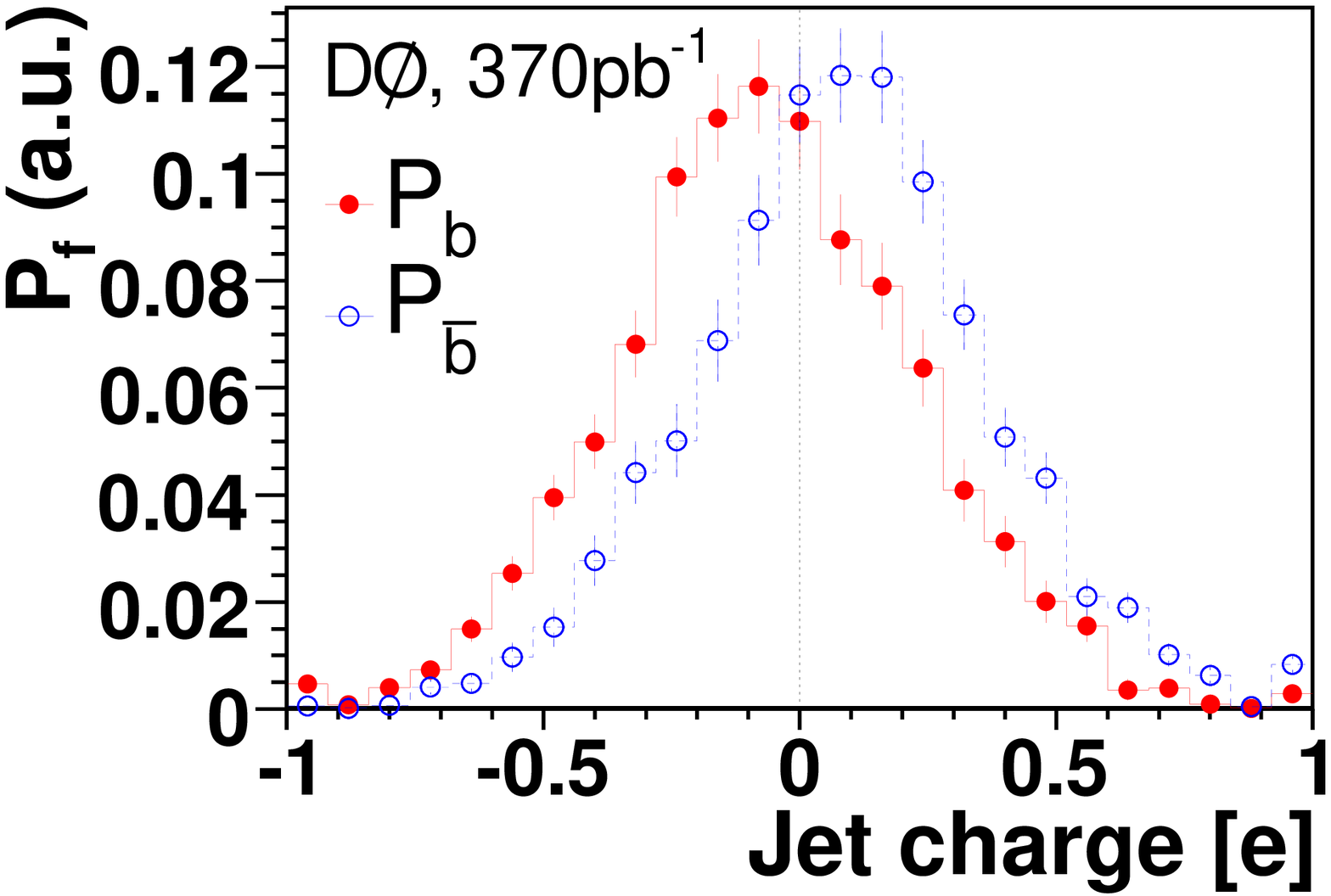}
  \includegraphics[width=1.0\textwidth]{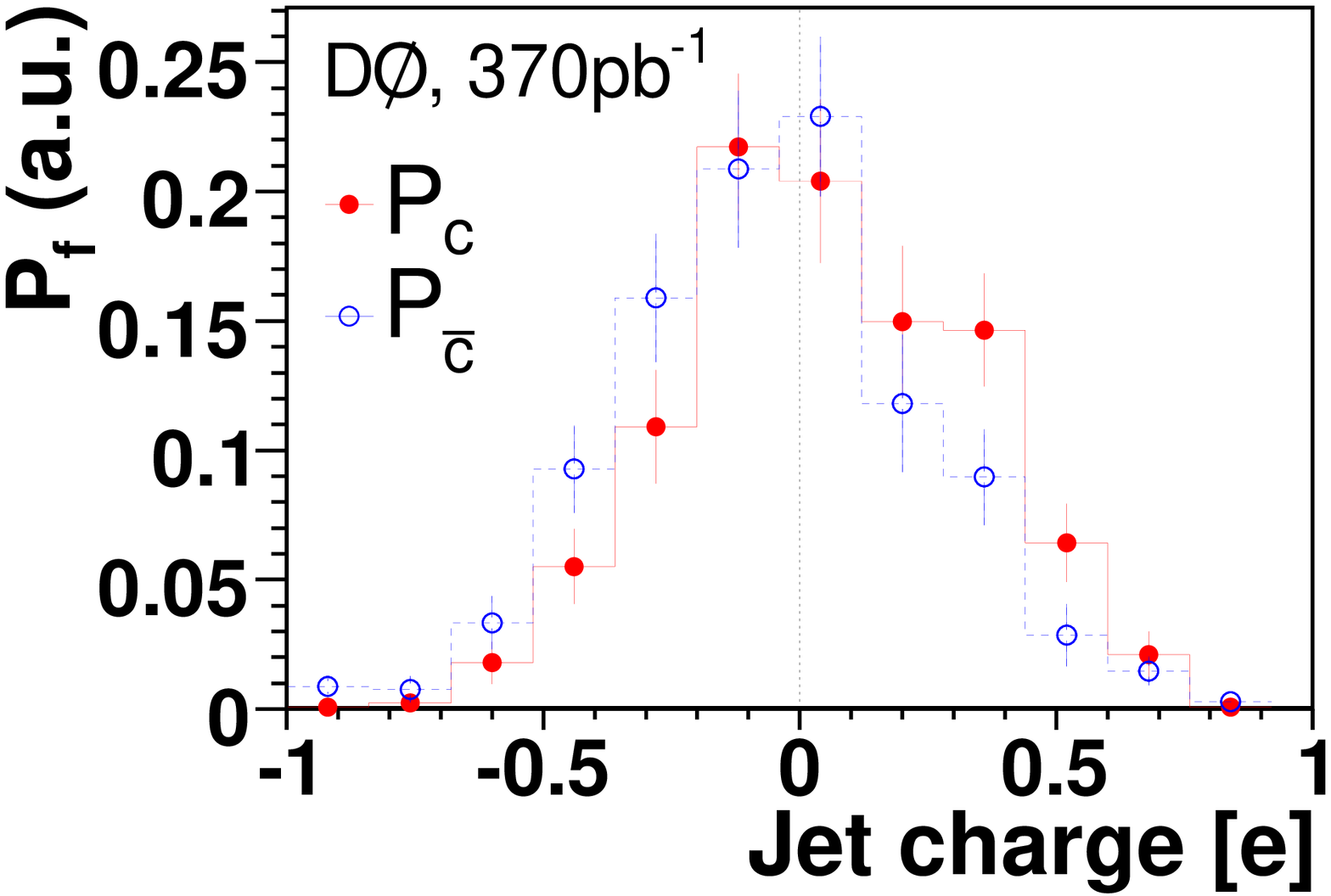}
  \caption{Jet charge templates for $b$- and $\bar{b}$-quark jets (top) and 
  $c$- and $\bar{c}$-quark jets (bottom) extracted from data.}
  \label{fig:final_templates}
\end{figure}

After applying the correction for the kinematical differences in the dijet and 
signal samples the final $b$-, $\bar{b}$-, $c-$ and $\bar{c}$-quark jet charge 
templates shown in Fig.~\ref{fig:final_templates} are obtained. 
 The jet charge templates are normalized to an area of one and can be seen as the 
probability density to measure a certain jet charge $Q_{\rm jet}$, given the type of quark 
($b$, $\bar{b}$, $c$ or $\bar{c}$) initiating the SVT-tagged jet. These templates are subsequently 
used to derive the expected charge templates for the SM top and the 
exotic quark.

\begin{figure}[]
  \centering
  \includegraphics[width=1.0\textwidth]{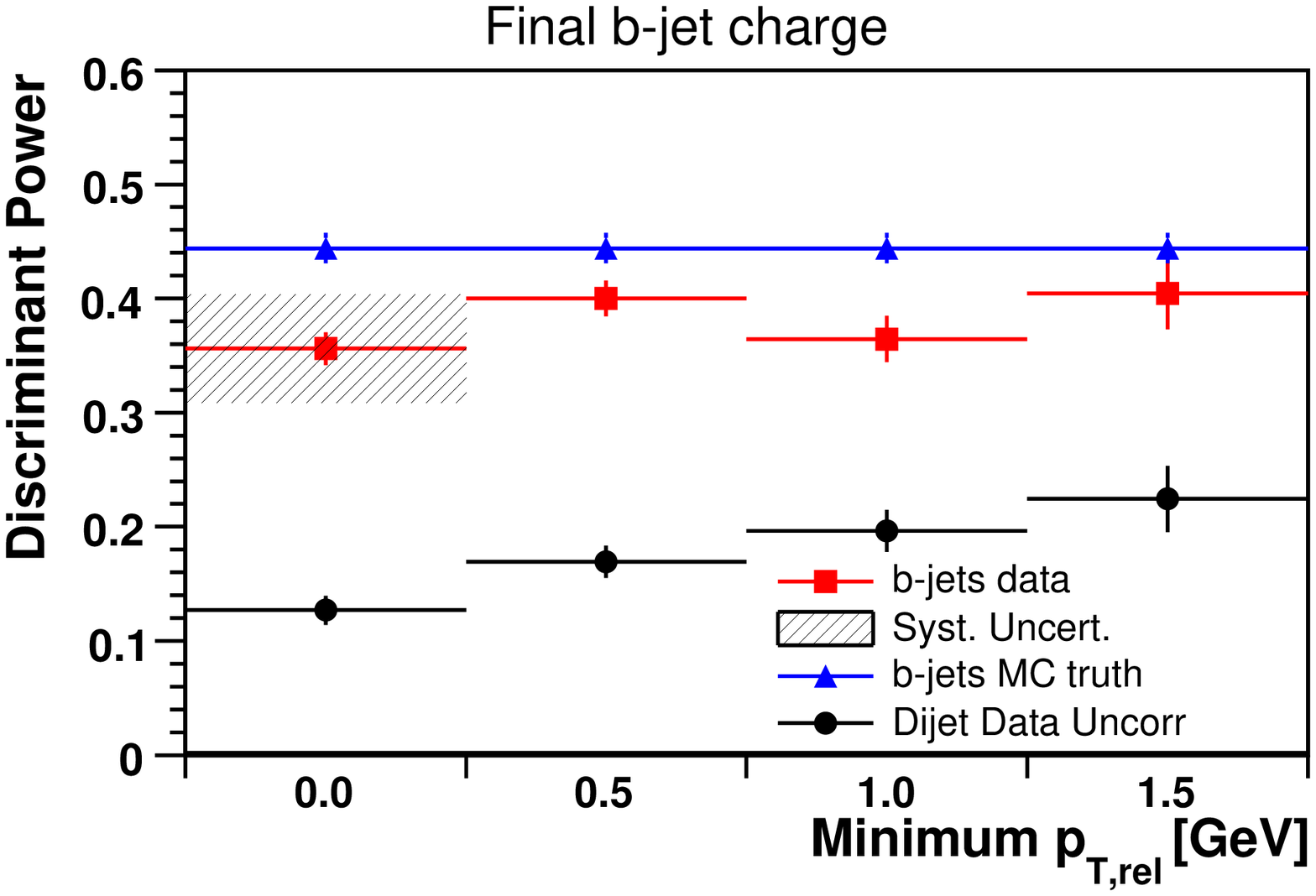}
  \caption{Discriminating power as function of the \ptrel~requirement for $b$-quark jets 
  after all corrections and in the tight dijet sample after selection. The discriminating 
  power in simulated events using the Monte Carlo history to find the true charge of 
  the quark initiating the jet is shown for comparison.}
  \label{fig:divfitresult}
\end{figure}

Figure~\ref{fig:divfitresult} shows the discriminating power of the final $b$-quark 
jet charge templates compared to the tight dijet sample after basic selections. 
The error band is the combined systematic uncertainty related to the extraction of the 
templates which is propagated to the final result. As required, the discriminating 
power of the extracted jet charge templates is independent of the minimum 
\ptrel~requirement on the tagging muon (remember that the final jet charge templates 
are derived with no requirement on the \ptrel~of the opposite tagging muon). 
As expected, the jet charge algorithm is less efficient for data than for Monte Carlo 
events where the Monte Carlo history is available to obtain the true type of quark initiating 
the jet.




\clearpage

\section{Top Quark Charge Observables}
\label{sec:topcharge}

In order to discriminate between the charge $2e/3$ SM top quark and 
the charge $4e/3$ exotic quark scenarios an event observable and an expectation 
of this observable for 
the two different hypotheses is needed. The charge $2e/3$ SM top quark and the 
charge $4e/3$ exotic quark scenarios will simply be refereed to as the SM and 
exotic scenarios respectively.

The observable is based on reconstructing the charge of the decay products of 
the top quark, assuming as mentioned earlier the decay $t \to bW^+$ (and similarly 
in the exotic charge scenario $Q \to bW^-$). Since there are two top quarks in each 
event, the top quark charge can be measured twice in each event. 

One top quark charge is constructed as the sum of the charge of the charged isolated 
lepton (referred to as the charged lepton if not otherwise specified) and the 
jet charge of the $b$-quark jet from the same top quark. The second top quark charge in the 
event is constructed as the sum of the second $b$-quark jet charge {\sl minus} the 
charge of the charged lepton. The two observables in each event are defined as:
\begin{eqnarray}
  \label{eq:top_charge_observables}
  Q_1 & = & |q_{\ell} + q_b| \\ \nonumber
  Q_2 & = & |-q_{\ell} + q_B| \\ \nonumber
\end{eqnarray}
where $q_{\ell}$ is the charge of the charged lepton, $q_b$ is the charge of the 
$b$-quark jet on the leptonic leg of the event (defined in Sec.~\ref{sec:overview}) 
and $q_B$ is the charge of the $b$-quark jet in the hadronic leg of the event.
The charges of the SVT-tagged jets $q_b$ and $q_B$ are obtained by applying 
the jet charge algorithm discussed earlier. The next section describes how the 
two $b$-quark jets are assigned to the two legs of the event.

\subsection{Associating SVT-Tagged Jets to the Correct $W$ Boson}
\label{subsec:hitfit}

Recall that the final state objects from the decay of \ttbar~pair in the 
\ljets~channel are two jets from the decay of the $W$ boson and a $b$-quark jet 
from the hadronic side of the event and a $b$-quark jet together with a charged 
lepton and a neutrino from the leptonic side of the event. In order to compute the 
top quark charge observables, the two $b$-quarks in the event needs to be assigned to 
two reconstructed jets. This is done using a kinematic fitting algorithm initially developed 
to measure the top quark mass~\cite{topmass_runI_PRL}. Below, the main features 
of the kinematic fit is described, more detailed information can be found in 
Ref.~\cite{hitfit}.

\subsubsection{Kinematic Fit}

The algorithm used performs a kinematic fit of top quark pair candidate 
events in the \ljets~topology. If the event contains the decay of a \ttbar~pair, 
with four jets (the issue regarding events with more than four jets is discussed 
below) in addition to the charged lepton and neutrino, the 
three jets (two light jets from the $W$ boson and a $b$-quark jet) forming the invariant mass of the 
``hadronic'' top quark (the top quark on the hadronic side of the event) is 
expected to be equal to the invariant mass formed by the charged lepton, 
neutrino and remaining $b$-quark jet (the leptonic side of the event). In 
addition to this constraint, two jets (out of three) on the hadronic side of 
the event is expected to form the invariant mass of the $W$ boson as is the charged 
lepton and the neutrino on the leptonic side of the event. Together with the overall 
energy-momentum conservation in the collision this over-constrains the problem.

The input variables to the kinematic fit are:
\begin{itemize}
  \item
  The measured energy (or momentum) of the four jets and the charged lepton ($3$ components 
  each) and their directions. The masses of the jets are fixed to zero except for the jets assigned as  
  $b$-quark jets which are given a mass of $4.6$~GeV.
  \item
  The $x$ and $y$ components of the measured missing transverse energy, \met, represents 
  the transverse momentum of the neutrino. Note that the neutrino momentum in the 
  longitudinal direction ($p^{\nu}_z$) cannot be inferred from momentum imbalance 
  in a similar way due to the spectator quarks in the colliding \ppbar~pair 
  which carries away a large fraction of the momentum in this direction which is mostly 
  unmeasured close to the beam direction.
\end{itemize}
The conclusion is that there are $17$ measured and one unmeasured variable 
(the neutrino momentum in the $z$ direction) under subject to three constraints 
when reconstructing the top quark event,
\begin{gather}
m_{W}^{\text{leptonic side}} = 80.4~{\rm GeV}, \\
m_{W}^{\text{hadronic side}} = 80.4~{\rm GeV}, \\
m_{t} = m_{\bar{t}},
\label{eq:fit_constraints}
\end{gather}
which makes the problem twice 
over-constrained\footnote{There are $13$ particles in the event, 
4 jets, 
two leptons, 
the \ppbar~pair, 
$W^{\pm}$, 
the \ttbar~pair 
and an additional pseudo-particle $X$ to obtain total four-momentum 
balance which gives 
$52~(13 \times 4)$ variables. 
The constraints come from the 
quarks, 
lepton and 
\met~giving 
$23~\left [ 5 \times 4 + 3~(p^{\nu}_x,p^{\nu}_y),{\rm mass of the neutrino} \right ]$ 
constraints. 
The \ppbar~pair gives an additional 
$ 8~(2 \times 4)$, 
four-momentum conservation in all $5$ vertices gives 
$20$, 
the known masses of $W^{\pm}$ gives 
$2$ 
and 
$m_t=m_{\bar{t}}$ gives one constraint. 
This leads in total to $54$ constraints for $52$ variables and 
thus twice over-constrained.}. In this analysis, which is not concerned with measuring 
the mass of the top quark, the top quark mass itself is used as an additional constraint 
and fixed to $175$~GeV (which is the top quark mass used in the generation of 
the simulated \ttbar~events). 

If the correspondence between jets and partons were known, a kinematic fit would not 
be needed. In general, this is not known. Therefore, the algorithm tries all $12$ possible 
permutations of jets and the ``best'' one can be chosen (there are $24$ permutations 
for four jets but the exchange of the two jet from the $W$ boson leaves the fit unchanged). 
To quantify ``best'', a ``$\chi^2$'' is defined which reflects to what degree it is possible 
to satisfy the 
given constraints for each jet permutation. The kinematic fit iteratively changes the 
measured and assigned kinematic variables based 
on their uncertainties and their impact on the $\chi^2$. For a given jet permutation the 
variables are pulled until the $\chi^2$ stops changing and the constraints are satisfied. 

For events with more than four jets the additional jets are assumed to 
originate from initial (ISR) or final state radiation (FSR). The number of permutations 
grows fast when including extra jets in the fit (note that if the extra jet is assumed 
to be FSR it must be merged with another jet in the event and 5 jets give 140 permutations, 
6 jets give 1020, etc.). As in the top mass analyses~\cite{topmass_runI_PRL}, this analysis 
uses only the four highest $p_T$ jets in the fit. Other jets are assumed to arise from 
ISR~\cite{hitfit}.

Consider now $\ttbar \to \ljets$ events where exactly two of the jets in the event are 
SVT-tagged. These $b$-tagged jets are denoted as $j_{SVT1}$, ..., $j_{SVTn}$, with $n = 2$
and the other jets $j_1$ ..., $j_n$, with $n \geq 2$. If $j_b$ and $j_B$ 
are denoted as the $b$-quark jets from the leptonic and hadronic leg of 
the event respectively, and $j_{w1}$ and $j_{w2}$ the two jets from the 
hadronic $W$ decay, then each permutation considered by the kinematic fit 
associates one jet ($j_{SVT1}$, $j_{SVT2}$, $j_1$ and $j_2$) to each of 
the jets $j_b$, $j_B$, $j_{w1}$ and $j_{w2}$. There are exactly 12 
combinations, but only two for which the SVT-tagged jets are associated 
to the two jets $j_b$ and $j_B$ from the $b$-quarks of the leptonic and hadronic 
legs:
\begin{enumerate}
\item $j_{SVT1}$ $\leftrightarrow$ $j_b$ and $j_{SVT2}$ $\leftrightarrow$ $j_B$, or
\item $j_{SVT2}$ $\leftrightarrow$ $j_b$ and $j_{SVT1}$ $\leftrightarrow$ $j_B$.
\end{enumerate}

For an event with more than two SVT-tagged jets the number of permutations for 
which the kinematic fit can associate SVT-tagged jets to $j_b$ or $j_B$ 
is multiplied. Note that only 1\% of \ttbar~events in Monte Carlo passing 
preselection have three or more SVT-tagged jets. 
\begin{quote}
  \item 
  Only permutations for which $j_b$ and $j_B$ are associated to SVT-tagged 
  jets are considered.
\end{quote}
Sometimes the input kinematic variables are very far from the constraints and the fit 
may therefore fail to pull the variables enough to satisfy the constraints or
it does not find a stable $\chi^2$ and fails to converge. The kinematic fit 
fails to converge in $17.6 \pm 0.6$\% of simulated \ttbar~events 
passing signal selections for any of the two permutations where the SVT-tagged 
jets are associated to $j_b$ and $j_B$, see Tab.~\ref{tab:conv_perms}. 
\begin{table}
\centering
\begin{tabular}{|c|c|c|}
\hline
Number of & Tagged & Data Pred. tagged \\
permutations & $t\bar{t}$ MC (\%) & $t\bar{t}$ MC (\%) \\ 
\hline
 $0$ & $17.6 \pm  0.6$ & $17.7 \pm  0.8$  \\ 

 $1$ & $ 5.6 \pm  0.3$ & $ 6.1 \pm  0.4$  \\ 

 $2$ & $71.9 \pm  1.4$ & $73.0 \pm  2.0$  \\ 

 $3$ & $ 0.2 \pm  0.1$ & $ 0.2 \pm  0.1$  \\ 

 $4$ & $ 1.9 \pm  0.2$ & $ 1.2 \pm  0.2$  \\ 

 $5$ & $ 0.20 \pm  0.05$ & $ 0.03 \pm  0.03$  \\ 

 $6$ & $ 2.6 \pm  0.2$ & $ 1.7 \pm  0.2$  \\ 

\hline
\end{tabular}
\caption{Number of permutations where the kinematic fit converges for 
simulated $\ttbar \to \ljets$ events. More than two permutations can arise in 
events where there are more than two SVT-tagged jets. In the rightmost column 
the SVT-tagged jets in the event was predicted by using the tagging 
efficiencies derived from data.}
\label{tab:conv_perms}
\end{table}
One should note that it is not seldom that the $W$ decay contains a $c$-quark 
jet, and the SVT-tagged jets do not correspond to 100\% of actual $b$-quark 
jets, see Tab.~\ref{tab:jet_flavor}.
\begin{table}
\centering
\begin{tabular}{|c|c|c|c|}
\hline
Jet flavor & Lowest $\chi^2$ & Double $b$-tagged & Data pred. $b$-tagged  \\
 & (\%) & events (\%) & events (\%) \\

\hline
 no match & $0.00 \pm 0.00$ & $3.40 \pm 0.16$ & $3.64 \pm 0.24$  \\

 u & $0.37 \pm 0.06$ & $0.45 \pm 0.06$ & $0.33 \pm 0.07$  \\

 d & $0.37 \pm 0.06$ & $0.47 \pm 0.06$ & $0.40 \pm 0.08$  \\

 s & $0.33 \pm 0.05$ & $0.45 \pm 0.06$ & $0.27 \pm 0.06$  \\

 c & $4.90 \pm 0.22$ & $7.16 \pm 0.24$ & $7.53 \pm 0.35$  \\

 b & $94.03 \pm 1.29$ & $88.07 \pm 1.10$ & $87.84 \pm 1.56$  \\

\hline
\end{tabular}
\caption{Jet flavor for the SVT-tagged jets in simulated $\ttbar \to \ljets$~events 
passing signal selections. The permutation 
giving the lowest $\chi^2$ in the kinematic fit gives as expected a higher 
fraction of $b$-quarks matched to the jets compared to any permutation (middle 
and last column). Using the tagging efficiency derived from data gives 
a similar result as for using the explicit SVT-tags in the Monte Carlo events. 
The true jet flavor is found by matching the closest parton within a cone of 
$\Delta R < 0.5$ to the jet axis. 
}\label{tab:jet_flavor}
\end{table}
Finally,  out of the permutations for which the kinematic fit assigns the 
SVT-tagged jet to $j_b$ and $j_B$, only the permutation with the lowest $\chi^2$ is 
used. Based on the Monte Carlo history, it is observed that the lowest $\chi^2$ 
permutation is the correct one in $83.8 \pm 1.8$\% of the cases. In the 
signal sample, 21 events are selected, out of which 16 have at least one converged
permutation where both SVT-tagged jets are assigned to $j_b$ and $j_B$. 
This is statistically consistent with the prediction $17.6 \pm 0.6$\% from 
simulated events.

\subsection{Expected Charge Templates in the Standard Model and Exotic Scenarios}
\label{subsec:templateconstruction}
The expected charge distribution templates in both scenarios 
are derived from simulated \ttbar~events and the jet charge templates derived from data 
earlier.


%

\subsubsection{Standard Model Charge Template}

The estimated charge distribution of the SM top and the exotic quark depends on 
how well the $b$- and $\bar{b}$-quark jets can be identified but 
also the assignment of the correct $b$-quark jet to correct 
$W$ boson in the event. Therefore, it is necessary that the 
kinematic fit is applied to the simulated events in an equivalent way as 
to data in order to extract the expected distributions.

\begin{figure}[]
  \centering
  \includegraphics[width=0.8\textwidth]{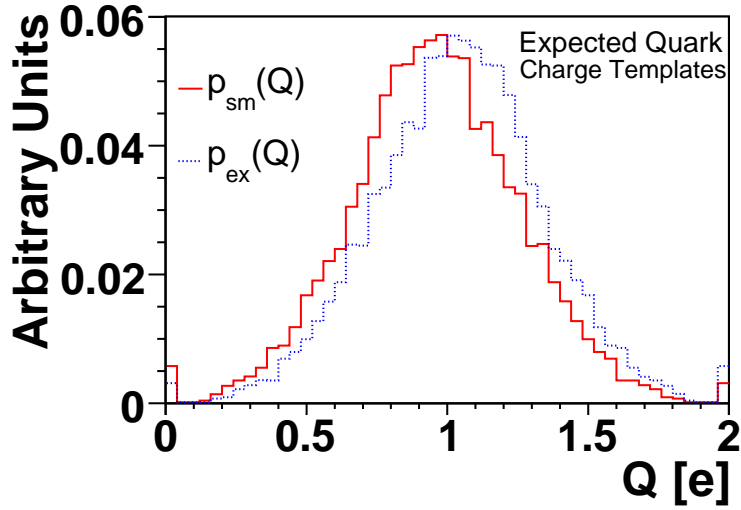}
  \caption{The predicted SM top ($p_{\rm sm}$) and exotic ($p_{\rm ex}$) quark charge 
  templates including backgrounds.}
  \label{fig:sm_ex_templates}
\end{figure}

The procedure to obtain the expected SM top quark charge distribution 
is explained below:
\begin{enumerate}
  \item 
  \label{enum:1}
  Apply signal selection criteria as defined in Sec.~\ref{sec:preselection} 
  on simulated \ttbar~events.
  \item 
  \label{enum:2}
  Fit the event using the constrained kinematic fit.
  \item 
  \label{enum:3}
  Select the lowest $\chi^2$ combination for which the jets $j_b$ and $j_B$
  are associated to SVT-tagged jets $j_{SVT1}$ and $j_{SVT2}$.
  \item 
  \label{enum:4}
  Determine the true flavor of the jets $j_{b}$ and $j_{B}$ 
  using jet-parton matching. The true flavor can be: $b$-, $\bar{b}$-,
  $c$- , $\bar{c}$- or light ($u$, $d$, $s$) jet.
  \item
  \label{enum:5}
  The jet charges $q_{b}$ and $q_{B}$ is set to one randomly chosen 
  value according to the probability density function $f(Q_{\rm jet})$ derived 
  from data depending on the flavor of the jet e.g. if the flavor is 
  $\bar{b}$, then the probability density function $f_{\bar{b}}(Q_{\rm jet})$. 
  If it is a light jet, then a random value from the corresponding
  jet charge probability density function is used, derived from Monte 
  Carlo\footnote{It is very seldom we need to sample the light jet 
  charge templates due to the low probability to tag a light jet 
  ($\approx1$\%), see Tab.~\ref{tab:jet_flavor}}. 
  \item 
  \label{enum:6}
  Compute the two observables $Q_1$ = $|q_{\ell}+q_{b}|$
  and $Q_2$ = $|-q_{\ell}+q_{B}|$.
  \item 
  \label{enum:7}
  Make an entry for $Q_1$ and another entry for $Q_2$ in a histogram
  to store the expected charge distribution for the SM top quark.
\end{enumerate}
To decrease the statistical uncertainty due to the sampling of the jet charge
distributions step~\ref{enum:5}, steps~\ref{enum:6} and~\ref{enum:7} 
are carried out 200 times per event.
The predicted SM top quark charge distribution referred to as $p_ {\rm sm}$ is 
shown in Fig.~\ref{fig:sm_ex_templates}.

\subsubsection{Exotic Scenario Charge Template}
The estimated charge distribution template for the exotic scenario is derived in 
a similar way as for the SM top quark. From Fig.~\ref{fig:diagram1} it can be 
seen that it can be obtained by simply replacing the above step~\ref{enum:6} 
by the following
\begin{enumerate}
\item[\ref{enum:6}]. The two expected observables are derived by permuting
  the jet charge of the SVT-tagged jet on the leptonic and hadronic 
  leg of the event: $Q_1$ = $|q_{\ell}+q_{B}|$ and 
  $Q_2$ = $|-q_{\ell}+q_{b}|$.
\end{enumerate}
The expected charge template for the exotic scenario is also shown in 
Fig.~\ref{fig:sm_ex_templates}. Note that with this procedure the charge 
template for the exotic scenario is a mirror distribution of the Standard 
Model top quark charge template.

\subsection{Backgrounds}
\label{susec:bkg}
The signal sample contains in addition to \ttbar~events also 
a small fraction of background processes, mostly $W$ boson production in 
association with a $b\bar{b}$ pair and two or more jets ($Wb\bar{b}jj$) 
but also some expected single top quark events. Details about the sample composition is described in 
Sec.~\ref{sec:preselection}. The observed charge in the signal 
sample is to be compared with a combination of the expected SM top quark and 
exotic quark charge templates described above and of the charge contribution 
from the background processes. Due to the large signal-to-background 
ratio in the signal sample ($\sim11$), the effect of adding 
the charge contribution from the background processes is small. The 
backgrounds considered are the $Wb\bar{b}jj$ and single top production 
while the other background processes are neglected. 
Two assumptions are made: {\it i)} the charge of the isolated high $p_T$ 
lepton is uncorrelated with the $b$-quark jets in $Wb\bar{b}jj$ events, 
{\it ii)} the $b$-quark jet charge in $Wb\bar{b}jj$ events are the same 
as for the $b$-quark jets in \ttbar~events. Due to the low expected 
contribution from the single top background ($\approx 0.3$ events) it is 
simply modeled by the same charge template as the $Wb\bar{b}jj$ background.

\begin{figure}[]
  \centering
  \includegraphics[width=0.8\textwidth]{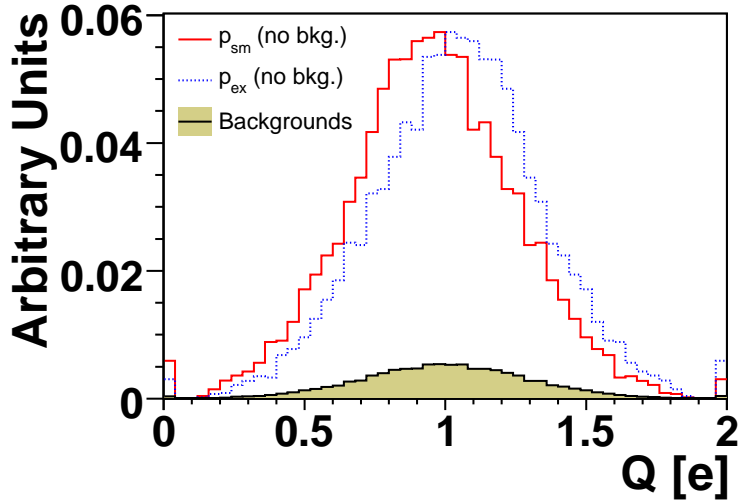}
  \caption{The combined background charge contribution compared to the SM top 
  and exotic quark charge templates.}
  \label{fig:wbbjjbkg}
\end{figure}

Figure~\ref{fig:wbbjjbkg} shows the background charge template contribution 
and the expected SM and exotic templates. 
The SM and exotic templates 
are normalized to an area of one and can be seen as the probability 
density functions to observe a certain charge $Q$ in the SM and exotic 
scenarios. In the rest of this thesis we denote these probability 
density functions by $p_{\rm sm}(Q)$ and $p_{\rm ex}(Q)$ for the SM 
top quark and the exotic quark scenario respectively.

\clearpage

\section{Systematic Uncertainties}
\label{sec:systematics}


In this section the sources of systematic uncertainties and the method to evaluate 
them are briefly described. A systematic uncertainty can affect the measurement in 
two ways: It can change the jet charge templates derived from data or it can affect 
the kinematic fit that 
assigns the SVT-tagged jets in the event to the correct $W$ boson. The result in the end 
is an uncertainty on the SM top and exotic quark templates. Thus, for each systematic 
uncertainty, the SM top and exotic quark charge templates 
are re-derived taking into account the uncertainty considered. The result is a set of 
varied SM top and exotic quark templates for each systematic which is taken into 
account when extracting the final result.

\subsubsection{Dependence on the Fragmentation Model}
\label{subsec:fragmodel}

\begin{figure}[]                
  \centering
  \includegraphics[width=0.8\textwidth]{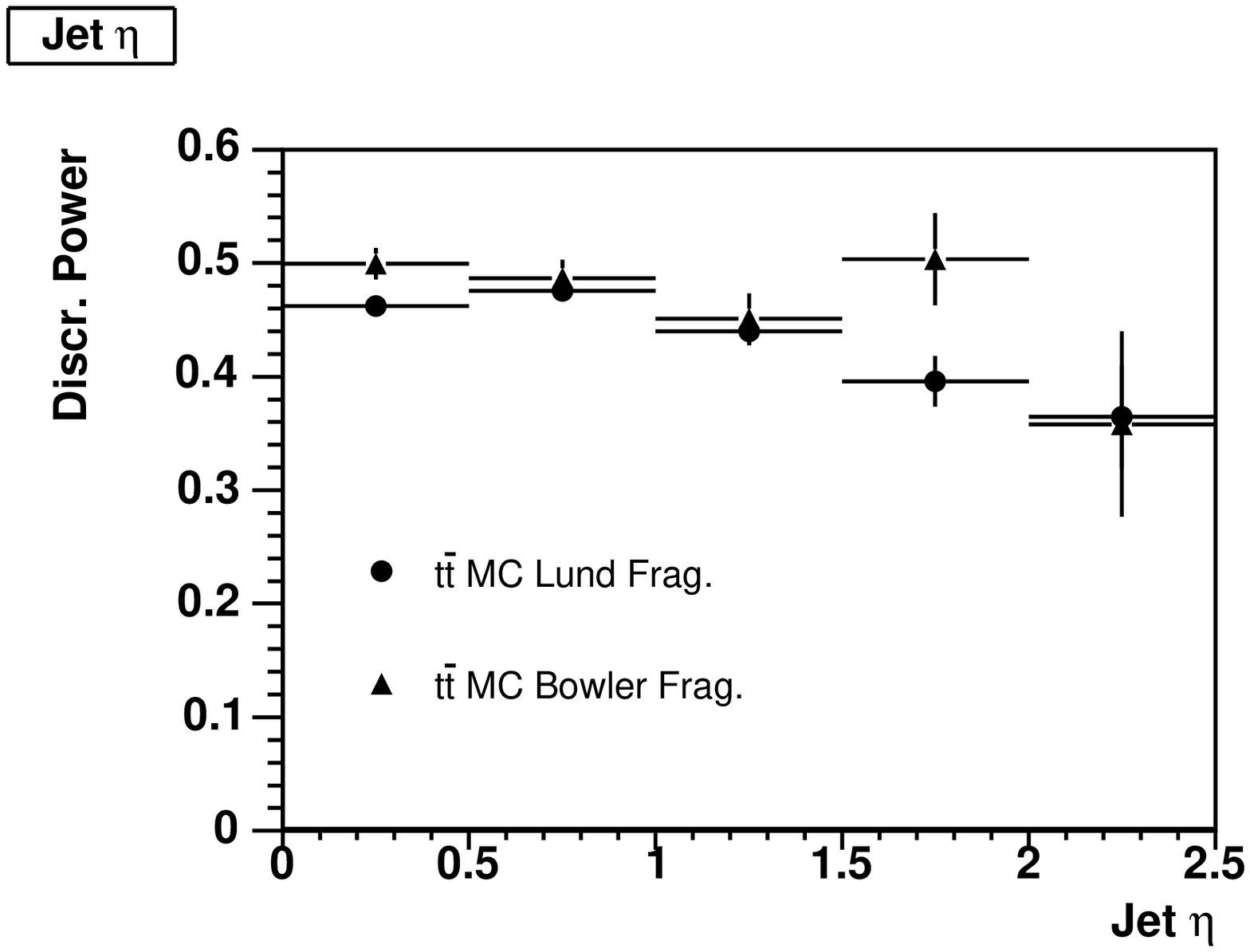}
  \includegraphics[width=0.8\textwidth]{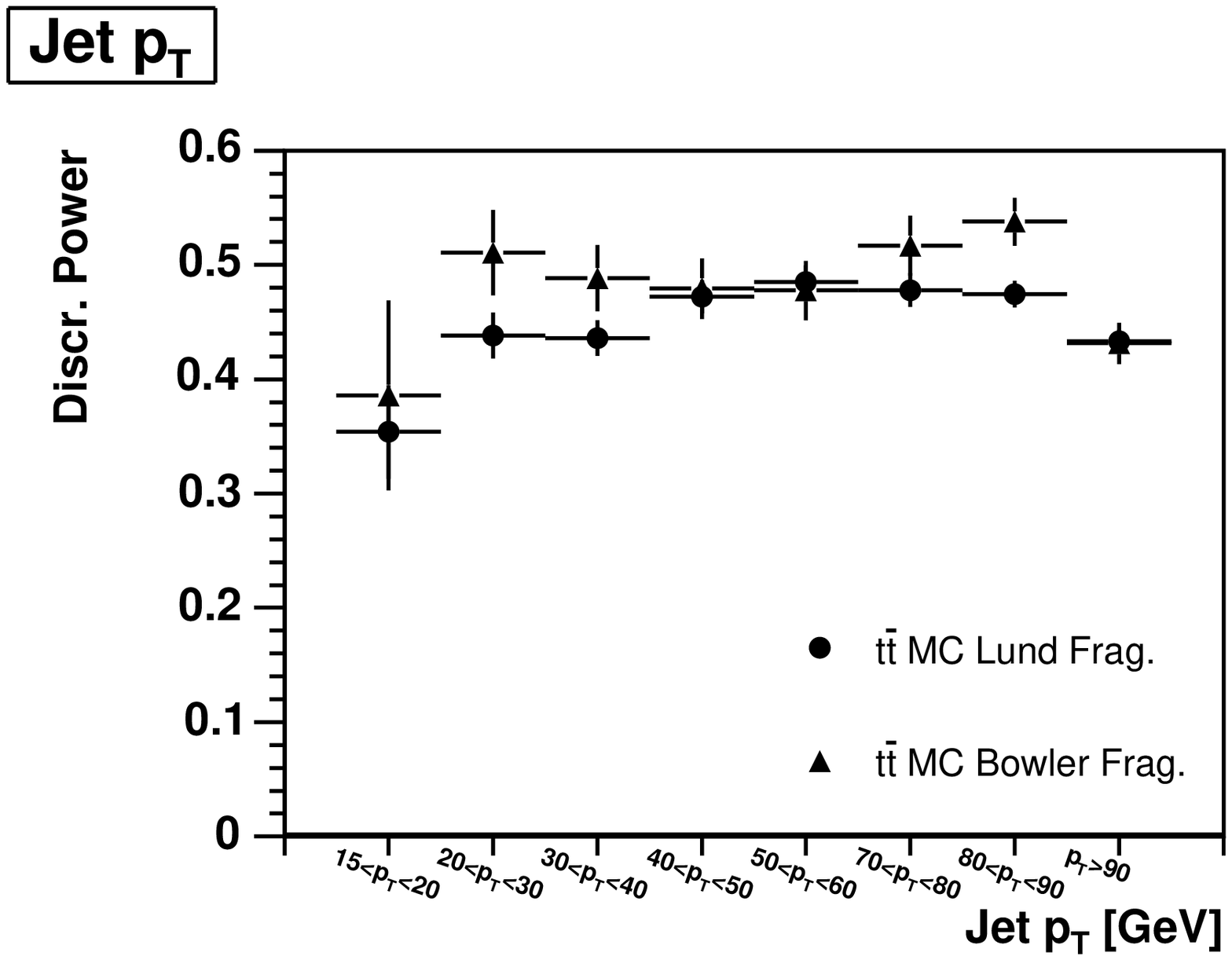}
  \caption{A comparison of the discriminating power in simulated \ttbar~events using 
  the Lund string and the Bowler fragmentation model as functions of $\eta$ (top) and 
  $p_{T}$ (bottom).}
  \label{fig:bowler}
\end{figure}

\begin{figure}[]
  \centering
  \includegraphics[width=0.8\textwidth]{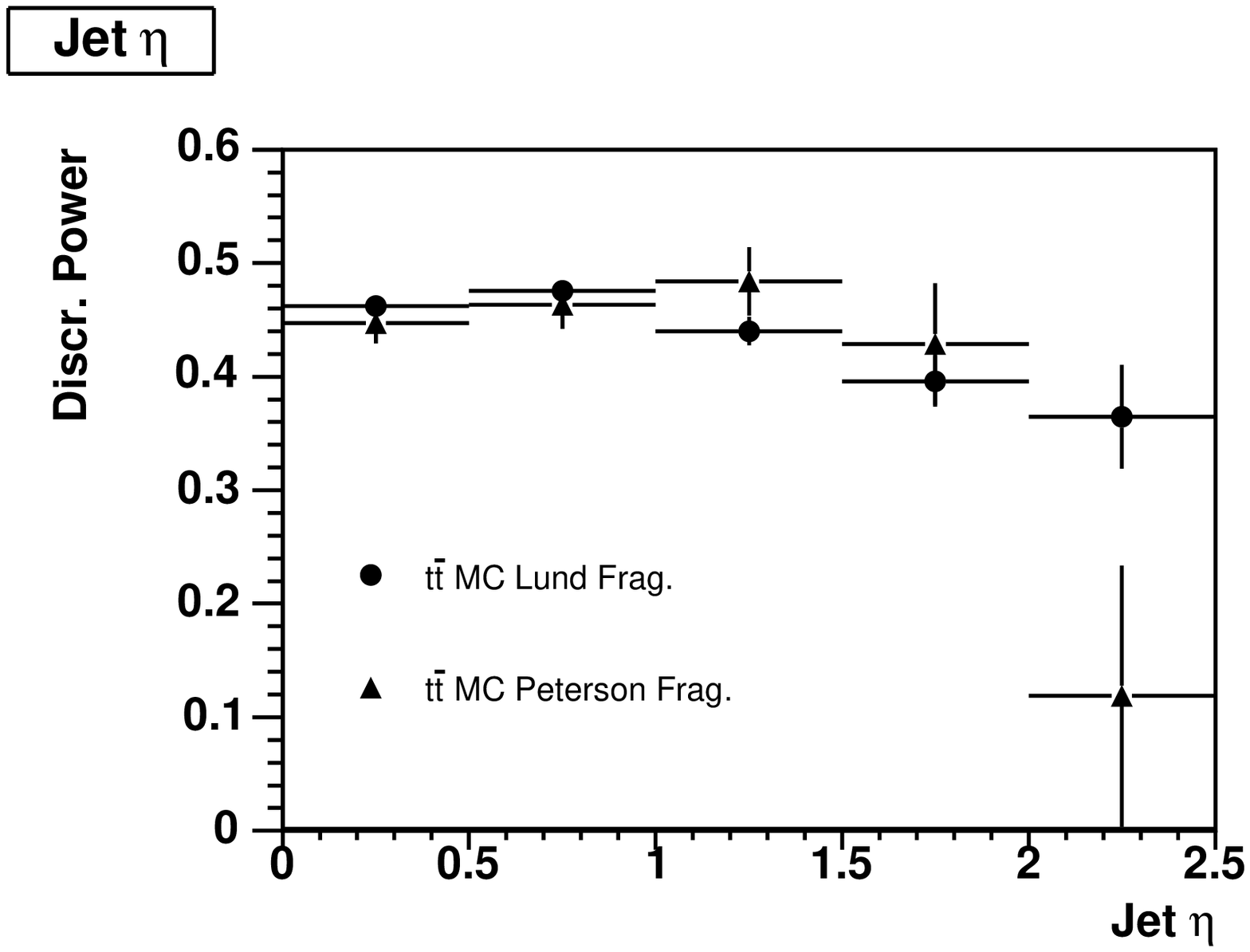}
  \includegraphics[width=0.8\textwidth]{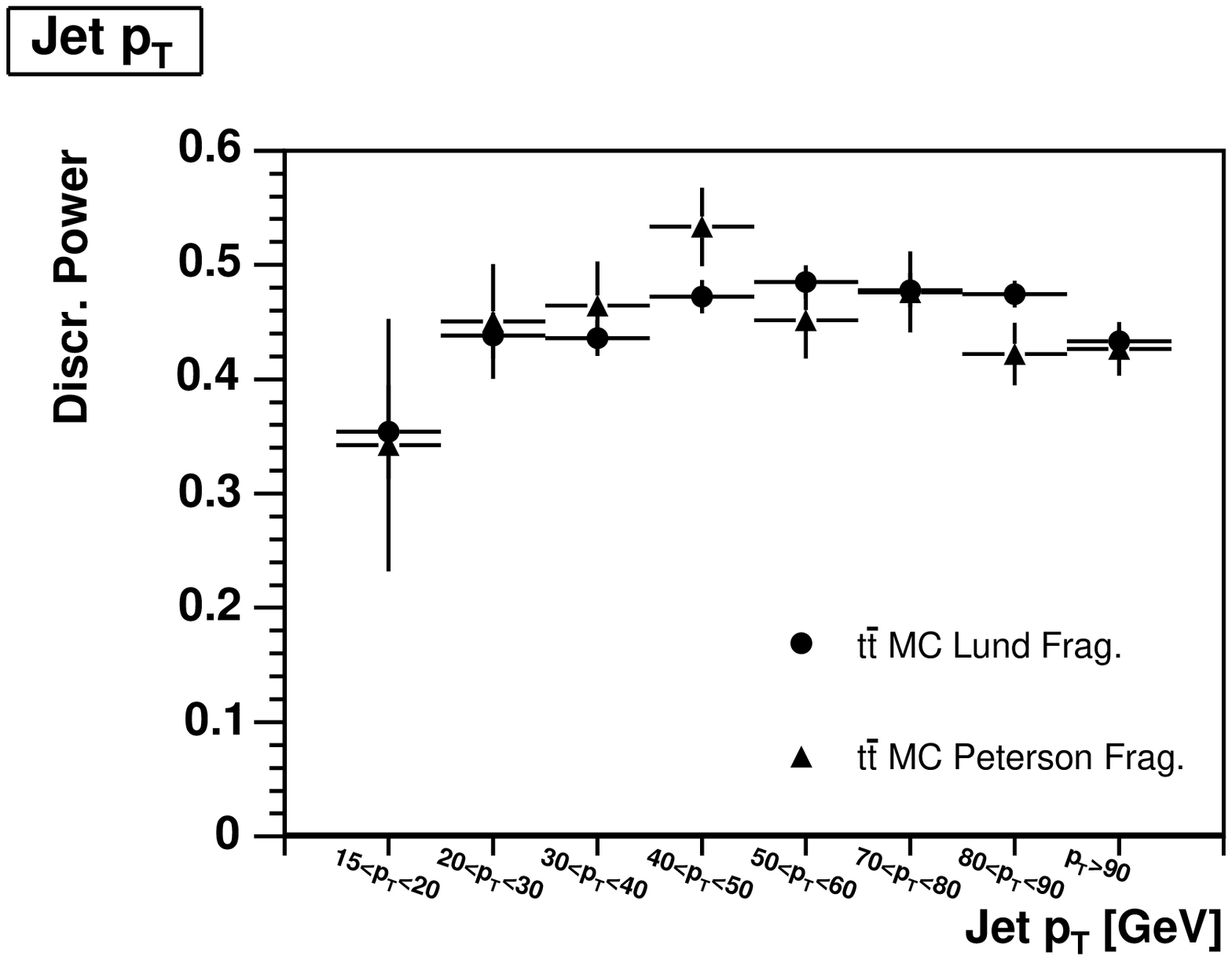}
  \caption{A comparison of the dicriminating power in simulated \ttbar~events for the Lund string 
  and Peterson fragmentation model as functions of $\eta$ (top) and  $p_{T}$ (bottom).}
  \label{fig:peterson}
\end{figure}

This analysis use as little input as possible from the Monte Carlo simulation in 
deriving the jet charge algorithm performance. For instance, the extraction of the correction 
function in Sec.~\ref{subsubsec:kincorr_der_func}, 
uses the ratio of Monte Carlo modeled variables as input. Nevertheless, the 
dependence of a different model for the fragmentation of $b$-quark jets has been 
investigated. As discussed in Sec.~\ref{subsec:topquarkmcsim}, the fragmentation 
of partons is modeled by the Lund string model. Figure~\ref{fig:bowler} 
and~\ref{fig:peterson} shows the difference in discriminating power between $b$- and 
$\bar{b}$-quark jet charge distributions compared to the alternative {\sc Bowler} 
and {\sc Peterson} fragmentation models respectively.
No systematic uncertainty due to the fragmentation model needs to be 
accounted for since the $b$- and $c$-quark jet charge templates are derived from 
data and as noted above, the kinematic corrections are ratios of Monte Carlo 
distributions, resulting in a large cancellation of any existing 
difference among the fragmentation models.


\subsubsection{Additional Jet Dependence}
\label{subsec:activitydepencence}
\begin{figure}[]
  \centering
  \includegraphics[width=0.8\textwidth]{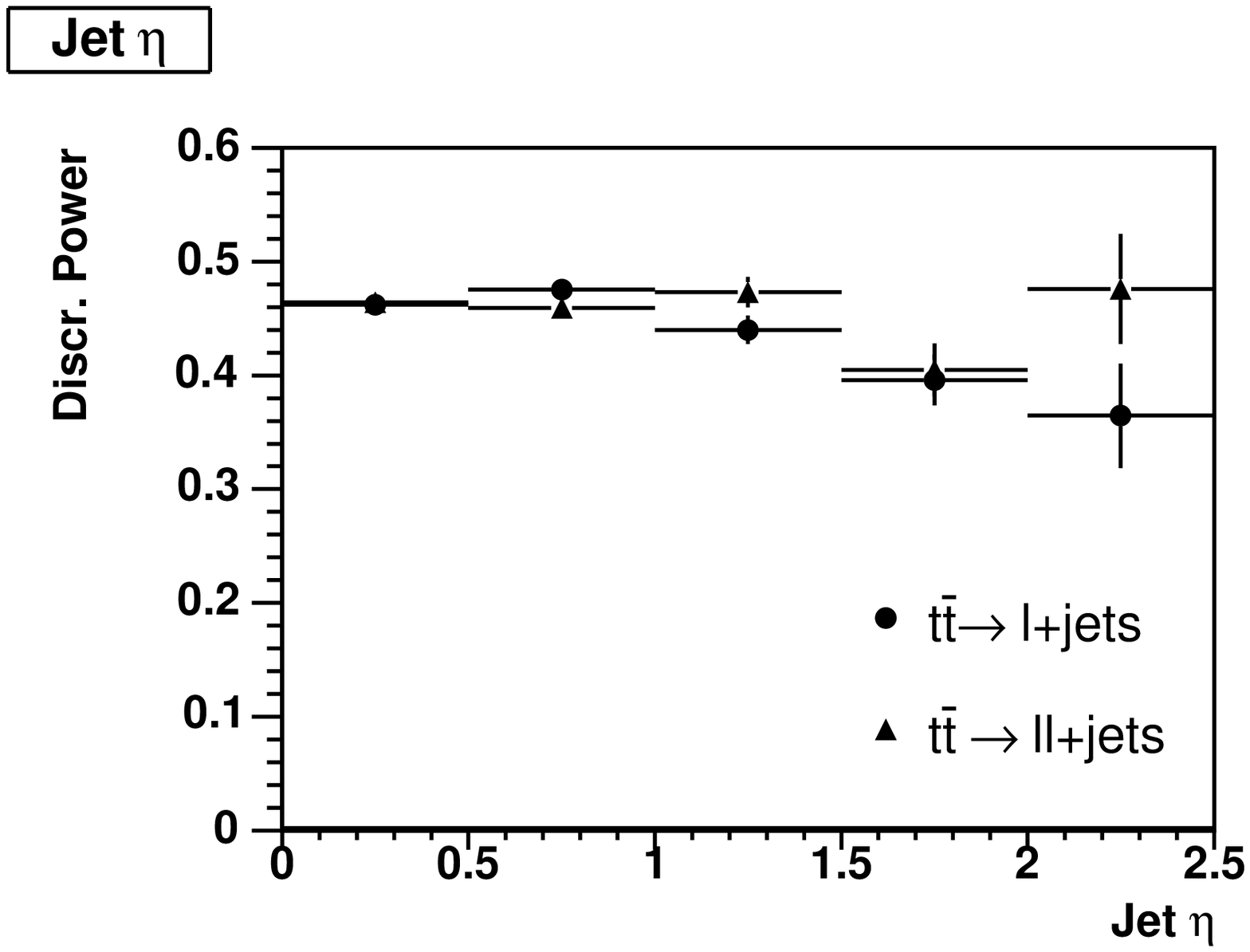}
  \includegraphics[width=0.8\textwidth]{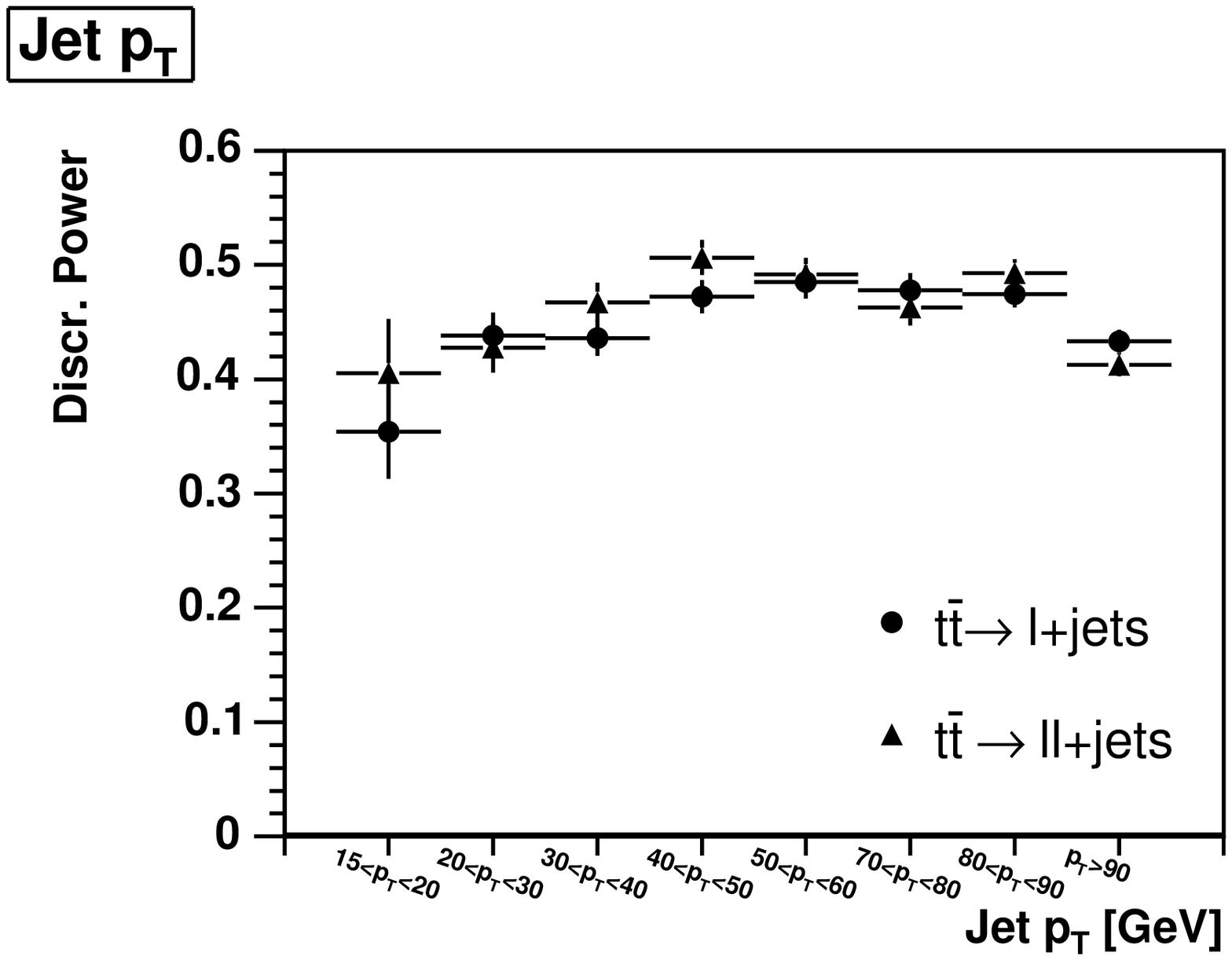}
  \caption{Comparison of the discriminating power between SVT-tagged $b$- and 
  $b$-quark jets in simulated $\ttbar \to \ell\text{+jets}$ and 
  $\ttbar \to \ell\ell\text{+jets}$ events as functions of jet $\eta$ 
  (upper plot) and $p_T$ (lower plot).}
  \label{fig:etabinsdilepton}
\end{figure}

The jet charge templates are derived from dijet samples where exactly two 
reconstructed jets are required while $\ttbar \to \ell\text{+jets}$ events have 
at least four jets. More activity in the event could affect the jet charge 
templates due to a higher number of tracks and possible overlap between tracks 
originating from different jets. This dependence is studied in 
Fig.~\ref{fig:etabinsdilepton}  by comparing the discriminating power for 
$b$- and $\bar{b}$-quark jets in simulated  $\ttbar \to \ell\text{+jets}$ and 
$\ttbar \to \ell\ell\text{+jets}$ events where only two jets are 
expected (additional jets can arise due to ISR and FSR). 
No sizable difference is found and thus no systematic uncertainty is considered.

\subsubsection{Tagging Efficiency in Data and Monte Carlo}
\label{subsec:tagprob_xcheck}
The $b$-tagging efficiencies are known to be different in simulated events 
compared to data. The method to determine the expected charge 
distributions for the SM top and exotic quark charge scenarios depends 
on the tagging efficiency for different jet flavors e.g. a SVT-tagged 
jet in a \ttbar~event is not always a $b$-quark jet but sometimes a $c$-quark jet 
from the $W$ boson decay. In addition the rate at which the various types 
of jets are SVT-tagged are different in data and simulated events. For this 
analysis, only the relative fraction of SVT-tagged jets of different flavors are 
important and not the absolute efficiencies.

To make sure that the procedure to find the flavor of the SVT-tagged jets 
in simulated \ttbar~events is correct, it is repeated but relying 
on the tagging efficiencies derived from data to predict which of the 
jets in the event that are SVT-tagged. The expected SM top and exotic 
quark charge templates were re-derived using this alternative procedure and the 
change was negligible.

\subsubsection{Dependence on the Primary Vertex Position}
\label{subsec:primvtx_dep} 

\begin{figure}[]
  \centering
  \includegraphics[width=0.8\textwidth]{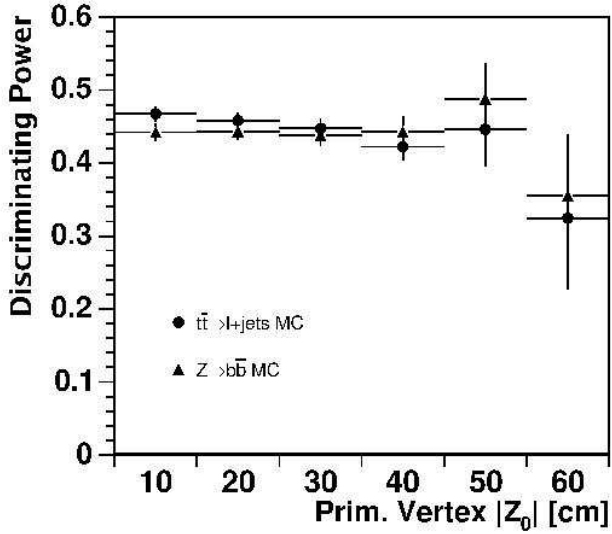}
  \caption{Discriminating power as a function of the PV 
  $z$-position for SVT-tagged $b$- and $b$-quark jets in 
  simulated $t\bar{t}\rightarrow \ell\text{+jets}$ and 
  \Zbbbar~events.}
\label{fig:dp_primvtxz0_ttbar_bbbar}
\end{figure}

The tracking efficiency and other tracking related variables depends 
on the geometry of the tracking detectors. Since samples with 
different PV distributions are used to extract the 
jet charge templates for jets in the signal sample, the discriminating 
power for $b$- and $\bar{b}$-quark jets was calculated with respect to 
the PV position. No significant dependence was found (see 
Fig.~\ref{fig:dp_primvtxz0_ttbar_bbbar}) which may be related to the fact 
that only SVT-tagged jets are considered. 
SVT-tagged jets are by definition associated with a set of well defined 
tracks (see Sec.~\ref{sec:bjets}).

\subsubsection{Fraction of $\mu$-Tagged Jets in the Dijet and Signal Sample}
\label{subsubsec:mutagbias}
The jet charge algorithm considers all tracks satisfying the standard quality cuts in 
Sec.~\ref{subsec:definition}. All jets are required to be tagged by the SVT 
algorithm but no veto is applied to reject $\mu$-tagged jets. The muon track usually 
has significantly higher transverse momentum than other tracks in the same jet, and 
the jet charge distribution for $b$-quark jets containing a muon is significantly 
different. The jet charge templates are derived on a different sample than the signal 
sample and therefore a possible bias may exists due to different fraction of $\mu$-tagged jets 
in the samples. Table~\ref{tab:mu_svt_tags} shows the fraction of the SVT-tagged jets 
that also has a muon associated to it.
\begin{table}
\centering
\begin{tabular}{|l|c|}
\hline
Sample & SVT- \& $\mu$-tagged jets (\%)  \\
\hline
Dijet data & $3.0 \pm 0.2$ \\
\ttbar~MC & $8.6 \pm 0.2$ \\
Signal sample & $11.9 \pm 5.0$ \\
\Zbbbar~MC & $3.2 \pm 0.2$ \\
\hline
\end{tabular}
\caption{ Fraction of SVT-tagged jets that have a muon associated to it in various samples. 
The difference between the dijet data and simulated \ttbar~events (passing signal 
selections) is taken into account as a systematic uncertainty. For comparison, 
the fractions in the signal sample and simulated \Zbbbar~events are shown.}
\label{tab:mu_svt_tags}
\end{table}
A systematic uncertainty due to this discrepancy is obtained by re-deriving 
the jet charge templates for $b$- and $c$-quark jets with a veto on any 
{\sl probe}-jet that has a muon matched to it within a cone of $\Delta R < 0.5$. 

\subsubsection{Fraction of $c$-Quark Jets in the Dijet Samples}
The uncertainty on the fraction of $c$-quark jets in the dijet samples 
originates from the uncertainty of the \ptrel~fit. The $b$- and $c$-quark 
jet charge distributions are re-derived varying the fraction of $c$-quark jets 
according to this uncertainty.

\subsubsection{Fraction of Muon Charge Sign Change in the Dijet and Monte Carlo Samples}
\label{subsubsec:cascsyst}
The amount of cascade decay passing the loose and tight dijet selections 
depends on the jet $p_T$. The simulated \Zbbbar~events and the dijet samples 
have similar but 
not equal jet $\eta$ and $p_T$ spectrum. Therefore, there is still the possibility 
that the fraction of cascade muons is different in the dijet samples and 
simulated \Zbbbar~events, that are used to extract the fraction of times the muon 
charge sign is different from the quark initiating the jet. This uncertainty is 
taken into account by re-deriving the jet charge templates by varying the fraction 
of times the muon charge changes sign, $x_{flip}$ (calculated in 
Sec.~\ref{subsec:muflip}), according to the uncertainty on the weighted average.

The measured charge of the muon can be different from the quark that initiated 
the jet simply because the muon track was poorly reconstructed. This effect exists 
also in the simulated events and we observe that it affects about $1$\% of the muons 
inside jets in simulated \Zbbbar~events. The rate at which the muon charge is 
measured incorrectly might be different in data and simulation. 
This potential difference between data and simulation is taken into account by 
varying how often the charge is misidentified between $0$\% and $3$\%.

\subsubsection{Statistical Uncertainty on the Kinematic Weighting}
\label{subsubsec:systematics_kinematics}
The procedure to weight the \ttbar~Monte Carlo events to reproduce the jet 
$p_T$ and $\eta$ spectrum of the dijet samples has a number of uncertainties 
associated with it. By using the $\pm1\sigma$ band in the fits, the statistical 
uncertainty can be evaluated. There are two weights that are applied, 
the jet $p_T$ and $\eta$ weights, both with a statistical uncertainty 
associated with them. Conservatively, the uncertainties are added in 
quadrature and the jet charge templates are re-derived varying this 
uncertainty.

\subsubsection{Statistical Uncertainty on the Kinematic Correction}
\label{subsubsec:systematics_kinematicsstat}
There is an additional systematic uncertainty related to the kinematic 
correction which arises from the limited statistics used in the correction. 
This systematic is evaluated by using the $\pm 1\sigma$ bands on the kinematic 
correction fit shown in Fig.~\ref{fig:kincorr_b_c_jets} and re-deriving the 
jet charge templates using these varied correction functions.

\subsubsection{Statistical Uncertainty on the Jet Charge Templates}

\begin{figure}[]
  \centering
  \includegraphics[width=0.8\textwidth]{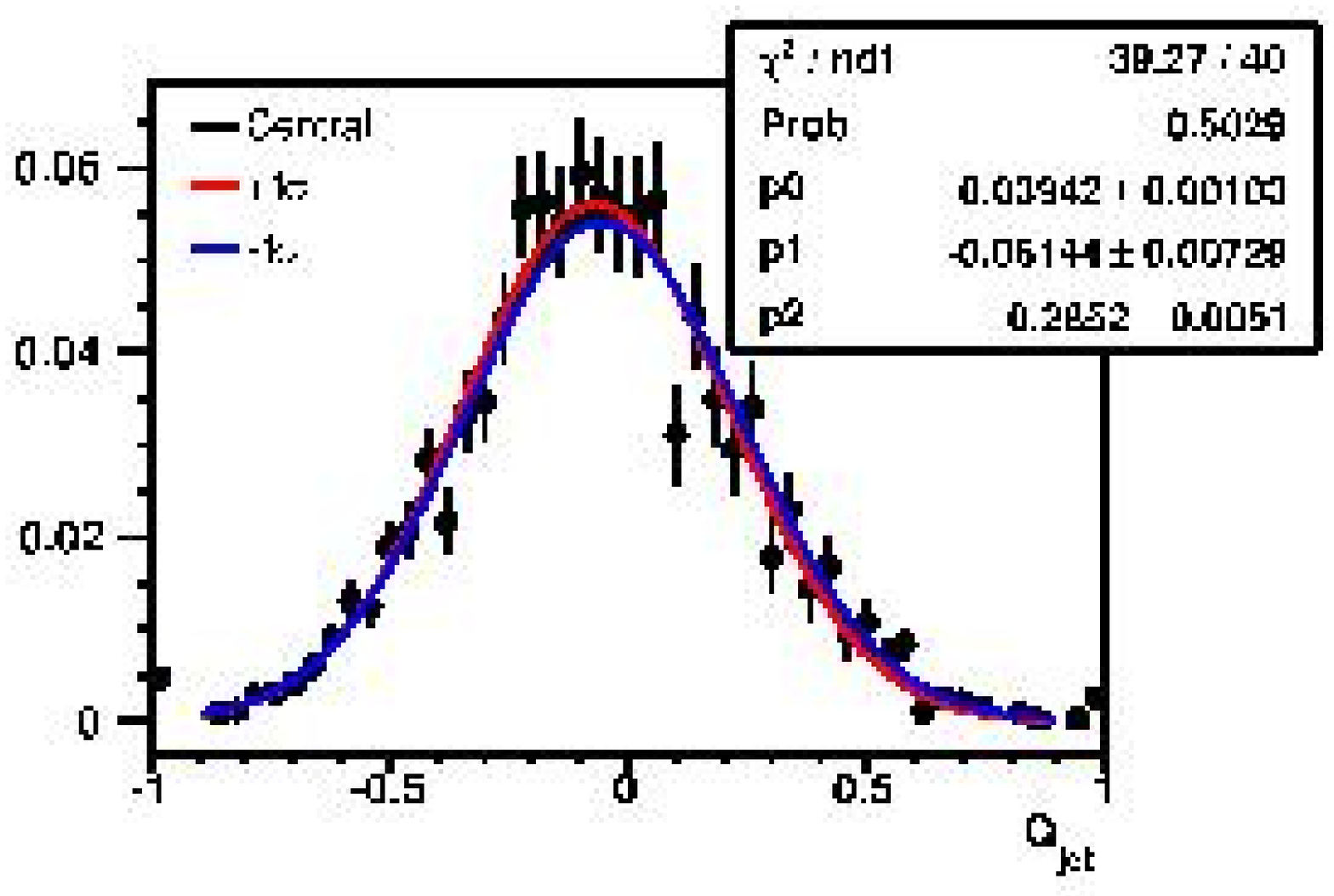}
  \includegraphics[width=0.8\textwidth]{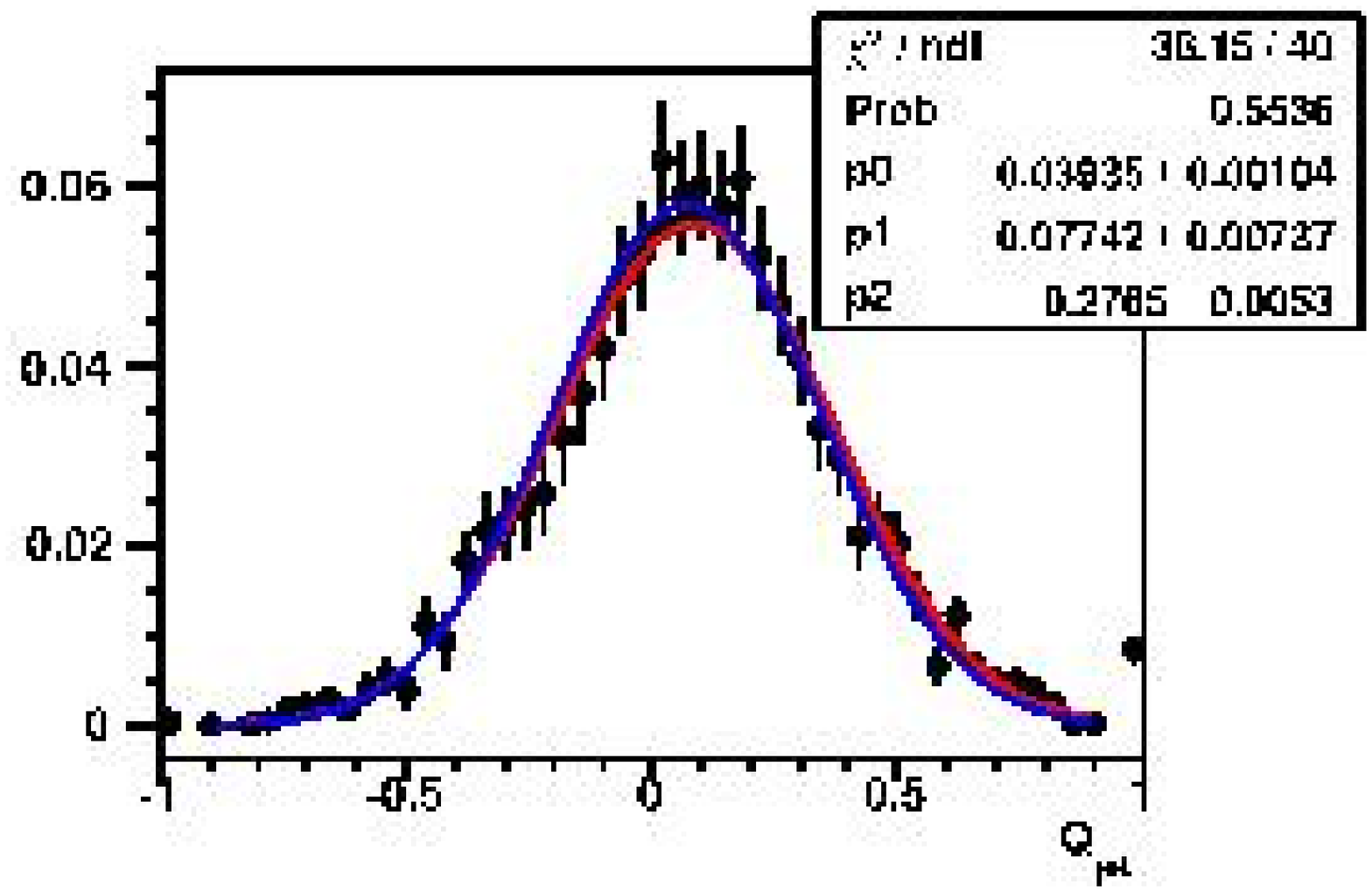}
  \caption{The $b$- and $\bar{b}$-quark jet charge distributions (top and 
  bottom respectively) fitted with a Gaussian function. The $\pm1\sigma$ jet 
  charge templates are derived by varying the 
  parameters of the Gaussian within their errors. The varied 
  jet charge templates are calculated from the difference between the varied 
  and central Gaussian.}
  \label{fig:systematics_bfrfit}
\end{figure}

\begin{figure}[]
  \centering
  \includegraphics[width=0.8\textwidth]{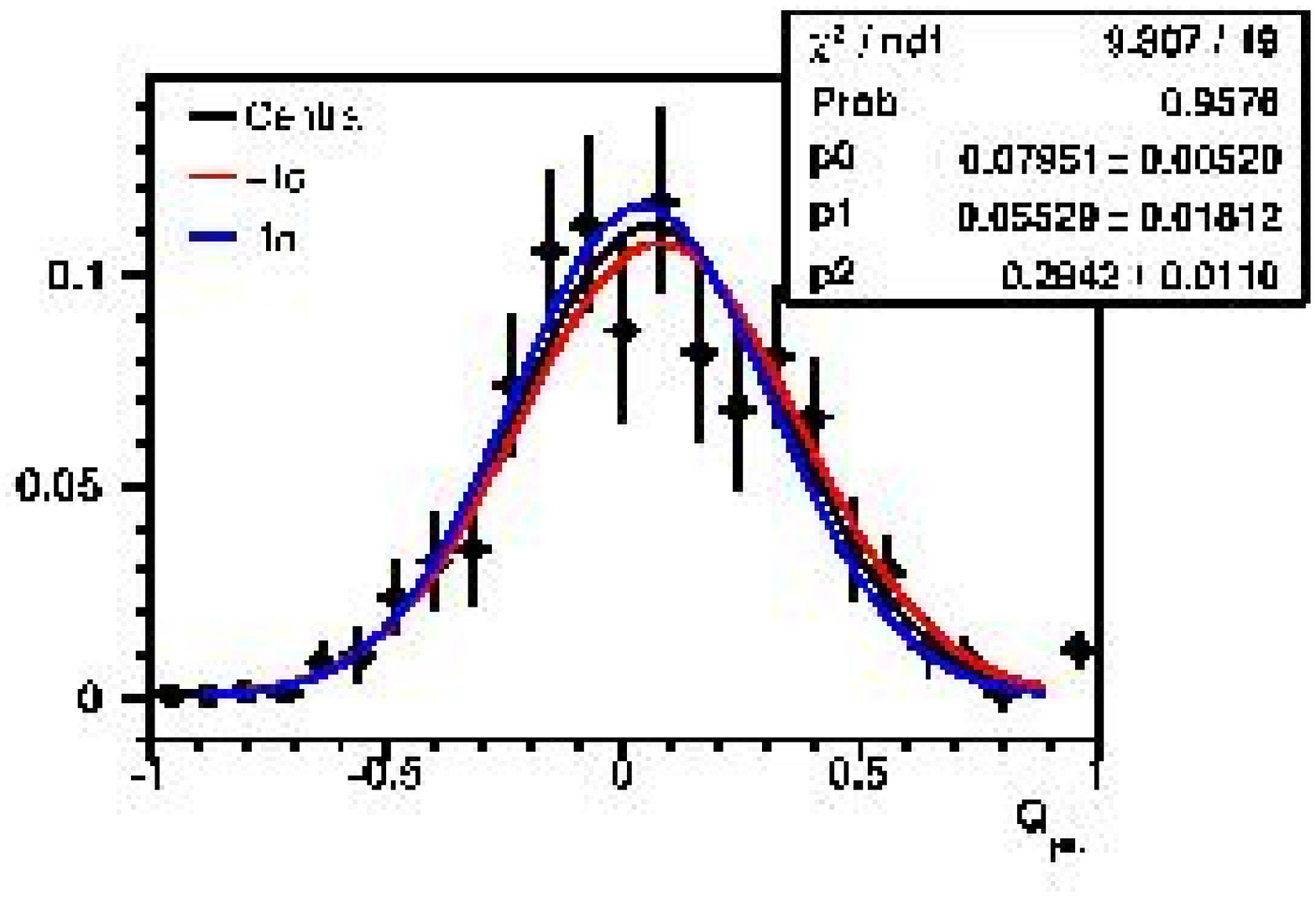}
  \includegraphics[width=0.8\textwidth]{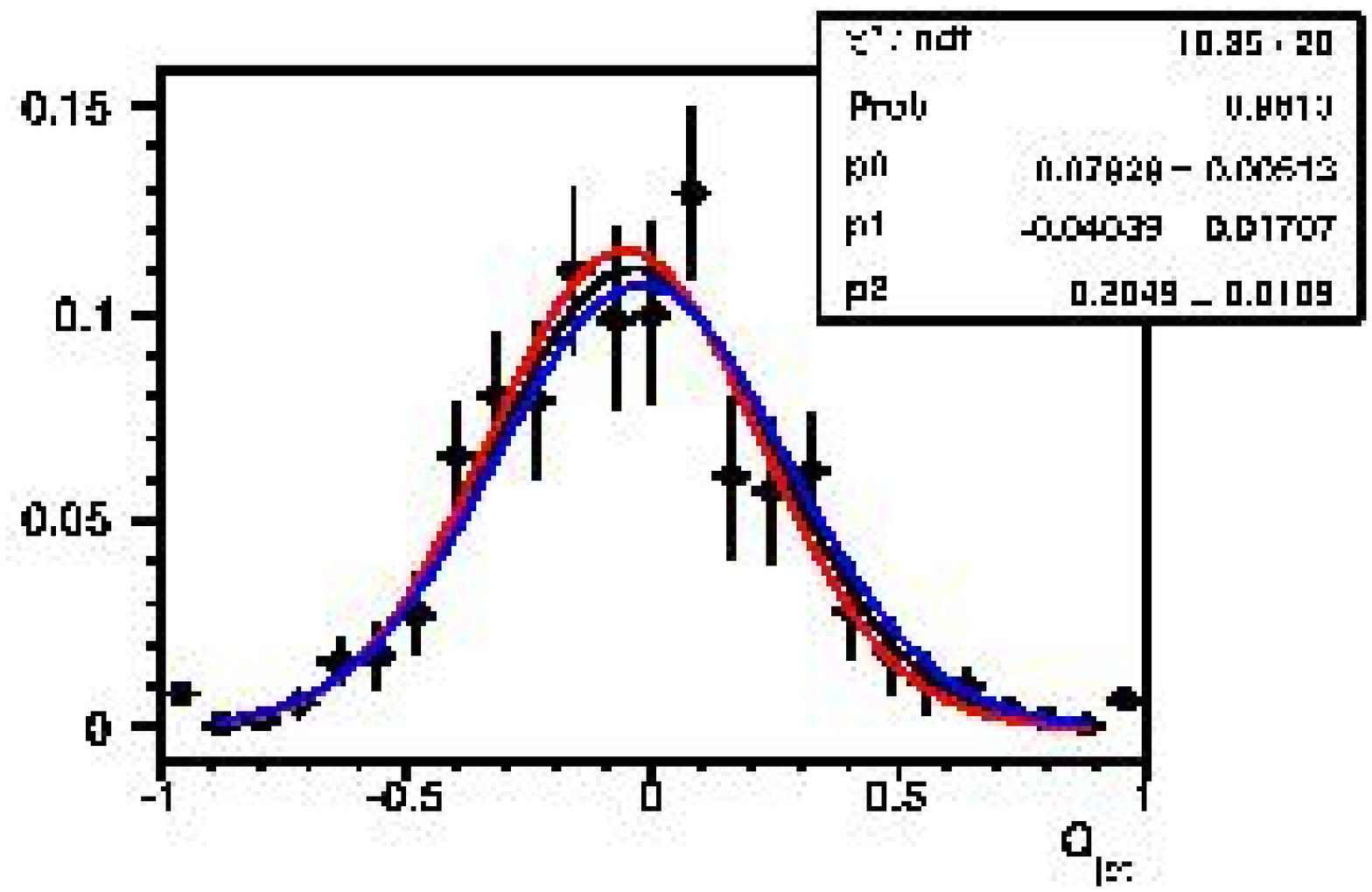}
  \caption{The $c$- and $\bar{c}$-quark jet charge distributions fitted with a 
  Gaussian function (top and bottom respectively). The $\pm1\sigma$ jet charge 
  templates are derived by varying the 
  parameters of the Gaussian within their errors. The varied 
  jet charge templates are calculated from the difference of the varied 
  to the central Gaussian.}
  \label{fig:systematics_cfrfit}
\end{figure}

The extraction of the jet charge template in 
Sec.~\ref{subsec:jetcharge_extraction} leads to a statistical uncertainty 
on the jet charge distributions. This uncertainty is evaluated by fitting 
the jet charge distributions with Gaussian functions and varying the fitted 
parameters within their uncertainties to produce $\pm 1 \sigma$ templates. The fit for $b$-
and $c$-quark jet charge distributions are shown in Fig.~\ref{fig:systematics_bfrfit} 
and \ref{fig:systematics_cfrfit} respectively.
The validity of this procedure was estimated by allowing each bin of the jet charge 
distributions to vary according to a Gaussian centered on the bin content 
and a width according to the error of that bin. After varying all bins 
in the jet charge templates the discriminating power is calculated and 
put in a histogram. This procedure is repeated $10,000$ times and the resulting 
distribution is fitted with a Gaussian where the width is compared to the 
difference in discriminating power between the $\pm 1\sigma$ varied jet charge 
distributions. The varied jet charge templates have a discriminating power 
that agrees with the width of the Gaussian and the statistical uncertainty 
on the templates are thus reasonable.

\subsubsection{Top Quark Mass Uncertainty}
The constrained kinematic fit uses the top quark mass as an additional constraint 
at a value of $175$~GeV. This value is chosen to be consistent with the mass 
used in the generation of the simulated events. Top quark mass is known only 
to a certain precision and constraining the mass in the fit to an alternative 
value might affect the fit performance. To take this into account, the SM top and 
exotic quark charge templates are re-derived using simulated \ttbar~events 
generated with different masses, while keeping the same top mass constraint 
in the fit. Events generated with a top quark mass of $170$ and 
$180$~GeV are used to evaluate this systematic uncertainty. The current world average 
uncertainty in the top mass is $\approx 2.3$~GeV~\cite{topmass_average} and therefore 
the SM top and exotic quark charge templates obtained from the $170$ and $180$~GeV 
samples are scaled corresponding to this uncertainty.

\subsubsection{Jet Energy Scale}
The jet energy scale correction is an attempt to correct the measured jet energies 
in the calorimeter back to the stable-particle level before interacting 
with the detector as described in Sec.~\ref{subsec:jes} and is applied 
to all jets. The jet energy scale correction is different for jets in data and in 
simulated events as shown in Fig.~\ref{fig:jes_data} and~\ref{fig:jes_mc}. The 
uncertainty from the jet energy scale is found by adding the uncertainties 
from data and simulation in quadrature, conservatively treating them totally 
uncorrelated. This uncertainty changes the jet energies in the event and 
may consequently affect the kinematic fit. The SM top 
and exotic quark charge templates are re-derived using the varied jet energy scale 
correction.

\subsubsection{Jet Energy Resolution}
The observed $p_T$ spectrum of jets in simulated events is smeared to match the 
resolution measured in data. The uncertainties 
on the parametrization of the jet energy introduces an uncertainty not taken 
into account by the jet energy scale. To account for this, the 
parametrization is varied according to the uncertainty and the SM top and exotic 
quark charge templates are re-derived.

\subsubsection{Jet Reconstruction and Identification Efficiency}
The efficiency to reconstruct a jet subject to all requirements in 
Sec.~\ref{sec:jetid} is higher in simulated events than in 
data~\cite{xsec_note240_mujets_topological}. The discrepancy is most prominent 
in the low $p_T$ region ($15-25$~GeV). Therefore, a $p_T$ dependent 
data-to-simulation scale factor is derived and jets in simulated events are 
removed to reproduce the jet reconstruction efficiency observed in data. This 
scale factor is varied according to its uncertainty to take into account 
any effect on the kinematic fit and the SM top and exotic quark charge 
templates are re-derived.

\subsubsection{Composition of the Signal Sample}
The signal sample composition discussed in Sec.~\ref{sec:preselection} has 
some uncertainty. Due to the large signal-to-background ratio this 
uncertainty has a small net effect on the analysis. It is evaluated 
by taking into account the statistical and systematic uncertainty 
when extracting the final result. A more detailed description is 
given in Sec~\ref{sec:results}.

\subsubsection{\bbbar~Production Mechanism}
\label{subsubsec:bbbar_prod_sys}
The dominant \bbbar~production mechanism in the dijet samples is assumed to be 
flavor creation. Other processes contributing to the production of \bbbar~pairs 
(flavor excitation and gluon splitting) destroy the correlation between the 
sign of the tagging muon and the quark initiating the {\sl probe}-jet. The 
fraction of flavor creation is sensitive to the 
azimuthal distance between the jets. Therefore, the jet charge templates 
are re-derived with the requirement of $\Delta \phi>2.65$ instead of 
$\Delta \phi>3.0$ on the azimuthal distance between the two jets in the 
dijet samples to take this uncertainty into account.

\subsubsection{\ttbar~Signal Modeling}
In the kinematic fit it is assumed that the four highest $p_T$ jets in the event are 
the result of the hadronization of the partons from the \ttbar~decay. As 
discussed before, additional jets can arise from ISR and FSR. When \ttbar~events 
are produced in association with a jet, the additional jet can be 
misinterpreted as a decay product from the \ttbar~pair. To assess the 
uncertainty in the modeling of these effects, events has been generated 
using a dedicated simulation of the production of \ttbar~events together 
with an additional parton using {\sc ALPGEN}. The fraction of such events 
is estimated not to be larger than $30$\%~\cite{top_mass_matrix_element_PRD}. 
The SM top and exotic quark charge templates are re-derived using 
this sample and systematic uncertainty of $30$\% of the difference between 
this and the default simulated \ttbar~sample is quoted.

\section*{}
To estimate the effect of the individual sources of systematic uncertainties, the 
discriminating power between the varied SM top and exotic quark charge 
templates is calculated and shown in Tab.~\ref{tab:dp_sys}.
\begin{table}
\centering
\begin{tabular}{|lc|}
\hline
Systematic uncertainty & Variation \\
\hline
Fraction of $c$-quark jets & $\pm0.003$ \\
Fraction of muon charge sign change & $\pm0.007$ \\
Stat. uncert. on the kinematic weighting & $\pm0.003$ \\
Stat. uncert. on the kinematic correction & $\pm0.014$ \\
\bbbar~production mechanism & $\pm0.008$ \\
Stat. uncert. $b$-quark jet charge templates & $\pm0.023$ \\
Stat. uncert. $c$-quark jet charge templates & $\pm0.002$ \\
Fraction of $\mu$-tagged jets & $\pm0.005$ \\
\ttbar~signal Modeling & $\pm0.007$ \\
Jet energy scale & $\pm0.003$ \\
Jet energy resolution & $\pm0.012$ \\
Jet reconstruction efficiency & $\pm0.009$ \\
Top quark mass uncertainty & $\pm0.002$ \\
\hline
\end{tabular}
\caption{The maximum shift in discriminating power between the varied 
SM top and exotic quark charge templates for the individual 
sources of systematic uncertainty.}
\label{tab:dp_sys}
\end{table}

\clearpage

\section{Results}
\label{sec:results}

The underlying models in this analysis are the two hypotheses of a SM top quark with 
a charge of $2e/3$ or an exotic quark with charge $4e/3$ referred 
to in the following as $H_{\rm sm}$ and $H_{\rm ex}$ respectively. 
A hypothesis test is performed in Sec.~\ref{subsec:lhratio} to assess 
the validity of the different hypotheses using a likelihood ratio test. 
However, the hypothesis test only provides the confidence in rejecting any 
of the two discrete scenarios, it does not assess the probability 
of a mixture of quarks with charge $2e/3$ and $4e/3$ in the signal sample.
Therefore, in Sec.~\ref{subsec:max_LH_fit} the fraction of quarks with 
the exotic charge $4e/3$ is estimated using the method of maximum likelihood.

\subsection{Discrimination Between Charge 2e/3 Top Quark and Charge 4e/3 Exotic Quark Production Scenarios}
\label{subsec:lhratio}

\begin{figure}[]
  \centering
  \includegraphics[width=1.0\textwidth]{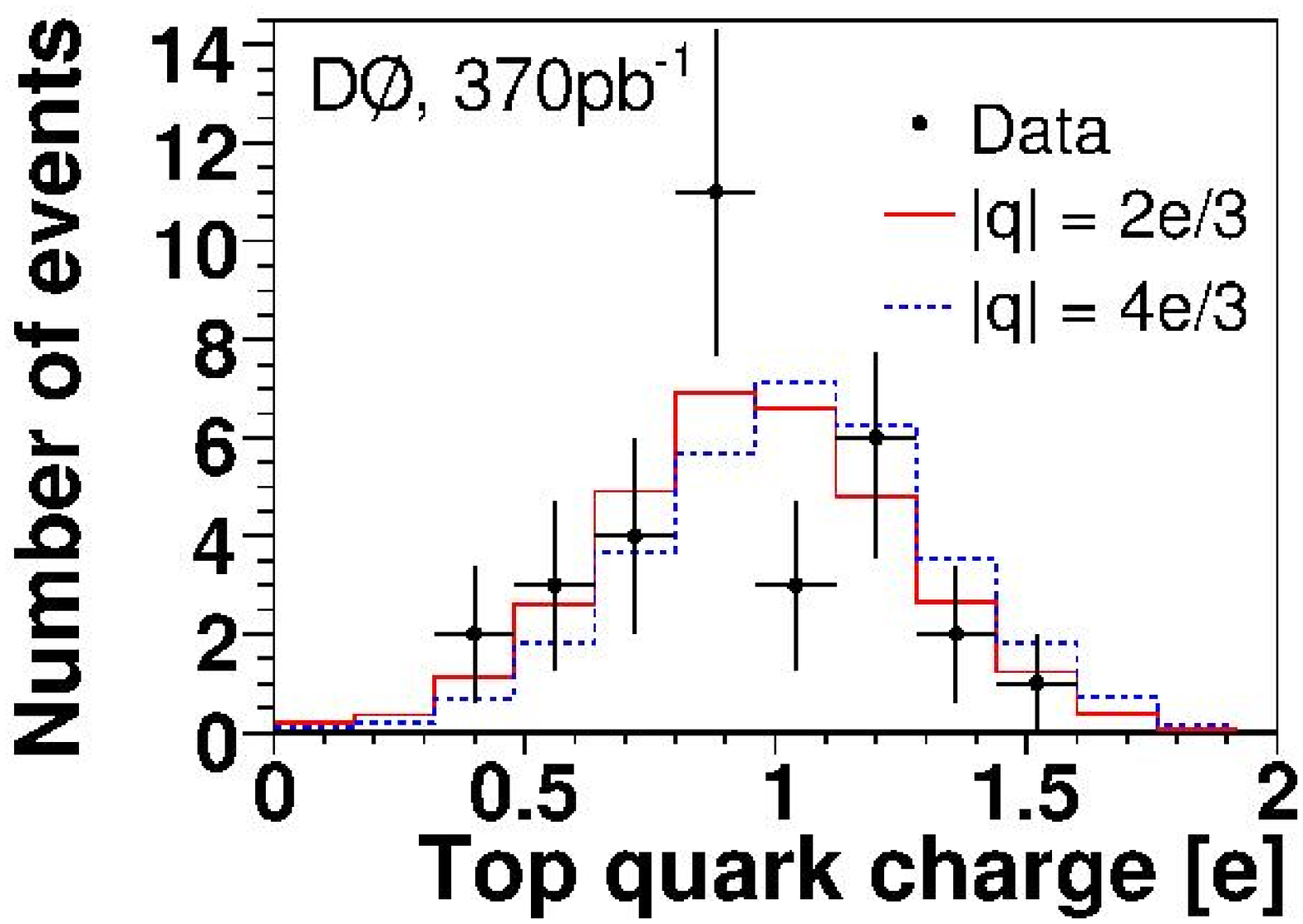}
  \includegraphics[width=1.0\textwidth]{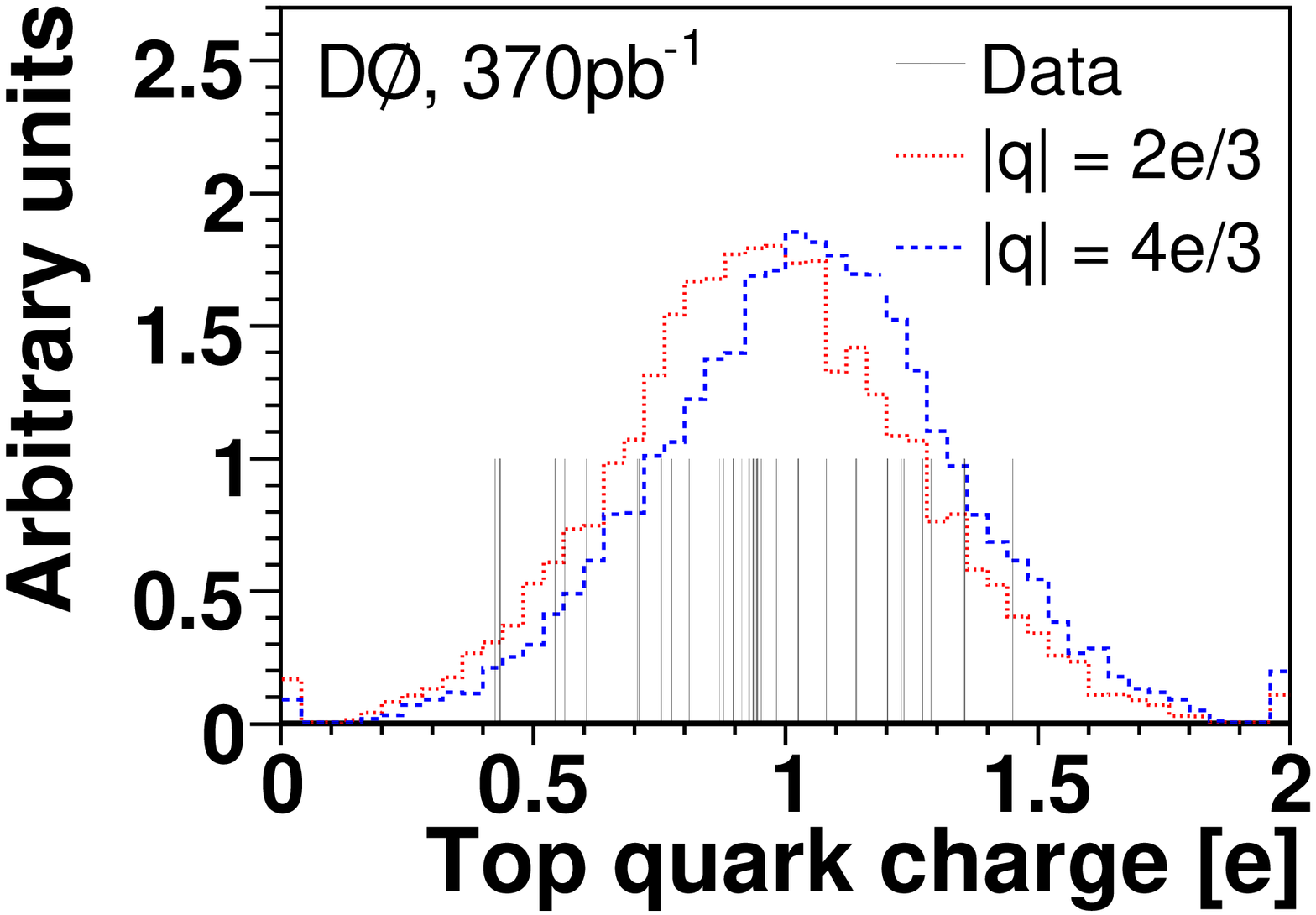}
  \caption{The 32 observed charges compared to the SM top and exotic quark charge templates 
  for binned (top) and unbinned (bottom) data.}
  \label{fig:topchargedata}
\end{figure}

In data a certain set of charges $Q_1$ and $Q_2$ are observed in the leptonic and 
hadronic leg of the \ljets~event respectively. There 
are 21 selected events in the signal sample. In 16 events, the 
kinematic fit converges in at least one of the permutations where 
the two jets $j_b$ and $j_B$ are associated to SVT-tagged jets. Each of the
16 events provides two observations of the charge, giving 16 observations 
of $Q_1$ and 16 observations of $Q_2$, i.e. a total of 32 charge 
observations. Figure~\ref{fig:topchargedata} shows the binned and 
unbinned distribution of measured charges in the 
data overlaid with the expected SM top and exotic quark charge templates. 
As can be seen from the figures, the observed data prefers the SM hypothesis. 
Below, a procedure to quantify the discrimination between the two hypotheses 
is described.

\subsubsection{Likelihood Ratio}

Often when discriminating between two scenarios it is useful to define a 
test statistics. If the set of 32 measured values 
$Q_1$ and $Q_2$ are denoted by a vector ${\bf q}^{\rm data}$ then the test statistics, 
denoted $\Lambda$, can be 
a single number or a vector with fewer components than ${\bf q}^{\rm data}$.
$\Lambda$ is a function of the data and its value reflects the level of agreement 
between the data and the hypothesis. Under assumption of each hypothesis 
different $\Lambda$ will be obtained for repeated experiments. 

The goodness is quantified by the $P$-value, which is the probability to 
find $\Lambda$ in the region of equal 
or lesser compatibility assuming a hypothesis $H$ than the level of 
compatibility observed with the real data $\Lambda^{\rm data}$.
In this analysis, the test statistics $\Lambda$ is the ratio of the 
likelihoods $\mathcal{L}$ for the two different hypotheses $H_{\rm sm}$ and 
$H_{\rm ex}$ given a vector of data ${\bf q}$
\begin{equation}
  \Lambda({\bf q}) = \frac{\mathcal{L}({\bf q}|H_{\rm sm})}{\mathcal{L}({\bf q}|H_{\rm ex})},
  \label{eq:lh_ratio_def}
\end{equation}
where the likelihood functions are defined as
\begin{gather}
  \mathcal{L}({\bf q}|H_{\rm sm}) = \prod_i p_{\rm sm}(q_i) \\
  \mathcal{L}({\bf q}|H_{\rm ex}) = \prod_i p_{\rm ex}(q_i).
  \label{eq:lh_def}
\end{gather}
Here the probability density functions (p.d.f.s) $p_{\rm sm}$ and $p_{\rm ex}$ 
are the SM top  and exotic quark charge templates derived in 
Sec.~\ref{subsec:templateconstruction} and the subscript $i$ runs from $1,2,..,32$. 

Under the assumption that the distribution of ${\bf q}$ follows a hypothesis $H$, 
the p.d.f. $g(\Lambda|H)$ will be determined for the test statistics. Here, the 
two hypotheses gives:
\begin{gather}
  \Lambda^{\rm sm} \equiv g(\Lambda|H_{\rm sm}) \\
  \Lambda^{\rm ex} \equiv g(\Lambda|H_{\rm ex}).
  \label{eq:g_def}
\end{gather}
Given the definition of the test statistic, the $P$-value for the exotic 
scenario is
\begin{equation}
  P = \int^{\infty}_{\Lambda^{\rm data}} \Lambda^{\rm ex}(\Lambda) d\Lambda .
  \label{eq:p_value}
\end{equation}

The likelihood ratio obtained from Eq.~\ref{eq:lh_ratio_def} with 
the set ${\bf q}^{\rm data}$ of 32 observed charges in the signal sample is,
\begin{equation}
  \Lambda^{\rm data} = 4.27 \nonumber
\end{equation}

\subsubsection{Generation of Pseudo-experiments}

The $P$-value is not the probability of the hypothesis but rather the 
probability under the assumption of the hypothesis of obtaining data 
as incompatible with the hypothesis as the data observed. Thus, in 
order to calculate the $P$-value for the exotic scenario defined in 
Eq.~\ref{eq:p_value}, the p.d.f. $\Lambda^{\rm ex}$ has to be computed.

This can be determined by the generation of 
pseudo-experiments which in an idealized world would correspond to 
building a set of new experiments and measure the charge under the 
assumption of $H_{\rm ex}$. The procedure is as follows:

\begin{enumerate}
  \item 
  \label{enum:sig}
  {\it Signal and background fraction} \newline
    The signal and background fractions are allowed to fluctuate according to 
	the statistical and systematic uncertainties: 
	The number of background events $N_{\rm bkg}$ is obtained from a Binomial 
	distribution with a mean at the prediction given by a Gaussian distribution with 
	a mean of $1.53$ (the central prediction) and a standard deviation equal to the total 
	uncertainty on the number of background events (see Fig.~\ref{fig:bkg_pred_ensembles}). 
	\begin{figure}[]
	  \centering
	  \includegraphics[width=0.7\textwidth]{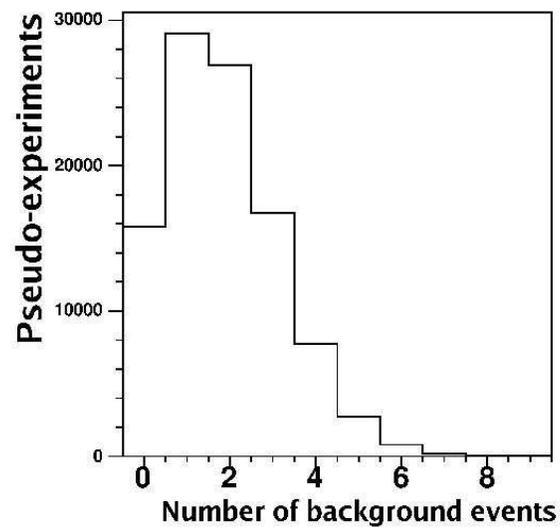}
	  \caption{The distribution of the number of background events for each pseudo-experiment.}
	  \label{fig:bkg_pred_ensembles}
    \end{figure} 
  \item
  \label{enum:sys}
  {\it Systematic uncertainties} \newline
	The various sources of systematic uncertainties affect the template 
	distributions and have to be taken into account. The method used here 
	allows the systematic uncertainties to change the default (or best 
	guess) template. All systematic uncertainties $j$ are assumed to follow 
	Gaussian distributions with widths $\sigma_j$. Each systematic uncertainty 
	is modeled by a free parameter $\nu_j$ called 
	{\sl nuisance parameter}~\cite{nuisance_params}. If 
	the default SM top quark template is characterized by a function $p_{\rm sm}(Q)$ and if 
	$p_{\rm sm}^{j+}(Q)$ and $p_{\rm sm}^{j-}(Q)$ are the $\pm 1 \sigma$ varied 
	templates for the source of error $j$, then the SM top quark template (and similar for the 
	exotic quark charge template) as a function of all systematic uncertainties can be written as:
	\begin{gather}
	  p_{\rm sm}(Q,\nu_j) = p_{\rm sm}(Q) + \sum_j \nu_j (p_{\rm sm}^{j \pm}(Q) - p_{\rm sm}(Q)), \\
	  p_{\rm ex}(Q,\nu_j) = p_{\rm ex}(Q) + \sum_j \nu_j (p_{\rm ex}^{j \pm}(Q) - p_{\rm ex}(Q)),
	  \label{eq:nuisance_params} 
	\end{gather}
	and similar for the exotic template. $\nu_j$ follow Gaussian distributions with mean zero and 
	standard deviation one. 
  \item
  \label{enum:pseudo}
  {\it Pseudo-measurements} \newline
	The number of pseudo-measurements has to be equal to the total number of observed 
	charges and thus $p_{\rm ex}(Q)$ is randomly sampled $32-N_{bkg}$ times and the background 
	template (given in Sec.~\ref{susec:bkg}) $N_{bkg}$ times to obtain a set of pseudo-measurements 
	${\bf q}^{\rm pseudo}$.
  \item
  \label{enum:lhratio}
  {\it Likelihood ratio} \newline
  	Given the set of pseudo-measurements ${\bf q}^{\rm pseudo}$, the likelihood ratio is 
	calculated using the generated $p_{\rm sm}(Q,\nu_j)$ and $p_{\rm ex}(Q,\nu_j)$ to 
	take into account the systematic uncertainties.

\end{enumerate}

\begin{figure}[]
  \centering
  \includegraphics[width=0.8\textwidth]{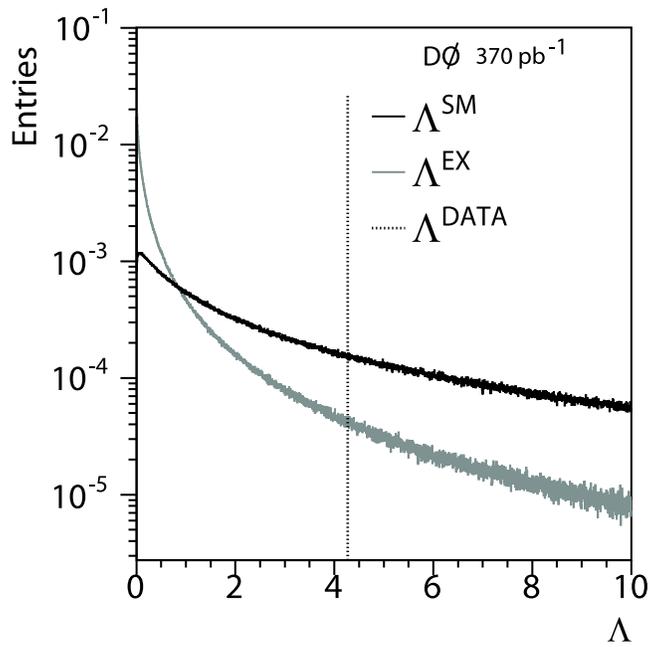}
  \caption{The distribution of $\Lambda^{\rm sm}$ and $\Lambda^{\rm ex}$ resulting from 
  the pseudo-experiments together with observed value of $\Lambda^{\rm data}$.}
  \label{fig:lambda}
\end{figure}

By repeating steps~\ref{enum:sig}-\ref{enum:lhratio} $100,000$ times (each time with 
a new background fraction and new set of nuisance parameters $\nu$), $\Lambda^{\rm ex}$ 
can be determined. In a similar way, the 
p.d.f. $\Lambda^{\rm sm}$ can be determined by sampling the $p_{\rm sm}(Q)$ charge
template in step~\ref{enum:sys}. Both $\Lambda^{\rm ex}$ and $\Lambda^{\rm sm}$ are 
shown in Fig.~\ref{fig:lambda} together with the observed $\Lambda^{\rm data}$ in data.

\subsubsection{Confidence Level}
\label{subsec:conflev}
From Fig.~\ref{fig:lambda} it is clear that the observed set of charges is more 
SM like than exotic like. The calculation of what confidence level $\alpha$ the 
observed charges excludes the hypothesis of 100\% quarks with the exotic $4e/3$ 
charge in the signal sample is simply done by observing that
\begin{equation}
  1 - \alpha = P.
\end{equation}
The confidence level is therefore extracted by calculating the $P$-value as 
defined in Eq.~\ref{eq:p_value}
\begin{equation}
  \int^{\infty}_{\Lambda^{\rm data}} \Lambda^{\rm ex}(\Lambda) d\Lambda = 0.078,
\end{equation}
and the corresponding confidence level is $\alpha = 92.2$\%.

The probability that the exotic hypothesis gives a set of charges as incompatible 
with the observed median of $\Lambda^{\rm sm}$ is also calculated. This can be 
thought of as an expected confidence level and is given by:
\begin{equation}
  1- \alpha^{\rm expected} =  \int^{\infty}_{\Lambda^{\rm sm}_{\rm Median}} \Lambda^{\rm ex}(\Lambda) d\Lambda = 0.088 
\end{equation}
which gives the expected confidence level $\alpha^{\rm expected} = 91.2$\% in good 
agreement with the observed $92.2$\%.

The effect of the systematic uncertainties on the confidence level can be seen 
in Tab.~\ref{tab:sysbreakdown} where the cumulative and individual effect of 
including systematic uncertainties are shown. 
%
%
%
%
%
%
%
%
%
%
%
%
%
%
%


\begin{table}
\centering
\begin{tabular}{|l|c|c|c|}
\hline
Systematic uncertainty & $\alpha$ & $\alpha^{\rm expected}$ & $\alpha^{\rm expected}_{\rm single}$ \\ 
\hline
Stat. uncert. & $95.82$  & $95.25$ & $95.25$ \\ 

Fraction of $c$-quark jets & $95.75$ & $95.23$ & $95.24$ \\ 

Fraction of muon charge sign changes & $95.66$ & $95.21$ & $95.26$ \\ 

Stat. uncert. on the kinematic weighting & $95.56$ & $95.21$ & $95.25$ \\ 

Stat. uncert. on the kinematic correction & $94.37$ & $94.11$ & $94.35$ \\ 

\bbbar~production mechanism & $93.70$ & $93.35$ & $94.55$ \\ 

Stat. uncert. $b$-quark jet charge templates & $93.47$ & $93.22$ & $95.18$ \\ 

Stat. uncert. $c$-quark jet charge templates & $93.33$ & $93.12$ & $95.23$ \\ 

Fraction of $\mu$-tagged jets & $93.28$ & $93.06$ & $95.19$ \\ 

\ttbar~signal modeling & $93.11$ & $92.89$ & $95.12$ \\ 

Jet energy scale & $93.08$ & $92.62$ & $95.15$ \\ 

Jet energy resolution & $92.82$ & $92.26$ & $95.16$ \\ 

Jet reconstruction efficiency & $92.44$ & $91.79$ & $95.17$ \\ 

Top quark mass uncertainty & $92.22$ & $91.17$ & $95.17$ \\ 

\hline

\end{tabular}
\caption{Breakdown of the systematic uncertainties and their cumulative effect ($\alpha$ and 
$\alpha^{\rm expected}$) on the confidence level.  $\alpha^{\rm expected}_{\rm single}$ is 
the effect on the confidence level allowing only one source of systematic uncertainty to enter the 
calculation.}
\label{tab:sysbreakdown}
\end{table}

\subsection{Fraction of Charge $2e/3$ Top Quark and Charge $4e/3$ Exotic Quarks}
\label{subsec:max_LH_fit}

The exclusion at $92$\% confidence level that the set of data 
${\bf q}^{\rm data}$ arose from the decay of a quark with the 
exotic $4e/3$ electric charge does not address the possibility that 
the set of data results from a mixture of a $2e/3$ and a $4e/3$ 
charge quark. In this section the fraction of exotic $4e/3$ 
charge quarks in the signal sample is estimated. 

\subsubsection{Maximum Likelihood Fit}

To extract as estimate of the fraction of exotic quarks $\rho$ the 
method of maximum likelihood is used.
If $p_{\rm mix}({\bf q}^{\rm data};\rho)$ is the p.d.f. from where the 
measurements ${\bf q}^{\rm data}$ arises, the estimated fraction of exotic 
quarks is the value of $\rho$ that maximizes the likelihood function, 
\begin{equation}
  \mathcal{L}(\rho) = \prod_i^{32} p_{\rm mix}(q_i^{\rm data}|\rho).
  \label{eq:lhfnc}
\end{equation}

The p.d.f.s $p_{\rm mix}({\bf q}^{\rm data}|\rho)$ for a specific 
fraction of exotic quarks $\rho$ is found by mixing the pure SM top quark 
charge template $p^{sm}(q)$ and the pure exotic quark charge template 
$p^{ex}(q)$ with the appropriate fraction.
\begin{equation}
  \label{eq:mixpdfs}
  p_{\rm mix}(q|\rho) = \rho \times p^{\rm ex}(q) + (1-\rho) \times p^{\rm sm}(q).
\end{equation}

For each value of $\rho$ (taken to be $0,0.05,...,1.0$) the unbinned 
likelihood that the data ${\bf q}^{\rm data}$ is consistent with 
the sum of signal and background is computed as described in 
Eq.~\ref{eq:lhfnc}. The $-\log{ \mathcal{L}}$ is 
minimized for the 32 observed charges in data and the result is 
a set of $-\log{ \mathcal{L}}$ points versus $\rho$. The result 
can be fitted with a parabola to find the maximum likelihood 
estimator of $\rho$.

\subsubsection{Interpretation of the Result to a Confidence Interval}
\label{subsubsec:confinterval}

For Gaussian errors, $-\log{ \mathcal{L}}$ has a parabolic shape and 
the confidence region (the probability that this region contains 
the true value of $\rho$) is given by,
\begin{equation}
  -\ln{\mathcal{L}(\rho)} \leq -\ln{\mathcal{L}_{\rm max}} + \Delta \ln{\mathcal{L}},
  \label{eq:confreg}
\end{equation}
and the $68$\% ($90$\%) standard deviation confidence intervals are 
given by $\Delta \mathcal{L} = 1/2 (2.71/2)$~\cite{PDG}.  

The result is also reported using a Bayesian approach where the prior 
p.d.f. $\pi(\rho)$ characterizes the prior knowledge of the true value 
of $\rho$. Using Bayes theorem,
\begin{equation}
  f(\rho|{\bf q}^{\rm data}) =  \frac{\mathcal{L}({\bf q}^{\rm data}|\rho) \pi(\rho)}{\int_{-\infty}^{\infty} \mathcal{L}({\bf q}^{\rm data}| \rho ') \pi(\rho ') d \rho '},
\end{equation}
the Bayesian confidence interval can be determined by 
computing the interval $[\rho_{\rm min},\rho_{\rm max}]$ which contain 
a given fraction $1-\alpha$ of the probability,
\begin{equation}
 1- \alpha =  \int_{\rho_{\rm min}}^{\rho_{\rm max}} f(\rho,|{\bf q}^{\rm data}) d {\bf q}^{\rm data}.
\end{equation}

In this analysis, a flat prior p.d.f. according to the physical region is used i.e. 
it is one in the interval $[0,1]$ and zero elsewhere. Using this, the 
calculation of the Bayesian interval is given by finding the interval 
$[\rho_{\rm min},\rho_{\rm max}]$ such that the probability of $1-\alpha$ is,
\begin{equation}
\label{eq:CL}
\frac{\int_{\rho_{ML}}^{\rho_{\rm max}}\mathcal{L}({\bf q}^{\rm data}|\rho)~d \rho}{\int_{0}^{1}\mathcal{L}({\bf q}^{\rm data}|\rho)~d \rho} = \frac{\int_{\rho_{\rm min}}^{\rho_{ML}}\mathcal{L}({\bf q}^{\rm data}|\rho)~d \rho}{\int_{0}^{1}\mathcal{L}({\bf q}^{\rm data}|\rho)~d \rho} = \frac{(1-\alpha)}{2},
\end{equation}
where $\rho_{ML}$ is the fraction at the minimum of the parabola 
(i.e. the fitted fraction). If $\rho_{ML}$ is outside the physical region 
a one-sided confidence interval is reported as,
\begin{equation}
\label{eq:CLlow}
\frac{\int_{0}^{\rho_{\rm max}}\mathcal{L}({\bf q}^{\rm data}|\rho)~d \rho}{\int_{0}^{1}\mathcal{L}({\bf q}^{\rm data}|\rho)~d \rho} = 1-\alpha,
\end{equation}
if $\rho_{ML}<0$ or,
\begin{equation}
\label{eq:CLhigh}
\frac{\int_{\rho_{\rm min}}^{1}\mathcal{L}({\bf q}^{\rm data}|\rho)~d \rho}{\int_{0}^{1}\mathcal{L}({\bf q}^{\rm data}|\rho)~d \rho} = 1-\alpha.
\end{equation}
if $\rho_{ML}>1$.

\subsubsection{Validation of the Fit Procedure}
\label{subsec:enstest}

Before applying the maximum likelihood fit to the selected signal 
sample the fit procedure is validated by performing pseudo-experiments 
where the input fraction of exotic quarks is known. Ideally, the 
fit returns the same fraction without any deviations together with 
information on the expected statistical sensitivity.

\begin{figure}[] 
  \centering
  \includegraphics[width=0.8\textwidth]{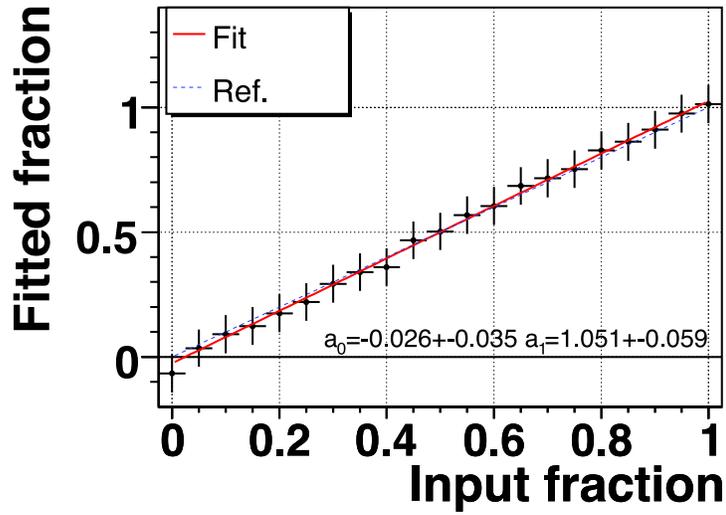} 
  \caption{The fitted versus input fraction of exotic quarks in the pseudo-experiments.}
  \label{fig:lhfitcal_fit}
\end{figure}

\begin{figure}[] 
  \centering
  \includegraphics[width=0.45\textwidth]{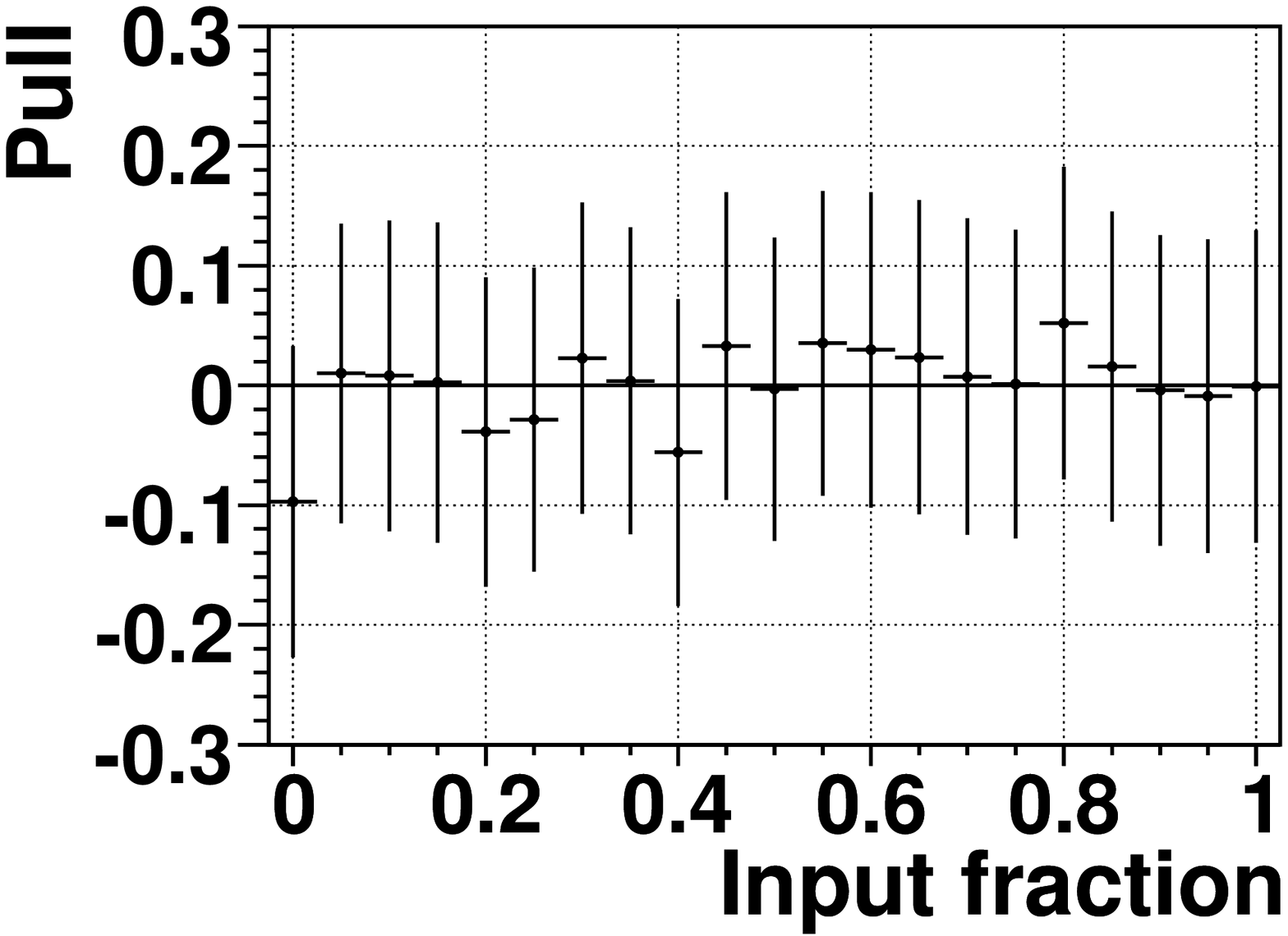}
  \includegraphics[width=0.45\textwidth]{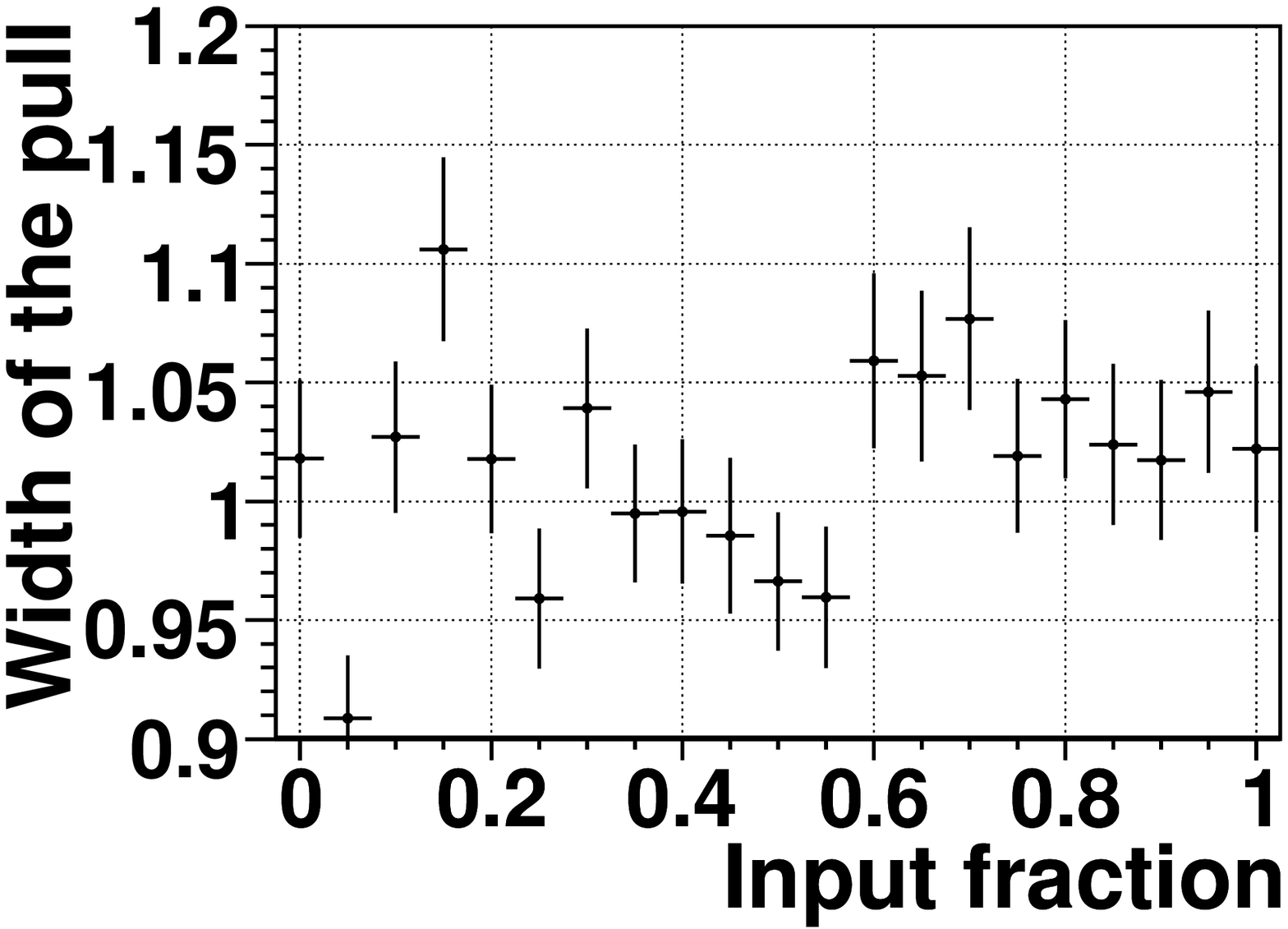}
  \caption{The pull (left) and the width of the fitted pull function (right) versus the input fraction 
  of exotic quarks.}
  \label{fig:lhfitcal_properties_pull}
\end{figure}

One pseudo-experiment is built by randomly sampling the p.d.f. 
$p_{\rm mix}(q|\rho)$ for a specific exotic fraction $\rho$ to 
obtain pseudo-data. In this procedure, the correct signal and 
background fractions are taken into account. 
A maximum likelihood fit is then performed with the same procedure 
as applied to the real data. 
By performing a large number of these pseudo-experiments for each 
input fraction $\rho = 0.0,0.05,..,1.0$ the performance of the fit 
can be evaluated.  
The fitted versus input fraction shown in 
Fig.~\ref{fig:lhfitcal_fit} is consistent with offset zero 
and slope one. The pull and the width of the fitted pull 
functions are shown in Fig.~\ref{fig:lhfitcal_properties_pull}.

\begin{figure}[] 
  \centering
  \includegraphics[width=0.8\textwidth]{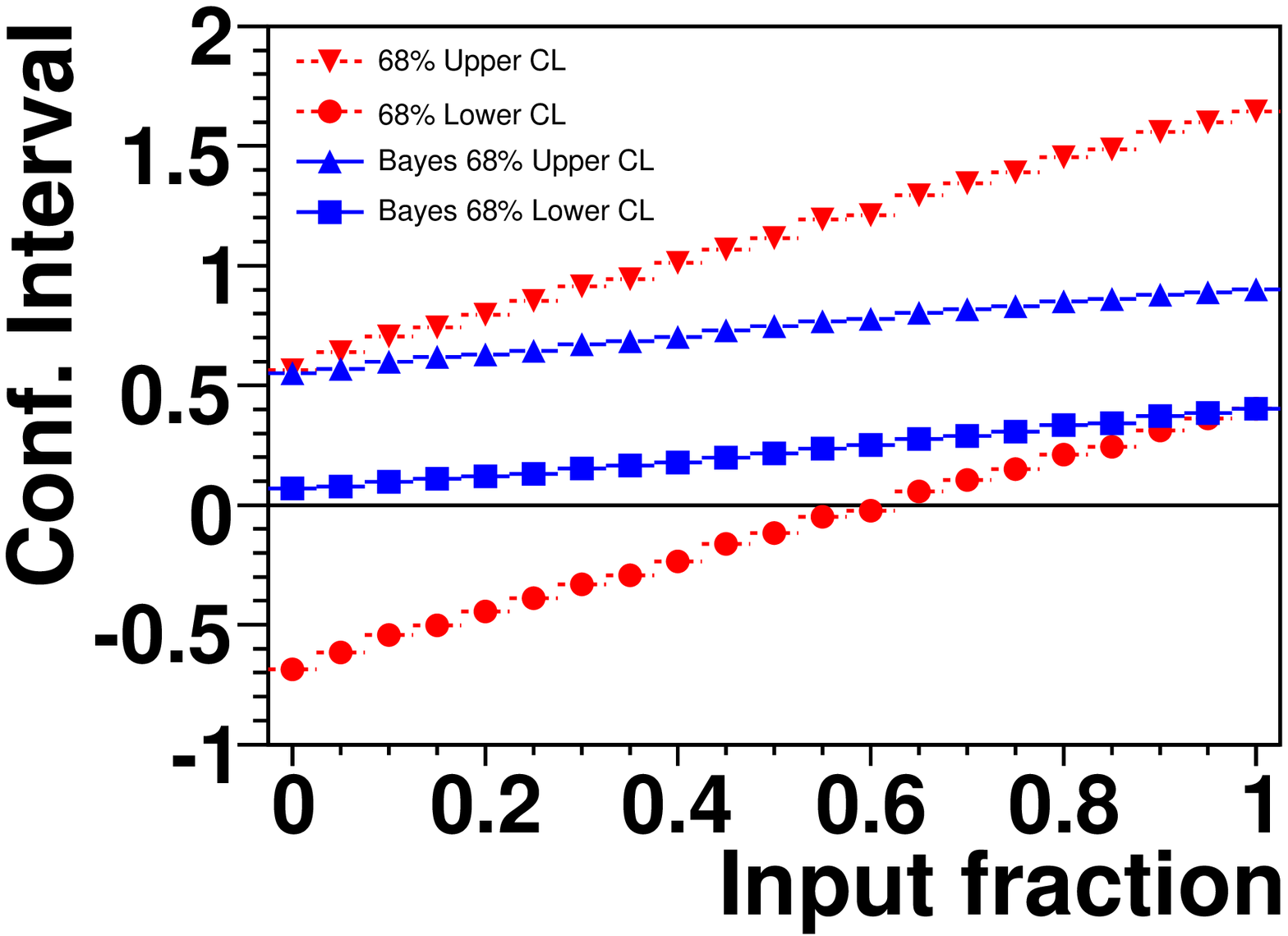} 
  \caption{The average $68$\% confidence intervals for the normal 
  (dashed) and Bayesian (solid) method. The red (blue) depicts the 
  upper (lower) limit.}
  \label{fig:expected_CI}
\end{figure}

For each pseudo-experiment, a confidence interval can be calculated. 
Figure~\ref{fig:expected_CI} shows the expected confidence 
intervals (taken as the mean of the confidence intervals for all 
pseudo-experiments) versus input fraction for both methods described 
above. 
The Bayesian mean value is constrained by the prior 
assumption of $\rho \in [0,1]$. The conclusion is that the expected 
width of the $68$\% confidence interval is approximately $\pm 0.6$.

\subsubsection{Systematic Uncertainties} 
\label{subsec:lhfit_syserr}
Sources of systematic uncertainties affect the fitted fraction 
in two ways: they can either change the fraction of signal and 
background or they can change the templates 
$p_{\rm mix}(q|\rho)$. 
The first is taken into account by allowing the predicted 
background fraction to vary according to its total 
uncertainty, constrained to the total number of events observed 
in data. The effect of the other systematic uncertainties 
are estimated by performing pseudo-experiment using the 
default p.d.f. $p_{\rm mix}(q|\rho)$ but with pseudo-data 
randomly sampled from the varied templates. The average observed 
shift in the fitted fraction for each systematic uncertainty 
(averaged over two input fractions) is 
added in quadrature to obtain the final shift from the systematic 
uncertainties. 
Table~\ref{tab:lhfit_syserr} shows the estimated systematic uncertainties.
\begin{table}[]
\centering
\begin{tabular}{lc}
\hline
Systematic uncertainty & $\Delta \rho$ \\ 
\hline
Fraction of $c$-quark jets & $\pm0.023$ \\ 
Fraction of muon charge sign change & $\pm0.027$ \\ 
Stat. uncert. on the kinematic weighting & $\pm0.029$ \\ 
Stat. uncert. on the kinematic correction & $\pm0.027$ \\ 
\bbbar~production mechanism & $\pm0.045$ \\ 
Stat. uncert. $b$-quark jet charge templates & $\pm0.045$ \\ 
Stat. uncert. $c$-quark jet charge templates & $\pm0.022$ \\ 
Fraction of $\mu$-tagged jets & $\pm0.031$ \\ 
\ttbar~signal Modeling & $\pm0.021$ \\ 
Jet energy scale & $\pm0.024$ \\ 
Jet energy resolution & $\pm0.035$ \\ 
Jet reconstruction efficiency & $\pm0.038$ \\ 
Top quark mass uncertainty & $\pm0.022$ \\ 
\hline
\hline
Total & $\pm0.111$ \\ 
\hline
 \end{tabular}
\caption{Summary of the observed shift in the fitted fraction for the 
systematic uncertainties.}
\label{tab:lhfit_syserr}
\end{table}

\subsubsection{Results from Data}
\label{subsec:lhfit_data}
\begin{figure}[] 
  \centering
  \includegraphics[width=1.0\textwidth]{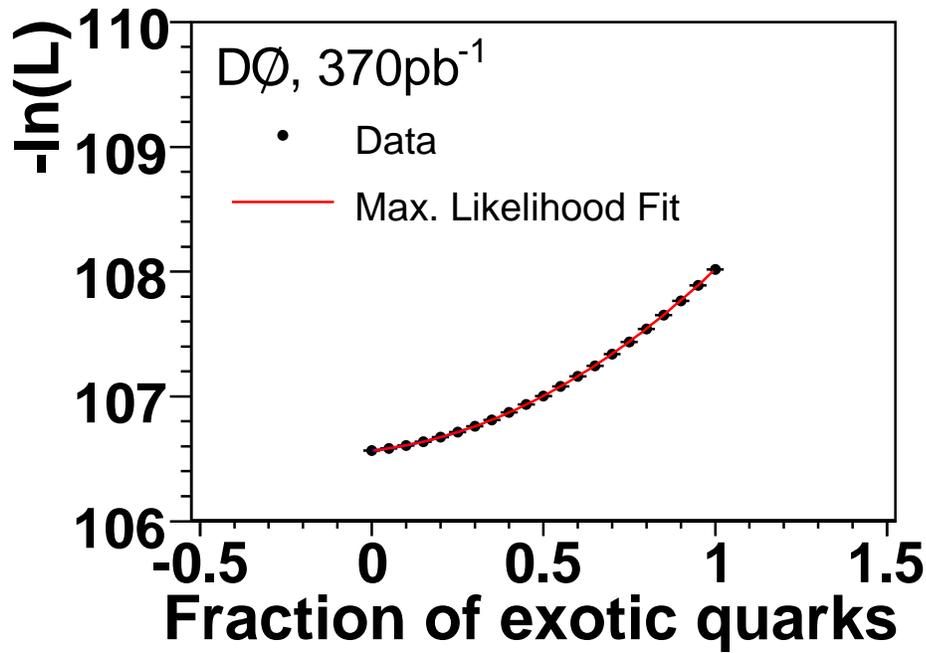} 
  \caption{The maximum likelihood fit applied to the observed charges ${\bf q}^{\rm data}$ 
  in the signal sample.}
  \label{fig:lhfit_result_stat}
\end{figure}

\begin{figure}[] 
  \centering
  \includegraphics[width=1.0\textwidth]{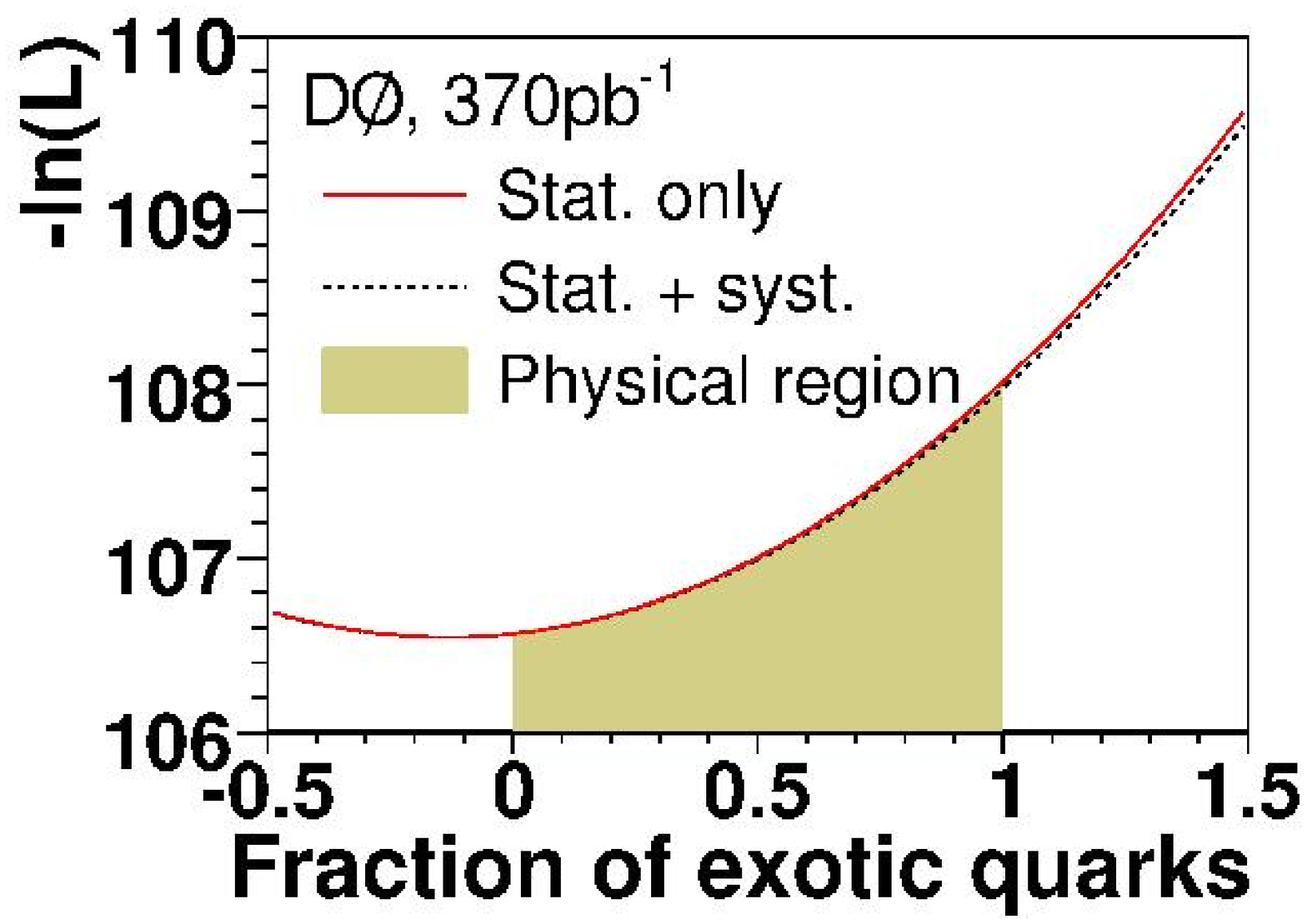} 
  \caption{The fitted $-\log{\mathcal{L}}$ with and without systematic 
  uncertainties.}
  \label{fig:lhfit_result_both}
\end{figure}

Applying the maximum likelihood fit to the observed charges 
${\bf q}^{\rm data}$ results in the fit shown in Fig.~\ref{fig:lhfit_result_stat}.  
The maximum likelihood estimator that minimizes the $-\log{\mathcal{L}}$ 
function is $\rho = -0.13\pm0.66\text{(stat)}\pm0.11\text{(syst)}$, in 
good agreement with the SM expectation. The effect of including systematic 
uncertainties in the $-\log{\mathcal{L}}$ function is shown in 
Fig.~\ref{fig:lhfit_result_both}. The systematic uncertainties are taken into 
account by convoluting the $-\log{ \mathcal{L}}$ function with a Gaussian centered 
on the minimum ($-\log{ \mathcal{L}}_{\rm max}$) and a width equal to the total 
systematic uncertainty calculated above. 
The upper limit on the fraction of exotic quarks in the signal sample is $0.80$ 
($0.52$) at $90$\% ($68$\%)C.L. The confidence intervals are shown in 
Tab.~\ref{tab:CI_stat} with statistical uncertainty only and in 
Tab.~\ref{tab:CI_sys} including systematic uncertainties.
\begin{table}
\centering
\begin{tabular}{l|c|c}
\hline
Probability & $\rho$ & $\rho$ (Bayesian) \\ 
\hline
$68$\%  & $-0.79 < \rho < 0.53$ & $0 < \rho < 0.52$ \\ 
$90$\%  & $-1.22 < \rho < 0.95$ & $0 < \rho < 0.79$ \\ 
\hline
\end{tabular}
\caption{Confidence intervals for statistical uncertainties only.}
\label{tab:CI_stat}
\end{table}
\begin{table}
\centering
\begin{tabular}{l|c|c}
\hline
Probability & $\rho$ & $\rho$ (Bayesian) \\ 
\hline
$68$\%  & $-0.80 < \rho < 0.54$ & $0 < \rho < 0.52$ \\ 
$90$\%  & $-1.24 < \rho < 0.97$ & $0 < \rho < 0.80$ \\ 
\hline
\end{tabular}
\caption{Confidence intervals including systematic uncertainties.}
\label{tab:CI_sys}
\end{table}



\cleardoublepage

\chapter{Conclusion and Outlook}
\label{ch:conclusions}

The first determination of the electric charge of the top quark is 
presented in this thesis. A data sample from \ppbar~collisions at 
$\sqrt{1.96}$~TeV recorded with the D\O\ detector at the Fermilab 
Tevatron collider corresponding to an integrated luminosity of 
$365$~pb$^{-1}$ was used. The top quark decays predominantly to 
$bW^+$ ($> 99.9$\%~\cite{PDG}. The predicted electric charge 
of the top quark is $2e/3$ in the SM. The top quark 
electromagnetic coupling is however, not measured and since the 
correlation between the charge sign of the decay products, i.e. 
$W^+$ and $b$ or $W^-$ and $\bar{b}$, was not known an exotic 
quark with electric charge $-4e/3$~\cite{exotic_top_paper} was 
not excluded. 

A pure sample of \ttbar~candidate events was selected in the 
\ljets~channel characterized by one high $p_T$ charged lepton 
from the $W$ boson decay and four jets out of which two are 
tagged using a $b$-tagging algorithm. A jet charge algorithm for 
$b$-quark jets based on the $p_T$ weighted sum of charges of the 
tracks associated with the jet was developed to distinguish 
between $b$- and $\bar{b}$-quark jets. The algorithm was calibrated using 
dijet data dominated by \bbbar~events. Template distributions to 
discriminate between the SM top and exotic quark charge 
scenarios were obtained by reconstructing the charge of the decay 
products: The charge of the high $p_T$ lepton and the $b$-tagged jets 
where the charge was extracted using the jet charge algorithm. A 
kinematic fit was performed to associate one of the $b$-quark jets 
to the charged lepton. 


A likelihood ratio test to discriminate between the SM and the
exotic charge hypothesis shows good agreement with the SM. The 
possibility that 100\% of the selected events are exotic quarks with electric 
charge $4e/3$ is ruled out to 92\% C.L. Nevertheless, the candidate heavy quarks 
could still be a mixture of two particles with the different charges. Using a Bayesian 
method to estimate the confidence intervals including both statistical and 
and systematic uncertainties, the fraction of exotic charges $\rho$ in the 
selected sample is
\begin{equation}
0 < \rho < 0.52~~ {\rm (68\%~ C.L.),} \nonumber
\end{equation}
\begin{equation}
0 < \rho < 0.80~~ {\rm (90\%~C.L.).}  \nonumber
\end{equation}
The fraction of exotic quarks in the selected sample is consistent with 
zero, the SM prediction.

The recorded integrated luminosity at the Tevatron is presently above $1$fb$^{-1}$ 
which together with another year of data taking should allow for an exclusion 
of the exotic charge scenario close to $95$\%. 

The Large Hadron Collider (LHC) at CERN is planned to start data taking 
during 2007. It will increase the luminosity with a factor of $\sim100$ and 
the center-of-mass to $14$~TeV. The top quark production cross-section is 
more than a factor of $100$ larger and the top quark pair 
production rate will be approximately 8~million per year~\cite{atlas_tdr} 
allowing for a possible determination of the top quark electromagnetic 
coupling~\cite{baur}. 

In general the top quark will play a major role in the physics program 
at LHC due to its the large mass, possibly playing a key role in electroweak 
symmetry breaking. The LHC is devoted to search for New Physics and 
any signature involving high $p_T$ leptons, jets and \met~are bound to have 
the top quark as a major background. In addition, the two jets from the 
hadronic decay of the W boson in a $\ttbar \to \ljets$ event may be used to calibrate 
the jet energy scale by reconstructing the invariant mass of the $W$ boson.

\cleardoublepage


\chapter*{Acknowledgements}
\addcontentsline{toc}{chapter}{Acknowledgements}


\begin{quote}
\item[]
I begin with thanking my supervisor Bengt Lund-Jensen for his decision to 
give me this position at KTH. Although sometimes far away from the analysis 
work he has always made sure that I was taken care of, not 
only research- and financial-wise, but also in my everyday life 
as a ``doktorand''. 
%

\item[]
When I started at KTH I was sent to Fermilab to work in the top working group. 
Christophe Cl\'ement became my supervisor and teammate for this 
project and has guided me from square one. Thank you Christophe: for your brains, 
for teaching me physics, for explaining all the details, for your company 
during long nights at Fermilab, for the lunch excursions, for the smooth 
collaboration and for your friendship.

\item[]
David Milstead has been of excellent help in my work and 
taking time to discuss and telling me how things really works. Thank you 
for getting me back on track when I was lost and for your cool english style. 
Thank you Sten Hellman for inspiring us to start this project.

\item[]
I would like to thank the whole D\O\ collaboration for running the experiment. 
In the top working group Regina Demina, Aurelio Juste, Chris Tully, Erich Varnes 
and Martijn Mulders made numerous comments and suggestions concerning this 
analysis. Sergey Burdin helped with $B$-physics related issues. In particular, 
I would like to mention the excellent people in the Swedish 
Consortium who made me feel welcome from the start: Jonas Strandberg, 
thank you for helping me with D\O\ code and answering all my questions 
about $b$-tagging. Thank you for your attempts to help me understand baseball.
Nils Gollub, thank you for the root-help, the many discussions and the fun times 
which hopefully continues in the coming years. Your proof-reading is much 
appreciated. Sara Strandberg always listened to my questions and helped 
me with everyday life at the lab. Thank you for lending me your car. 
I'm also very grateful to Barbro \AA sman who not only helped me with my travel 
arrangements to Fermilab and the D\O\ administration but also allowed me to drive her sports car 
(until Nils borrowed it and it broke down). 

\item[]
My good friends Paul DiTuro and Alejandro Daleo helped me to stay in shape 
at Fermilab and got me out from the lab occassionally. I will never forget 
the crazy nights in Chicago, not only (!) because of the ``TVR''-scar. 
Christian Schwanenberger was the only one putting up with my ice hockey babble. 

\item[]
Thank you everybody in the particle physics group at KTH. Tore Ersmark, 
my roommate, has been my excellent computer support. 
Thank you Jens Lund for the 
laughs and (inappropriate) jokes and Sara Bergenius Gavler 
who kept organizing despite the workaholic crew. I would also 
like to thank Mark Pearce, Janina \"Ostling, Jan Conrad, Erik Fredenberg, 
Karl-Johan Grahn, Silvio Orsi, Cecilia Marini Bettolo and Petter Hofvenberg 
for their nice company and lively discussions.  

\item[]
A big thanks goes to Christopher Engman who provided lodging and to Hanna K\"allman 
and Therese Nordlund who are always there for me. Svalan made sure I 
thought about other things than physics from time to time, thank you. 

\item[]
Without the full support and love of my parents, Kaj and Thory, this thesis 
would not have been possible. Thank you for raising me and allowing me to 
find my interests. I love you. 

\item[]
I also would like to thank my brother and best friend Bj\"orn for all the discussions, the good times, 
the trips and general ``hets''. You are the best. 

\item[]
Finally, without my wonderful Anna this thesis would have been a lot shorter. She 
has always supported me and accepted my absence many evenings and weekends. 
You mean the world to me!

\end{quote}

\cleardoublepage












\addcontentsline{toc}{chapter}{Bibliography}

%
%


\begin{thebibliography}{99}
%
\bibitem{discovery}
  CDF Collaboration,  F.\ Abe {\em et al.},    Phys. Rev. Lett. {\bf 74}, 2626 (1995);
  D\O\ Collaboration, S.\ Abachi {\em et al.}, Phys. Rev. Lett. {\bf 74}, 2632 (1995).

\bibitem{top_review}
  P. C. Bhat, H. Prosper, and S. S. Snyder, Int. J. Mod. Phys. A {\bf 13}, 5113 (1998).

\bibitem{exotic_top_paper}
  D. Chang, W. FG. Chang, and E. Ma, Phys. Rev. D {\bf 59}, 091503 (1999); {\bf 61}, 037301 (2000);
  D. Choudhury, T. M. Tait, C. E. Wagner, Phys. Rev., D {\bf 65} (2002) 053002.

\bibitem{CDF_4th_quark_search}
  CDF Collaboration, CDF Conference Note 8495 (2006). 

\bibitem{PDG}
  S. Eidelman {\em et al.}, Phys. Lett. B 592, 1 (2004); Particle Data Group web site:
  \verb+http://pdg.lbl.gov+.

\bibitem{lep_higgs}
  G. Abbiendi {\em et al.}, Phys. Lett. B {\bf 565} 61-75 (2003).

\bibitem{griffiths}
  D. Griffiths, Introduction to Elementary Particles, Wiley, (1987).

\bibitem{frank_mod_article}
  F. Wilczek, Int. J. Mod. Phys. {\bf A13}, 863 (1997).

\bibitem{EWprecLetterB}
  LEP Collaboration, Phys. Lett. B {\bf 276} 247 (1992).

\bibitem{EWphysicsReport2004}
  G. Altarelli, M. Grunewald, Phys. Rep. 403-404 189-201 (2004).

\bibitem{top_prod_theory}
  C. Cacciari {\em et al.}, JHEP 0404:068 (2004);
  N. Kidonakis, R. Vogt, Phys. Rev. D {\bf 68}, 114014 (2003).

\bibitem{top_single_production}
  D\O\ Collaboration, V.M. Abazov {\em et al.}, Phys. Lett. B {\bf 622}:265-276 (2005);
  CDF Collaboration, D. Acosta {\em et al.}, Phys. Rev. D {\bf 71}, 012005 (2005).

\bibitem{top_single_prod_theory}
  S. Cortese, R. Petronzio, Phys. Lett. B {\bf 253}, 494-498 (1991);
  D.A. Dicus, S.S.D. Willenbrock, Phys. Rev. D {\bf 34}, 155 (1986).

\bibitem{tt_quarkonium}
  I.I. Bigi {\em et al.}, Phys.Lett. {\bf B} 181, 157 (1986).

\bibitem{topmass_average}
  CDF and D\O\ Collaborations, hep-ex/0603039.

\bibitem{technicolor}
  Christopher T. Hill, Stephen J. Parke, Phys. Rev. D {\bf 49}, 4454-4462 (1994).

\bibitem{d0_collaboration_top_results_online_archive}
  D\O\ Collaboration, V.M. Abazov {\em et al.}, Top Physics Online Archive, available at: 
  \verb+http://www-d0.fnal.gov/Run2Physics/top/index.html+

\bibitem{CKM_org}
  N. Cabibbo, Phys. Rev. Lett. {\bf 10}, 531-533 (1963);
  M. Kobayashi, T. Maskawa, Prog. Theor. Phys. {\bf 49}, 652-657 (1973).

\bibitem{Vtb}
  D\O\ Collaboration, V.M. Abazov {\em et al.}, Phys. Lett. {\bf B} 639, 616 (2006).

\bibitem{W_hel_d0}
  D\O\ Collaboration, V.M. Abazov {\em et al.}, hep-ex/0609045, submitted to Phys. Rev. Lett.(2006).

\bibitem{W_hel_cdf_f0}
  CDF Collaboration, A. Abulencia {\em et al.}, Phys. Rev. {\bf D} 73, 111103 (2006).

\bibitem{W_hel_d0_f0}
  D\O\ Collaboration, V.M. Abazov {\em et al.}, Phys. Lett. {\bf B} 617, 1 (2005).

\bibitem{topcondensate}
  C.T. Hill, Phys. Lett. {\bf B266}, 419 (1991).

\bibitem{topasstechnicolor}
  C.T. Hill, Phys. Lett. {\bf B345}, 483 (1995).

\bibitem{ttbarresonance}
  CDF Collaboration, T. Affolder {\em et al.}, Phys. Rev. Lett. 85, 2062-2067 (2000);
  D\O\ Collaboration, V.M. Abazov {\em et al.}, Phys. Rev. Lett. 92, 221801 (2004);
  D\O\ Collaboration, V.M. Abazov {\em et al.}, D\O\ Conference Note 4880-CONF (2005).

\bibitem{spincorr}
  D\O\ Collaboration, B. Abbott {\em et al.}, Phys. Rev. Lett. 85, 256-261 (2000).

\bibitem{EWWG_summer05_combination}
  The LEP Electroweak Working Group, hep-ex/0511027 (2005); Lep Electroweak Working Group web site: \verb+http://lepewwg.web.cern.ch+

\bibitem{d0_detector}
  D\O\ Collaboration, V.M. Abazov {\em et al.}, Nucl. Instrum. and Meth. A~{\bf 565}, 463-537 (2006).
  
\bibitem{kleinknecht}
  K. Kleinknecht, ``Detectors for particle radiation'', Cambridge.

\bibitem{pt_res}
  R. Hopper, G. Landsberg, D\O\ Note 4230 (2003).


\bibitem{muon_detector}
  D\O\ Collaboration, V.M. Abazov {\em et al.}, Nucl. Instrum. Meth. A~{\bf 552}, 372 (2005).

\bibitem{d0_lumi}
  T. Edwards {\em et al.}, FERMILAB-TM-2278-E (2004).

\bibitem{inel_ppbar_xsec}
  S. Klimenko, J. Konigsberg, T. M. Liss, FERMILAB-FN-0741 (2003).






\bibitem{top_group}
  D\O\ top working group, web site: \texttt{http://www-d0.fnal.gov/Run2Physics/top/}.

\bibitem{alpgen_manual}
  M. L. Mangano {\em et al.}, JHEP {\bf 07}, 001 (2003).

\bibitem{xsec_240_ljets_topo_publ}
  D\O\ Collaboration, V.M. Abazov {\em et al.}, Phys. Lett. B 626, 45 (2005).




\bibitem{kalman_fitter}
  H. Greenlee, D\O\ Note 4303 (2004).
\bibitem{pv_alg_p14}
  A. Garcia. Bellido,  D\O\ Note 4320 (2004); 
  M. Narain, F. Stichelbaut,  D\O\ Note 3560 (2004). 
\bibitem{pv_prob_p14}
  A. Schwartzmann, M. Narain, D\O\ Note 4042 (2002). 



\bibitem{p14muonid}
    C. Cl\'ement {\em et al.}, D\O\ Note 4350 (2004). 
    
\bibitem{xsec_note240_mujets_topological}
  C. Clement {\em et al.}, D\O\ Note 4667 (2004).



\bibitem{emlhood_p14_org}
  D. Whiteson {\em et al}, D\O\ Note 4184 (2003). 

\bibitem{emlhood_p14}
  J. Kozminski {\em et al}, D\O\ Note 4449 (2004). 

\bibitem{xsec_note240_ejets_topological}
  C. Clement {\em et al.}, D\O\ Note 4662 (2004).



\bibitem{cone_algorithm}
  G. Blazey {\em et al}, D\O\ Note 3750 (2000). 
\bibitem{T42_1}
  U. Bassler, G. Bernardi, D\O\ Note 4124 (2002). 
\bibitem{T42_2}
  J.R. Vlimant {\em et al}, D\O\ Note 4146 (2003). 
\bibitem{T42_3}
  G. Bernardi, E. Busato, J.R. Vlimant, D\O\ Note 4335 (2004). 

\bibitem{jes_group}
  D\O\ jet energy scale working group, web page: \texttt{http://www-d0.fnal.gov/phys\_id/jes/d0\_private/jes.html}.



%

\bibitem{jes_nim}
  D\O\ Collaboration, B.~Abbot {\it et al.}, Nucl. Instrum. Meth. A~{\bf 424}, 352 (1999).

\bibitem{jer_jetcorr5.3}
  M. Agelou {\em et al}, D\O\ Note 4775 (2005). 
\bibitem{dl} Decay length $L_{xy}$ is defined as the distance 
  from the primary to the secondary vertex in the plane transverse to the beamline.
  Decay length significance is defined as $L_{xy}/\sigma_{L_{xy}}$, where
  $\sigma_{L_{xy}}$ is the uncertainty on $L_{xy}$.
\bibitem{bid_group}
  D\O\ $b$-identification Working Group webpage: \texttt{http://www-d0.fnal.gov/phys\_id/bid/d0\_private/bid.html}.
\bibitem{SVT}
  D. Boline {\em et al}, D\O\ Note 4796 (2005);
  A. Schwartzman, M. Narain, D\O\ Note 4080 (2003);
  A. Schwartzman, M. Narain, D\O\ Note 4081 (2003).


\bibitem{xsec_240_ljets_btag_publ}
  D\O\ Collaboration, V. Abazov {\em et al.}, Phys. Lett. B 626, 35 (2005).


\bibitem{xsec_note360_ljets_btag}
  C. Cl\'ement {\em et al.}, on the {\it Measurement of the \ttbar~cross section in the \ljets~channel at sqrt(s) = 1.96 TeV using lifetime tagging}, see: \texttt{http://www-d0.fnal.gov/Run2Physics/top/private/Confs05/\\ljets\_btag\_note\_v2\_1.pdf}.




\bibitem{pdf_fit_cteq}
  J. Huston {\em et al.}, Eur.Phys.J. C12 375-392 (2000).

\bibitem{pythia_manual}
  T. Sj\"ostrand {\em et al.}, Comp. Phys. Commun. {\bf 135}, 238 (2001).

\bibitem{tuneA}
  CDF Collaboration, T. Affolder {\em et al.}, Phys. Rev. D {\bf 65}, 092002 (2002).


\bibitem{lund_model}
  B. Andersson, G. Gustafson, G. Ingelman and T. Sj\"ostrand, Phys. Rep. {\bf 97} 31 (1983);
  B. Andersson, ``The Lund Model'', Cambridge University Press (1998).

\bibitem{bowler}
  M.G. Bowler, Z. Phys. C11, 169 (1981).

\bibitem{peterson}
  C. Peterson {\em et al.}, Phys. Rev. D {\bf 27} 105 (1983). 

\bibitem{ref_evtgen}
  D. J. Lange, Nucl. Instrum. Meth. A~{\bf 462}, 152 (2001).

\bibitem{d0gstar}
  D\O\ Collaboration, see: \texttt{http://www-d0.fnal.gov/computing/MonteCarlo/\\simulation/d0gstar.html}.

\bibitem{ref_geant}
  R. Brun, F. Carminati, CERN Program Library Long Writeup W5013 (unpublished) (1993).



\bibitem{topanalyze}
  M. Klute, L. Phaf and D. Whiteson, D\O\ Note 4122 (2003).

\bibitem{root}
  ROOT, ``An Object-Oriented Data Analysis Framework'', see: \texttt{http://root.cern.ch}.


\bibitem{topphysics_winter2004_note}
  The D\O\ Top Working group, D\O\ Note 4419 (2004).











\bibitem{hitfit}
  D\O\  Collaboration, A. Abachi {\em et al.},  Phys.Rev.Lett. {\bf 79}, 1197 (1997);
  Scott Snyder, D\O\ Doctoral Thesis, State Univeristy of Stony Brook, FERMILAB-THESIS-1995-27 (1995). 









\bibitem{top_trigger_package}
  M. Agelou {\em et al.}, D\O\ Note 4512 (2004).







\bibitem{jet_charge_aleph}
  ALEPH Collaboration, D. Buskulic {\em et al.}, Phys. Lett., B {\bf 356} (1995).

\bibitem{jet_charge_cdf}
  CDF Collaboration, T. Affolder {\em et al}, Phys. Rev., D {\bf 61}, 072005 (2000);
  CDF Collaboration, R. Affolder {\em et al}, Phys. Rev., D {\bf 60},  072003 (1999).

\bibitem{jet_charge_delphi}
  DELPHI Collaboration, J. Abdallah {\em et al.}, hep-ex/0412004, CERN-PH-EP-2004-062.

\bibitem{jet_charge_l3}
  L3 Collaboration, Acciarri, M. {\em et al.}, Phys. Lett., B {\bf 439} 225 (1998).

\bibitem{jet_charge_opal}
  OPAL Collaboration, G. Abbiendi {\em et al.}, Phys. Lett., B {\bf 546} 29-47 (2002).





\bibitem{rick_field}
  R. D. Field, Phys. Rev. D {\bf 65}, 094006 (2002).

\bibitem{cdf_b_production_runI}
  CDF Collaboration, D. Acosta {\em et al.}, Phys. Rev. D {\bf 71}, 092001 (2005).

\bibitem{dzero_thesis}
  Daniel Abraham Wijngaarden, PhD thesis , FERMILAB-THESIS-2005-14 (2005).






\bibitem{topmass_runI_PRL}
  D\O\ Collaboration, S. Abachi {\em et al.}, Phys. Rev. Lett. {\bf 79}, 1197 (1997).











\bibitem{top_mass_matrix_element_PRD}
  D\O\ Collaboration, V.M. Abazov {\em et al.}, submitted to Phys. Rev. {\bf D}, hep-ex/0609053 (2006). 



\bibitem{nuisance_params}
E.T. Jaynes, "Probability Theory", Cambridge.




\bibitem{atlas_tdr}
  Atlas Collaboration, ATLAS TDR 14, CERN/LHCC 99-14 (1999).
  
\bibitem{baur}
U. Baur, Phys. Rev. D{\bf 64} 094019 (2001);  
U. Baur, A. Juste, L. H. Orr, D. Rainwater, {\em Probing electroweak top quark couplings at hadron and lepton colliders}, to appear in the proceedings of 8th DESY Workshop on Elementary Particle Theory: Loops and Legs in Quantum Field Theory, Eisenach, Germany, 23-28 Apr 2006, hep-ph/0606264, (2006).











  














 
  







%

\end{thebibliography}




\end{document}